# Subspace-selective unitary manipulation based on the Hilbert-space symmetric structures in the multiple-quantum operator algebra spaces in the quantum-computing speedup theory

Xijia Miao

Chinese name: 缪希茄

Somerville, Massachusetts, USA

## Extended Abstract

A large and important problem in the quantum-computing speedup theory [Ref[1]][1] is how to make use of the fundamental quantum-computing-speedup resources to speed up essentially quantum computing and quantum simulating. In the present work the author makes a great effort toward solving this important problem. The multiple-quantum operator algebra space is positioned as the central place where one makes use of the fundamental quantum-computing-speedup resources to speed up essentially quantum computing and quantum simulating. It may be considered as the standard theoretical equipment to describe any unitary quantum dynamical processes including the unitary time-evolutional processes of quantum spin systems in the Heisenberg picture and/or the Dirac picture. This is similar to that the Hilbert space is the standard theoretical equipment to describe quantum-mechanically the time-dependent quantum states of any quantum system in a unitary time-evolutional process in the Schrödinger picture. The base-operator expansion principle can be generally used in the finite-dimensional multiple-quantum operator algebra space. According to the quantum-computing speedup theory the symmetrical structures and properties of quantum systems are considered as the fundamental

[1] [Ref[1]] X. Miao, (*i*) *The duality-character solution-information-carrying unitary propagators*, arXiv.org: 2012.13250 [quant-ph] (2020); (*ii*) *The unitary dynamical state-locking process, the HSSS quantum search process, and the quantum-computing speedup theory*, arXiv.org:1612.05969 [quant-ph] (2016); (*iii*) *The universal quantum driving force to speed up a quantum computation – The unitary quantum dynamics*, arXiv:1105.3573 [quant-ph] (2011)

quantum-computing-speedup resources which are responsible for exponentially speeding up quantum computing and also quantum simulating, e.g., quantum simulating for the unitary time-evolutional processes of quantum spin systems. They may be characterized via these different kinds of basic quantum spaces: the multiple-quantum operator algebra spaces, the density operator spaces, and/or the Hilbert spaces of the quantum systems. Therefore, the fundamental quantum-computing-speedup resources of a quantum system may exist in these different kinds of basic quantum spaces of the quantum system. Here the multiple-quantum operator algebra spaces are the linear operator spaces, while the Hilbert spaces and the density operator spaces are the quantum-state spaces. Making use of the fundamental quantum-computing-speedup resources of quantum spin systems to achieve essential quantum-computing speedup needs to use a variety of the multiple-quantum-transition operators of the multiple-quantum operator algebra spaces of the spin systems. One large part of the present work therefore are devoted to describing and investigating a variety of the multiple-quantum-transition operators.

Recognize that the multiple-quantum operator algebra space is the central place to make use of the fundamental quantum-computing-speedup resources to achieve essential quantum-computing speedup. Then the fundamental quantum-computing-speedup resources which are original from the corresponding Hilbert spaces must be explicitly taken into account in the multiple-quantum operator algebra spaces. In this paper the subspace-selective unitary manipulation aims to harness the fundamental quantum-computing-speedup resources original from the corresponding Hilbert space to speed up quantum computing and quantum simulating in the multiple-quantum operator algebra space. Theoretically it is based on the symmetrical structures of the Hilbert space which are characterized by the direct-sum decomposition of the Hilbert space. It plays a key role in achieving concretely essential quantum-computing speedup. As an important application, it can be used to realize concretely and efficiently the search-space dynamical reduction of the $HSSS$ unstructured quantum search algorithm ([Ref[1]] and [Ref[2]][2]) in the multiple-quantum operator algebra space. In this paper two different kinds of the Hermitian operators of the multiple-quantum operator algebra space are proposed to act as bridge to connect the corresponding Hilbert space to the multiple-quantum operator algebra space in the aspect of the symmetrical structures and properties. They are the Pseudo-Diagonal Hermitian (PDH) operators and the subspace-selective Multiple-Quantum-Transition (MQT) Hermitian operators, respectively. These two kinds of the Hermitian operators and their generated unitary operators are described and investigated in detail. They are very important to realize concretely and efficiently the subspace-selective unitary manipulation in the multiple-quantum operator algebra space. As the specific subspace-selective unitary manipulation, in this paper the Hilbert-space-enlarging processes (and their inverses) are deliberately designed to make use of the fundamental quantum-computing-speedup resources original from the Hilbert space to achieve essential quantum-computing speedup in the multiple-quantum operator algebra space. They may be performed in the multiple-quantum operator algebra space. They work also based on the direct-sum subspaces of the Hilbert space, but they are able to take into account the tensor-product symmetrical structure of the Hilbert space of the composite quantum system under study, which is also the fundamental quantum-computing-speedup resource. Therefore, it is expected that they have important applications in quantum computing and quantum simulating. In this paper the Hilbert-space-enlarging processes are described and investigated in detail in the $n-$spin$-1/2$ systems. The subspace-selective MQT Hermitian operators and their generated unitary operators and the subspace-selective PDH operators and their generated unitary operators may be used to construct concretely and realize efficiently the Hilbert-space-enlarging processes.

[2] [Ref[2]] X. Miao, *Efficient dynamical reduction from the exponentially large unstructured search space of a search problem to a polynomially small subspace in an n-qubit spin system*, Unpublished work (PDF document) (2012)

The Hilbert-space-enlarging processes may be used to harness the fundamental quantum-computing-speedup resources original from the Hilbert space to achieve essential quantum-computing speedup via the way that the Hilbert-space-enlarging processes and their inverses can selectively change at will the occupied direct-sum subspaces of the Hilbert space and can adjust at will the dimensional sizes of these occupied direct-sum subspaces in the multiple-quantum operator algebra space. Rather than the dimensional size of the whole Hilbert space, the dimensional sizes of the direct-sum subspaces of the Hilbert space are really related to the symmetrical structures of the Hilbert space. They constitute an important physical quantity and also one important aspect to reflect the symmetrical structures and properties of the quantum system under study. The dimensional size of any direct-sum subspace is a natural number and owns an infinitely high precision. More importantly, the inverse of dimensional size of a direct-sum subspace may act as a discrete variable which owns an infinite-high precision and can take an extremely small discrete value. These dimensional sizes of the subspaces and especially their inverses (i.e., the discrete variables) may be considered as important resource which may be harnessed to realize essential quantum-computing speedup for quantum simulating and quantum computing.

# Contents

# 1. Introductory considerations

The quantum-computing speedup theory [Ref[1]] considers the symmetrical structures and properties of quantum systems as the fundamental quantum-computing-speedup resources which are responsible for exponentially speeding up quantum computing and quantum simulating. At present a large and important problem in the quantum-computing speedup theory is how to make use of the fundamental quantum-computing-speedup resources to speed up essentially quantum computing and quantum simulating. Here the author makes a great effort toward solving this important problem in the present theoretical research work in this paper. The theoretical research work is mainly divided into the Part I and the Part II in this paper. The Part I investigates mainly the multiple-quantum operator algebra spaces. In the meantime the relationships also are analyzed theoretically among the multiple-quantum operator algebra spaces, quantum simulating the unitary time-evolutional processes, and the fundamental quantum-computing-speedup resources which exist in these different kinds of basic quantum spaces: the multiple-quantum operator algebra space, the density operator space, and the Hilbert space. It concludes that the multiple-quantum operator algebra space [4] must be positioned as the central place where the fundamental quantum-computing-speedup resources are exploited to speed up essentially quantum computing and quantum simulating. The Part II investigates mainly the subspace-selective unitary manipulation based on the Hilbert-space symmetrical structures. Recognize that the multiple-quantum operator algebra space is the central place to make use of the fundamental quantum-computing-speedup resources to achieve essential quantum-computing speedup. Then those fundamental quantum-computing-speedup resources which are original from the corresponding Hilbert space must be explicitly taken into account in the multiple-quantum operator algebra space. How the resources original from the Hilbert space can be explicitly taken into account in the multiple-quantum operator algebra space is an important problem in the quantum-computing speedup theory. The subspace-selective unitary manipulation [7] is able to solve this important problem. It aims to harness the fundamental quantum-computing-speedup resources which are original from the corresponding Hilbert space to speed up essentially quantum computing and quantum simulating in the multiple-quantum operator algebra space.

In this Introductory considerations (this Section One) several research subjects are introduced, respectively. The first three subjects, i.e., "The fundamental quantum-computing-speedup resources", "The multiple quantum transitions and the multiple-quantum-transition operators", and "The subspace-selective unitary manipulation", are the main theoretical research subjects in this paper. The last two subjects, i.e., "The Hilbert-space-symmetrical-structure-($HSSS$)-based unstructured quantum search algorithm" and "Programmable quantum simulating for the unitary time-evolutional processes", are the future applications of the present theoretical research work.

## The fundamental quantum-computing-speedup resources

According to the quantum-computing speedup theory [Ref[1]] both the unitary quantum dynamics and the quantum-mechanical symmetry are considered as the two pillars to build efficient quantum-computing processes to solve hard computational problems. The quantum-computing speedup theory considers the symmetrical structures and properties of quantum systems (i.e., the quantum-mechanical symmetry) as the fundamental quantum-computing-speedup resources which are responsible for exponentially speeding up quantum computing and also quantum simulating. The symmetric structures and properties of a quantum system such as a quantum spin system may be characterized through these different kinds of basic quantum spaces in quantum mechanics: the multiple-quantum operator algebra space (or the Liouville operator algebra space), the density operator space, and/or the Hilbert space of the quantum system. Here, from the point of view of quantum mechanics, the multiple-quantum operator algebra spaces (or the Liouville operator algebra spaces) [4] are the linear operator spaces, while the Hilbert spaces [1, 2, 3] and the density operator spaces [3, 6, 5] are the quantum-state spaces. Therefore, the fundamental quantum-computing-speedup resources of a quantum system may exist in these different kinds of basic quantum spaces of the quantum system. One important and large problem in the quantum-computing speedup theory at present is how to correctly and concretely make use of the fundamental quantum-computing-speedup resources (i.e., the symmetric structures and properties of quantum systems) to speed up essentially quantum computing and quantum simulating. In the present theoretical research work in this paper the author makes a great effort toward solving this important problem. The multiple-quantum operator algebra space (Ref.[4] and [Ref[3]]) must be considered as the central place where one makes use of the fundamental quantum-computing-speedup resources to speed up essentially quantum computing and also quantum simulating. As a consequence, one is able to solve correctly and concretely the first step of the $H$ilbert-$S$pace-$S$ymmetrical-$S$tructure-($HSSS$)-based unstructured quantum search algorithm [Ref[1]], where the first step includes the efficient search-space dynamical reduction [Ref[2]]. The search-space dynamical reduction ([Ref[2]] and [Ref[1]]) of the $HSSS$ unstructured quantum search algorithm must be performed in the multiple-quantum operator algebra space. Ultimately the whole $HSSS$ unstructured quantum search algorithm should be realized in the multiple-quantum operator algebra space (or the Liouville operator algebra space).

The fundamental quantum-computing-speedup resources of a quantum system may exist in these three different kinds of basic quantum spaces mentioned above in quantum mechanics. These basic quantum spaces have their own fundamental and inherent attributes:

*The Hilbert space* [1,2,3]: The simplest one; Physical property: the quantum-state space; Accommodate quantum (pure) states which are represented by vectors or wavefunctions; No multiplication operations exist

*The density operator space* [6,3,5]: More complex than the Hilbert space; Physical property: the quantum-state space; Accommodate quantum mixed states which are represented by the density operators (or matrices); No multi-plication operations exist

*The multiple-quantum operator algebra space (or the Liouville operator algebra space)* [4]: The most complicated one; Physical property: the linear operator space; Accommodate linear operators and quantum dynamical variables; Obey multiplication operations

The symmetrical structures and properties of a quantum system are considered as the fundamental and inherent attributes of the quantum system. Consequently the fundamental quantum-computing-speedup resources must be the fundamental and inherent attributes of the quantum system (the key point).

Why is the multiple-quantum operator algebra space considered as the central place where one makes use of the fundamental quantum-computing-speedup resources to speed up essentially quantum computing and quantum simulating? In the quantum-computing speedup theory [Ref[1]] both the unitary quantum dynamics and the quantum-mechanical symmetry are thought of as the two pillars to build an efficient quantum-computing process which has an exponential quantum-computing speedup over its classical counterpart. Moreover, the unitary quantum dynamics is fundamental in quantum physics and is considered as the universal quantum driving force to essentially speed up quantum computing [Ref[1]]. Actually, it is well known in quantum mechanics that any quantum-symmetry operations of quantum systems (including the geometric-symmetric, the dynamical-symmetric, the permutational-symmetric operations and so on) can be represented by the unitary transformations [2, 3, 17], while the latter each can be realized by the unitary operators (or propagators) which are generated by the dynamical variables including the quantum-system Hamiltonians or the Hermitian operators [2]. Especially the unitary time-evolutional propagators (representing the unitary time-evolutional dynamics) play the central role in describing, calculating, and quantum simulating the unitary time-evolutional processes of quantum spin systems [4]. Actually, for quantum spin systems any quantum-symmetry operations also can be realized by the unitary time-evolutional propagators. Therefore, the unitary quantum dynamics including the unitary time-evolutional dynamics must be considered as the central place in quantum computing and quantum simulating. This is the direct reason why the multiple-quantum operator algebra space is considered as the central place to make use of the fundamental quantum-computing-speedup resources to achieve essential quantum-computing speedup. Here the unitary propagators and the unitary operators that govern the unitary quantum dynamical processes are generated by the dynamical variables including the quantum-system Hamiltonians and/or the Hermitian operators. Now examine the fundamental and inherent attributes of these three different kinds of basic quantum spaces: the multiple-quantum operator algebra spaces, the density operator spaces, and the Hilbert spaces. It can be found that only the multiple-quantum operator algebra spaces are the linear operator spaces, while the Hilbert spaces and the density operator spaces are the quantum-state spaces, and more importantly, only the multiple-quantum operator algebra spaces are able to accommodate the linear operators including the dynamical variables such as the quantum-system Hamiltonians and the Hermitian operators. These are the main reasons why the multiple-quantum operator algebra space must be considered as the central place where the fundamental quantum-computing-speedup resources are exploited to speed up essentially quantum computing and quantum simulating.

Recognize that the multiple-quantum operator algebra space is the central place where the fundamental quantum-computing-speedup resources are exploited to speed up essentially quantum computing and quantum simulating. Then the fundamental quantum-computing-speedup resources which are original from the corresponding Hilbert space must be explicitly taken into account in the multiple-quantum operator algebra space. This is an important problem that needs to be solved in the quantum-computing speedup theory. In this paper quite a large part of the theoretical research work are devoted to solving this important problem (Read the subject "The subspace-selective unitary manipulation" later in this Section One).

Here something must be corrected in the quantum-computing speedup theory [Ref[1]] due to that the multiple-quantum operator algebra space is positioned as the central place to make use of the fundamental quantum-computing-speedup resources to achieve essential quantum-computing speedup. As an example, the $HSSS$ unstructured quantum search algorithm [Ref[1]] emphasizes importance of the (tensor-product) symmetric structure of the Hilbert space in the quantum-computing speedup theory. This is completely correct. However, the multiple-quantum operator algebra space is disregarded and this is not consistent with the spirit of the quantum-computing speedup theory. The $HSSS$ unstructured quantum search algorithm should be realized in the multiple-quantum operator algebra space (or the Liouville operator algebra space). In particular, the search-space dynamical reduction ([Ref[2]] and [Ref[1]]) of the $HSSS$ unstructured quantum search algorithm must be performed in the multiple-quantum operator algebra space.

In this paper the Part I including the Sections 2 and 3 are mainly devoted to investigating the multiple-quantum operator algebra spaces. And in the meantime the relationships are also analyzed theoretically among the multiple-quantum operator algebra spaces, quantum simulating the unitary time-evolutional processes, and the fundamental quantum-computing-speedup resources. Here the Section 3 is devoted to investigating the multiple-quantum operator algebra spaces. In the quantum-computing speedup theory the multiple-quantum operator algebra space may be considered as the standard theoretical equipment to describe any unitary quantum dynamical processes including the unitary time-evolutional processes of quantum spin systems in the Heisenberg picture and/or the Dirac picture. This is similar to that the Hilbert space is the standard theoretical equipment to describe quantum-mechanically the time-dependent quantum states of any quantum system in any unitary time-evolutional process in the Schrödinger picture. There are the relationships among the multiple-quantum operator algebra spaces, quantum simulating the unitary time-evolutional processes, and the fundamental quantum-computing-speedup resources that exist in these three different kinds of basic quantum spaces mentioned above. These relationships are theoretically analyzed in the Section 3. A general quantum-mechanical description for the unitary time-evolutional processes of a quantum system is given in the Section 2.

The base-operator expansion principle is introduced in the Subsection 3.1 for the multiple-quantum operator algebra spaces (or the Liouville operator algebra spaces). It can be used generally in the multiple-quantum operator algebra spaces of quantum spin systems [4], which are the finite-dimensional linear operator spaces. If the Liouville operator algebra space (or the multiple-quantum operator algebra space) is an infinite-dimensional linear operator space, then situation becomes complicated. That is, the base-operator expansion for an operator may or may not exist in an infinite-dimensional linear operator space. And this situation will not be discussed in this paper. The operator basis sets of the multiple-quantum operator algebra spaces are introduced in the Subsection 3.2. With the help of the base-operator expansion principle any unitary time-evolutional process of a quantum spin system may be determined (or realized) in the multiple-quantum operator algebra space. It is well known that any unitary time-evolutional process of a quantum spin system can be determined (or realized) in the Hilbert space [1, 2, 3] of the spin system, and it also can be determined in the density operator space [5, 6, 3] of the spin system. These two methods are perhaps the unique two methods that have been used most extensively in quantum physics to determine (or realize) any unitary time-evolutional processes. Here the multiple-quantum operator algebra space provides the third determination (or realization) method (See the Subsection 3.1), which are different from these two methods, to determine any unitary time-evolutional processes.

# The multiple quantum transitions and the multiple-quantum-transition operators

The concept of multiple quantum transitions in the quantum-simulating and quantum-computing research field (Ref.[7], [Ref[3]], [Ref[8]][3], Ref.[8], Ref.[4][4]) is original from nuclear magnetic resonance spectroscopy (See, for example, [Ref[5]], [Ref[6]][5], and [Ref[4]]; See also the classical scientific literatures and therein: Ref.[5], Ref.[11]) Quantum transition is a fundamental concept in quantum mechanics [2]. It dates back to the Bohr's quantum-physical model of hydrogen atom in 1913 and it is closely related to the atomic spectroscopy. Magnetic resonance spectroscopy (See, e.g., Ref.[11], Ref.[12]) and a variety of atomic spectroscopy (See, e.g., Ref.[13]) are closely related to the quantum transitions (or jumpings) between discrete quantum-system energy levels. Multiple quantum transitions in nuclear spin systems in nuclear magnetic resonance spectroscopy are the specific aspect of the quantum transitions in quantum mechanics [2].

[3] [Ref[8]] X. Miao, *universal construction of quantum computational networks in superconducting Josephson junctions*, http://arxiv.org/abs/quant-ph/0003113 (2000)

[4] This paper (Molec. Phys. 98, 625 (2000)) is the author's first paper that is related to the concept of multiple quantum transitions and the MQT operators in the quantum-computing research field. It was first submitted to journal (J.Chem.Phys.) to publish in 1998, but unfortunately since then it had been rejected to publish until the year 2000.

[5] [Ref[6]] X. Miao, et al., *Application of the product operator formalism to spin (I=1/2) systems under a radio-frequency irradiation*, Molec. Phys. 90, 499 (1997)

The multiple quantum transitions and the multiple-quantum-transition operators which can induce the multiple quantum transitions in the spin systems have been extensively used in nuclear magnetic resonance spectroscopy [5, 11].

Why are the multiple quantum transitions and the multiple-quantum-transition operators introduced into the quantum-computing research field? Initially the author thought that the unitary quantum dynamics and especially the unitary time-evolutional dynamics of quantum spin systems in the physical essence may be better understood from the quantum-computing research field. And the author hoped to find opportunity to have their applications and solve some important problems in the research field. Here these applications should refer to those related to the multiple-quantum transitions and the multiple-quantum-transition operators, which of course involve the unitary time-evolutional processes and the multiple-quantum operator algebra spaces of quantum spin systems [4]. Now one of the most important reasons for this is stated as follows: Once it is recognized that the multiple-quantum operator algebra space (Ref.[4] and [Ref[3]]) is the central place where the fundamental quantum-computing-speedup resources are exploited to speed up essentially quantum computing and quantum simulating, it becomes inevitable to investigate and make use of the multiple quantum transitions and the multiple-quantum-transition operators of quantum spin systems in the quantum-simulating and quantum-computing research field.

Harnessing the fundamental quantum-computing-speedup resources of quantum spin systems to essentially speed up quantum computing and quantum simulating needs to employ a variety of the multiple-quantum transition operators of the multiple-quantum operator algebra spaces of the quantum spin systems. Therefore, one large part of the present work in this paper are devoted to describing, investigating, constructing, and realizing a variety of the multiple-quantum transition operators. Here an $n-$spin$-1/2$ system which consists of $n$ spin$-1/2$ particles, each one of which has spin quantum number $I = 1/2$, is a simple and typical quantum spin system and has been frequently employed for the present work in this paper.

The Section 3 is devoted to investigating the multiple-quantum operator algebra spaces and the Subsection 3.1 is devoted to introducing the base-operator expansion principle in the multiple-quantum operator algebra spaces. In the Subsection 3.2 several typical operator basis sets of the multiple-quantum operator algebra space are described in detail. The base operators of these operator basis sets are related to a variety of multiple-quantum transition operators. These operator basis sets may have important applications in quantum computing and quantum simulating. The Cartesian product operator basis set [5] of an $n-spin-1/2$ system (the Subsection 3.2.1) may be best used in the programmable quantum simulating for the unitary time-evolutional processes [8]. The multiple-quantum-transition (MQT) product operators of an $n-spin-1/2$ system (the Subsection 3.2.1 and the Section A), which constitute a complete set of base operators of the multiple-quantum operator algebra space, can be used to express any $p-$order quantum transition operators and characterize the multiple-quantum operator algebra space. They could be more useful for the theoretical treatments in quantum computing and quantum simulating. The Hermitian pseudospin operators (See the Subsection 3.2.2) may be used to set up connection between the multiple-quantum operator algebra space and a usual $n-$qubit quantum system (e.g., an $n-$pseudospin$-1/2$ system) in quantum computing and quantum simulating.

The research work in the Section 4 is closely related to the subspace-selective unitary manipulation. It is described in the subject "The subspace-selective unitary manipulation" later in this Section One.

The Section 5 is devoted to describing the selective and subspace-selective multiple quantum transition operators. The $p-$order selective-quantum-transition operators and the Hermitian selective-$|p|-$quantum-transition operators are first defined formally (Subsection 5.1.1). They are basic not only in theory but also in experimental implementation. Theoretically the Hermitian selective-$|p|-$quantum-transition operators [7] may be further used to construct any subspace-selective MQT operators. A subspace-selective MQT operator [7] is closely related to the direct-sum subspaces of the Hilbert space of the spin system under study. As introduced in the Subsection 5.1.2, a subspace-selective $|p|-$quantum-transition operator with a given order $p \neq 0$ may be constructed by using these Hermitian selective-$|p|-$quantum-transition operators. It involves two chosen direct-sum subspaces of the Hilbert space. Similarly, as introduced in the Subsection 5.1.3, a subspace-selective zero-quantum-transition operator may be constructed also by employing these Hermitian selective-zero-quantum-transition operators. It involves only one chosen direct-sum subspace of the Hilbert space. Generally, any subspace-selective MQT operators are defined on the basis of the symmetric structures and properties of the Hilbert space which is specified by the direct-sum decomposition. The subspace-selective MQT unitary operators may be employed to realize the subspace-selective unitary manipulation [7] in the multiple-quantum operator algebra space. Therefore, in the Subsection 5.2 the subspace-selective MQT operators including the subspace-selective MQT Hermitian operators and unitary operators are described in detail, concretely constructed and realized, and have been intensively investigated. The theoretical research work in the Subsection 5.2 is one main contribution to this paper. The Section 5 is one important part of the subspace-selective unitary manipulation.

The Section 6 is an important part of the subspace-selective unitary manipulation. The research work in the Section 6 is described in the subject "The subspace-selective unitary manipulation" later in this Section One.

Finally, the Section A in this paper is devoted to introducing generally the multiple quantum transitions and the multiple-quantum transition operators. It is relatively closely related to nuclear magnetic resonance spectroscopy [5, 11]. The formal definitions for the multiple-quantum transition operators with any quantum-transition orders are first introduced in the Section A. On the basis of these formal definitions a theoretical framework for the multiple-quantum transition operators is set up, which may be more suitable for applications in quantum simulating and quantum computing. Quantum computing and quantum simulating stress to be mathematical logical, strict, and quantitative. And this theoretical framework meets this point. Then the zero-quantum transition operators, the invariant subspaces under the zero-quantum operator set (or subspace), and the MQT product operators of an $n - spin - 1/2$ system (the Subsection 3.2.1) are discussed and analyzed theoretically, respectively.

## The subspace-selective unitary manipulation

As shown in the previous subject "The fundamental quantum-computing-speedup resources", according to the quantum-computing speedup theory the multiple-quantum operator algebra space is considered as the central place to make use of the fundamental quantum-computing-speedup resources to achieve essential quantum-computing speedup. And these fundamental quantum-computing-speedup resources may exist in these different kinds of basic quantum spaces: the multiple-quantum operator algebra space, the density operator space, and/or the Hilbert space. Now recognize that the multiple-quantum operator algebra space is the central place where making use of the fundamental quantum-computing-speedup resources to achieve essential quantum-computing speedup. Then those fundamental quantum-computing-speedup resources which are original from the corresponding Hilbert space must be explicitly taken into account in the multiple-quantum operator algebra space. When the fundamental quantum-computing-speedup resources are exploited to speed up quantum computing and quantum simulating, how those resources which are original from the corresponding Hilbert space can be explicitly taken into account in the multiple-quantum operator algebra space — this is an important problem in the quantum-computing speedup theory. The subspace-selective unitary manipulation [7] is able to solve this important problem.

In this paper the Part II including the Sections 5 and 6 are devoted to investigating the subspace-selective unitary manipulation based on the Hilbert-space symmetrical structures. In addition the research work in the Section 4 is related to the subspace-selective unitary manipulation. Therefore, these three Sections 4, 5, and 6 are the core part of the theoretical research work in the current research subject.

Initially the subspace-selective unitary manipulation [7] works on the Hilbert space of a quantum spin system. It is closely related to the subspace-selective MQT operators and the subspace-selective multiple-quantum-transition processes between different direct-sum subspaces of the Hilbert space of the quantum spin system. It also have been used in the $HSSS$ unstructured quantum search algorithm [Ref[10]]. In this paper the subspace-selective unitary manipulation aims to harness the fundamental quantum-computing-speedup resources which are original from the corresponding Hilbert space to speed up essentially quantum computing and quantum simulating in the multiple-quantum operator algebra space. Theoretically it is based on the symmetrical structures and properties of the Hilbert space which are characterized by the direct-sum decomposition of the Hilbert space.

The multiple-quantum operator algebra space is a linear operator space, while its corresponding Hilbert space is a quantum-state space. How to set up connection between the two completely different basic quantum spaces so that the fundamental quantum-computng-speedup resources original from the corresponding Hilbert space can be exploited to essentially speed up quantum computing and quantum simulating in the multiple-quantum operator algebra space. According to quantum mechanics [1,3] any linear operators are defined on the Hilbert space of a quantum system, where the Hilbert space is a linear vector space. Then the linear operator space of a quantum system can connect to its corresponding Hilbert space on which any linear operators of the linear operator space are defined. Consequently the symmetrical structures and properties of the corresponding Hilbert space may connect to the counterparts of the multiple-quantum operator algebra space. Therefore, theoretically it becomes possible to make use of the fundamental quantum-computing-speedup resources original from the corresponding Hilbert space to speed up essentially quantum computing and quantum simulating in the multiple-quantum operator algebra space. The Pseudo-Diagonal Hermitian (PDH) operators (the Section 4) and the subspace-selective MQT Hermitian operators (the Subsection 5.2) are purposefully designed to set up connection between the corresponding Hilbert space and the multiple-quantum operator algebra space in the aspect of symmetrical structures and properties. They are very important to realize concretely and efficiently the subspace-selective unitary manipulation in the multiple-quantum operator algebra space.

The Section 4 is devoted to investigating the pseudo-diagonal Hermitian operators. In this Section 4 a general pseudo-diagonal Hermitian operator is first defined. Then it is described and investigated in detail. The pseudo-diagonal Hermitian operators are the specific Hermitian operators of the multiple-quantum operator algebra space. They are perhaps the simplest non-diagonal Hermitian operators through which the connection in the aspect of symmetrical structures and properties may be conveniently set up between the corresponding Hilbert space and the multiple-quantum operator algebra space. With the aid of the pseudo-diagonal Hermitian operators the symmetric structures and properties original from the corresponding Hilbert space may be easily taken into account explicitly in the multiple-quantum operator algebra space. Therefore, the pseudo-diagonal Hermitian operators and their generated unitary operators may have important applications in the subspace-selective unitary manipulation in the multiple-quantum operator algebra space. Especially they have important applications in the Hilbert-space-enlarging processes (the Section 6) which are one specific kind of the subspace-selective unitary manipulation.

The Section 6 is devoted to investigating a Hilbert-space-enlarging process. The so-called Hilbert-space-enlarging process is a unitary quantum dynamical process that changes one Hilbert subspace to another with a larger dimensional size in the Hilbert space of the quantum system under study. It may be performed in the multiple-quantum operator algebra space. Its inverse process is a Hilbert-space-shrinking process that changes one Hilbert subspace to another with a smaller dimensional size. Here the Hilbert subspace should refer to a direct-sum subspace of the Hilbert space which the quantum system really occupies. It may be said intuitively that a Hilbert-space-enlarging process changes the occupied Hilbert subspace from a small to a large dimensional size.

In the Section 6 the Hilbert-space-enlarging processes are described in detail, constructed and realized concretely, and have been intensively investigated in the $n - spin - 1/2$ systems. They are deliberately designed to make use of the fundamental quantum-computing-speedup resources original from the corresponding Hilbert space to achieve essential quantum-computing speedup in the multiple-quantum operator algebra space. As a specific kind of the subspace-selective unitary manipulation, the Hilbert-space-enlarging processes also work on the basis of the direct-sum subspaces of the Hilbert space, but they are able to take into account the tensor-product symmetrical structure of the Hilbert space of a composite quantum system under the specific (direct-sum-decomposition) Hilbert-space symmetrical structure. According to the quantum-computing speedup theory [Ref[1]] the tensor-product symmetrical structure of the Hilbert space is also the fundamental quantum-computing-speedup resource. Therefore, it is expected that the Hilbert-space-enlarging processes have important applications in quantum computing and quantum simulating.

A general Hilbert-space-enlarging process may be expressed as a sequence of the basic building blocks which may be conveniently chosen as the subspace-selective MQT unitary operators (the Subsection 5.2) or the subspace-selective unitary operators generated by the pseudo-diagonal Hermitian operators (the Section 4). Therefore, the subspace-selective MQT unitary operators and the subspace-selective unitary operators generated by the PDH operators as well as the subspace-selective MQT Hermitian operators and PDH operators (the dynamical variables) may be used to realize efficiently the Hilbert-space-enlarging processes.

Both the Hilbert-space-enlarging processes and their inverses (the Hilbert-space-shrinking processes) constitute one specific kind of the subspace-selective unitary manipulation in the multiple-quantum operator algebra space. They together can selectively change at will the occupied direct-sum subspaces of the Hilbert space. More importantly, the dimensional sizes of these occupied direct-sum subspaces can be adjusted at will by the suitable Hilbert-space-enlarging processes and their inverses in the multiple-quantum operator algebra space. Here the dimensional sizes of the direct-sum subspaces of the Hilbert space constitute an important physical quantity [7]. Rather than the dimensional size of the whole Hilbert space, the dimensional sizes of the direct-sum subspaces of the Hilbert space are related to the symmetrical structures of the Hilbert space [7]. They constitute one important aspect to reflect the symmetrical structures and properties of the Hilbert space of the quantum system under study. Therefore, the Hilbert-space-enlarging processes may be used to harness the fundamental quantum-computing-speedup resources original from the Hilbert space to achieve essential quantum-computing speedup via the way that the Hilbert-space-enlarging processes and their inverses can selectively change at will the occupied direct-sum subspaces of the Hilbert space and can adjust at will the dimensional sizes of the occupied direct-sum subspaces in the multiple-quantum operator algebra space. This is an extraordinary and important application for the Hilbert-space-enlarging processes in quantum computing and quantum simulating.

The dimensional size of any direct-sum subspace of the Hilbert space is a natural number and owns an infinitely high precision. More importantly, the inverse of dimensional sizes of the direct-sum subspaces of the Hilbert space may constitute a discrete variable which also owns an infinite-high precision and can take an extremely small discrete value. Here the Hilbert-space-enlarging processes and their inverses can really play a crucial role in that they can selectively change at will the occupied direct-sum subspaces of the Hilbert space and can adjust at will the dimensional sizes of the occupied direct-sum subspaces in the multiple-quantum operator algebra space. Consequently it becomes possible that the inverse of dimensional size of the occupied direct-sum subspace can really act as a discrete variable which owns an infinite-high precision and can take any extreme-small discrete value (which corresponds to an exponentially large dimensional size). These direct-sum-subspace dimensional sizes and especially their inverses (i.e., the discrete variables) may be considered as important resource which may be harnessed to achieve essential quantum-computing speedup in quantum computing and quantum simulating. The relevant work will be reported in future.

# The Hilbert-space-symmetrical-structure-(HSSS)-based unstructured quantum search algorithm

According to the quantum-computing speedup theory [Ref[10]][6] a quantum-computing speedup process obeys not only the fundamental quantum-physical laws, i.e., the unitary quantum dynamics and the quantum-mechanical symmetry, but also the mathematical-logical principle of a computational problem to be solved. This dual character of quantum-computing speedup reflects the interaction between the unitary quantum dynamics (quantum physics) and the mathematical-logical principle (mathematics) in the quantum-computing speedup process. It is essential in the quantum-computing speedup theory. The $H$ilbert-$S$pace-$S$ymmetrical- $S$tructure-($HSSS$)-based unstructured quantum search process [Ref[10]] can best illustrate the dual character of quantum-computing speedup. The dual character of quantum-computing speedup may be best reflected by the basic $SIC$ unitary operators (or propagators) of the $HSSS$ unstructured quantum search process (See [Ref[10]] and [Ref[3]][7]):

$$U_{\lambda}^{sic}(a_m^s, \theta_m) = \exp(-i\theta_m a_m^s I_{m\lambda}),\ 1 \leq m \leq n,\ \lambda = x, y, z \tag{1.1}$$

where $I_{mx}$, $I_{my}$, and $I_{mz}$ are the spin operators of the $m-$th spin$-1/2$ particle of the $n-$qubit spin$-1/2$ system. Here the most important is the so-called duality-character quantity $a_m^s$ which takes the discrete values $a_m^s = \pm 1$ and is called the double-valued mathematical-logical number (or the unit number [Ref[3]]). It represents information of the solution to the unstructured search problem. This basic $SIC$ unitary operator is the $s$olution-$i$nformation-$c$arrying

[6] [Ref[10]] X. Miao, *The duality-character solution-information-carrying unitary propagators*, arXiv.org: 2012.13250 [quant-ph] (2020)

[7] [Ref[3]] X. Miao, *Universal construction for the unsorted quantum search algorithms,* https://arXiv.org/abs/quant-ph/0101126 (2001)

($SIC$) unitary propagator (or operator). It carries the information of the $m-$th component state $|s_m\rangle$ of the solution state $|S\rangle$ ($|S\rangle = |s_n\rangle ... |s_m\rangle ... |s_1\rangle$) of the search problem [Ref[10]], where $|s_m\rangle = |0_m\rangle$ or $|1_m\rangle$. Therefore, from the point of view of quantum physics the basic $SIC$ unitary operator of (1.1) is the unitary propagator that governs the unitary time-evolutional process of the quantum system under study, while from the point of view of mathematics it carries the information of the solution to the unstructured search problem and hence the mathematical-logical principle of the search problem must be obeyed in the unitary time-evolutional process. The duality-character quantity $a_m^s$ may be explained as a hidden variable in the $HSSS$ unstructured quantum search process [Ref[10]] which is a deterministic unitary quantum dynamical process. Consequently one can differentiate the $HSSS$ unstructured quantum search process from the conventional quantum computing based on the orthodox quantum mechanics.

According to the quantum-computing speedup theory [Ref[1]] the symmetric structures and properties of quantum system (i.e., the quantum-mechanical symmetry) are considered as the fundamental quantum-computing-speedup resources which are responsible for exponentially speeding up quantum computing. Therefore, both the dual character of quantum-computing speedup and the fundamental quantum-computing-speedup resources are responsible for achieving an exponential quantum-computing speedup for an efficient quantum-computing algorithm to solve a hard computational problem over its classical counterpart. From the point of view of pure quantum mechanics both the unitary quantum dynamics and the quantum symmetry are considered as the two pillars to build an efficient quantum-computing process to solve a hard computational problem in the quantum-computing speedup theory [Ref[10]].

If now one wants to prepare the basic $SIC$ unitary operator of (1.1) by starting from the multiple-quantum operator algebra space and the density operator space (or equivalently the Hilbert space) of the $n-$qubit spin$-1/2$ system, respectively, then it can be found that one can directly obtain the correct result of (1.1) only in the multiple-quantum operator algebra space. Therefore, the basic $SIC$ unitary operator can be prepared directly and correctly only in the multiple-quantum operator algebra space instead of the density operator space (or equivalently the Hilbert space). This preparation for the basic $SIC$ unitary operator is the first step of the $HSSS$ unstructured quantum search algorithm [Ref[1]], which includes the efficient search-space dynamical reduction [Ref[2]]. Therefore, this result better illustrates that the multiple-quantum operator algebra space must be positioned as the central place where the fundamental quantum-computing-speedup resources are exploited to speed up essentially quantum computing and quantum simulating.

The $HSSS$ unstructured quantum search process [Ref[10]] which solves an unstructured search problem consists of the two successive steps that the first step is the search-space dynamical reduction and the second the dynamical quantum-state-difference amplification. Below only the first step is discussed. How to realize efficiently the second step will be reported in future. Here the ultimate goal of the first step is set to prepare efficiently the basic $SIC$ unitary operator (or propagator) of (1.1) simply due to that the basic $SIC$ unitary operator is simplest and most elementary. The core part of the first step is the search-space dynamical reduction. Therefore, the search-space dynamical reduction is usually referred to as the first step.

Exponentially large unstructured search space is a huge obstacle to any quantum search algorithm to solve efficiently an unstructured search problem. The search-space dynamical reduction therefore is the number-one priority for the whole $HSSS$ unstructured quantum search process [Ref[1]]. It eliminates dynamically the exponentially large unstructured search space (i.e., the math Hilbert space) and purges the classical-physical effect [Ref[10]]. And in the meantime a polynomially large reduced search space is generated after the search-space dynamical reduction. In particular, the smallest reduced search space may be two-dimensional. It may carry only the information ($a_m^s$) of the $m-$th component state $|s_m\rangle$ of the solution state $|S\rangle$ of the search problem. In this special case, by starting from the final result of the efficient search-space dynamical reduction [Ref[2]] one can further prepare efficiently the basic $SIC$ unitary operators (or propagators) of (1.1). This preparation is governed mainly by the pure quantum-mechanical unitary dynamical process and hence it is relatively secondary.

The search-space dynamical reduction [Ref[10]] can be expressed as a sequence of the duality-character oracle operations $\{C_S(\theta)\}$ of the unstructured search problem with the original (unreduced) unstructured search space (e.g., the $2^n-$dimensional math Hilbert space) and the relevant quantum-mechanical unitary operators. In the search-space dynamical reduction the core components are the duality-character oracle operations $\{C_S(\theta)\}$. The duality-character oracle operation $C_S(\theta)$ is a reversible (or unitary) selective diagonal operator. It may be written in the exponential operator form ([Ref[1]], [Ref[3]])

$$C_S(\theta) = \exp(-i\theta D_S), \tag{1.2}$$

where the oracle diagonal operator $D_S = |S\rangle\langle S|$ can be expressed as

$$D_S = \bigotimes_{m=1}^{n} \left(\frac{1}{2}E_m + a_m^s I_{mz}\right) \tag{1.3}$$

and the corresponding candidate solution state $|S\rangle$ may be written in the form

$$|S\rangle = \bigotimes_{m=1}^{n} \left(\frac{1}{2}|T_m) + a_m^s |S_m)\right) \tag{1.4}$$

with $|T_m) = |0_m\rangle + |1_m\rangle$ and $|S_m) = \frac{1}{2}(|0_m\rangle - |1_m\rangle)$, here $|0_m\rangle$ and $|1_m\rangle$ are the two usual computational base vectors of the $m-$th spin$-1/2$ particle of the $n-$qubit spin$-1/2$ system, and $E_m$ and $I_{mz}$ are the unity operator and the $z-$component spin operator of the $m-$th spin$-1/2$ particle, respectively.

It can be seen from (1.3) that the oracle diagonal operator $D_S$ indeed uses the tensor-product symmetrical structure of the $n-$qubit spin$-1/2$ system. This indicates that the tensor-product symmetrical structure is necessary to express the oracle diagonal operator $D_S$ and the duality-character oracle operation $C_S(\theta)$, and hence it is really necessary for the $HSSS$ unstructured quantum search algorithm. Obviously, the most important quantity in (1.3) and (1.4) is the duality-character double-valued logical number set $\{a_m^s\}$ (here $a_m^s = \pm 1$ in value with $1 \leq m \leq n$) which characterizes completely the duality-character oracle operation $C_S(\theta)$ [Ref[10]].

The tensor-product symmetrical structure of the Hilbert space of the $n-$qubit spin$-1/2$ system can be seen from the candidate solution state $|S\rangle$ of (1.4) to the unstructured search problem. The Hilbert space provides quite a large convenience to describe and explain how the $HSSS$ unstructured quantum search algorithm works [Ref[1]]. Therefore, it is plausible to emphasize importance of the symmetrical structure of the Hilbert space in the $HSSS$ unstructured quantum search algorithm and this is also completely correct, but the multiple-quantum operator algebra space is disregarded and this is not consistent with the spirit of the quantum-computing speedup theory. However, these descriptions and explanations can be seamlessly translated to the corresponding multiple-quantum operator algebra space (or the Liouville operator algebra space).

Now according to the quantum-computing speedup theory the multiple-quantum operator algebra space is positioned as the central place to make use of the fundamental quantum-computing-speedup resources to achieve essential quantum-computing speedup. Then this means that the $HSSS$ unstructured quantum search algorithm should be performed in the multiple-quantum operator algebra space and especially the search-space dynamical reduction must be realized in the multiple-quantum operator algebra space. Here the tensor-product symmetrical structure of the Hilbert space is still necessary for the search-space dynamical reduction. However, the efficient search-space dynamical reduction [Ref[2]] works on the basis of the direct-sum subspaces of the Hilbert space of the $n-$qubit spin$-1/2$ system. According to the quantum-computing speedup theory the subspace-selective unitary manipulation works on the basis of the direct-sum subspaces of the Hilbert space and may be performed in the multiple-quantum operator algebra space. Moreover, it is able to take into account the tensor-product symmetrical structure of a composite quantum system under study. Therefore, it is expected that the subspace-selective unitary manipulation can play key role in realizing correctly and concretely the efficient search-space dynamical reduction and preparing efficiently the basic $SIC$ unitary operators of (1.1) for the $HSSS$ unstructured quantum search algorithm. The relevant work will be reported in future.

# Programmable quantum simulating for the unitary time-evolutional processes

Both the unitary quantum dynamics and the fundamental quantum-computing-speedup resources play as well the key role in essentially speeding up quantum simulating any quantum systems. Especially the unitary time-evolutional propagators (i.e., the unitary time-evolutional dynamics) play the central role in describing, calculating, and quantum simulating the unitary time-evolutional processes of quantum spin systems [8, 4]. Programmable quantum simulating for the unitary time-evolutional processes (Ref.[8], [Ref[8]]) is an alternative to the Feynman's quantum simulating which was described for the first time by Feynman [14]. In theory the programmable quantum simulating is initially related to exact and analytical calculations of the unitary time-evolutional processes in a variety of large and complex quantum spin systems including the strongly-coupled spin$-1/2$ systems [Ref[5]] and the coupled multiple-spin$-1/2$ systems under radio-frequency pulse ([Ref[6]], [Ref[7]][8]) in nuclear magnetic resonance spectroscopy. The exact and analytical calculations are based on the method [Ref[11]][9] mainly original from the Lie group and Lie algebra ([Ref[12]][10], [Ref[13]][11], [Ref[14]][12]), here the core point is that the unitary time-evolutional propagator of any quantum spin system is decomposed completely into an ordered product of a series of elementary propagators. In experiment the programmable quantum simulating may be implemented by employing nuclear magnetic resonance selective-pulse experimental techniques [Ref[4]] (i.e., implementation of the one-spin rotation operators and the two-spin unitary operators).

The programmable quantum simulating is divided into the two steps [8]. The first step is that the unitary time-evolutional propagator of any quantum spin system is decomposed thoroughly into an ordered product of a series of elementary propagators. Then the second step is that every elementary propagator is further expressed as a sequence of the single-spin (or one-body) rotation operators and the two-spin (or two-body) elementary propagators. Consequently the whole unitary time-evolutional propagator is finally decomposed thoroughly into an ordered product of a series of the single-spin rotation operators and the two-spin elementary propagators (which may be diagonal). As a typical example, consider an $n-spin-1/2$ system. Any spin Hamiltonian $H_s$ of the $n-spin-1/2$ system can be expanded in terms of the complete set $\{B_s\}$ of base operators of the $4^n-$dimensional multiple-quantum operator algebra space of the spin system (See (3.1) in the Section 3 below): $H_s=\sum_{k=0}^{4^n-1}a_kB_k$, where the base operators $\{B_s\}$ may be taken as the Hermitian Cartesian product operators (See Ref.[5] or the Cartesian product operators (3.10) in the Section 3). Here every Hermitian Cartesian product operator $B_s$ may generate a unitary exponential operator $R_s(B_s)=\exp(-ib_sB_s)$ with real parameter $b_s$. This unitary operator $R_s(B_s)$ is the so-called elementary propagator (Ref.[8] and [Ref[5]]). These unitary propagators $\{R_s(B_s)\}$ are the simplest and most elementary propagators of the $n-spin-1/2$ system. Now the unitary time-evolutional propagator $U_s(t)=\exp(-iH_st/\hbar)$ which is generated by the spin

[8] [Ref[7]] X. Miao, *An explicit criterion for existence of the Magnus solution for a coupled spin system under a time-dependent radiofrequency pulse*, arXiv: 1204.4872 [quant-ph] (corrected Ed.) and Phys. Lett. A 271, 296 (2000) (published Ed.)

[9] [Ref[11]] X. Miao, Unpublished work, 1990

[10] [Ref[12]] Z. Yan and Y. Xu, *Lie group and Lie algebra*, the Higher Education Press, Beijing, 1985 (Chinese)

[11] [Ref[13]] J. Wei and E. Norman, *Lie algebraic solution of linear differential equations*, J. Math. Phys. 4, 575 (1963)

[12] [Ref[14]] W. Magnus, *On the exponential solution of differential equations for a linear operator*, Comm. Pure Appl. Math. 7, 649 (1954)

Hamiltonian $H_s$ can be decomposed thoroughly into an ordered product of a series of elementary propagators $\{R_r(B_r)\}$:

$$U_s(t) = \exp\left(-i\frac{t}{\hbar}\sum_{k=0}^{4^n-1} a_k B_k\right) = \prod_r R_r(B_r) \tag{1.5}$$

where the symbol $\prod$ stands for the ordered product. This is the first step of the programmable quantum simulating [8].

The second step of the programmable quantum simulating [8] is that every elementary propagator $R_s(B_s) = \exp(-ib_sB_s)$ is completely decomposed into a sequence of the single-spin (or one-body) rotation operators and the two-spin (or two-body) elementary propagators. This step is efficient. As shown by (3.10) in the Section 3, almost all the Cartesian product operators $\{B_s\}$ of the $n-spin-1/2$ system may be considered as many-spin interaction terms with spin number greater than 2, and every many-spin interaction term $B_s$ generates a many-spin elementary propagator $R_s(B_s)$ . These many-spin elementary propagators $R_s(B_s)$ usually can not be directly implemented in experiment. Therefore, every many-spin elementary propagator $R_s(B_s)$ must be further expressed as a sequence of the single-spin rotation operators and the two-spin elementary propagators, here the single-spin rotation operators and the two-spin elementary propagators can be implemented directly in experiment [Ref[4]]. This can be done by the two steps $(i)$ and $(ii)$ below. $(i)$ Any non-diagonal Cartesian product operator $B_s$ can be efficiently converted into a diagonal Cartesian product operator by a sequence of the single-spin rotation operators $\{R_\lambda(\theta)\}$ [8]. As a typical example, consider a non-diagonal product operator $B_s = I_{kx}I_{ly}...I_{jz}...I_{my}$ (See Ref.[5] or the Cartesian product operators (3.10) in the Section 3). Then it can be converted into the diagonal product operator $I_{kz}I_{lz}...I_{jz}...I_{mz}$ by a sequence of the single-spin rotation operators:

$$I_{kx}I_{ly}...I_{jz}...I_{my}\xrightarrow{\exp(i\pi I_{ky}/2)}\xrightarrow{\exp(-i\pi I_{lx}/2)}......\xrightarrow{\exp(-i\pi I_{mx}/2)}I_{kz}I_{lz}...I_{jz}...I_{mz}$$

where the single-spin rotation operators are given by $R_{q\lambda}(\theta) = \exp(-i\theta I_{q\lambda})$ with $\lambda = x, y, z$ and $q = k, l, m,$ *etc.* Here for any non-diagonal product operator in an $n-spin-1/2$ system the number of the single-spin rotation operators $\{R_{q\lambda}(\theta)\}$ in the sequence is not more than $n$. $(ii)$ Any $(m+1)$-spin diagonal elementary propagator that is generated by a diagonal product operator $2^m I_{k_1z}I_{k_2z}...I_{k_mz}I_{k_{m+1}z}$ $(n \geq m+1 > m \geq 2)$ can be recursively expressed as [8]

$$\begin{aligned}&\exp\left(-i\theta 2^m I_{k_1z}I_{k_2z}...I_{k_mz}I_{k_{m+1}z}\right)\\ &= V_m \exp\left(-i\theta 2^{m-1} I_{k_1z}I_{k_2z}...I_{k_{m-1}z}I_{k_{m+1}z}\right) V_m^+\end{aligned} \tag{1.6}$$

where the unitary operator $V_m$ is given by

$$\begin{aligned}V_m = &\exp\left(-i\frac{\pi}{2}I_{k_{m+1}x}\right)\exp\left(-i\pi I_{k_mz}I_{k_{m+1}z}\right)\\ &\times\exp\left(i\frac{\pi}{2}I_{k_{m+1}x}\right)\exp\left(-i\frac{\pi}{2}I_{k_{m+1}y}\right)\end{aligned} \tag{1.7}$$

The operator recursive relation (1.6) shows clearly that the $(m+1)$-spin diagonal elementary propagator $\exp\left(-i\theta 2^{m} I_{k_1 z} I_{k_2 z} ... I_{k_m z} I_{k_{m+1} z}\right)$ $(m \geq 2)$ can be expressed as the ordered product of the $m-$spin diagonal elementary propagator $\exp\left(-i\theta 2^{m-1} I_{k_1 z} I_{k_2 z} ... I_{k_{m-1} z} I_{k_{m+1} z}\right)$ and the unitary operators $V_m$ and $V_m^{+}$, while the formula (1.7) shows that the unitary operator $V_m$ (and $V_m^{+}$) is a sequence of three single-spin rotation operators and one two-spin diagonal elementary propagator. The operator recursive relation (1.6) can be used repeatedly so that the $(m+1)$-spin diagonal elementary propagator $\exp(-i\theta 2^{m} I_{k_1 z} I_{k_2 z} ... \times I_{k_m z} I_{k_{m+1} z})$ $(m \geq 2)$ finally can be efficiently decomposed into a sequence of the single-spin rotation operators and the two-spin diagonal elementary propagators. By combining the two steps $(i)$ and $(ii)$ together it can be shown that any many-spin elementary propagator $R_s\,(B_s)$ can be efficiently decomposed into a sequence of the single-spin rotation operators and the two-spin diagonal elementary propagators. It therefore concludes that the second step of the programmable quantum simulating is efficient.

The above theoretical analysis shows that in computational complexity the programmable quantum simulating is decided by the first step of (1.5) that the unitary time-evolutional propagator $U_s\,(t) = \exp\left(-iH_s t/\hbar\right)$ is decomposed thoroughly into an ordered product of a series of elementary propagators $\{R_s\,(B_s)\}$. The first step is usually hard for a general quantum spin system. The symmetrical structures and properties of the multiple-quantum operator algebra spaces of quantum spin systems [4] can provide an essential help to speed up the first step. Beside this it is expected that the other fundamental quantum-computing-speedup resources also may help speed up the first step.

# PART I. The multiple-quantum operator algebra spaces and the unitary time-evolutional processes

## 2. The unitary time-evolutional processes of quantum systems

Before the multiple-quantum operator algebra space is investigated in detail in the next Section, this Section is devoted to describing generally the unitary time-evolutional processes of quantum systems such as quantum spin systems in quantum mechanics. The unitary time-evolutional process of a quantum system may be performed in the Hilbert space of the quantum system. It also may be performed in the density operator space of the quantum system. These two methods are perhaps the unique two methods that have been used most extensively in quantum physics to determine (or realize) any unitary time-evolutional processes of quantum systems. Note that these realization (or determination) methods have nothing to do with any description ways (or pictures) to describe a unitary time-evolutional process in quantum mechanics. The latter are just what one wants to discuss in this Section. For example, in the Hilbert space one may have the three different basic description ways (or pictures) to describe any unitary time-evolutional process of a quantum system, as can be seen below.

The multiple-quantum operator algebra space [4] may provide the third realization (or determination) method, that is, any unitary time-evolutional process of a quantum system such as a quantum spin system may be performed (or realized) in the multiple-quantum operator algebra space, as shown in the next Section. This, of course, is another thing different from what one wants to discuss in this Section.

Quantum mechanically the time-evolutional process of a quantum system is a unitary quantum dynamical process. It describes how the quantum system changes unitarily from one instant of time to another. It is described by one or more equations of motion in quantum mechanics. There are a number of different ways to describe the time-evolutional process quantum mechanically, but for any description ways there is the same time evolution for the expectation value of the relevant dynamical variable [3, 2]. Among these description ways there are the three basic description ways named the Schrödinger picture, the Heisenberg picture, and the Dirac (or interaction) picture in quantum mechanics [2], respectively. In each basic picture the time-evolutional process is governed by one or two equations of motion.

In wave mechanics the time-evolutional process of a quantum system is governed by the Schrödinger wave equation and this description way is called the Schrödinger picture. Here the Schrödinger wave equation may be written as [2]

$$i\hbar\frac{\partial}{\partial t}\Psi\left(t\right)=H\Psi\left(t\right) \tag{2.1}$$

where $\Psi\left(t\right)$ is quantum state of the quantum system and it is time-dependent wave function or state vector and $H$ is Hamiltonian operator of the quantum system which is usually Hermitian. Below for simplicity suppose that the Hamiltonian $H$ is time-independent unless stated otherwise.

Quantum states that characterize completely a quantum system in quantum mechanics are the fundamental elements in the Schrödinger picture. The time-evolutional process therefore is that the quantum states change in the course of time. Quantum states may be represented by the wave functions (or state vectors) or by the density operators (or matrices) more generally. Correspondingly the unitary time-evolutional process obeys the Schrödinger wave equation (2.1) or the Liouville-von Neumann equation. Here the Liouville-von Neumann equation may be written as [1, 5]

$$i\hbar\frac{d}{dt}\rho\left(t\right)=\left[H,\rho\left(t\right)\right] \tag{2.2}$$

where $\rho\left(t\right)$ is time-dependent density operator (or matrix). Generally the density operator (or matrix) $\rho\left(t\right)$ represents the mixed state of a quantum system. As special case, it also can represent the quantum state $\Psi\left(t\right)$ (called the pure state) via $\rho\left(t\right)=\left|\Psi\left(t\right)\right\rangle\left\langle\Psi\left(t\right)\right|$, where $\Psi\left(t\right)$ is the normalized wave function (or state vector) of the quantum system in the Schrödinger wave equation (2.1). And in this special case the Liouville-von Neumann equation (2.2) is equivalent to the Schrödinger wave equation (2.1).

In the Schrödinger picture quantum states ($\rho(t)$ or $\Psi(t)$) are fundamental and are in central position for describing the time-evolutional process of any quantum system. This is consistent with the orthodox quantum mechanics. Hence it is natural to employ directly the time-dependent quantum states to describe the time-evolutional process of any quantum system [3]. The unitary time-evolutional propagator $U(t) = \exp(-iHt/\hbar)$ of the quantum system then can be derived from the time-dependent quantum states, if the latter are given (See, for example, Ref.[3]). However, in this situation it is hard to say the Schrödinger picture is a direct and explicit way to describing a unitary quantum dynamical process such as the unitary time-evolutional process of a quantum system. There are a large number of methods to solve the Schrödinger wave equation (2.1) and the Liouville-von Neumann equation (2.2), respectively. In the Schrödinger picture any dynamical variables are not involved explicitly in the motion equations (2.1) and (2.2) except the Hamiltonian $H$ of the quantum system itself which is also a dynamical variable.

The unitary time-evolutional propagators (representing the unitary time-evolutional dynamics) are fundamental and really in central position for describing, calculating, and simulating the unitary time-evolutional processes of a variety of quantum spin systems [4]. Naturally they are fundamental and in core position in any description ways or pictures (See below) to describing the unitary time-evolutional processes of the quantum spin systems. According to the quantum-computing speedup theory [Ref[1]] the unitary quantum dynamics is fundamental and is considered as the universal quantum driving force to essentially speed up quantum computing. Then generally the unitary time-evolutional propagators also are fundamental and in core position in any description ways (or pictures) to describing and quantum simulating the unitary time-evolutional processes of any quantum systems including the quantum spin systems. It may be thought that the unitary time-evolutional propagators which are generated by the Hamiltonians of the quantum systems drive the quantum systems to evolve in time. Hence the unitary time-evolutional propagators may be intuitively considered as the universal quantum driving force to drive the time evolution of the quantum systems. And this reflects the spirit of the quantum-computing speedup theory that the unitary quantum dynamics is the universal quantum driving force to essentially speed up quantum computing and also quantum simulating.

In order to emphasize in the Schrödinger picture that the unitary time-evolutional propagators play the central role in describing and quantum simulating the unitary time-evolutional process of a quantum system, it is necessary to rewrite appropriately the motion equations (2.1) and (2.2) in the manners that can reflect importance of the unitary time-evolutional propagators, respectively. The solution to the Schrödinger wave equation (2.1) may be formally written as $[1, 2]$

$$\Psi(t) = U(t)\,\Psi(0) \tag{2.3}$$

and the formal solution to the Liouville-von Neumann equation (2.2) may be

given by [1, 5, 4]

$$\rho\left(t\right)=U\left(t\right)\rho\left(0\right)U^{+}\left(t\right) \tag{2.4}$$

where the unitary time-evolutional propagator $U\left(t\right)$ is explicitly written as

$$U\left(t\right)=\exp\left(-iHt/\hbar\right) \tag{2.5}$$

for the time-independent Hamiltonian $H$; and if the Hamiltonian $H = H\left(t\right)$ is time-dependent, then the propagator $U\left(t\right)$ may be formally written as (See, e.g., Refs.[5, 4])

$$U\left(t\right)=\hat{T}\exp\left\{-\frac{i}{\hbar}\int_{0}^{t}H\left(t'\right)dt'\right\} \tag{2.6}$$

where $\hat{T}$ is the Dyson time-ordering operator [18] and $U\left(t_0\right) = \mathbf{E}$ (the unity operator) with the initial time $t_0 = 0$. The time-evolutional propagator $U\left(t\right)$ is unitary, whether or not the Hamiltonian $H$ depends on the time $t$. The formula (2.3) is equivalent to the motion equation (2.1), while the formula (2.4) is equivalent to the motion equation (2.2).

The formal solutions (2.3) and (2.4) each may reflect importance of the unitary time-evolutional propagator $U\left(t\right)$. It can be seen from (2.4) or (2.3) that in the Schrödinger picture the time-evolutional process of a quantum system is described through the way that the time-evolutional propagator $U(t)$ drives the quantum state ($\rho\left(t_0\right)$ or $\Psi\left(t_0\right)$) to evolve in time.

The Schrödinger picture alone may not be complete to describe the time-evolutional process of a quantum system. Beside the Schrödinger picture there is also the Heisenberg picture to describe the unitary time-evolutional process of a quantum system in quantum mechanics. In matrix mechanics the unitary time-evolutional process of a quantum system is governed by the Heisenberg motion equation [2]:

$$i\hbar\frac{d}{dt}A\left(t\right)=\left[A\left(t\right),H\right] \tag{2.7}$$

where the time-dependent dynamical variable (or operator) $A\left(t\right)$ is defined by

$$A\left(t\right)=U^{+}\left(t\right)AU\left(t\right) \tag{2.8}$$

with any time-independent dynamical variable $A$. Due to $U(0) = E$ (the unity operator) it follows from (2.8) that the initial condition of the equation (2.7) is $A(0) = A$ at the initial time $t_0 = 0$. The formula (2.8) also is the formal solution to the Heisenberg motion equation (2.7). This is easy to verify. First, it is easy to obtain $i\hbar\left(dU\left(t\right)/dt\right) = HU\left(t\right)$ by differentiating the unitary propagator $U(t)$ of (2.5) with the time variable $t$. Next, by making the time derivative on both sides of the equation (2.8) and then using $i\hbar\left(dU\left(t\right)/dt\right) = HU\left(t\right)$ it can be found that the Heisenberg motion equation (2.7) holds. Therefore, the equation (2.8) is equivalent to the equation (2.7).

Now the unitary time-evolutional process of the quantum system may be described by the Heisenberg motion equation (2.7) or its formal solution (2.8). This is the so-called Heisenberg picture. Obviously, in the Heisenberg picture any quantum states of a quantum system are not involved explicitly in the motion equation (2.7) or (2.8).

It can be seen from (2.8) that in the Heisenberg picture the time-evolutional process of a quantum system is described through the way that the unitary time-evolutional propagator $U(t)$ drives the dynamical variable $A$ to evolve in time.

There is an important property in the Heisenberg picture, that is, if the dynamical variable (or operator) $A$ evolves in time in accordance with the equation (2.8), then function $f(A)$ of the dynamical variable $A$ evolves in time in accordance with the following equation [3]:

$$f(A(t)) = f\left(U^{+}(t) AU(t)\right) = U^{+}(t) f(A) U(t) \tag{2.9}$$

As a special case of (2.9), consider that the dynamical variable $A$ is a Hermitian operator and the function $f(A)$ is the exponential operator function:

$$f(A) = \exp(-iA\tau_0/\hbar) \tag{2.10}$$

with real number $\tau_0$. Obviously, the exponential operator $\exp(-iA\tau_0/\hbar)$ is unitary. It is a dynamical variable too [2]. Now suppose that the dynamical variable $A$ evolves in time in accordance with (2.8). Then according to the equation (2.9) the operator function $f(A(t))$ evolves in time in accordance with the following equation:

$$f(A(t)) = \exp(-iA(t)\tau_0/\hbar) = U^{+}(t) \exp(-iA\tau_0/\hbar) U(t) \tag{2.11}$$

There is the property in the Heisenberg picture [3] that the operator $A(t)$ is Hermitian at any time $t$, if the operator $A$ is Hermitian. Then it follows from (2.11) that the exponential operator $\exp(-iA(t)\tau_0/\hbar)$ is unitary at any time $t$.

Besides the Schrödinger picture and the Heisenberg picture mentioned above there is a third basic description way named the interaction picture (or the Dirac picture) [2]. For simplicity, suppose that the Hamiltonian of a quantum system is divided into the two parts:

$$H(t) = H_0 + H_1(t) \tag{2.12}$$

where the first part $H_0$ is time-independent and the second part $H_1(t)$ is allowed to depend explicitly on the time $t$. The time-independent component Hamiltonian $H_0$ in (2.12) may generate a unitary propagator (or operator) $U_0(t) = \exp(-iH_0t/\hbar)$. Now with the help of the unitary propagator $U_0(t)$ the interaction picture (or the Dirac picture) is defined by [2]

$$A_I(t) = U_0^{+}(t) AU_0(t) \tag{2.13}$$

$$\rho_I(t) = U_0^{+}(t) \rho(t) U_0(t) \text{ or } \Psi_I(t) = U_0^{+}(t) \Psi(t) \tag{2.14}$$

where the density operator $\rho(t)$ and the state vector $\Psi(t)$ are given in the Liouville-von Neumann equation (2.2) and the Schrödinger wave equation (2.1), respectively. Therefore, in the interaction picture the time-dependent dynamical variable $A_I(t)$ is given by (2.13), while the time-dependent density operator $\rho_I(t)$ or state vector $\Psi_I(t)$ is given in (2.14). There are two equations of motion in the interaction picture, one of which governs the time evolution of the dynamical variable $A_I(t)$ and another governs the time evolution of the density operator $\rho_I(t)$ or the state vector $\Psi_I(t)$.

The motion equation that the dynamical variable $A_I(t)$ obeys can be obtained by differentiating the equation (2.13) with respect to the time $t$ and is explicitly written as [2]

$$i\hbar\frac{d}{dt}A_I(t) = [A_I(t), H_{0I}] \tag{2.15}$$

where the Hamiltonian $H_{0I} = H_0$. The initial condition for the motion equation (2.15) is given by $A_I(0) = A$ with the initial time $t_0 = 0$. Obviously, the formula (2.13) is the formal solution to the motion equation (2.15). Here the unitary propagator $U_0(t)$ (or the Hamiltonian $H_{0I}$) is the quantum driving force that drives the dynamical variable $A$ to evolve in time.

There is an important property in the Dirac picture (i.e., the interaction picture) that if the dynamical variable $A$ evolves in time in accordance with the equation (2.13), then function $f(A)$ of the dynamical variable $A$ evolves in time in accordance with the following equation [3]:

$$f(A_I(t)) = f\left(U_0^+(t)AU_0(t)\right) = U_0^+(t)f(A)U_0(t). \tag{2.16}$$

Let the dynamical variable $A$ be a Hermitian operator and the function $f(A)$ be the unitary exponential operator of (2.10). Then it can be found from (2.16) that the operator function $f(A_I(t))$ evolves in time in accordance with the following equation:

$$f(A_I(t)) = \exp(-iA_I(t)\tau_0/\hbar) = U_0^+(t)\exp(-iA\tau_0/\hbar)U_0(t) \tag{2.17}$$

where $A_I(t)$ is a Hermitian operator and $\exp(-iA_I(t)\tau_0/\hbar)$ is a unitary exponential operator at any time $t$.

The motion equation that the density operator $\rho_I(t)$ (or the state vector $\Psi_I(t)$) obeys can be derived from the first (or the second) equation in (2.14). By differentiating the first equation in (2.14) with respect to the time $t$ and then making use of the motion equation (2.2) and the Hamiltonian $H(t)$ of (2.12) it can be found that the motion equation of the density operator $\rho_I(t)$ is written as [1, 5]

$$i\hbar\frac{d}{dt}\rho_I(t) = [H_I(t), \rho_I(t)]; \tag{2.18}$$

and in similar fashion, by differentiating the second equation in (2.14) and then making use of the motion equation (2.1) and the Hamiltonian of (2.12) it can be found that the motion equation for the state vector $\Psi_I(t)$ is given by [2]

$$i\hbar\frac{\partial}{\partial t}\Psi_I(t) = H_I(t)\Psi_I(t). \tag{2.19}$$

Here the interaction Hamiltonian $H_I(t)$ in (2.18) and (2.19) is written as

$$H_I(t) = U_0^+(t) H_1(t) U_0(t). \tag{2.20}$$

The initial conditions for the motion equations (2.18) and (2.19) are given by $\rho_I(0) = \rho(0)$ and $\Psi_I(0) = \Psi(0)$ with the initial time $t_0 = 0$, respectively.

The formal solutions to the motion equations (2.18) and (2.19) are given respectively by

$$\rho_I(t) = U_I(t) \rho_I(0) U_I^+(t) \tag{2.21}$$

and

$$\Psi_I(t) = U_I(t) \Psi_I(0) \tag{2.22}$$

where the unitary propagator $U_I(t)$ is given by

$$U_I(t) = \hat{T} \exp\left(-\frac{i}{\hbar}\int_0^t H_I(t')\, dt'\right) \tag{2.23}$$

Moreover, the total time-evolutional propagator $U(t)$, which is generated by the time-dependent Hamiltonian $H(t)$ of (2.12) and may be expressed as (2.6), is the product of the unitary propagators $U_0(t)$ in (2.13) and $U_I(t)$ of (2.23),

$$U(t) = U_0(t) U_I(t) \tag{2.24}$$

This can be verified as follows. Note that $\rho_I(0) = \rho(0)$ and $\rho(0)$ is arbitrary. With the help of (2.21) and the first equation in (2.14) it can be found that there is the formula $\rho(t) = U_0(t) U_I(t) \rho(0) U_I^+(t) U_0^+(t)$, and then by comparing this formula with (2.4) it is verified that the equation (2.24) holds. Similarly, noting that $\Psi_I(0) = \Psi(0)$ and $\Psi(0)$ is arbitrary, with the aid of (2.22) and the second equation in (2.14) it is found that there is the formula $\Psi(t) = U_0(t) U_I(t) \Psi(0)$, and then by comparing this formula with (2.3) it is confirmed that the equation (2.24) holds.

The unitary time-evolutional process (2.13) of the dynamical variable $A_I(t)$, the unitary time-evolutional process (2.21) of the density operator $\rho_I(t)$ (or (2.22) of the state vector $\Psi_I(t)$), and the decomposition (2.24) of the total time-evolutional propagator $U(t)$ together show that in the interaction picture the unitary time-evolutional process of a quantum system is described through the way that the unitary time-evolutional propagator $U_0(t)$ drives the dynamical variable $A$ to evolve in time and in the meantime the unitary time-evolutional propagator $U_I(t)$ drives the quantum state ($\rho_I(t_0)$ or $\Psi_I(t_0)$) to evolve in time. Therefore, not only the dynamical variables ($A(t)$) but also the quantum states ($\rho_I(t)$ or $\Psi_I(t)$) are involved explicitly in the Dirac picture.

According to quantum mechanics [2, 3] there is the same time evolution of the matrix elements of the dynamical variable (or operator) $A$ under study for any description ways (or pictures). This is the theoretical basis to set up connection between one description picture with another in quantum mechanics. This is also the theoretical basis for that beside the aforementioned three basic pictures there may be a number of different description pictures to describe the unitary time-evolutional process of a quantum system. However, each one of all these description pictures falls in one of the three basic pictures mentioned above: the quantum states evolve in time alone, the dynamical variables evolve in time alone, or both the dynamical variables and the quantum states evolve in time at the same time.

In a linear operator space the inner (or scalar) product of a pair of linear operators $A$ and $B$ may be defined by $(A, B) = Tr\,(A^{+}B)$ [1,3]. Particularly the expectation value $\langle A\rangle$ of the dynamical variable $A$ under study may be written as the inner product $(\rho, A)$, i.e., $\langle A\rangle = (\rho, A) = Tr\,(\rho A)$ [1,3], where the density operator $\rho$ is Hermitian and belongs to the density operator space. A quantum system may be described completely by a density operator of the density operator space [6,3,5]. This is just like that a quantum system is completely described by a state vector of the Hilbert space. In physics the density operator space is a quantum-state space and is used to describe quantum (mixed) states of a quantum system, but in mathematics it still may be considered as a specific linear operator space (i.e., a linear space of operators). Hence here the inner product (e.g., $(\rho, A)$) related to the density operator $\rho$ is still defined on the linear operator space [3]. In the Hilbert space of a quantum system the matrix element of the dynamical variable (or operator) $A$ between a pair of state vectors $\Psi_\alpha$ and $\Psi_\beta$ may be written as the inner product $(\Psi_\alpha, A\Psi_\beta)$ [1,3] (or $\langle\Psi_\alpha|A|\Psi_\beta\rangle$ [2]). Particularly the matrix element $(\Psi, A\Psi) = \langle A\rangle$ is the expectation value $\langle A\rangle$ of the dynamical variable $A$.

Now consider the case that quantum states are represented by the density operators ($\rho$) of the density operator space. In the Schrödinger picture the dynamical variable $A$ under study, which is usually different from the Hamiltonian of the quantum system, is fixed or time-independent, while the density operator $\rho\,(t)$ is time-dependent and is given by (2.4). Then the time dependence of the expectation value $\langle A\,(t)\rangle$ of the dynamical variable $A$ may be expressed as $\langle A\,(t)\rangle = Tr\,(\rho\,(t)\,A)$. In the Heisenberg picture the quantum states ($\rho_H\,(t)$) are fixed or time-independent (i.e., $\rho_H\,(t) = \rho_H\,(0)$), while the dynamical variable $A\,(t)$ under study is time-dependent. Then the time dependence of the expectation value $\langle A\,(t)\rangle$ may be written as $\langle A\,(t)\rangle = Tr\,(\rho_H\,(0)\,A\,(t))$, where $A\,(t)$ is given by (2.8) and the initial conditions are $\rho_H\,(0) = \rho\,(0)$ and $A\,(0) = A$. In the Dirac picture both the density operator $\rho_I\,(t)$ and the dynamical variable $A_I\,(t)$ under study are time-dependent. Then the time dependence of the expectation value $\langle A\,(t)\rangle$ is given by $\langle A\,(t)\rangle = Tr\,(\rho_I\,(t)\,A_I\,(t))$, where $A_I\,(t)$ and $\rho_I\,(t)$ are given by (2.13) and (2.21), respectively, and the initial conditions are $\rho_I\,(0) = \rho\,(0)$ and $A_I\,(0) = A$. Then for all these three basic pictures mentioned above there is the same time dependence (or time evolution) of the expectation value $\langle A\,(t)\rangle$ [3,2], indicating that there are the following equalities:

$$\langle A\,(t)\rangle = Tr\,(\rho\,(t)\,A) = Tr\,(\rho_H\,(0)\,A\,(t)) = Tr\,(\rho_I\,(t)\,A_I\,(t)) \tag{2.25}$$

where the initial conditions are $\rho_I\,(0) = \rho_H\,(0) = \rho\,(0)$ and $A_I\,(0) = A\,(0) = A$ at the initial time $t_0 = 0$. In fact, by making use of the property of trace of operator [2], i.e., $Tr\,(ABC) = Tr\,(BCA)$ for operators $A$, $B$, and $C$, it can be confirmed that all these equalities in (2.25) hold indeed. These equalities in (2.25) set up connection between any two of these three basic pictures, when quantum states are represented by the density operators ($\rho$) of the density operator space.

In similar fashion, if quantum states are represented by the state vectors of the Hilbert space, then it can be found that for all these three basic pictures mentioned above there is the same time evolution of the expectation value $\langle A(t)\rangle$ [3, 2]:

$$\langle A(t)\rangle = (\Psi(t), A\Psi(t)) = (\Psi_H(0), A(t)\Psi_H(0)) = (\Psi_I(t), A_I(t)\Psi_I(t)) \tag{2.26}$$

where the initial conditions are $\Psi_I(0) = \Psi_H(0) = \Psi(0)$ and $A_I(0) = A(0) = A$ at the initial time $t_0 = 0$. Actually, with the help of the property of the inner product: $(A\Psi_\alpha, \Psi_\beta) = (\Psi_\alpha, A^+\Psi_\beta)$ [1, 2, 3] it can be verified that all these equalities in (2.26) hold. Therefore, these equalities in (2.26) set up connection between any pair of these three basic pictures, when quantum states are represented by the state vectors of the Hilbert space.

# 3. The multiple-quantum operator algebra spaces

The quantum-computing speedup theory [Ref[10]] considers that both the unitary quantum dynamics and the quantum-mechanical symmetry are the two pillars to build efficient quantum-computing processes. It also considers the symmetrical structures and properties of quantum systems as the fundamental quantum-computing-speedup resources which are responsible for essentially speeding up quantum computing. Moreover, the fundamental quantum-computing-speedup resources also are responsible for essentially speeding up quantum simulating the unitary time-evolutional processes of any quantum systems including the quantum spin systems. The symmetric structures and properties of a quantum system such as a quantum spin system [4] may be characterized through these different kinds of basic quantum spaces: the multiple-quantum operator algebra space (or the Liouville operator algebra space) [4], the density operator space [3,6,5], and/or the Hilbert space [1,2,3] of the quantum system. Therefore, the fundamental quantum-computing-speedup resources may exist in these different kinds of basic quantum spaces. As shown in the Section One in this paper, the quantum-computing speedup theory further considers that the multiple-quantum operator algebra space should be positioned as the central place where one makes use of the fundamental quantum-computing-speedup resources (i.e., the symmetric structures and properties of quantum system) to speed up essentially quantum computing and quantum simulating.

There are the relationships among the multiple-quantum operator algebra spaces, quantum simulating the unitary time-evolutional processes, and the fundamental quantum-computing-speedup resources that exist in these different kinds of basic quantum spaces mentioned above. According to the quantum-computing speedup theory [Ref[1]] the unitary quantum dynamics is fundamental in quantum physics and is considered as the universal quantum driving force to essentially speed up quantum computing and quantum simulating, and as shown in the previous Section 2, especially the unitary time-evolutional propagators, which represent the unitary time-evolutional dynamics, play the central role in describing, calculating, and quantum simulating the unitary time-evolutional processes of quantum systems such as quantum spin systems [4]. Therefore, the unitary quantum dynamics including the unitary time-evolutional dynamics is the direct reason why the multiple-quantum operator algebra space should be considered as the central place to make use of the fundamental quantum-computing-speedup resources to achieve essential quantum-computing speedup. Moreover, according to quantum mechanics the Hamiltonians of the quantum systems that generate the unitary time-evolutional propagators are the dynamical variables. Now among these different kinds of basic quantum spaces, i.e., the multiple-quantum operator algebra spaces, the density operator spaces, and the Hilbert spaces, only the multiple-quantum operator algebra spaces can accommodate the dynamical variables including the Hamiltonians of the quantum systems. Consequently the multiple-quantum operator algebra space (or the Liouville operator algebra space) must be positioned as the central place where one makes use of the fundamental quantum-computing-speedup resources to speed up essentially quantum computing and also quantum simulating such as quantum simulating for the unitary time-evolutional processes of quantum spin systems.

There are a number of different ways to describe the unitary time-evolutional process of a quantum system in quantum mechanics [2, 3] and among these different description ways there are the three basic description ways (or pictures), that is, the Schrödinger picture, the Heisenberg picture, and the Dirac picture, as can be seen in the previous Section 2. Therefore, the unitary time-evolutional process of the quantum system may be simulated in a number of different ways. Among these different quantum-simulating ways there are the three basic quantum-simulating ways to simulate any unitary time-evolutional process, which correspond to the three basic description pictures, respectively. Therefore, quantum simulating any unitary time-evolutional process of a quantum system may be carried out in the Schrödinger picture, the Heisenberg picture, or the Dirac picture.

It is expected that the fundamental quantum-computing-speedup resources can play a key role in essentially speeding up quantum simulating any unitary time-evolutional processes. When the equation of motion (2.1) or (2.3), i.e., the Schrödinger wave equation, is solved (exactly or approximately) with the quantum-simulating method in the Schrödinger picture, the symmetrical structures and properties of the Hilbert space [1,2,3] may be used to simplify the unitary time-evolutional propagators $U(t)$ in (2.3). Then it is expected that the symmetrical structures and properties of the Hilbert space can play a key role in essentially speeding up solving the Schrödinger wave equation with the quantum-simulating method. In analogous way, when the Liouville-von Neumann equation (2.2) or (2.4) is solved with the quantum-simulating method in the Schrödinger picture, the symmetrical structures and properties of the density operator space [5,6,3] may be used to speed up solving the Liouville-von Neu-mann equation. According to the quantum-computing speedup theory [Ref[1]] these symmetrical structures and properties of the Hilbert space and the density operator space are considered as the fundamental quantum-computing-speedup resources. In the Schrödinger picture quantum simulating the unitary time-evolutional process is performed in a quantum-state space, i.e., in the Hilbert space or in the density operator space. Then it is said that the fundamental quantum-computing-speedup resources are original from the quantum-state space, i.e., the Hilbert space or the density operator space. The symmetrical structures and properties of a quantum system are considered as the fundamental and inherent attributes of the quantum system. Consequently the fundamental quantum-computing-speedup resources which are original from these quantum-state spaces of the quantum system are the fundamental and inherent attributes of the quantum system.

Quantum simulating any unitary time-evolutional process of a quantum system also may be carried out in the Heisenberg picture or in the Dirac picture. In the Heisenberg picture the unitary time-evolutional process is described by the Heisenberg motion equation (2.7). Consequently the dynamical variable $A$ under study evolves in time in accordance with the Heisenberg equation (2.7) or (2.8) in which any quantum states are not involved explicitly. However, here it must be pointed out that generally the dynamical variable $A$ under study in the Heisenberg equation (2.7) has nothing to do with the symmetrical structures and properties of the quantum system. Therefore, whether or not one can make full use of the fundamental quantum-computing-speedup resources of the quantum system to speed up essentially solving the Heisenberg equation (2.7) with the quantum-simulating method is also dependent upon the dynamical variable $A$ under study. Here the fundamental quantum-computing-speedup resources should refer to those original from the Hilbert space of the quantum system. It is still expected that the symmetrical structures and properties of the Hilbert space can play a key role in speeding up solving the Heisenberg equation with the quantum-simulating method. The case related to the dynamical variable $A$ in the Dirac picture is similar to the present case in the Heisenberg picture.

According to the quantum-computing speedup theory the symmetrical structures and properties of the multiple-quantum operator algebra space [4] are also considered as the fundamental quantum-computing-speedup resources. It is said that the fundamental quantum-computing-speedup resources are original from the multiple-quantum operator algebra space. Just like those resources original from the Hilbert space or the density operator space the fundamental quantum-computing-speedup resources original from the multiple-quantum operator algebra space of quantum system such as quantum spin system [4] are also the fundamental and inherent attributes of the quantum system. The symmetric structures and properties of the multiple-quantum operator algebra spaces have been used to simplify greatly exact determination of the unitary time-evolutional processes of general quantum spin systems [4] and hence greatly speed up quantum simulating these unitary time-evolutional processes [8].

The essential difference between quantum simulating a unitary time-evolutional process in the multiple-quantum operator algebra space and that one in the Hilbert space or in the density operator space is that quantum simulating in the multiple-quantum operator algebra space is performed in a linear operator space, while the latter is performed in a quantum-state space.

The theoretical analysis above shows that the fundamental quantum-computing-speedup resources may exist in these different kinds of basic quantum spaces, that is, the multiple-quantum operator algebra space, the density operator space, and/or the Hilbert space of the quantum system under study. The multiple-quantum operator algebra space of any quantum spin system [4] is a linear operator space, while its corresponding Hilbert space is a quantum-state space. Recognize that the multiple-quantum operator algebra space is the central place where the fundamental quantum-computing-speedup resources are exploited to speed up essentially quantum computing and quantum simulating. Then it is worth devoting this whole Section 3 to investigating in detail the multiple-quantum operator algebra spaces. As for the fundamental quantum-computing-speedup resources which are original from the corresponding Hilbert spaces, the next Sections are devoted to their detailed investigation.

In quantum mechanics any dynamical variables are represented by linear operators. In the Heisenberg motion equation (2.7) the dynamical variable $A$ under study, the unitary propagator $U(t)$ and the Hamiltonian $H$ of the quantum system all are the dynamical variables and can be represented by the linear operators. Therefore, if the Heisenberg equation (2.7) is solved with the quantum-simulating method in the multiple-quantum operator algebra space (or the Liouville operator algebra space), at least in method it becomes easier to making use of the fundamental quantum-computing-speedup resources original from the multiple-quantum operator algebra space to speed up essentially solving the Heisenberg equation, since the dynamical variable $A$, the unitary propagator $U(t)$, and the Hamiltonian $H$ in the Heisenberg equation all are in the same multiple-quantum operator algebra space. Then it is expected that the symmetrical structures and properties of the multiple-quantum operator algebra space can play a key role in speeding up solving the Heisenberg equation (2.7) or (2.8).

## 3.1. The base-operator expansion principle

The spin Hamiltonian operator $H_s$ of a general spin system that generates the unitary time-evolutional propagator $U(t) = \exp(-iH_s t/\hbar)$ may be expanded in terms of the base operators $\{B_s\}$ of the finite-dimensional linear operator space of the spin system [4],

$$H_s = \sum_{s=0}^{N\times N-1} a_s B_s, \tag{3.1}$$

where $\{a_s\}$ are expansion coefficients. Here all these $N\times N$ linearly-independent base operators $\{B_s\}$ span the linear operator space with dimension $N\times N$. The inner product (or scalar product) of any pair of base operators $B_s$ and $B_t$ of the linear operator space may be defined by $(B_s, B_t) = Tr\,(B_s^{+} B_t)$. This finite-dimensional $(N\times N)$ linear operator space may be called the multiple-quantum operator algebra space (or the Liouville operator algebra space) [4].

Here it is particularly pointed out that the Liouville operator space in Ref.[5] is generally treated as the density operator space [6, 3] and is used as a quantum-state space to describe quantum (mixed) states ($\rho$) of spin systems in nuclear magnetic resonance spectroscopy. In contrast, the Liouville operator space in Ref.[4] and in this paper is considered as a linear operator space and is used to describe linear operators (e.g., the Hamiltonian operators) of spin systems. Moreover, in this paper the density operator space [6, 3, 5] which is comprised of all the density operators $\{\rho\}$ (or matrices) of a quantum system such as a spin system is generally treated as a quantum-state space and is used to describe quantum (mixed) states ($\rho$) of the quantum system.

In quantum mechanics it is postulated that any physical quantities including any dynamical variables are represented by linear operators on vector space [1, 3, 2]. All linear operators on a vector space form a linear space. A theoretical proof for this theorem may be seen in textbook (See, e.g., Ref.[3]). In this paper suppose that any linear operator $A$ on an $N-$dimensional vector space $\{\psi_j\}$ is defined by the linear transformation $A\psi_k = \sum_{j=0}^{N-1} a_{jk}\psi_j$, where $\psi_k$ takes every one of all the $N$ base vectors $\{\psi_j\}$ of the $N-$dimensional vector space and all these $N^2$ scalars $\{a_{jk}\}$ form a unique $N \times N$ matrix [3,2,1]. This means that the linear operator $A$ is completely characterized by the $N \times N$ matrix. Now this theorem also may be understood from the viewpoint of matrix as follows. All linear operators on an $N-$dimensional vector space correspond one-to-one to all $N \times N$ matrices [3,2,1]. All the $N \times N$ matrices form a linear space of matrices with dimension $N \times N$. This linear space of the $N \times N$ matrices is closed under multiplication operation, because the product of any two $N \times N$ matrices of the linear space of matrices is still an $N \times N$ matrix and belongs to the same linear space of matrices. Correspondingly all the linear operators on the $N-$dimensional vector space form a linear operator space with dimension $N \times N$. Moreover, this linear operator space is closed under multiplication operation. Finally, in parallel with inner product of matrices the inner product of linear operators $A$ and $B$ of the linear operator space may be defined by $(A, B) = Tr\,(A^{+}B)$ [1, 3]. For simplicity, in this paper it is assumed that the linear operator spaces under investigation are finite-dimensional.

Just like that a vector space may be completely described by a complete set of base vectors (i.e., a vector basis set), a linear operator space may be completely described by a complete set of base operators (i.e., an operator basis set). There are $N$ orthogonal base vectors which form a complete set of base vectors for an $N-$dimensional vector space, while there are $N^2$ orthogonal base operators which form a complete set of base operators for the corresponding $N \times N-$dimensional linear operator space. Obviously, there is not one-to-one correspondence between all the vector basis sets of the $N-$dimensional vector space and all the operator basis sets of the corresponding $N \times N-$dimensional linear operator space. Therefore, even given a complete set of base vectors for the $N-$dimensional vector space one is still allowed to choose arbitrarily a complete set of base operators for the corresponding $N \times N-$dimensional linear operator space.

Just like that any vector of a vector space can be expanded in terms of a complete set of base vectors of the vector space, any operator of a linear operator space may be expanded in terms of a complete set of base operators of the linear operator space. This is the so-called base-operator expansion in the linear operator space. A typical instance for this base-operator expansion is shown by (3.1), where the Hamiltonian operator $H_s$ of a spin system is expanded in terms of a complete set of base operators $\{B_s\}$ of the multiple-quantum operator algebra space of the spin system [4].

Generally, an arbitrary operator $Q$ of the multiple-quantum operator algebra space can be expanded in terms of the complete set of base operators $\{B_s\}$,

$$Q = \sum_{k=0}^{N^2-1} a_k B_k, \tag{3.2}$$

where the expansion coefficient $a_k$ can be determined by the inner products:

$$a_k = \frac{(B_k, Q)}{(B_k, B_k)} = \frac{Tr\left(B_k^+ Q\right)}{Tr\left(B_k^+ B_k\right)},$$

here the orthogonal relations for the base operators $\{B_s\}$ have already been used, that is, $(B_k, B_l) = Tr\left(B_k^+ B_l\right) = Tr\left(B_k^+ B_k\right)\delta_{kl}$ with the $\delta$ symbol: $\delta_{kl} = 0$ if $k \neq l$ and $\delta_{kl} = 1$ if $k = l$. As an example, the operator $Q$ in the base-operator expansion (3.2) may be the dynamical variable under study in the motion equation (2.8) in the Heisenberg picture or in the motion equation (2.13) in the Dirac picture.

The base-operator expansion (3.2) (and also (3.1)) is applied to the multiple-quantum operator algebra space of a general spin system [4]. It can be applied as well to any finite-dimensional linear operator space of a quantum system, if any complete set of base operators of the linear operator space replaces the complete set of base operators $\{B_s\}$ in (3.2) of the multiple-quantum operator algebra space. Any operator of the finite-dimensional linear operator space of a quantum system can be expanded in terms of a complete set of base operators of the linear operator space. This base-operator expansion principle in the linear operator space corresponds to the eigenfunction expansion principle in the Hilbert space in quantum mechanics [1, 2, 3]. The base-operator expansion principle may be considered as the fundamental mathematical principle in quantum mechanics. Generally it is useful and suitable for dealing with quantum-mechanical problems of the quantum systems with finite-dimensional linear operator spaces [1, 3]. It has been extensively used to study the unitary time-evolutional processes of general spin systems [4].

In quantum mechanics the density operator space [6, 3, 5] of a quantum system is a quantum-state space and is used to describe the quantum (mixed) states of the quantum system. However, from the point of view of mathematics the density operator space may be thought of as a specific linear operator space (i.e., a linear space of operators). It accommodates only the density operators ($\rho$) which are positive and Hermitian operators and satisfy $Tr\,(\rho) = 1$. In order to describe completely and generally the unitary quantum dynamics such as the unitary time-evolutional processes of quantum systems, a more general and complicated linear operator space such as the multiple-quantum operator algebra space (or the Liouville operator algebra space) [4] is necessary which can accommodate the quantum-system Hamiltonians and dynamical variables (or operators), general Hermitian operators, and the unitary time-evolutional propagators and general unitary operators, and so on. For simplicity, in this paper this linear operator space is limited to be finite-dimensional. This finite-dimensional linear operator space is still called the multiple-quantum operator algebra space or the Liouville operator algebra space. From the point of view of quantum physics the density operator space is completely different from the multiple-quantum operator algebra space, that is, the former describes quantum states, while the latter describes linear operators that act on quantum states. However, from the point of view of mathematics the multiple-quantum operator algebra space contains the whole density operator space. The essential difference between the two operator spaces from the point of view of mathematics is that the multiple-quantum operator algebra space is closed under multiplication operation, while the density operator space is not[13].

The multiple-quantum operator algebra space (or the Liouville operator algebra space) is the central place where the fundamental quantum-computing-speedup resources are exploited to speed up essentially quantum computing and quantum simulating. The base-operator expansion principle can be generally used in the multiple-quantum operator algebra space. Therefore, in quantum computing and quantum simulating the multiple-quantum operator algebra space may be considered as the standard theoretical equipment to describe any unitary quantum dynamical process, which may be any unitary time-evolutional process of quantum spin system in the Heisenberg picture or the Dirac picture. This is similar to that the Hilbert space is the standard theoretical equipment to describe quantum-mechanically the time-dependent quantum states of a quantum system in any unitary time-evolutional process in the Schrödinger picture.

With the help of the base-operator expansion principle the unitary time-evolutional process of a quantum spin system may be determined (or realized) in the multiple-quantum operator algebra space. Suppose that the unitary time-evolutional process of the quantum spin system is governed by the unitary propagator $U\,(t) = \exp\,(-iH_s t/\hbar)$ which is generated by the spin Hamiltonian $H_s$ of (3.1). Let the operator $Q$ of (3.2) be any dynamical variable of the spin system. It has the base-operator expansion (3.2), that is, it can be expanded in terms of the complete set $\{B_s\}$ of base operators of the multiple-quantum operator algebra space of the spin system. Then the unitary time-evolutional process of

[13] There is the relation $\rho^2 = \rho$ for the density operator $\rho$ of the density operator space of a pure-state quantum system. This relation may distinguish a pure-state quantum system from a mixed-state one ($\rho^2 \neq \rho$), but it has nothing to do with the definition of a density operator (or matrix) and the symmetry of quantum system, and it does not yet change the fact that $\rho$ always represents quantum state.

the spin system may be expressed as

$$U\left(t\right)^{+}QU\left(t\right)=\sum_{k=0}^{N^{2}-1}a_{k}U\left(t\right)^{+}B_{k}U\left(t\right) \tag{3.3a}$$

This is a unitary time-evolutional process of the spin system in the Heisenberg picture or the Dirac picture. Now this unitary time-evolutional process can be determined by determining the unitary time-evolutional process of any base operator $B_k$ of the operator basis set $\{B_s\}$ in the multiple-quantum operator algebra space [4],

$$U\left(t\right)^{+}B_{k}U\left(t\right)=\sum_{s=0}^{N^{2}-1}b_{s}B_{s} \tag{3.3b}$$

This exact method to determine the unitary time-evolutional process of (3.3a) needs to employ the base-operator expansion principle and the unitary time-evolutional processes of the base operators $\{B_s\}$ in the multiple-quantum operator algebra space. This determination method can be used not only by the unitary time-evolutional processes, but it also can be available for any unitary transformation [2,3,17] in the multiple-quantum operator algebra space. Suppose that the unitary transformation may be written as

$$WQW^{+}=\sum_{k=0}^{N^{2}-1}a_{k}WB_{k}W^{+} \tag{3.4a}$$

where $W$ is any unitary transformation acting on the operator $Q$ of (3.2) which may be any linear operator, e.g., a dynamical variable or a Hermitian operator and so on. Obviously, by determining the unitary transformation $W$ on any base operator $B_k$ in the multiple-quantum operator algebra space,

$$WB_{k}W^{+}=\sum_{s=0}^{N^{2}-1}c_{s}B_{s}, \tag{3.4b}$$

one is able to determine the unitary transformation of (3.4a). The unitary transformation (3.4b) may be called the operator-basis unitary transformation in the multiple-quantum operator algebra space.

There may be various methods to determine the unitary time-evolutional process (3.3b) of any base operator $B_k$ or the unitary transformation (3.4b) of any base operator $B_k$ in the multiple-quantum operator algebra space. One important method is to employ the decomposition of the unitary time-evolutional propagator $U(t)$ (or the unitary transformation $W$) into an ordered product of a series of elementary propagators to determine (3.3b) (or (3.4b)) [4, 8].

The unitary time-evolutional process of a quantum spin system can be determined (or realized) in the Hilbert space [2,3,1] of the spin system. It also can be determined in the density operator space [5,6,3] of the spin system. These two methods have been used most extensively in quantum physics. Here the multiple-quantum operator algebra space [4] can provide the third method to determine (or realize) the unitary time-evolutional process.

The symmetric structures and properties of the multiple-quantum operator algebra spaces of general spin systems are basically investigated in Ref. [4]. Generally the multiple-quantum operator algebra space of any quantum spin system contains a diagonal operator subspace which may be called as usual the longitudinal magnetization and spin order (LOMSO) operator subspace. All the diagonal operators of the LOMSO operator subspace commute mutually. The LOMSO operator subspace should be the largest commuting operator subspace of the multiple-quantum operator algebra space. Beside the LOMSO operator subspace there are the non-diagonal operator subspaces in the multiple-quantum operator algebra space. The general non-diagonal operator subspaces may include the zero-quantum operator subspace and the even-order multiple-quantum operator subspace. The basic properties of these diagonal and non-diagonal operator subspaces are described in Ref. [4]. The zero-quantum operator subspace contains the LOMSO operator subspace, while the even-order multiple-quantum operator subspace contains both the zero-quantum operator subspace and the LOMSO operator subspace. These general multiple-quantum operator subspaces and the LOMSO operator subspace have been used to simplify the complete decomposition of the unitary time-evolutional propagator of a general spin system into an ordered product of a series of elementary propagators. Consequently the unitary time-evolutional process of the spin system may be exactly calculated and simulated more quickly.

A general linear operator space is usually much complicated in quantum mechanics. The base-operator expansion of an operator may or may not exist in an infinite-dimensional linear operator space. Therefore, it is necessary to constrain the linear operator spaces under investigation in this paper. For simplicity, suppose that any linear operator space under investigation is comprised of all the linear operators that act on the finite-dimensional Hilbert space of a quantum system with discrete energy spectrum. This means that in this paper the multiple-quantum operator algebra spaces (or the Liouville operator algebra spaces) under investigation are finite-dimensional linear operator spaces. Moreover, any member of the multiple-quantum operator algebra space under investigation is a bounded operator. Below the multiple-quantum operator algebra spaces are further investigated extensively.

## 3.2. The complete sets of base operators

The multiple-quantum operator algebra space of any quantum spin system [4] is finite-dimensional and may be spanned by an operator basis set. According to the base-operator expansion principle any operator of the spin system can be expanded in terms of a complete set of base operators (i.e., an operator basis set) of the multiple-quantum operator algebra space of the spin system, as shown by the base-operator expansion (3.2). There are a number of choices of the operator basis set for the base-operator expansion (3.2) of the multiple-quantum operator algebra space. It is important to choose suitably an operator basis set for the multiple-quantum operator algebra space so that the symmet-ric structures and properties of the spin system can be conveniently taken into account. Therefore, a good choice of the operator basis set usually needs to consider a detailed quantum spin system under study. In quantum mechanics the most important linear operators are Hermitian operators and unitary operators. The Hermitian operators are often chosen to form an operator basis set for the multiple-quantum operator algebra space, partly because they are simpler and can generate simply and directly the unitary exponential operators, and more importantly they can better reflect symmetry of quantum system, i.e., the symmetric structures and properties of quantum system. Of course, beside the Hermitian operators other linear operators also are possibly chosen to form an operator basis set. Below three different kinds of the operator basis sets of the multiple-quantum operator algebra spaces are introduced in quantum spin systems. The base operators of these operator basis sets are closely related to the multiple-quantum transition operators of the spin systems.

### 3.2.1. Tensor products of Cartesian spin operators

Tensor (or direct) products of Cartesian spin operators [5] of a multiple-spin system may be simply called the Cartesian product operators (or briefly product operators). They are perhaps most suitable to treat theoretically an $n-$spin$-1/2$ system which consists of $n$ spin$-1/2$ particles, each one of which has the spin angular momentum quantum number $I = 1/2$. The Cartesian spin operators of a single spin$-1/2$ particle (e.g., the $k-$th spin$-1/2$ particle of the $n - spin - 1/2$ system) are simply given by the spin angular momentum operators $I_{kx}$, $I_{ky}$, and $I_{kz}$ (in unit $\hbar = 1$). (The spin angular momentum operators $J_{kx}, J_{ky}, J_{kz}$ of a single spin$-1/2$ particle [2] are really given by $J_{k\mu} = \hbar\sigma_{k\mu}/2 = \hbar I_{k\mu}$ with $\mu = x, y, z$.) They obey the basic commutation rules of angular momentum operators: $[I_{k\alpha}, I_{k\beta}] = iI_{k\gamma}$, where $\alpha, \beta, \gamma = x, y, z$ and cyclic permutations. With the help of the tensor product method the Cartesian product operators of the $n-$spin$-1/2$ system may be constructed by the tensor product of the Cartesian spin operators $\{I_{kx}, I_{ky}, I_{kz}\}$ plus the unity operators $E_k$ $(k = 1, 2, ..., n)$ of all the $n$ individual spin$-1/2$ particles of the $n-$spin$-1/2$ system. These Cartesian product operators may be chosen to form a complete set of base operators of the multiple-quantum operator algebra space of the $n - spin - 1/2$ system. This is perhaps the most straightforward choice for the operator basis set of the multiple-quantum operator algebra space. A great advantage for this choice is that these Cartesian product operators are Hermitian and can act as the generating operators to directly generate the unitary exponential operators.

First of all, consider a single spin$-1/2$ particle. This is a single-spin$-1/2$ system. It is well-known in quantum mechanics [2] that a single spin$-1/2$ particle with spin quantum number $I = 1/2$ is associated with a discrete two-dimensional Hilbert space. The Liouville operator algebra space of the single-spin$-1/2$ system, which corresponds to the two-dimensional Hilbert space, is therefore four-dimensional. Then one may simply choose these three Cartesian spin operators $I_x$, $I_y$, $I_z$, and the unity operator $E$ of the single spin$-1/2$ particle together to form a complete set $\{E, I_x, I_y, I_z\}$ of base operators for the Liouville operator algebra space. This operator basis set $\{E, I_x, I_y, I_z\}$ is Hermitian as a whole in the sense that every base operator in the operator basis set is Hermitian. Any spin Hamiltonian operator $H_s$ of the single-spin$-1/2$ system then may be expanded in terms of the base operators $\{E, I_x, I_y, I_z\}$,

$$H_s = \alpha_0 E + \alpha_x I_x + \alpha_y I_y + \alpha_z I_z. \tag{3.5}$$

Here all these expansional coefficients $\alpha_0$, $\alpha_x$, $\alpha_y$, and $\alpha_z$ are real due to that every operator in (3.5) is Hermitian. The unitary time-evolutional propagator that is generated by the spin Hamiltonian $H_s$ of (3.5) may be generally written as

$$U_s(\tau) = \exp(-iH_s\tau/\hbar) = \exp[-i(\alpha_0 E + \alpha_x I_x + \alpha_y I_y + \alpha_z I_z)\tau/\hbar] \tag{3.6a}$$

where $\tau$ is the time interval. The unitary time-evolutional process of the single-spin$-1/2$ system then is generally described by this single-spin unitary propagator $U_s(\tau)$.

The unitary spin rotation operators $R_\lambda(\theta)$ with $\lambda = x, y, z$ of the single spin$-1/2$ particle, which are generated by the Hermitian spin operators $I_x$, $I_y$, and $I_z$, respectively, may be written as

$$R_\lambda(\theta) = \exp(-i\theta I_\lambda), \ \lambda = x, y, z \tag{3.6b}$$

Obviously, these spin rotation operators $R_\lambda(\theta)$ of the single spin$-1/2$ particle are the special cases of the general single-spin unitary propagator (or operator) $U_s(\tau)$ of (3.6a). These single-spin rotation operators $R_\lambda(\theta)$ and more generally the single-spin unitary operator $U_s(\tau)$ may represent the spin-selective excitation pulses which are selectively applied to a single selected spin$-1/2$ particle in a multiple-spin$-1/2$ system. They are the basic building blocks [8] which may be used to realize quantum simulating the unitary time-evolutional process of a general multiple-spin$-1/2$ system. They may be directly realized in nuclear magnetic resonance selective-pulse experiments (See, e.g., [Ref[4]] and Ref.[11]).

The single-spin Hamiltonian $H_s$ of (3.5) can be expanded in terms of the base operators $\{E, I_x, I_y, I_z\}$ of the Liouville operator algebra space of the single-spin$-1/2$ system. This is a simplest application of the base-operator expansion principle. The base-operator expansion principle can be used not only for a single-spin$-1/2$ system but also for a general spin system. Consider a two-spin$-1/2$ system which consists of two individual spin$-1/2$ particles. The two-spin$-1/2$ system is associated with a discrete four-dimensional Hilbert space, which is the tensor product of the component two-dimensional Hilbert spaces of the two individual spin$-1/2$ particles [2]. Correspondingly the multiple-quantum operator algebra space (or the Liouville operator algebra space) of the two-spin$-1/2$ system is a sixteen-dimensional linear operator space. A straightforward choice for the complete set of base operators of the multiple-quantum operator algebra space is the Cartesian product operators of the two-spin$-1/2$ system. Then any spin Hamiltonian of the two-spin$-1/2$ system can be expanded in terms of these Cartesian product operators.

Below the Cartesian product operators are explicitly constructed with the help of the tensor product method for the $two-spin-1/2$ system. For convenience, here the unity operator $E$ is temporarily renamed $I_o$, i.e., $E=I_o$. Then the operator basis set $\{E,I_x,I_y,I_z\}$ of any individual spin$-1/2$ particle of the $two-spin-1/2$ system may be rewritten as $\{I_{ko},I_{kx},I_{ky},I_{kz}\}$ with $k=1,2$. Now, by starting from the operator basis set $\{E,I_x,I_y,I_z\}$ of an individual spin$-1/2$ particle, the Cartesian product operators $\{B_s\}$ of the $two-spin-1/2$ system may be constructed explicitly by the tensor (or direct) product of these two operator basis sets $\{I_{ko},I_{kx},I_{ky},I_{kz}\}$ with $k=1,2$ of the two individual spin$-1/2$ particles of the $two-spin-1/2$ system:

$$\begin{aligned}\{B_s\} &= \{I_{1o},I_{1x},I_{1y},I_{1z}\}\bigotimes\{I_{2o},I_{2x},I_{2y},I_{2z}\}\\ &\overset{Def}{\equiv}\{I_{1o}\bigotimes I_{2o},I_{1o}\bigotimes I_{2x},I_{1o}\bigotimes I_{2y},I_{1o}\bigotimes I_{2z};\\ &I_{1x}\bigotimes I_{2o},I_{1x}\bigotimes I_{2x},I_{1x}\bigotimes I_{2y},I_{1x}\bigotimes I_{2z};\\ &I_{1y}\bigotimes I_{2o},I_{1y}\bigotimes I_{2x},I_{1y}\bigotimes I_{2y},I_{1y}\bigotimes I_{2z};\\ &I_{1z}\bigotimes I_{2o},I_{1z}\bigotimes I_{2x},I_{1z}\bigotimes I_{2y},I_{1z}\bigotimes I_{2z}\}\\ &=\{I_o,I_{2x},I_{2y},I_{2z};I_{1x},I_{1x}I_{2x},I_{1x}I_{2y},I_{1x}I_{2z};\\ &I_{1y},I_{1y}I_{2x},I_{1y}I_{2y},I_{1y}I_{2z};I_{1z},I_{1z}I_{2x},I_{1z}I_{2y},I_{1z}I_{2z}\}\end{aligned}\tag{3.7a}$$

where the tensor product operator $B_s=I_{1\lambda}\bigotimes I_{2\mu}$ with $\lambda,\mu=o,x,y,z$ may be rewritten as $B_s=(I_{1\lambda}\bigotimes I_{2o})(I_{1o}\bigotimes I_{2\mu})=I_{1\lambda}I_{2\mu}$ and the latter is the direct matrix product of $I_{1\lambda}\equiv I_{1\lambda}\bigotimes I_{2o}$ (matrix) and $I_{2\mu}\equiv I_{1o}\bigotimes I_{2\mu}$ (matrix), and especially the total unity operator $I_o=I_{1o}\bigotimes I_{2o}=I_{1o}I_{2o}$. In (3.7a) the symbol $\overset{Def}{\equiv}$ in $A\overset{Def}{\equiv}B$ means that $A$ is defined by $B$. For convenience in the theoretical treatments and applications every product operator $B_s$ of (3.7a) may be multiplied by a suitable factor. Then the product operators $\{B_s\}$ of (3.7a) may be rewritten in the general form $\{B_s\}=\{2^{r-1}I_{1\lambda}I_{2\mu}\}$ with the real factor $2^{r-1}$ and $\lambda,\mu=o,x,y,z$. If the factor $r=1$ for every product operator $2^{r-1}I_{1\lambda}I_{2\mu}$, then the product operators $\{2^{r-1}I_{1\lambda}I_{2\mu}\}$ are reduced to (3.7a). If the factor $r$ is the number of the single-spin operators $\{I_{kx},I_{ky},I_{kz}\}$ appearing in the product operator $2^{r-1}I_{1\lambda}I_{2\mu}$, then the product operators $\{2^{r-1}I_{1\lambda}I_{2\mu}\}$ may be reduced to the form [5]

$$\{B_s\}=\{I_o/2,I_{1\xi},I_{2\eta},2I_{1\xi}I_{2\eta}\}\,;\ \xi,\eta=x,y,z.\tag{3.7b}$$

The product operators $\{B_s\}=\{2^{r-1}I_{1\lambda}I_{2\mu}\}$ form a complete set of base operators of the sixteen-dimensional multiple-quantum operator algebra space of the two-spin$-1/2$ system. This is an operator basis set which consists of the orthogonal and Hermitian product operators. There are the orthogonal relations for the product operators $\{B_s\}=\{2^{r-1}I_{1\lambda}I_{2\mu}\}$:

$$(B_s,B_t)=Tr\left(B_s^+B_t\right)=Tr\left(B_s^+B_s\right)\delta_{st}\tag{3.8}$$

It is easy to prove these orthogonal relations. This is based on $(i)$ every spin operator $I_{k\xi}$ is traceless: $Tr\,(I_{k\xi}) = 0$ for $\xi = x, y, z$ and $k = 1, 2$; $(ii)$ $Tr\,(I_{1\lambda} \bigotimes I_{2\mu}) = Tr\,(I_{1\lambda})\, Tr\,(I_{2\mu})$; and $(iii)$ $I_{k\xi}^2 = I_{ko}/4$ for $\xi = x, y, z$ and $I_{k\alpha} I_{k\beta} = \frac{1}{2} i I_{k\gamma}$ for $\alpha, \beta, \gamma = x, y, z$ and cyclic permutations. The factor $2^{r-1}$ does not affect the orthogonality (3.8) of the product operators $\{B_s\} = \{2^{r-1} I_{1\lambda} I_{2\mu}\}$. Of course, it can affect the normalization constant $Tr\,(B_s^+ B_s)$. There is $Tr\,(B_s^+ B_s) = 1$ [5] for any product operator $B_s$ of (3.7b).

All the product operators $\{B_s\} = \{2^{r-1} I_{1\lambda} I_{2\mu}\}$ are Hermitian. Then each one (e.g., $I_{1\lambda} I_{2\mu}$) of these product operators $\{B_s\}$ of (3.7) may generate a unitary exponential operator $R_{\lambda\mu}\,(J_{\lambda\mu})$,

$$R_{\lambda\mu}\,(J_{\lambda\mu}) = \exp\left(-i J_{\lambda\mu} I_{1\lambda} I_{2\mu}\right) \text{ for } \lambda, \mu = o, x, y, z \tag{3.9}$$

where $J_{\lambda\mu}$ is real interacting parameter. Except the unity operator $R_{oo}\,(J_{oo}) = \exp\,(-i J_{oo})\, I_o$ with the phase factor $\exp\,(-i J_{oo})$, these unitary operators of (3.9) may be divided into two different kinds. One kind are the unitary operators $R_{k\lambda}\,(J_{k\lambda}) = \exp\,(-i J_{k\lambda} I_{k\lambda})$ with $\lambda = x, y, z$ and $k = 1, 2$. This kind are generated by the single-spin operators $I_{k\lambda}$ which may represent the single-spin interactions $(J_{k\lambda} I_{k\lambda})$ between the $k-$th spin$-1/2$ particle and the external electromagnetic wave field (or pulse). They are really the single-spin rotation operators of (3.6b). Therefore, the single-spin rotation operator $R_{k\lambda}\,(J_{k\lambda})$ represents the spin-selective excitation pulse that is applied to only the $k-$th spin$-1/2$ particle of the two-spin$-1/2$ system. Another kind are the unitary operators $R_{\lambda\mu}\,(J_{\lambda\mu})$ which are generated by the two-spin interaction operators $J_{\lambda\mu} I_{1\lambda} I_{2\mu}$ with $\lambda, \mu = x, y, z$ of the two-spin$-1/2$ system. Note that any non-diagonal two-spin interaction operator $J_{\lambda\mu} I_{1\lambda} I_{2\mu}$ with $\lambda, \mu = x, y, z$ can be transformed to the diagonal two-spin interaction operator $J_{\lambda\mu} I_{1z} I_{2z}$ by a unitary transformation which is composed of the suitable single-spin rotation operators of (3.6b) [8]. The first kind $\{R_{k\lambda}\,(J_{k\lambda})\}$ and the second kind $\{R_{\lambda\mu}\,(J_{\lambda\mu})\}$ of the unitary operators (3.9) are the one-spin and the two-spin elementary propagators, respectively. They are the basic building blocks which can be used for quantum simulating the unitary time-evolutional process of a general spin$-1/2$ system [8]. They may be conveniently realized in nuclear magnetic resonance selective-pulse experiments (See, e.g., [Ref[4]], Ref.[11]).

Generally, the Cartesian product operators for any $n-$spin$-1/2$ system may be explicitly constructed as well with the help of the tensor product method that is already used in the $two - spin - 1/2$ system. There is the spin operator basis set $\{I_{ko}, I_{kx}, I_{ky}, I_{kz}\}$ for any individual spin$-1/2$ particle ($i.e.$, the $k-$th spin$-1/2$ particle with $k = 1, 2, ..., n$) of the $n$ spin$-1/2$ particles of the $n - spin-1/2$ system. By starting from the spin operator basis set $\{I_{ko}, I_{kx}, I_{ky}, I_{kz}\}$ of an individual spin$-1/2$ particle, the Cartesian product operators $\{B_s\}$ of any $n - spin - 1/2$ system may be constructed explicitly by the tensor product of these $n$ spin operator basis sets $\{I_{ko}, I_{kx}, I_{ky}, I_{kz}\}$ with $k = 1, 2, ..., n$ of the $n$ spin$-1/2$ particles of the $n - spin - 1/2$ system:

$$\{B_s\} = \{I_{1o}, I_{1x}, I_{1y}, I_{1z}\} \bigotimes \{I_{2o}, I_{2x}, I_{2y}, I_{2z}\} \bigotimes ... \bigotimes \{I_{no}, I_{nx}, I_{ny}, I_{nz}\}$$

$$= \{I_o, I_{k_1\lambda_1}, I_{k_1\lambda_1}I_{k_2\lambda_2}, I_{k_1\lambda_1}I_{k_2\lambda_2}I_{k_3\lambda_3}, ..., I_{k_1\lambda_1}I_{k_2\lambda_2}...I_{k_l\lambda_l}, ..., I_{1\lambda_1}I_{2\lambda_2}...I_{n\lambda_n}\} \quad (3.10a)$$

or

$$\{B_s\} = \{I_o/2, I_{k_1\lambda_1}, 2I_{k_1\lambda_1}I_{k_2\lambda_2}, 2^2 I_{k_1\lambda_1}I_{k_2\lambda_2}I_{k_3\lambda_3}, ..., 2^{l-1}I_{k_1\lambda_1}I_{k_2\lambda_2}...I_{k_l\lambda_l}, ..., 2^{n-1}I_{1\lambda_1}I_{2\lambda_2}...I_{n\lambda_n}\} \quad (3.10b)$$

where $\lambda_1, \lambda_2, ..., \lambda_l = x, y, z$; $k_1, k_2, ..., k_l = 1, 2, ..., n$ and $k_1 < k_2 < ... < k_l$; $1 \leq l \leq n$; and the total unity operator $I_o = I_{1o} \bigotimes I_{2o} \bigotimes ... \bigotimes I_{no}$. This is the tensor product method to construct explicitly the Cartesian product operators $\{B_s\}$ for any $n - spin - 1/2$ system. The product operators $\{B_s\}$ of (3.10b) are given in Ref.[5], which obey the orthogonal relations: $Tr\,(B_s B_t) = 2^{n-2}\delta_{st}$.

There are $4^n$ orthogonal and Hermitian product operators $\{B_s\}$ of (3.10) for the $n - spin - 1/2$ system. They form a complete set of base operators of the $4^n-$dimensional multiple-quantum operator algebra space of the $n - spin - 1/2$ system. This complete set may be called the Cartesian-product-operator basis set or simply the product-operator basis set. Obviously, this product-operator basis set $\{B_s\}$ is Hermitian as a whole. Now any operator, e.g., the spin Hamiltonian $H_s$ of the $n - spin - 1/2$ system, of the multiple-quantum operator algebra space may be expanded in terms of these $4^n$ orthogonal product operators $\{B_s\}$ of (3.10) in accordance with the base-operator expansion principle. And this base-operator expansion is still given formally by (3.1) with $N = 2^n$ and the base operators $\{B_s\}$ given by (3.10a) or (3.10b).

Since the Cartesian-product-operator basis set $\{B_s\}$ of (3.10) is Hermitian as a whole, every Hermitian product operator $B_s$ of the basis set $\{B_s\}$ may generate a unitary exponential operator (or propagator) $R_s\,(B_s) = \exp\,(-ib_s B_s)$ with real parameter $b_s$. These unitary propagators $\{R_s\,(B_s)\}$ are the simplest and most elementary propagators of the $n - spin - 1/2$ system. Hence they are called the elementary propagators ([Ref[5]], Ref.[8]). Now the time-evolutional process of the $n - spin - 1/2$ system is governed by the unitary time-evolutional propagator $\exp\,(-iH_s t/\hbar)$, where the spin Hamiltonian $H_s$ of the $n - spin - 1/2$ system may be generally expressed as the base-operator expansion (3.1) in which the base operators $\{B_s\}$ are given by the Hermitian product-operator basis set of (3.10). The programmable quantum simulating for the unitary time-evolutional process of the $n - spin - 1/2$ system then may be carried out by the two steps [8]: The first step is that the unitary propagator $\exp\,(-iH_s t/\hbar)$ is decomposed into a sequence of the elementary propagators $\{R_s\,(B_s)\}$, and the second step is that every elementary propagator $R_s\,(B_s)$ is expressed as a sequence of the single-spin elementary propagators of (3.6b) and the two-spin elementary propagators of (3.9). As shown in Ref. [8], the second step is efficient. However, the first step is usually hard. The symmetrical structures and properties of the multiple-quantum operator algebra space of the spin system may be helpful for speeding up the first step [4].

The symmetrical structures and properties of the multiple-quantum operator algebra space [4] of the $n - spin - 1/2$ system may be basically characterized with the help of the product-operator basis set $\{B_s\}$ of (3.10). Among these $4^n$ orthogonal product operators $\{B_s\}$ there are $2^n$ orthogonal diagonal product operators $\{\tilde{B}_s\}$ which may be given by [4]

$$\{\tilde{B}_s\} = \{E/2, I_{k_1z}, 2I_{k_1z}I_{k_2z}, 2^2 I_{k_1z}I_{k_2z}I_{k_3z}, ...,$$
$$2^{l-1}I_{k_1z}I_{k_2z}...I_{k_lz}, ..., 2^{n-1}I_{1z}I_{2z}...I_{nz}\} \quad (3.11)$$

where $k_1, k_2, ..., k_l = 1, 2, ..., n$ and $k_1 < k_2 < ... < k_l$; $1 \leq l \leq n$; and the total unity operator $E = I_o$. These $2^n$ orthogonal diagonal product operators $\{\tilde{B}_s\}$ span a commuting operator subspace of the multiple-quantum operator algebra space of the $n-$spin$-1/2$ system. This is just the $2^n-$dimensional diagonal operator subspace [4] called the LOMSO operator subspace. In addition to these $2^n$ diagonal product operators $\{\tilde{B}_s\}$ there are the $4^n - 2^n$ non-diagonal product operators in the product-operator basis set $\{B_s\}$ of (3.10). These non-diagonal product operators are the multiple-quantum transition operators (See Ref.[5] and the Section A). By combining with the diagonal product operators $\{\tilde{B}_s\}$ of (3.11) these multiple-quantum-transition (MQT) operators may form different MQT operator subspaces of the multiple-quantum operator algebra space [4]. These MQT operator subspaces may include the zero-quantum operator subspace and the even-order multiple-quantum operator subspace. They are more complicated than the LOMSO operator subspace.

As a typical example, consider the multiple-quantum operator algebra space of the two-spin$-1/2$ system. The product-operator basis set $\{B_s\}$ of the multiple-quantum operator algebra space is comprised of sixteen product operators of (3.7). As shown by (3.11), there are four orthogonal diagonal product operators $\{\tilde{B}_s\}$ of the product-operator basis set $\{B_s\}$:

$$\{\tilde{B}_s\} = \{E/2, I_{1z}, I_{2z}, 2I_{1z}I_{2z}\} \quad (3.12a)$$

They form an operator basis subset of the LOMSO operator subspace of the multiple-quantum operator algebra space. The other twelve product operators are non-diagonal operators. Among them these four product operators $\{2I_{1x}I_{2x},\ 2I_{1x}I_{2y}, 2I_{1y}I_{2x}, 2I_{1y}I_{2y}\}$ each are a linear combination of the zero- and two-quantum transition operators and the rest eight product operators $\{I_{1x}, I_{1y}, I_{2x},\ I_{2y}, 2I_{1x}I_{2z}, 2I_{1y}I_{2z}, 2I_{1z}I_{2x}, 2I_{1z}I_{2y}\}$ each are a linear combination of the one-quantum transition operators [5]. All these zero-, one-, and two-quantum transition operators are the multiple-quantum transition operators. Then in the multiple-quantum operator algebra space of the two-spin$-1/2$ system the LOMSO operator subspace is spanned by the diagonal product-operator basis subset $\{\tilde{B}_s\}$ of (3.12a); the zero-quantum operator subspace is spanned by the product operator basis subset:

$$\{B_s^{zq}\} = \{E/2, I_{1z}, I_{2z}, 2I_{1z}I_{2z};$$
$$(2I_{1x}I_{2x} + 2I_{1y}I_{2y})/2, (2I_{1y}I_{2x} - 2I_{1x}I_{2y})/2\}, \quad (3.12b)$$

where $(2I_{1x}I_{2x} + 2I_{1y}I_{2y})/2$ and $(2I_{1y}I_{2x} - 2I_{1x}I_{2y})/2$ are the Hermitian zero-quantum transition operators; and the even-order multiple-quantum operator subspace is spanned by the product operator basis subset:

$$\{B_s^{emq}\} = \{E/2, I_{1z}, I_{2z}, 2I_{1z}I_{2z}; (2I_{1x}I_{2x} + 2I_{1y}I_{2y})/2, (2I_{1y}I_{2x} - 2I_{1x}I_{2y})/2;$$

$$(2I_{1x}I_{2x} - 2I_{1y}I_{2y})/2, (2I_{1y}I_{2x} + 2I_{1x}I_{2y})/2\}, \tag{3.12c}$$

where $(2I_{1x}I_{2x} - 2I_{1y}I_{2y})/2$ and $(2I_{1y}I_{2x} + 2I_{1x}I_{2y})/2$ are the Hermitian two-quantum transition operators [5]. It is easy to find that the zero-quantum operator subspace $\{B_s^{zq}\}$ contains the LOMSO operator subspace $\{\tilde{B}_s\}$, while the even-order multiple-quantum operator subspace $\{B_s^{emq}\}$ contains the LOMSO operator subspace $\{\tilde{B}_s\}$ and the zero-quantum operator subspace $\{B_s^{zq}\}$.

Now any operator (e.g., the spin Hamiltonian $H_s$) of the multiple-quantum operator algebra space of the two-spin$-1/2$ system may be expanded in terms of the product operators $\{B_s\}$ of (3.7),

$$H_s = H_0 + H_1 + H_{02}, \tag{3.13}$$

where

$$H_0 = \alpha_0 E + \Omega_1 I_{1z} + \Omega_2 I_{2z} + J_{z,z} I_{1z} I_{2z} \tag{3.14a}$$

$$H_1 = \sum_{k=1}^{2} (\omega_{kx} I_{kx} + \omega_{ky} I_{ky})$$

$$+ J_{x,z} I_{1x} I_{2z} + J_{z,x} I_{1z} I_{2x} + J_{y,z} I_{1y} I_{2z} + J_{z,y} I_{1z} I_{2y} \tag{3.14b}$$

$$H_{02} = J_{x,x} I_{1x} I_{2x} + J_{x,y} I_{1x} I_{2y} + J_{y,x} I_{1y} I_{2x} + J_{y,y} I_{1y} I_{2y} \tag{3.14c}$$

Here $\{I_{k\xi}\}$ are the one-spin interaction operators for $k = 1, 2$ and $\xi = x, y, z$ and $\{2I_{1\xi}I_{2\eta}\}$ the two-spin interaction operators for $\xi, \eta = x, y, z$. And $H_0$ is a diagonal operator of the LOMSO operator subspace $\{\tilde{B}_s\}$ of (3.12a), $H_1$ is a linear combination of the single-quantum transition operators, and $H_{02}$ is a linear combination of the zero- and double-quantum transition operators and belongs to the even-order multiple-quantum operator subspace $\{B_s^{emq}\}$ of (3.12c).

The Cartesian product operators (3.10) of an $n-$spin$-1/2$ system mentioned above are constructed by the tensor product of the Cartesian spin-operator basis sets of the $n$ individual spin$-1/2$ particles of the $n-$spin$-1/2$ system. Beside the Cartesian product operators there also may be different types of product operators (See, e.g., Ref.[5]). Below one type of product operators which are different from the previous Cartesian product operators are constructed for the $n-$spin$-1/2$ system. The construction still employs the tensor-product method and is in accordance with quantum mechanics and reflects the point that quantum computing and quantum simulating stress to be mathematical logical, strict, and quantitative. This type of product operators are named the multiple-quantum-transition (MQT) product operators or simply the MQT product operators so that one can differentiate them from the previous Cartesian product operators (3.10). Theoretically the MQT product operators could be more useful. From this type of product operators the multiple-quantum operator algebra space may be understood more easily and the multiple-quantum-transition operators may be described more clearly (See also the Section A).

According to quantum mechanics a single spin$-1/2$ particle (in external magnetic field along the $z-$axis) owns only two discrete spin energy levels [2] and is associated with a two-dimensional Hilbert space. The two discrete spin energy levels may be represented by the two eigenbase vectors $|\alpha\rangle$ and $|\beta\rangle$ of the $z-$component spin operator $I_z$ of the spin angular momentum operator $\mathbf{I}$ (in unit $\hbar = 1$) of the single spin$-1/2$ particle, respectively. Let $|0\rangle$ and $|1\rangle$ be these two eigenbase vectors $|\alpha\rangle$ and $|\beta\rangle$, respectively. The two eigenbase vectors $|\alpha\rangle$ and $|\beta\rangle$ (or $|0\rangle$ and $|1\rangle$) may be respectively written as [2]

$$|0\rangle = |\alpha\rangle = \begin{pmatrix} 1 \\ 0 \end{pmatrix}, \ |1\rangle = |\beta\rangle = \begin{pmatrix} 0 \\ 1 \end{pmatrix}, \tag{3.15}$$

where $|\alpha\rangle$ and $|\beta\rangle$ obey the eigenvalue equations $I_z|\alpha\rangle = \frac{1}{2}|\alpha\rangle$ and $I_z|\beta\rangle = -\frac{1}{2}|\beta\rangle$, respectively. Then, in the representation defined by the spin operator $I_z$, the spin operators $\{I_x, I_y, I_z\}$ and the unity operator $E$ of the single spin$-1/2$ particle may be respectively expressed as [2]

$$I_x = \frac{1}{2}\sigma_x = \frac{1}{2}\begin{pmatrix} 0 & 1 \\ 1 & 0 \end{pmatrix} = \frac{1}{2}\left(|0\rangle\langle 1| + |1\rangle\langle 0|\right), \tag{3.16a}$$

$$I_y = \frac{1}{2}\sigma_y = \frac{1}{2}\begin{pmatrix} 0 & -i \\ i & 0 \end{pmatrix} = \frac{1}{2i}\left(|0\rangle\langle 1| - |1\rangle\langle 0|\right), \tag{3.16b}$$

$$I_z = \frac{1}{2}\sigma_z = \frac{1}{2}\begin{pmatrix} 1 & 0 \\ 0 & -1 \end{pmatrix} = \frac{1}{2}\left(|0\rangle\langle 0| - |1\rangle\langle 1|\right), \tag{3.16c}$$

$$E = \begin{pmatrix} 1 & 0 \\ 0 & 1 \end{pmatrix} = \left(|0\rangle\langle 0| + |1\rangle\langle 1|\right), \tag{3.16d}$$

where $\sigma_x$, $\sigma_y$, and $\sigma_z$ are the Pauli spin operators (or matrices). It is easy to find from (3.16) that these four orthogonal operators $|0\rangle\langle 0|$, $|1\rangle\langle 1|$, $|0\rangle\langle 1|$, and $|1\rangle\langle 0|$ may be respectively expressed as

$$|0\rangle\langle 0| = \frac{1}{2}E + I_z, \ \ |1\rangle\langle 1| = \frac{1}{2}E - I_z, \ \ |0\rangle\langle 1| = I_x + iI_y, \ \ |1\rangle\langle 0| = I_x - iI_y \tag{3.17}$$

These four orthogonal operators $\{|0\rangle\langle 0|, |1\rangle\langle 1|, |0\rangle\langle 1|, |1\rangle\langle 0|\}$ also form a complete set of base operators of the Liouville operator algebra space of the single-spin$-1/2$ system. The operator basis set $\{|0\rangle\langle 0|, |1\rangle\langle 1|, |0\rangle\langle 1|, |1\rangle\langle 0|\}$ is clearly different from the Cartesian spin operator basis set $\{E, I_x, I_y, I_z\}$. A significant difference is that the former is not Hermitian as a whole, while the latter is. However, they are equivalent to one another in the sense that one operator basis set can be changed to another by the operator-basis linear transformation of (3.16) or (3.17) (or equivalently by the base-operator expansions).

Now any operator of the Liouville operator algebra space of the single-spin$-1/2$ system can be expanded in terms of the base operators $\{|0\rangle\langle 0|, |1\rangle\langle 1|, |0\rangle\langle 1|, |1\rangle\langle 0|\}$. As an example, the single-spin Hamiltonian $H_s$ of (3.5) may be expressed as

$$H_s = \beta_{00}|0\rangle\langle 0| + \beta_{11}|1\rangle\langle 1| + \beta_{01}|0\rangle\langle 1| + \beta_{10}|1\rangle\langle 0| \tag{3.18}$$

where the expansional coefficients are given by $\beta_{00} = \alpha_0 + \frac{1}{2}\alpha_z$, $\beta_{11} = \alpha_0 - \frac{1}{2}\alpha_z$, $\beta_{01} = \frac{1}{2}\alpha_x + \frac{1}{2i}\alpha_y$, and $\beta_{10} = \frac{1}{2}\alpha_x - \frac{1}{2i}\alpha_y$. It can be seen that these expansional coefficients may be complex due to that the operator basis set $\{|0\rangle\langle 0|, |1\rangle\langle 1|, |0\rangle\langle 1|, |1\rangle\langle 0|\}$ is not Hermitian as a whole.

As a simple application of the operator basis set $\{|0\rangle\langle 0|, |1\rangle\langle 1|, |0\rangle\langle 1|, |1\rangle\langle 0|\}$, the specific Hermitian operator $H_2^{pd}$ of a single-spin$-1/2$ system is constructed by

$$H_2^{pd} = (\alpha|0\rangle + \beta|1\rangle)(\alpha^*\langle 0| + \beta^*\langle 1|) \tag{3.19}$$

where the coefficients $\alpha$ and $\beta$ may be complex. The Hermitian operator $H_2^{pd}$ is a pseudo-diagonal Hermitian operator (See the Section 4 below). The unitary propagator that is generated by the pseudo-diagonal Hermitian operator $H_2^{pd}$ is given by

$$U_2^{pd}(\tau) = \exp\left(-iH_2^{pd}\tau/\hbar\right) = \exp\left[-i(\alpha|0\rangle + \beta|1\rangle)(\alpha^*\langle 0| + \beta^*\langle 1|)\tau/\hbar\right] \tag{3.20}$$

The Hermitian operator $H_2^{pd}$ can be easily expanded in terms of the base operators $\{|0\rangle\langle 0|, |1\rangle\langle 1|, |0\rangle\langle 1|, |1\rangle\langle 0|\}$ :

$$H_2^{pd} = \alpha\alpha^*|0\rangle\langle 0| + \beta\beta^*|1\rangle\langle 1| + \alpha\beta^*|0\rangle\langle 1| + \beta\alpha^*|1\rangle\langle 0|. \tag{3.21a}$$

Obviously, it is a non-diagonal Hermitian operator. Now it also can be expanded in terms of the Cartesian spin operator basis set $\{E, I_x, I_y, I_z\}$,

$$H_2^{pd} = a_0'E + a_x'I_x + a_y'I_y + a_z'I_z. \tag{3.21b}$$

Here with the help of the base-operator expansions (3.21a), (3.18), and (3.5) it can be found that the expansional coefficients $a_0'$, $a_x'$, $a_y'$, and $a_z'$ in (3.21b) are determined by $a_0' + \frac{1}{2}a_z' = |\alpha|^2$, $a_0' - \frac{1}{2}a_z' = |\beta|^2$, $\frac{1}{2}a_x' + \frac{1}{2i}a_y' = \alpha\beta^*$, and $\frac{1}{2}a_x' - \frac{1}{2i}a_y' = \beta\alpha^*$.

It is shown above that the Cartesian spin operator basis set $\{E, I_x, I_y, I_z\}$ is equivalent to the operator basis set $\{|0\rangle\langle 0|, |1\rangle\langle 1|, |0\rangle\langle 1|, |1\rangle\langle 0|\}$ and vice versa for an individual spin$-1/2$ particle. Here for convenience the operator basis set $\{|0\rangle\langle 0|, |1\rangle\langle 1|, |0\rangle\langle 1|, |1\rangle\langle 0|\}$ of the $j-$th spin$-1/2$ particle of the $n$ individual spin$-1/2$ particles of the $n-$spin$-1/2$ system is written as $\{S_j^{-k_j,+l_j}\} = \{|0_j\rangle\langle 0_j|, |1_j\rangle\langle 1_j|, |0_j\rangle\langle 1_j|, |1_j\rangle\langle 0_j|\}$, where the base operator $S_j^{-k_j,+l_j}$ of the $j-$th spin$-1/2$ particle is defined by

$$S_j^{-k_j,+l_j} = |k_j\rangle\langle l_j| \text{ for } k_j, l_j = 0, 1 \text{ and } j = 1, 2, ..., n. \tag{3.22}$$

Obviously, there is the operator identity $\left(S_j^{-k_j,+l_j}\right)^+ = |l_j\rangle\langle k_j| = S_j^{-l_j,+k_j}$. There are the base-operator expansions for these base operators $\{S_j^{-k_j,+l_j}\}$ in terms of the Cartesian spin operator basis set $\{E_j, I_{jx}, I_{jy}, I_{jz}\}$, respectively,

$$\begin{cases} S_j^{-0,+0} = |0_j\rangle\langle 0_j| = \frac{1}{2}E_j + I_{jz}, \ S_j^{-1,+1} = |1_j\rangle\langle 1_j| = \frac{1}{2}E_j - I_{jz}, \\ S_j^{-0,+1} = |0_j\rangle\langle 1_j| = I_{jx} + iI_{jy}, \ S_j^{-1,+0} = |1_j\rangle\langle 0_j| = I_{jx} - iI_{jy} \end{cases} \tag{3.23}$$

They are really the base-operator expansions of (3.17) which are applied to the base operators $S_j^{-k_j,+l_j}$ of the $j-$th spin$-1/2$ particle for $k_j, l_j = 0, 1$.

It is known above that, by starting from the Cartesian spin operator basis set $\{E, I_x, I_y, I_z\}$ (or $\{I_{ko}, I_{kx}, I_{ky}, I_{kz}\}$) of an individual spin$-1/2$ particle, with the help of the tensor product method the Cartesian product operators (3.10) of the $n-$spin$-1/2$ system may be constructed by the tensor product of these $n$ spin operator basis sets $\{I_{ko}, I_{kx}, I_{ky}, I_{kz}\}$ of the $n$ individual spin$-1/2$ particles of the $n-spin-1/2$ system. In analogous way, by starting from the operator basis set $\{|0\rangle\langle 0|, |1\rangle\langle 1|, |0\rangle\langle 1|, |1\rangle\langle 0|\}$ (or $\{S_j^{-k_j,+l_j}\}$) of an individual spin$-1/2$ particle, with the help of the tensor product method the product operators $\{O_S\}$ of the $n-$spin$-1/2$ system also may be constructed by the tensor product of these $n$ operator basis sets $\{S_j^{-k_j,+l_j}\}$ of the $n$ individual spin$-1/2$ particles of the $n-spin-1/2$ system,

$$\begin{aligned}\{O_S\} &= \{S_1^{-k_1,+l_1} \bigotimes S_2^{-k_2,+l_2} \bigotimes ... \bigotimes S_n^{-k_n,+l_n}\} \\ &= \{S_1^{-k_1,+l_1} S_2^{-k_2,+l_2} ... S_n^{-k_n,+l_n}\} \qquad (3.24)\end{aligned}$$

There are only $4^n$ product operators $\{O_S\}$ which are mutually orthogonal. All these product operators $\{O_S\}$ also form a complete set of the orthogonal base operators of the $4^n-$dimensional multiple-quantum operator algebra space of the $n-$spin$-1/2$ system. The product operator basis set $\{O_S\}$ may be called the multiple-quantum-transition ($MQT$) product operator basis set or simply the $MQT$ product operator basis set so as to differentiate it from the previous Cartesian product operator basis set $\{B_s\}$ of (3.10). The MQT product-operator basis set $\{O_S\}$ is not Hermitian as a whole. However, any operator of the multiple-quantum operator algebra space of the $n-$spin$-1/2$ system can be expanded in terms of the MQT product operators $\{O_S\}$.

With the help of (3.22) any MQT product operator $O_S$ of (3.24) may be rewritten as

$$O_S = S_1^{-k_1,+l_1} S_2^{-k_2,+l_2} ... S_n^{-k_n,+l_n} = (|k_1\rangle |k_2\rangle ... |k_n\rangle)(\langle l_n| ... \langle l_2| \langle l_1|) \qquad (3.25)$$

where $k_j, l_j = 0, 1$ for $j = 1, 2, ..., n$. Let $|K\rangle = |k_1\rangle |k_2\rangle ... |k_n\rangle$ and $|L\rangle = |l_1\rangle |l_2\rangle ... |l_n\rangle$. Note that $|0_j\rangle$ and $|1_j\rangle$ are the two eigenbase vectors $|\alpha\rangle$ and $|\beta\rangle$ of the $z-$component spin operator $I_{jz}$ of the $j-$th spin$-1/2$ particle of the $n-spin-1/2$ system, respectively. Then it can prove that the tensor-product vectors $|K\rangle$ and $|L\rangle$ are just the orthonormal eigenbase vectors of the total spin operator $I_z = \sum_{j=1}^n I_{jz}$ of the $n-spin-1/2$ system (See the Section A). Therefore, $|K\rangle$ and $|L\rangle$ are the orthonormal tensor-product base vectors of the Hilbert space of the $n-spin-1/2$ system. By using the tensor-product base vectors $|K\rangle$ and $|L\rangle$ the MQT product operator $O_S$ of (3.25) may be simply written as $O_S = |K\rangle\langle L|$. All the $2^n$ orthonormal tensor-product base vectors $\{|K\rangle\}$ form a complete set of base vectors of the $2^n-$dimensional Hilbert space of the $n-spin-1/2$ system, while all the $4^n$ orthogonal MQT product operators $\{O_S\} = \{|K\rangle\langle L|\}$ form a complete set of the orthogonal base operators of the

corresponding $4^n-$dimensional multiple-quantum operator algebra space of the spin system.

With the help of the MQT product operators $\{O_S\}$ the symmetrical structures and properties of the multiple-quantum operator algebra space of the $n-$spin$-1/2$ system may be characterized more clearly and conveniently. In the MQT product operator basis set $\{O_S\}$ there are $2^n$ orthogonal diagonal operators:

$$\left\{\tilde{O}_S\right\} = \{S_1^{-k_1,+k_1}...S_2^{-k_j,+k_j}...S_n^{-k_n,+k_n}\} \text{ for } k_j = l_j = 0, 1 \text{ and } j = 1, 2, ..., n. \tag{3.26}$$

Note that these diagonal operators are still called the MQT diagonal product operators in unified form, although they are not the multiple-quantum transition operators. These $2^n$ diagonal product operators $\{\tilde{O}_S\}$ each belong to the LOMSO operator subspace of the multiple-quantum operator algebra space of the $n-$spin$-1/2$ system. Moreover, they also can form a complete set of base operators of the LOMSO operator subspace. Though the MQT product operator basis set $\{O_S\}$ is not Hermitian as a whole, the diagonal product operator basis subset $\{\tilde{O}_S\}$ is Hermitian as a whole. Obviously, for every one of these $2^n$ diagonal product operators $\{\tilde{O}_S\}$ there is the relation $p = \sum_{j=1}^{n}(-k_j + l_j) = 0$ owing to the relations $k_j = l_j$ for $j = 1, 2, ..., n$.

Beside these $2^n$ diagonal product operators $\{\tilde{O}_S\}$ the other $4^n - 2^n$ orthogonal MQT product operators $\{O_S\}$ of (3.24) are non-diagonal operators and they are really the multiple-quantum transition operators. Generally, it can prove (See the Section A) that any MQT product operator $O_S = S_1^{-k_1,+l_1}S_2^{-k_2,+l_2}...$ $S_n^{-k_n,+l_n}$ of (3.24) is a $p-$order quantum transition operator, here the quantum-transition order $p$ is determined by $p = \sum_{j=1}^{n}(-k_j + l_j)$ where $k_j \neq l_j$ for one individual spin$-1/2$ particle at least in these $n$ spin$-1/2$ particles of the $n-$spin$-1/2$ system. Therefore, it can be seen that any MQT product operator $O_S = S_1^{-k_1,+l_1}S_2^{-k_2,+l_2}...S_n^{-k_n,+l_n}$ is a zero-order quantum transition operator, if the quantum-transition order $p = \sum_{j=1}^{n}(-k_j + l_j) = 0$ where $k_j \neq l_j$ for one individual spin$-1/2$ particle at least in the $n-$spin$-1/2$ system. The number of the linearly-independent zero-order quantum transition operators of the $n-$spin$-1/2$ system is $N_{zq} = \begin{pmatrix} 2n \\ n \end{pmatrix} - 2^n$. Generally, any MQT product operator $O_S = S_1^{-k_1,+l_1}S_2^{-k_2,+l_2}...S_n^{-k_n,+l_n}$ is a $p-$order quantum transition operator with order $p \neq 0$, if the quantum-transition order $p$ is given by $p = \sum_{j=1}^{n}(-k_j + l_j) \neq 0$ where $k_j \neq l_j$ for one individual spin$-1/2$ particle at least in the $n-$spin$-1/2$ system. The highest quantum-transition order $|p|$ is given by $|p| = n$ for an $n-spin-1/2$ system. Among these $4^n - 2^n$ non-diagonal MQT product operators $\{O_S\}$ of (3.24) the numbers $N_{\pm p}$ $(p \geq 1)$ of the linearly-independent $\pm p-$order quantum transition operators of the $n-spin-1/2$ system are given by

$$N_p = N_{-p} = \begin{pmatrix} 2n \\ n-p \end{pmatrix}, \ p = 1, 2, ..., n$$

In fact it can be found [5] that $\frac{1}{2}N_{zq}$ and $N_p$ $(p \geq 1)$ are the (largest) numbers of the zero-order and the $p-$order quantum transitions of the $n - spin - 1/2$ system, respectively.

As shown by (3.25), every diagonal product operator $\tilde{O}_S$ of (3.26) may be rewritten as

$$\tilde{O}_S = S_1^{-k_1,+k_1} S_2^{-k_2,+k_2} ... S_n^{-k_n,+k_n} = |K\rangle\langle K| \quad (3.27)$$

These $2^n$ diagonal product operators $\{\tilde{O}_S\} = \{|K\rangle\langle K|\}$ each belong to the LOMSO operator subspace. Moreover, they can form a complete set of base operators of the LOMSO operator subspace. Obviously, every diagonal product operator $\tilde{O}_S = |K\rangle\langle K|$ of (3.27) is Hermitian. It can directly generate a unitary exponential operator:

$$R_K^Z(\theta) = \exp\left(-i\theta S_1^{-k_1,+k_1} S_2^{-k_2,+k_2} ... S_n^{-k_n,+k_n}\right) = \exp\left(-i\theta |K\rangle\langle K|\right) \quad (3.28)$$

Evidently $R_K^Z(\theta)$ is a unitary diagonal operator and belongs to the LOMSO operator subspace.

It can be found that every $p-$order quantum transition operator $O_S = |K\rangle\langle L|$ of (3.25) with $|K\rangle \neq |L\rangle$ is not Hermitian (even though the order $p = 0$). Therefore, it cannot be used directly to generate a unitary exponential operator, indicating that an elementary propagator which is unitary can not be generated directly by the non-diagonal MQT product operator $O_S$ of (3.24) (i.e., the non-Hermitian multiple-quantum transition operator $O_S$). The non-Hermitian multiple-quantum transition operators $\{|K\rangle\langle L|\}$ may be related to the Hermitian pseudospin operators $\{Q_x^{KL}, Q_y^{KL}, Q_z^{KL}\}$ in the Subsection 3.2.2 later, while the latter are Hermitian and can generate directly the elementary propagators. The MQT product operators $\{O_S\}$ are further investigated in the Section A below.

### 3.2.2. The Hermitian pseudospin operators

First of all, a general definition is given for the Hermitian pseudospin operators of a quantum system such as a general spin system. Theoretically the Hermitian pseudospin operators are suited to treat a multi-level and multi-spin system which may contain many spin$-I$ particles with spin quantum number $I \geq 1/2$. Suppose that $\{|K\rangle\}$ are the spin energy eigenbase vectors of the spin Hamiltonian $H_s$ of a general multi-level spin system and the energy eigenvalue equation is given by $H_s |K\rangle = E_K |K\rangle$ with spin energy eigenvalue $E_K$. All these energy eigenbase vectors $\{|K\rangle\}$ constitute a complete set of the orthonormal base vectors of the Hilbert space of the spin system. Let $|K\rangle$ and $|L\rangle$ be any pair of the spin energy eigenbase vectors of the spin system. Then corresponding to the two spin energy eigenbase vectors $|K\rangle$ and $|L\rangle$ one may define respectively the three Hermitian pseudospin operators $Q_x^{KL}$, $Q_y^{KL}$, and $Q_z^{KL}$ by (See, e.g., the definition (3.30) in [Ref[10]])

$$\begin{cases} Q_x^{KL} = \frac{1}{2}(|K\rangle\langle L| + |L\rangle\langle K|), \ Q_y^{KL} = \frac{1}{2i}(|K\rangle\langle L| - |L\rangle\langle K|), \\ Q_z^{KL} = \frac{1}{2}(|K\rangle\langle K| - |L\rangle\langle L|), \ E^{KL} = |K\rangle\langle K| + |L\rangle\langle L| \end{cases} \quad (3.29)$$

By comparing (3.29) with (3.16) it can be found that the Hermitian pseudospin operators $Q_x^{KL}, Q_y^{KL}, Q_z^{KL}$ of the pair of spin energy levels $|K\rangle$ and $|L\rangle$ corre-spond to the Cartesian spin operators $I_x, I_y, I_z$ of a single spin$-1/2$ particle, respectively, and moreover, the energy diagonal operator $E^{KL}$ corresponds to the unity operator $E$ of the single spin$-1/2$ particle. In fact, for a single spin$-1/2$ particle these pseudospin operators $Q_x^{KL}$, $Q_y^{KL}$, and $Q_z^{KL}$ are exactly equal to the spin operators $I_x$, $I_y$, and $I_z$, respectively, and the diagonal operator $E^{KL}$ is exactly equal to the unity operator $E$. The two spin energy levels $|K\rangle$ and $|L\rangle$ may form a two-level system $\{|K\rangle, |L\rangle\}$ of the spin system. These pseudospin operators $Q_x^{KL}$, $Q_y^{KL}$, and $Q_z^{KL}$ and their related two-level systems $\{|K\rangle, |L\rangle\}$ have been studied continually and have extensive applications in diverse research fields.[14]

These Hermitian pseudospin operators $Q_x^{KL}$, $Q_y^{KL}$, and $Q_z^{KL}$ obey the basic commutation rules of angular momentum operators: $\left[Q_\alpha^{KL}, Q_\beta^{KL}\right] = iQ_\gamma^{KL}$, where $\alpha, \beta, \gamma = x, y, z$ and cyclic permutations. Beside these basic commutation rules they also obey many other commutation rules (See Ref. [5] in detail, where $Q_\mu^{KL}$ with $\mu = x, y, z$ are replaced with the single-transition operators $I_\mu^{(KL)}$). These three pseudospin operators $\{Q_x^{KL}, Q_y^{KL}, Q_z^{KL}\}$ and the diagonal operator $E^{KL}$ in (3.29) together form a complete set of the orthogonal base operators of the Liouville operator algebra space of the two-level system $\{|K\rangle, |L\rangle\}$. This complete set $\{E^{KL}, Q_x^{KL}, Q_y^{KL}, Q_z^{KL}\}$ correspond one-to-one to the Cartesian spin operator basis set $\{E, I_x, I_y, I_z\}$ of an individual spin$-1/2$ particle. Therefore, the two-level system $\{|K\rangle, |L\rangle\}$ behaves like a single spin$-1/2$ particle. However, a two-level system $\{|K\rangle, |L\rangle\}$ is usually not a tensor-product subsystem such as an individual spin$-1/2$ particle in the composite $n - spin - 1/2$ system $(n > 1)$.

Consider an $n - spin - 1/2$ system. There are totally $N = 2^n$ spin energy levels and hence there are totally $\begin{pmatrix} N \\ 2 \end{pmatrix}$ different pairs $\{|K\rangle, |L\rangle\}$ of the spin energy levels for the $n$−spin−$1/2$ system. Then it can be found from (3.29) that there are totally $3\begin{pmatrix} N \\ 2 \end{pmatrix} = \frac{3}{2}N(N-1)$ Hermitian pseudospin operators $\{Q_x^{KL}, Q_y^{KL}, Q_z^{KL}\}$. The total number $\frac{3}{2}N(N-1)$ of the pseudospin operators is generally far larger than the dimensional size $N^2 = 4^n$ of the multiple-quantum operator algebra space of the $n - spin - 1/2$ system with the spin number $n > 1$. Therefore, these Hermitian and traceless pseudospin operators $\{Q_x^{KL}, Q_y^{KL}, Q_z^{KL}\}$ of (3.29) plus the total unity operator $E$ can form at least one complete set $\{E, Q_x^{KL}, Q_y^{KL}, Q_z^{KL}\}$ of the linearly-independent base operators of the multiple-quantum operator algebra space. This operator basis set $\{E, Q_x^{KL}, Q_y^{KL}, Q_z^{KL}\}$ evidently consists of only $4^n - 1$ linearly-independent

[14] In quantum mechanics, See, for example, E. Merzbacher, *Quantum mechanics*, Chapt. 13, Wiley, New York, 1970; In NMR spectroscopy, See, for example, R. R. Ernst, et al., *Principles of nuclear magnetic resonance in one and two dimensions*, Oxford University Press, Oxford, 1987; In optical resonance of atomic systems and the atomic spectroscopy, See, for example, L. Allen, et al., *optical resonabce and two-level atoms*, Dover, New York, 1987 (originally published by Wiley, 1975)

pseudospin operators $\{Q_x^{KL}, Q_y^{KL}, Q_z^{KL}\}$ and the unity operator $E$ and moreover, it is Hermitian as a whole.

Now any operator of the multiple-quantum operator algebra space, e.g., the spin Hamiltonian $H_s$ of the spin system under study, may be expanded in terms of the Hermitian pseudospin operators $\{E, Q_x^{KL}, Q_y^{KL}, Q_z^{KL}\}$ [4],

$$H_s = \alpha_0 E + \sum_{K,L} \left(\alpha_{KL}^x Q_x^{KL} + \alpha_{KL}^y Q_y^{KL} + \alpha_{KL}^z Q_z^{KL}\right), \tag{3.30}$$

where these expansional coefficients may be real due to that the spin Hamiltonian $H_s$ is Hermitian, and the coefficient $\alpha_0 = 0$ if the Hamiltonian $H_s$ is traceless. It is allowed to run the sum $\sum_{K,L}$ in (3.30) over all the Hermitian pseudospin operators $\{Q_x^{KL}, Q_y^{KL}, Q_z^{KL}\}$ of (3.29).

The Hermitian pseudospin operators $\{Q_x^{KL}, Q_y^{KL}, Q_z^{KL}\}$ may connect to the multiple-quantum transition operators and hence may be used to characterize the symmetrical structures and properties of the multiple-quantum operator algebra space. The first scheme is that the Hermitian pseudospin operators $\{Q_x^{KL}, Q_y^{KL}, Q_z^{KL}\}$ may be related to the MQT product operators $\{O_S\}$ of (3.25) which include non-Hermitian multiple-quantum transition operators in an $n-spin-1/2$ system. Here consider simply an $n-spin-1/2$ system which may be, for example, a non-interacting $n-$spin$-1/2$ system or a $zz-$interacting $n-spin-1/2$ system whose interaction terms are $I_{kz}I_{lz}$, etc. In the $n-spin-1/2$ system the spin Hamiltonian $H_s$ commutes with the total spin operator $I_z$. Hence both the spin Hamiltonian $H_s$ and the total spin operator $I_z$ have the common eigenbase vectors. Then in the spin system an energy quantum transition is just a multiple quantum transition, as shown in the Section A. The spin energy eigenbase vectors $\{|K\rangle\}$ (i.e., the common eigenbase vectors) then may be given by the tensor-product base vectors $\{|\Phi_l^z\rangle\}$ of (A1.31) in the Section A. Therefore, according to (A1.32) in the Section A there is $|K\rangle = |m_1\rangle |m_2\rangle ... |m_n\rangle$, where the eigenbase vector $|m_j\rangle$ (here $|m_j\rangle \overset{Def}{\equiv} |I_j, m_j\rangle$ with spin quantum number $I_j = 1/2$) and the eigenvalue $m_j = \pm 1/2$ satisfy the eigenvalue equation $I_{jz}|m_j\rangle = m_j |m_j\rangle$ of the spin operator $I_{jz}$ of the $j-$th spin$-1/2$ particle of the $n-spin-1/2$ system for $j = 1, 2, ..., n$. Let $|m_j\rangle = |1/2\rangle = |0_j\rangle$ and $|m_j\rangle = |-1/2\rangle = |1_j\rangle$ for $j = 1, 2, ..., n$. Then the spin energy eigenbase vectors $|K\rangle$ and $|L\rangle$ may be respectively written as $|K\rangle = |k_1\rangle |k_2\rangle ... |k_n\rangle$ and $|L\rangle = |l_1\rangle |l_2\rangle ... |l_n\rangle$, where $k_j, l_j = 0, 1$ and $j = 1, 2, ..., n$. It can be found that the tensor-product base vectors $|K\rangle$ and $|L\rangle$ are also the orthonormal eigenbase vectors of the total spin operator $I_z = \sum_{j=1}^n I_{jz}$ of the $n-spin-1/2$ system. Therefore, the base operators $|K\rangle\langle L|$ and $|L\rangle\langle K|$ in the pseudospin operators $\{Q_x^{KL}, Q_y^{KL}, Q_z^{KL}\}$ of (3.29) are also the MQT product operators $\{O_S\}$ of (3.25) and may be respectively written as

$$|K\rangle\langle L| = S_1^{-k_1,+l_1} S_2^{-k_2,+l_2} ... S_n^{-k_n,+l_n} \tag{3.31a}$$

$$|L\rangle\langle K| = S_1^{-l_1,+k_1} S_2^{-l_2,+k_2} ... S_n^{-l_n,+k_n} \tag{3.31b}$$

where according to (3.22) the operators $S_j^{-k_j,+l_j} = |k_j\rangle\langle l_j|$ for $k_j, l_j = 0, 1$ and $j = 1, 2, ..., n$. It can be found from (3.31) that there is the relation $|K\rangle\langle L| = (|L\rangle\langle K|)^+$ due to the operator identities $\left(S_j^{-k_j,+l_j}\right)^+ = S_j^{-l_j,+k_j}$ for $j = 1, 2, ..., n$. Let $p = \sum_{j=1}^n (-k_j + l_j)$. Then $-p = \sum_{j=1}^n (-l_j + k_j)$. Therefore, it can be found from (3.31) that the base operator $|K\rangle\langle L|$ is a $p-$order quantum transition operator $Q_p$, while $|L\rangle\langle K|$ is a $-p-$order quantum transition operator $Q_{-p}$. Both the base operators $|K\rangle\langle L|$ and $|L\rangle\langle K|$ of (3.31) evidently are not Hermitian, if $|K\rangle \neq |L\rangle$. As shown in the Section A, a Hermitian $|p|-$quantum transition operator $Q_{|p|}$ may be generated by $Q_{|p|} = \frac{1}{2}\left(Q_p + Q_p^+\right)$ or by $Q_{|p|} = \frac{1}{2i}\left(Q_p - Q_p^+\right)$. Now one has $Q_p = |K\rangle\langle L|$. Then $Q_p^+ = (|K\rangle\langle L|)^+ = |L\rangle\langle K| = Q_{-p}$. Therefore, it can be found that

$$Q_{|p|} = \frac{1}{2}\left(Q_p + Q_{-p}\right) = \frac{1}{2}(|K\rangle\langle L| + |L\rangle\langle K|) = Q_x^{KL} \tag{3.32a}$$

$$Q_{|p|} = \frac{1}{2i}\left(Q_p - Q_{-p}\right) = \frac{1}{2i}(|K\rangle\langle L| - |L\rangle\langle K|) = Q_y^{KL} \tag{3.32b}$$

These formulae show clearly that the Hermitian pseudospin operators $Q_x^{KL}$ and $Q_y^{KL}$ of (3.29) are really the Hermitian $|p|-$quantum transition operators $Q_{|p|}$ and belong to the multiple-quantum operator algebra space of the $n-spin-1/2$ system. In analogous way it can be shown that the base operators $|K\rangle\langle K|$ and $|L\rangle\langle L|$ in the Hermitian pseudospin operators $Q_z^{KL}$ and the energy diagonal operator $E^{KL}$ in (3.29) are also the diagonal MQT product operators $\{\tilde{O}_S\}$ of (3.26) and therefore, both the operators $Q_z^{KL}$ and $E^{KL}$ belong to the LOMSO operator subspace of the multiple-quantum operator algebra space.

There is the second scheme which connects the Hermitian pseudospin operators $\{Q_x^{KL}, Q_y^{KL}, Q_z^{KL}\}$ to the multiple-quantum transition operators and the multiple-quantum operator algebra space. This scheme is described below. As pointed out previously, a two-level system $\{|K\rangle, |L\rangle\}$ is usually not a tensor-product subsystem such as any single spin$-1/2$ particle in an $n - spin - 1/2$ system. If a two-level system $\{|K\rangle, |L\rangle\}$ is able to act as a tensor-product subsystem like an individual spin$-1/2$ particle, then thing becomes simple. Without lost generality, here consider a composite quantum spin system, i.e., an $n - spin - I$ system which consists of $n$ $spin - I$ particles with spin quantum number $I > 1/2$. A single spin$-I$ particle in external magnetic field owns $2I+1$ discrete spin energy levels with $2I+1 > 2$ [2]. Suppose that $|r\rangle$ and $|t\rangle$ are a pair of the spin energy eigenbase vectors (i.e., a pair of discrete spin energy levels) of the single spin$-I$ particle. Corresponding to the pair of the energy eigenbase vectors $|r\rangle$ and $|t\rangle$ the three Hermitian pseudospin operators $Q_x^{rt}$, $Q_y^{rt}$, and $Q_z^{rt}$ and the energy diagonal operator $E^{rt}$ are still defined by (3.29). Moreover, the pair of the energy eigenbase vectors $|r\rangle$ and $|t\rangle$ also form a two-level system $\{|r\rangle, |t\rangle\}$ of the single spin$-I$ particle. Evidently the two-level system $\{|r\rangle, |t\rangle\}$ and its related pseudospin operators $Q_x^{rt}$, $Q_y^{rt}$, and $Q_z^{rt}$ and diagonal operator $E^{rt}$ all are related only to the single spin$-I$ particle alone. The pseudospin operator basis set $\{E^{rt}, Q_x^{rt}, Q_y^{rt}, Q_z^{rt}\}$ of the two-level system $\{|r\rangle, |t\rangle\}$ of the single spin$-I$ particle is isomorphic to the Cartesian spin operator basis set $\{E, I_x, I_y, I_z\}$ of a single spin$-1/2$ particle. Therefore, the two-level system $\{|r\rangle, |t\rangle\}$ of the single spin$-I$ particle behaves like a single spin$-1/2$ particle and may be treated as a single pseudospin$-1/2$ $'$particle$'$. It may be said that the single pseudospin$-1/2$ $'$particle$'$ is created from the single spin$-I$ particle alone.

The composite $n-spin-I$ system consists of $n$ $spin-I$ particles with spin quantum number $I > 1/2$. Now each individual spin$-I$ particle of the composite $n-spin-I$ system may create one pseudospin$-1/2$ $'$particle$'$ at least. Note that each individual spin$-I$ particle is a tensor-product subsystem of the composite $n-spin-I$ system. Then the whole composite $n-spin-I$ system may create one composite $n-$pseudospin$-1/2$ system at least. Here the composite $n-$pseudospin$-1/2$ system evidently consists of the $n$ pseudospin$-1/2$ $'$particles$'$, each one of which is created by an individual spin$-I$ particle of the composite $n-spin-I$ system. Therefore, any single pseudospin$-1/2$ $'$particle$'$, i.e., the two-level system $\{|r\rangle, |t\rangle\}$ of any individual spin$-I$ particle of the composite $n-spin-I$ system, is a tensor-product subsystem of the composite $n-$pseudospin$-1/2$ system. Here the pseudospin operator basis set $\{E^{rt}, Q_x^{rt}, Q_y^{rt}, Q_z^{rt}\}$ is a complete set of the orthogonal base operators of the Liouville operator algebra space of the single pseudospin$-1/2$ $'$particle$'$ associated with the two-level system $\{|r\rangle, |t\rangle\}$.

The concept of the pseudospin$-1/2$ $'$particle$'$ (i.e., a two-level system) is fundamental and general in quantum mechanics (See References in the footnote on page 50). It can be applied to not only a single spin$-I$ particle with spin quantum number $I > 1/2$ but also a single multi-level atomic particle and so on. A pseudospin$-1/2$ $'$particle$'$ may act as a quantum bit (i.e., qubit [Ref[9]][15]) in quantum computing and quantum simulating.

Theoretically a composite $n-$pseudospin$-1/2$ system may be treated just like an $n-spin-1/2$ system. It may be created from the composite $n-spin-I$ system with spin quantum number $I > 1/2$ or any other composite quantum systems (See, e.g., Ref.[8], [Ref[8]]). Any linear operators of the $n-$pseudospin$-1/2$ system may be defined on the Hilbert space of the $n-$pseudospin$-1/2$ system (i.e., an $N-$dimensional vector space with dimension $N = 2^n$). The total $z-$component pseudospin operator $I_z$ of the $n-$pseudospin$-1/2$ system is defined by $I_z = \sum_j^n Q_{jz}^{rt}$, where $Q_{jz}^{rt}$ is the $z-$component pseudospin operator (i.e., $Q_z^{KL}$ in (3.29)) of the $j-$th pseudospin$-1/2$ $'$particle$'$ of the $n-$pseudospin$-1/2$ system. The multiple-quantum-transition (MQT) operators of the $n-$pseudospin $-1/2$ system then may be defined as well in accordance with the formal definition (A1.5) or (A1.6) in the Section A. The symmetrical structures and properties of the multiple-quantum operator algebra space of the $n-$pseudospin$-1/2$ system therefore may be characterized by using these MQT operators.

The symmetrical structures and properties of the multiple-quantum operator algebra space of a general spin system also may be specified by the Hermitian pseudospin operators of (3.29). For simplicity, below consider an $n-spin-1/2$ system in which the spin Hamiltonian $H_s$ commutes with the total spin operator $I_z$. The $4^n-$dimensional multiple-quantum operator algebra space of the $n-spin-1/2$ system may be spanned by one pseudospin operator basis set $\{E, Q_x^{KL}, Q_y^{KL}, Q_z^{KL}\}$ of (3.29). In the pseudospin operator basis set there are only the $2^n-1$ linearly-independent diagonal pseudospin operators $\{Q_z^{KL}\}$ and the total unity operator $E$. All these $2^n$ linearly-independent diagonal base operators $\{E, Q_z^{KL}\}$ belong to the LOMSO operator subspace of the multiple-quantum operator algebra space of the $n-spin-1/2$ system and moreover, they span the LOMSO operator subspace. The total number of the diagonal pseudospin operators $\{Q_z^{KL}\}$ of (3.29) is generally far larger than the dimensional size $2^n$ of the LOMSO operator subspace. It is known from (3.32) that the non-diagonal pseudospin operators $Q_x^{KL}$ and $Q_y^{KL}$ are the Hermitian $|p|-$quantum transition operators, i.e., $Q_x^{KL}=\frac{1}{2}(Q_p+Q_{-p})$ and $Q_y^{KL}=\frac{1}{2i}(Q_p-Q_{-p})$. Here the $p-$order quantum transition operators $Q_p$ may be given by the MQT product operators $\{O_s\}=\{|K\rangle\langle L|\}$ of (3.31). Note that if $Q_p$ is a zero-order or an even-order quantum transition operator, then $Q_{-p}$ is still a zero-order or an even-order quantum transition operator. Therefore, the Hermitian $|p|-$quantum transition operators $Q_x^{KL}$ and $Q_y^{KL}$ are zero-order quantum transition operators, if $Q_p$ is a zero-order quantum transition operator; and $Q_x^{KL}$ and $Q_y^{KL}$ are even-order quantum transition operators, if $Q_p$ is an even-order quantum transition operator. Therefore, the multiple-quantum operator subspaces including the zero-quantum operator subspace and the even-order quantum operator subspace of the multiple-quantum operator algebra space may be characterized by the Hermitian $|p|-$quantum transition operators $\{Q_x^{KL}, Q_y^{KL}\}$. The related work may further refer to the Section A and the Part II in this paper.

[15] [Ref[9]] B. Schumacher, *Quantum coding*, Phys. Rev. A 51, 2738 (1995)

Every one of the Hermitian pseudospin operators $\{Q_x^{KL}, Q_y^{KL}, Q_z^{KL}\}$ of a quantum spin system may generate a unitary exponential operator (See, e.g., Refs [5, 4]):

$$R_\mu^{KL}(\theta)=\exp\left(-i\theta Q_\mu^{KL}\right),\ \mu=x,y,z \tag{3.33}$$

where $R_\mu^{KL}(\theta)$ is called the pseudospin rotation operator. The pseudospin rotation operator $R_\mu^{KL}(\theta)$ is also an elementary propagator due to that the Hermitian pseudospin operator $Q_\mu^{KL}$ is a base operator. Suppose that the Hermitian pseudospin operator $Q_\mu^{KL}$ that generates the pseudospin rotation operator $R_\mu^{KL}(\theta)$ is any member of the pseudospin operator basis set $\{E, Q_x^{KL}, Q_y^{KL}, Q_z^{KL}\}$ of the $4^n-$dimensional multiple-quantum operator algebra space of the $n-spin-1/2$ system. Then in addition to the identity operator $R_o(\theta)=\exp(-i\theta)E$ with the phase factor $\exp(-i\theta)$ there are only the $4^n-1$ elementary propagators $\{R_\mu^{KL}(\theta)\}$ of (3.33) which are generated by the $4^n-1$ linearly-independent pseudospin operators $\{Q_x^{KL}, Q_y^{KL}, Q_z^{KL}\}$, respectively. As shown previously, the total number of these Hermitian pseudospin operators $\{Q_x^{KL}, Q_y^{KL}, Q_z^{KL}\}$ of (3.29) of the $n-spin-1/2$ system is generally far larger than the number $4^n$ of the base operators of the pseudospin operator basis set $\{E, Q_x^{KL}, Q_y^{KL}, Q_z^{KL}\}$. Here any pseudospin rotation operator $R_\mu^{KL}(\theta)$ that is generated by one of these Hermitian pseudospin operators $\{Q_x^{KL}, Q_y^{KL}, Q_z^{KL}\}$ of (3.29) of the $n - spin - 1/2$ system may be considered as a candidate of the elementary propagators.

Now a comparison is made between the pseudospin rotation operators $R_\mu^{KL}(\theta)$ of (3.33) and the spin rotation operators $R_{k\mu}(\theta) = \exp(-i\theta I_{k\mu})$ of (3.6b) with $\mu = x, y, z$. Both the unitary operators $R_\mu^{KL}(\theta)$ and $R_{k\mu}(\theta)$ are able to act as the elementary propagators. The spin-selective excitation pulse may be represented by the spin rotation operator $R_{k\mu}(\theta)$ which is selectively applied to the $k-$th spin$-1/2$ particle of the $n - spin - 1/2$ system. The spectral-line-selective excitation pulse [11] may be represented by the pseudospin rotation operator $R_\mu^{KL}(\theta)$. The spectral-line-selective energy quantum transition between two selected spin energy levels $|K\rangle$ and $|L\rangle$ of the $n - spin - 1/2$ system may be directly induced by the pseudospin rotation operator $R_\mu^{KL}(\theta)$ $(\mu = x, y)$. Both the pseudospin rotation operators $R_\mu^{KL}(\theta)$ and the spin rotation operators $R_{k\mu}(\theta)$ may be directly realized in nuclear magnetic resonance selective-pulse experiments (See, e.g., [Ref[4]][16] and See, generally, Ref.[11]). They can act as the basic building blocks in quantum simulating and quantum computing.

The unitary transformation that any unitary pseudospin rotation operator $R_\lambda^{kl}(\theta)$ of (3.33) with $K = k$ and $L = l$ acts on any operator (e.g., the spin Hamiltonian $H_s$ of (3.30)) of the multiple-quantum operator algebra space may be expressed as

$$\left(R_\lambda^{kl}(-\theta)\right)^+ H_s \left(R_\lambda^{kl}(-\theta)\right) = R_\lambda^{kl}(\theta) H_s \left(R_\lambda^{kl}(\theta)\right)^+$$

$$= \alpha_0 E + \sum_{K,L}(\alpha_{KL}^x R_\lambda^{kl}(\theta) Q_x^{KL} \left(R_\lambda^{kl}(\theta)\right)^+$$

$$+\alpha_{KL}^y R_\lambda^{kl}(\theta) Q_y^{KL} \left(R_\lambda^{kl}(\theta)\right)^+ + \alpha_{KL}^z R_\lambda^{kl}(\theta) Q_z^{KL} \left(R_\lambda^{kl}(\theta)\right)^+) \qquad (3.34)$$

where the operator identities $\left(R_\lambda^{kl}(-\theta)\right)^+ = R_\lambda^{kl}(\theta) = \exp\left(-i\theta Q_\lambda^{kl}\right)$ are obtained from (3.33) and the Hermitian pseudospin operator $Q_\lambda^{kl}$ is given by (3.29) with $K = k$ and $L = l$ for $\lambda = x, y, z$. The basic unitary transformations in the multiple-quantum operator algebra space that the unitary pseudospin rotation operator (i.e., an elementary propagator) $R_\lambda^{kl}(\theta)$ with $\lambda = x, y, z$ acts on any pseudospin operators $Q_\mu^{KL}$ with $\mu = x, y, z$ may be written as

$$\left(R_\lambda^{kl}(-\theta)\right)^+ Q_\mu^{KL} \left(R_\lambda^{kl}(-\theta)\right) = R_\lambda^{kl}(\theta) Q_\mu^{KL} \left(R_\lambda^{kl}(\theta)\right)^+ \qquad (3.35)$$

Once the basic unitary transformations of (3.35) are exactly obtained, the unitary transformation of (3.34) can be exactly calculated. The basic unitary transformations of (3.35) can be exactly determined (See, e.g., Ref.[5]).

There are a number of methods to calculate exactly the basic unitary transformations of (3.35) in the multiple-quantum operator algebra space. Below a general method is simply introduced that can make transition from the vector-basis unitary transformations in the Hilbert space to the operator-basis unitary

[16] [Ref[4]] X. Miao and R. Freeman, *Spin-echo modulation experiments with soft Gaussian pulses*, J. Magn. Reson. A 119, 90 (1996)

transformations in the multiple-quantum operator algebra space. First of all, the pseudospin rotation operator $R_{\lambda}^{KL}(\theta)$ $(\lambda = x, y, z)$ of (3.33) can be generally written in the operator-expansion form (See, e.g., Ref.[7])

$$R_{\lambda}^{KL}(\theta) = E + \left(-1 + \cos\frac{1}{2}\theta\right)(|K\rangle\langle K| + |L\rangle\langle L|) - i2Q_{\lambda}^{KL}\sin\frac{1}{2}\theta \tag{3.36}$$

With the help of this operator expansion it is easy to calculate exactly the basic unitary transformations of (3.35). This method involves only the multiple-quantum operator algebra space. Note that the pseudospin operators $\{Q_{\lambda}^{KL}\}$ of (3.29) of a spin system under study are defined on the basis of the energy eigenbase vectors $|K\rangle$ and $|L\rangle$ of the Hilbert space of the spin system. Then a general method to calculate exactly the basic unitary transformations of (3.35) may be that the basic unitary transformations for the pseudospin rotation operators $R_{\lambda}^{kl}(\theta) = \exp\left(-i\theta Q_{\lambda}^{kl}\right)$ $(\lambda = x, y, z)$ in the Hilbert space of the spin system may be employed to exactly calculate the basic unitary transformations of (3.35) in the multiple-quantum operator algebra space. These basic unitary transformations for the pseudospin rotation operators $R_{\lambda}^{kl}(\theta)$ in the Hilbert space can be easily calculated exactly with the aid of the operator expansion (3.36) of the pseudospin rotation operator $R_{\lambda}^{kl}(\theta)$. Consequently they are given respectively by

$$\exp\left(-i\theta Q_{x}^{kl}\right)|k\rangle = |k\rangle\cos\frac{1}{2}\theta - i\,|l\rangle\sin\frac{1}{2}\theta \tag{3.37a}$$

$$\exp\left(-i\theta Q_{x}^{kl}\right)|l\rangle = |l\rangle\cos\frac{1}{2}\theta - i\,|k\rangle\sin\frac{1}{2}\theta \tag{3.37b}$$

$$\exp\left(-i\theta Q_{y}^{kl}\right)|k\rangle = |k\rangle\cos\frac{1}{2}\theta + |l\rangle\sin\frac{1}{2}\theta \tag{3.37c}$$

$$\exp\left(-i\theta Q_{y}^{kl}\right)|l\rangle = |l\rangle\cos\frac{1}{2}\theta - |k\rangle\sin\frac{1}{2}\theta \tag{3.37d}$$

$$\exp\left(-i\theta Q_{z}^{kl}\right)|k\rangle = \exp(-i\theta/2)\,|k\rangle\,,\ \ \exp\left(-i\theta Q_{z}^{kl}\right)|l\rangle = \exp(i\theta/2)\,|l\rangle \tag{3.37e}$$

and if $|j\rangle \neq |k\rangle, |l\rangle$, then there are the identical transformations:

$$\exp\left(-i\theta Q_{\lambda}^{kl}\right)|j\rangle = |j\rangle \text{ for } |j\rangle \neq |k\rangle, |l\rangle \text{ and } \lambda = x, y, z \tag{3.37f}$$

Here $|k\rangle$, $|l\rangle$, and $|j\rangle$ are any spin energy eigenbase vectors of the Hilbert space of the spin system. Now with the help of these basic unitary transformations of (3.37) one can exactly calculate any basic unitary transformations of (3.35) in the multiple-quantum operator algebra space. As an example, consider the basic unitary transformation $\left(R_{x}^{kl}(-\theta)\right)^{+} Q_{x}^{lj}\left(R_{x}^{kl}(-\theta)\right)$ with $|j\rangle \neq |k\rangle, |l\rangle$ in the multiple-quantum operator algebra space. As shown in (3.29), the pseudospin operator $Q_{x}^{lj} = \frac{1}{2}(|l\rangle\langle j| + |j\rangle\langle l|)$. Then the basic unitary transformation $\left(R_{x}^{kl}(-\theta)\right)^{+} Q_{x}^{lj}\left(R_{x}^{kl}(-\theta)\right)$ is given by

$$\left(R_{x}^{kl}(-\theta)\right)^{+} Q_{x}^{lj}\left(R_{x}^{kl}(-\theta)\right) = R_{x}^{kl}(\theta)\,Q_{x}^{lj}\left(R_{x}^{kl}(\theta)\right)^{+}$$

$$= \frac{1}{2}\left(R_x^{kl}(\theta)\,|l\rangle\langle j| + |j\rangle\langle l|\left(R_x^{kl}(\theta)\right)^+\right)$$
$$= Q_x^{lj}\cos\frac{1}{2}\theta + Q_y^{kj}\sin\frac{1}{2}\theta$$

where the second equality is obtained due to the identical transformation (3.37f) and the last equality holds due to the basic unitary transformation (3.37b) in the Hilbert space.

### 3.2.3. The single-matrix-element base operators

In the linear operator space with dimension $N \times N$ all linear operators on an $N-$dimensional Hilbert space correspond one-to-one to all $N \times N$ matrices. Then one may choose the operator basis set for the linear operator space via the $N \times N$ matrices. This method may be used to choose the operator basis set for the multiple-quantum operator algebra space of a general spin system. As an important application, here this method is used to choose the simplest operator basis set in theory which is comprised of all the single-matrix-element base operators. The single-matrix-element base operators usually may be defined in the energy representation. Of course, they also can be defined in any other representation than the energy representation. Consider a general spin system whose Hilbert space is $N-$dimensional. Then all the $N$ spin energy eigenbase vectors $\{|k\rangle\}$ of the spin system form a complete set of the orthonormal base vectors of the $N-$dimensional Hilbert space of the spin system. Suppose that $|k\rangle$ and $|l\rangle$ are any spin energy eigenbase vectors of the $N-$dimensional Hilbert space. Corresponding to the orthonormal energy eigenbase vectors $|k\rangle$ and $|l\rangle$ one may define the single-matrix-element (SME) base operators $\{B_{kl}\}$ by

$$B_{kl} = |k\rangle\langle l| \text{ for } k \neq l,\ B_{kk} = |k\rangle\langle k| \text{ for } k = l \tag{3.38}$$

where $k, l = 0, 1, ..., N-1$. It can be found from the definition (3.38) that there are only $N^2$ orthogonal SME base operators $\{B_{kl}\}$. Then all these $N^2$ orthogonal SME base operators $\{B_{kl}\}$ with $0 \leq k, l \leq N-1$ form a complete set of base operators of the $N^2-$dimensional multiple-quantum operator algebra space which corresponds to the $N-$dimensional Hilbert space.

The matrix element $(B_{kl})_{ij}$ with row $i$ and column $j$ of the representation matrix $\left((B_{kl})_{ij}\right)$ of the SME base operator $B_{kl}$ of (3.38) in the energy representation is given by

$$(B_{kl})_{ij} = \langle i|B_{kl}|j\rangle = \langle i|k\rangle\langle l|j\rangle = \delta_{ik}\delta_{lj} \tag{3.39}$$

where the orthonormal base vectors $|i\rangle$ and $|j\rangle$ belong to the complete set $\{|k\rangle\}$ of the energy eigenbase vectors of the $N-$dimensional Hilbert space. It can be found from (3.39) that, in the representation matrix $\left((B_{kl})_{ij}\right)$ of the SME base operator $B_{kl}$, only one matrix element $(B_{kl})_{ij} = 1$ (one) for the row $i = k$ and the column $j = l$ and any other matrix elements $(B_{kl})_{ij} = 0$ (zero) for the row $i \neq k$ and/or the column $j \neq l$. Any matrix $(A_{ij})$ is called a single-element matrix, if all the elements of the matrix $(A_{ij})$ are zero except one element equal to one. Then it can be found from (3.39) that any base operator $B_{kl}$ of (3.38) owns an $N \times N$ single-element representation matrix $\left((B_{kl})_{ij}\right)$ in the energy representation. Therefore, the base operators $\{B_{kl}\}$ are called the single-matrix-element (SME) base operators.

The SME base operators $\{B_{kl}\}$ are simplest in theory. They are not always Hermitian, but they are quite useful in the theoretical treatments. All these orthogonal SME base operators of (3.38) form an operator basis set $\{B_{kl}\}$ of the multiple-quantum operator algebra space of the spin system under study. Therefore, any operator $A$ of the multiple-quantum operator algebra space may be expanded in terms of the complete set $\{B_{kl}\}$ of the orthogonal SME base operators,

$$A = \sum_{k,l=0}^{N-1} a_{kl} B_{kl}, \tag{3.40}$$

where $a_{kl}$ is an expansional coefficient and may be real or complex. The representation matrix $(A_{kl})$ of the operator $A$ in the energy representation is given by $(A_{kl}) = (a_{kl})$ , indicating that the expansional coefficient $a_{kl}$ in (3.40) is just the element $A_{kl}$ of the representation matrix $(A_{kl})$ of the operator $A$. The number of the orthogonal SME base operators of the operator basis set $\{B_{kl}\}$ is equal to the dimensional size $N^2$ of the multiple-quantum operator algebra space. Moreover, the operator basis set $\{B_{kl}\}$ exactly contains the $N^2$ orthogonal SME base operators $\{B_{kl}\}$.

As a typical example, the MQT product operators $\{O_S\} = \{|K\rangle\langle L|\}$ of (3.25) each are the SME base operator $B_{kl}$ of (3.38), if the base vectors $\{|K\rangle\}$ are the spin energy eigenbase vectors of the $N-$dimensional Hilbert space $(N = 2^n)$ of an $n - spin - 1/2$ system. Moreover, the non-diagonal MQT product operators $\{O_S\}$ are the multiple-quantum transition operators.

# 4. The Pseudo-Diagonal Hermitian (PDH) operators and their generated unitary operators

The research work in this Section is one main contribution to this paper. As shown in the previous Sections, on the one hand, the fundamental quantum-computing-speedup resources of a quantum system may exist in these different kinds of basic quantum spaces: the multiple-quantum operator algebra space, the density operator space, and the Hilbert space; on the other hand, according to the quantum-computing speedup theory the multiple-quantum operator algebra space must be considered as the central place where the fundamental quantum-computing-speedup resources are exploited to speed up essentially quantum computing and quantum simulating. Recognize that the multiple-quantum operator algebra space is the central place. Then the fundamental quantum-computing-speedup resources which are original from the symmetrical structures and properties of the corresponding Hilbert space must be considered explicitly in the multiple-quantum operator algebra space. This is an important problem that need to be solved in this Section and the next Sections.

In quantum mechanics any linear operators are defined on the Hilbert space of a quantum system. Then the linear operator space of a quantum system may connect to the corresponding Hilbert space of the quantum system on which any linear operators of the linear operator space are defined. Consequently the symmetrical structures and properties of the Hilbert space may connect to the counterpart of the corresponding multiple-quantum operator algebra space.

The pseudo-diagonal Hermitian operators are the specific Hermitian operators of the multiple-quantum operator algebra space. They are perhaps one kind of the simplest non-diagonal Hermitian operators through which a connection in the aspect of symmetrical structures and properties may be easily set up between the Hilbert space and its corresponding multiple-quantum operator algebra space. With the help of the pseudo-diagonal Hermitian operators the symmetric structures and properties which are original from the corresponding Hilbert space may be easily considered explicitly in the multiple-quantum operator algebra space. The pseudo-diagonal Hermitian operators have important applications in the subspace-selective unitary manipulation in the next Sections. In this Section a general pseudo-diagonal Hermitian operator of quantum spin system is described and investigated in detail.

## 4.1. Definition of a general pseudo-diagonal Hermitian operator

Below a general definition is first given for any pseudo-diagonal Hermitian operators. According to the base-operator expansion principle any spin Hamiltonian $H_s$ of an $n-spin-1/2$ system may be expanded in terms of a complete set of base operators of the multiple-quantum operator algebra space of the $n-spin-1/2$ system. Here the complete set of base operators may be chosen as the Hermitian Cartesian product operator basis set $\{B_s\}$ of (3.10) or the Hermitian pseudospin operator basis set $\{E, Q_x^{KL}, Q_y^{KL}, Q_z^{KL}\}$ of (3.29). It also may be chosen as the MQT product operator basis set $\{O_S\} = \{|R\rangle \langle T|\}$ of (3.25). As far as the representation matrices are concerned, these pseudospin operators $\{Q_x^{KL}, Q_y^{KL}, Q_z^{KL}\}$ and the MQT product operators $\{|R\rangle \langle T|\}$ are far simpler than the Cartesian product operators $\{B_s\}$. Especially the MQT product operators $\{|R\rangle \langle T|\}$ each may have a single-element representation matrix and hence they are simplest in theory. Therefore, the two operator basis sets $\{|R\rangle \langle T|\}$ and $\{E, Q_x^{KL}, Q_y^{KL}, Q_z^{KL}\}$ each are more suitable to act as a stepping stone to set up connection in the aspect of symmetrical structures and properties between the Hilbert space of a quantum spin system and its corresponding multiple-quantum operator algebra space.

It is known in the Subsection 3.2.1 that all these $4^n$ MQT product operators $\{O_S\} = \{|R\rangle \langle T|\}$ of (3.25) constitute a complete set of the orthogonal base operators of the multiple-quantum operator algebra space of the $n-$spin$-1/2$ system. Any spin Hamiltonian $H_s$ of the $n-spin-1/2$ system then may be expanded in terms of the MQT product operators $\{|R\rangle \langle T|\}$,

$$H_s = \sum_{R,T} \beta_{RT} |R\rangle \langle T|, \tag{4.1}$$

where all the $2^n$ orthonormal tensor-product base vectors $\{|R\rangle\}$ form a complete set of base vectors of the $2^n-$dimensional Hilbert space of the $n-spin-1/2$ system. Note that the MQT product-operator basis set $\{|R\rangle\langle T|\}$ is not Hermitian as a whole. Then in the base-operator expansion (4.1) any expansional coefficient $\beta_{RT}$ may be real or complex. However, because the spin Hamiltonian $H_s$ is Hermitian, these expansional coefficients $\{\beta_{RT}\}$ in (4.1) must ensure that the base-operator expansion (4.1) is Hermitian so that the time-evolutional propagator $U_s(t) = \exp(-iH_s t/\hbar)$ is unitary that is generated by the spin Hamiltonian $H_s$ with the base-operator expansion (4.1). Actually the spin Hamiltonian $H_s$ of (4.1) also may be expanded in terms of the Hermitian pseudospin operator basis set $\{E, Q_x^{KL}, Q_y^{KL}, Q_z^{KL}\}$ of (3.29) or the Cartesian product operator basis set $\{B_s\}$ of (3.10). For the pseudospin operator basis set $\{E, Q_x^{KL}, Q_y^{KL}, Q_z^{KL}\}$ this base-operator expansion is just given by (3.30). For the Cartesian product operator basis set $\{B_s\}$ it is still given formally by (3.1) with the base operators $\{B_s\}$ given by (3.10). These three different base-operator expansions for the same spin Hamiltonian $H_s$ are equivalent to each other theoretically.

Suppose that any expansional coefficient $\beta_{RT}$ in the base-operator expansion (4.1) of the spin Hamiltonian $H_s$ is taken as $\beta_{RT} = \epsilon_s \alpha_R \alpha_T^*$ for $R, T = 0, 1, 2, ..., N-1$, where the sign $\epsilon_s = \pm 1$, the coefficient $\alpha_R$ may be real or complex, and $N = 2^n$ is the dimensional size of the Hilbert space of the $n-spin-1/2$ system. Then by substituting these expansional coefficients $\{\beta_{RT}\} = \{\epsilon_s \alpha_R \alpha_T^*\}$ into (4.1) it can be found that the spin Hamiltonian $H_s$ of (4.1) may be reduced to the specific form

$$H_s = \epsilon_s \left( \sum_{R=0}^{N-1} \alpha_R |R\rangle \right) \left( \sum_{T=0}^{N-1} \alpha_T^* \langle T| \right) \tag{4.2}$$

where for convenience all the $N = 2^n$ orthonormal tensor-product base vectors of the Hilbert space of the $n - spin - 1/2$ system are denoted by $\{|R\rangle\}$ with $R = 0, 1, ..., N-1$. It is easy to prove that the spin Hamiltonian $H_s$ of (4.2) is always Hermitian for any expansional coefficients $\{\alpha_R\}$ and the sign $\epsilon_s = \pm 1$. Though the spin Hamiltonian $H_s$ of (4.2) is the special one of the general spin Hamiltonian $H_s$ of (4.1), it is more suitable for the present purpose.

The Hermitian spin Hamiltonian (4.2) is the specific form of the general spin Hamiltonian $H_s$ of (4.1), but it can reflect more clearly the Hilbert space of the $n-spin-1/2$ system. Inspired by the spin Hamiltonian (4.2) a general pseudo-diagonal Hermitian operator $H_N^{pd}$ that acts on an $N-$dimensional vector space may be defined by

$$H_N^{pd} = \epsilon_{pd} \left( \sum_{k=0}^{N-1} \alpha_k |k\rangle \right) \left( \sum_{k'=0}^{N-1} \alpha_{k'}^* \langle k'| \right) \tag{4.3}$$

where the sign $\epsilon_{pd} = \pm 1$ and any expansional coefficient $\alpha_k$ may be real or complex. In (4.3) the vector basis set $\{|k\rangle\}$ with $k = 0, 1, ..., N-1$ may stand for any complete set of base vectors of the $N-$dimensional vector space. Theoretically it is allowed that the base-vector expansion $\sum_{k=0}^{N-1} \alpha_k |k\rangle$ in (4.3) is not considered as any state vector of a quantum system. The pseudo-diagonal Hermitian operator $H_N^{pd}$ is positive if the sign $\epsilon_{pd} = +1$ and it is negative if the sign $\epsilon_{pd} = -1$. Moreover, $H_N^{pd} = 0$ only when every expansional coefficient $\alpha_k = 0$ in (4.3).

The Pseudo-Diagonal Hermitian (PDH) operators $H_N^{pd}$ of (4.3) are independent upon any detailed vector basis set of the $N-$dimensional vector space. They make sense for any dimensional size $N$ of the $N-$dimensional vector space. They are independent of any detailed direct-sum decomposition of a Hilbert space, when the $N-$dimensional vector space is the Hilbert space.

The PDH operator $H_N^{pd}$ may reflect clearly the $N-$dimensional vector space (or the Hilbert space of a quantum system). A great advantage for the specific form of (4.3) of the PDH operator $H_N^{pd}$ is that the Hermitian property of the PDH operator $H_N^{pd}$ always can be ensured no matter what these expansional coefficients $\{\alpha_k\}$ are taken in (4.3). Another great advantage is that the specific form of (4.3) keeps unchanged, when any unitary transformation acts on the PDH operator $H_N^{pd}$. These advantages make greatly convenient for the PDH operator $H_N^{pd}$ to be used extensively in future.

Consider the special case that the base-vector expansion $\sum_{k=0}^{N-1} \alpha_k |k\rangle$ in the PDH operator $H_N^{pd}$ of (4.3) is separately equal to an energy eigenvector of a quantum system (or an eigenvector of some dynamical variable $\Omega$). Then it can be found that the PDH operator $H_N^{pd}$ is a diagonal operator (or matrix) in the energy representation and moreover, it is really a single-matrix-element (SME) diagonal operator (apart from a constant). However, the PDH operator $H_N^{pd}$ also may be a non-diagonal operator (or matrix) in other representation than the energy representation. This means that the PDH operator $H_N^{pd}$ may be a SME diagonal operator (or matrix) in one representation, but it also may be a non-diagonal operator (or matrix) in another representation. This is the reason why the operators $H_N^{pd}$ of (4.3) are called the pseudo-diagonal Hermitian operators in unified form. Generally speaking, in a linear operator space the real SME diagonal operators should be the simplest Hermitian operators. Then the PDH operators $H_N^{pd}$ are perhaps the simplest non-diagonal and Hermitian operators in addition to the Hermitian pseudospin operators of (3.29).

For convenience in the theoretical treatments and applications in future the PDH operator $H_N^{pd}$ of (4.3) may be written briefly as

$$H_N^{pd} = \left( \sum_{k=0}^{N-1} \alpha_k |k\rangle \right) \langle h.c.| \tag{4.4}$$

where the right part $\langle h.c.|$ stands for the Hermitian conjugate (or the transpose conjugate) of the left part $\sum_{k=0}^{N-1} \alpha_k |k\rangle$ in round brackets, that is, $\langle h.c.| = \left(\sum_{k=0}^{N-1} \alpha_k |k\rangle\right)^+ = \sum_{k=0}^{N-1} \alpha_k^* \langle k|$, and the sign $\epsilon_{pd} = \pm 1$ is dropped temporarily without lost generality. The PDH operators $H_N^{pd}$ of (4.4) are quite popular and important. Many important operators in diverse research fields (See, e.g.,

Refs.[2,3,5], [Ref[3]]) also take the specific form of (4.4) of the PDH operators $H_N^{pd}$. The Hamiltonian operators (or the dynamical variables) that have been used extensively in the future work often take the specific form of (4.3) or (4.4).

Obviously, the spin Hamiltonian $H_s$ of (4.2) is a special PDH operator $H_N^{pd}$. Some instances for the PDH operator $H_N^{pd}$ include the specific Hermitian operator $H_2^{pd}$ of a single-spin$-1/2$ system in the Subsection 3.2.1 and the specific Hermitian operator $H_4^{pd}$ of a two-spin$-1/2$ system in the Subsection A1.4 of the Section A. It is easy to understand from these instances that the PDH operator $H_N^{pd}$ may be expanded in terms of a different operator basis set of the multiple-quantum operator algebra space.

Consider the spin Hamiltonian $H_s$ of a general spin system under study which is a dynamical variable and belongs to the multiple-quantum operator algebra space. According to quantum mechanics the spin Hamiltonian $H_s$ is diagonal in the energy representation and it may be generally expressed as

$$H_s = Diag\left(E_0, E_1, ..., E_{N-1}\right) = \sum_{k=0}^{N-1} E_k \left|\psi_k\right\rangle \left\langle\psi_k\right| \tag{4.5}$$

where $\left|\psi_k\right\rangle$ and $E_k$ are the $k-$th normalized energy eigenbase vector and energy eigenvalue, respectively, and obey the energy eigenvalue equation $H_s \left|\psi_k\right\rangle = E_k \left|\psi_k\right\rangle$. The Hamiltonian (4.5) may be called the energy diagonal operator (or matrix) in the energy representation. It also may be rewritten as $H_s = \sum_{k=0}^{N-1} H_k$, where $H_k = E_k \left|\psi_k\right\rangle \left\langle\psi_k\right|$ is the $k-$th component energy operator (or matrix). Evidently every component energy operator $H_k$ with $k = 0, 1, ..., N-1$, which is a dynamical variable and belongs to the multiple-quantum operator algebra space too, is a Hermitian diagonal operator (or matrix) in the energy representation and hence owns the specific form of (4.3).

Now consider another representation that is defined by the total spin operator $I_z$ of the total spin angular momentum operator $\mathbf{I}$ of the spin system under study (See the Section A). It is different from the energy representation. Here for convenience it is called the Spin$-I_z$ representation. The eigenvalue equation for the total spin operator $I_z$ may be written as $I_z \left|M, K_M\right\rangle = M \left|M, K_M\right\rangle$, where $\left|M, K_M\right\rangle$ is the orthonormal degenerate eigenbase vector, the total spin magnetic quantum number $M$ is the eigenvalue, and the index $K_M$ distinguishes between different degenerate eigenbase vectors $\{\left|M, K_M\right\rangle\}$. All the orthonormal eigenbase vectors $\{\left|M, K_M\right\rangle\}$ of the total spin operator $I_z$ form a complete set of base vectors of the Hilbert space of the spin system under study (See the Section A).

Consider the case that the spin Hamiltonian $H_s$ does not commute with the total spin operator $I_z$ and they do not have the common eigenbase vectors. According to the eigenfunction expansion principle in quantum mechanics, every energy eigenbase vector (e.g., $\left|\psi_k\right\rangle$) in the spin Hamiltonian $H_s$ of (4.5) always can be expanded in terms of the complete set $\{\left|M, K_M\right\rangle\}$ of the eigenbase vectors of the total spin operator $I_z$ (Ref.[7], Ref.[8], and the Section A),

$$|\psi_k\rangle = \sum_M \sum_{K_M} C_{k,M,K_M} |M, K_M\rangle \tag{4.6}$$

Then in the Spin$-I_z$ representation the component energy operator $H_k$ may be explicitly written as

$$H_k = E_k |\psi_k\rangle \langle\psi_k|$$
$$= E_k \left( \sum_M \sum_{K_M} C_{k,M,K_M} |M, K_M\rangle \right) \left( \sum_M \sum_{K_M} C_{k,M,K_M} |M, K_M\rangle \right)^+ \tag{4.7}$$

Here the energy eigenvalue $E_k$ may be positive, zero, negative. Let $E_k = \epsilon_k |E_k|$ where the sign $\epsilon_k = +1$ if $E_k \geq 0$ and $\epsilon_k = -1$ if $E_k < 0$. Then the component energy operator $H_k$ of (4.7) may be rewritten as

$$H_k = \epsilon_k \left( \sum_M \sum_{K_M} \sqrt{|E_k|} C_{k,M,K_M} |M, K_M\rangle \right) \left( \sum_M \sum_{K_M} \sqrt{|E_k|} C^*_{k,M,K_M} \langle M, K_M| \right) \tag{4.8}$$

This is general expression for the component energy operator $H_k$ in the Spin$-I_z$ representation. Obviously, the component energy operator $H_k$ of (4.8) is not a diagonal but a non-diagonal Hermitian operator (or matrix) in the Spin$-I_z$ representation. By comparing the expression (4.8) with the definition (4.3) of a PDH operator $H_N^{pd}$ it can be found that the component energy operator $H_k$ of (4.8) is indeed a pseudo-diagonal Hermitian operator. Finally, for the case that the spin Hamiltonian $H_s$ commutes with the total spin operator $I_z$ and they have the common eigenbase vectors, the pseudo-diagonal Hermitian operator of (4.8) may be reduced to a Hermitian diagonal operator, i.e., the component energy operator (or matrix) $H_k$ in the energy representation.

## 4.2. The basic properties of the PDH operators

There are some basic properties for a $PDH$ operator $H_N^{pd}$ which takes generally the specific form of (4.3) or is briefly written in the specific form of (4.4). These basic properties for the PDH operator $H_N^{pd}$ of (4.4) (or (4.3)) are described as follows:

$(i)$ The PDH operator $H_N^{pd}$ of (4.3) is always Hermitian for any expansional coefficients $\{\alpha_k\}$ and the sign $\epsilon_{pd} = \pm 1$

$(ii)$ Any unitary transformation $W$ acting on the PDH operator $H_N^{pd}$ may be generally expressed as

$$W H_N^{pd} W^+ = \left( \sum_{k=0}^{N-1} \alpha_k W |k\rangle \right) \langle h.c.| \tag{4.9}$$

Here the right part $\langle h.c.| = \sum_{k=0}^{N-1} \alpha_k^* \langle k| W^+ = \left( \sum_{k=0}^{N-1} \alpha_k W |k\rangle \right)^+$ is just the Hermitian conjugate of the left part $\sum_{k=0}^{N-1} \alpha_k W |k\rangle$ in round brackets. Therefore, the generated operator $W H_N^{pd} W^+$ is still a PDH operator $H_N^{pd}$. This means

that the specific form of (4.4) (or (4.3)) and the norm for a PDH operator $H_N^{pd}$ keep unchanged, when the PDH operator $H_N^{pd}$ is acted on by any unitary transformation. The operator $WH_N^{pd}W^+$ may be directly realized in experiment, if the PDH operator $H_N^{pd}$ represents some dynamical variable.

(*iii*) Any operator $Q$ left acting on the PDH operator $H_N^{pd}$ may be expressed as

$$QH_N^{pd} = \left(\sum_{k=0}^{N-1} \alpha_k Q\,|k\rangle\right)\left(\sum_{k=0}^{N-1} \alpha_k^* \langle k|\right) \tag{4.10a}$$

and any operator $Q^+$ right acting on the PDH operator $H_N^{pd}$ may be written as

$$H_N^{pd}Q^+ = \left(\sum_{k=0}^{N-1} \alpha_k\,|k\rangle\right)\left(\sum_{k=0}^{N-1} \alpha_k^* \langle k|\,Q^+\right) \tag{4.10b}$$

Obviously, both the generated operators $QH_N^{pd}$ and $H_N^{pd}Q^+$ may not be the PDH operators. Though the generated operator $QH_N^{pd}Q^+$ is still a PDH operator, the norm of the PDH operator $QH_N^{pd}Q^+$ may be different from the original norm of the PDH operator $H_N^{pd}$ if the operator $Q$ is not unitary. Generally the operator $QH_N^{pd}Q^+$ for any non-unitary operator $Q$ can not be directly realized in experiment.

(*iv*) There is the identity for any value $A$:

$$\left(A\sum_{k=0}^{N-1} \alpha_k\,|k\rangle\right)\langle h.c.| = |A|^2\left(\sum_{k=0}^{N-1} \alpha_k\,|k\rangle\right)\langle h.c.| \tag{4.11}$$

where the value $A$ may be real or complex. This identity directly leads to the selective-phase-inversion identity:

$$\left(-\alpha_j\,|j\rangle + \sum_{k\neq j,k=0}^{N-1} \alpha_k\,|k\rangle\right)\langle h.c.| = \left(\alpha_j\,|j\rangle - \sum_{k\neq j,k=0}^{N-1} \alpha_k\,|k\rangle\right)\langle h.c.| \tag{4.12}$$

where the selective phase inversion for a single base vector $|j\rangle$ is defined as $|j\rangle \rightarrow -|j\rangle$ with $0 \leq j \leq N-1$.

## 4.3. The unitary-invariant norms of the PDH operators

The *PDH* operators $H_N^{pd}$ may act as bridge to connect the Hilbert space with its corresponding multiple-quantum operator algebra space in the aspect of symmetrical structures and properties. Their norms may be related to both the Hilbert space and its corresponding multiple-quantum operator algebra space. In this Subsection the unitary-invariant norms [15, 16] of a general PDH operator $H_N^{pd}$ are derived in detail. In quantum mechanics the Hilbert space of a quantum system is a vector space with scalar product (or inner product). The scalar product of any pair of vectors $|\psi\rangle$ and $|\phi\rangle$ in the vector space may be written as $(\psi,\phi)$ or $\langle\psi|\phi\rangle$ . Particularly the scalar product $\langle\psi|\psi\rangle$ of the vector $|\psi\rangle$ is non-negative. The non-negative number $\sqrt{\langle\psi|\psi\rangle}$ is called the norm or the length of the vector $|\psi\rangle$ in the vector space. Here for convenience denote $||\psi|| \equiv |||\psi\rangle|| = \sqrt{\langle\psi|\psi\rangle}$. Then $||\psi||^2 = \langle\psi|\psi\rangle$ is the square norm of the vector $|\psi\rangle$. Suppose that $\{|\psi_k\rangle\}$ is an orthonormal vector basis set of the $N-$dimensional vector space. Then any vector $|\Psi\rangle$ of the $N-$dimensional vector space may be expanded in terms of the base vectors $\{|\psi_k\rangle\}$ :

$$|\Psi\rangle = \sum_{k=0}^{N-1} a_k |\psi_k\rangle \tag{4.13}$$

where $\{a_k\}$ are expansional coefficients. Then the square norm $||\Psi||^2$ of the vector $|\Psi\rangle$ may be given by

$$||\Psi||^2 = \left|\left|\sum_{k=0}^{N-1} a_k |\psi_k\rangle\right|\right|^2 = \sum_{k=0}^{N-1} |a_k|^2 \tag{4.14}$$

The norm $||\Psi||$ is the length of the vector $|\Psi\rangle$ in the $N-$dimensional vector space (i.e., the $N-$dimensional Hilbert space). There is an important property for the norm $||\Psi||$ that the norm is invariant, when the vector $|\Psi\rangle$ undergoes any unitary transformation $U$ (i.e., $U|\Psi\rangle$). That is, $||U|\Psi\rangle|| = ||\Psi||$ . Therefore, the norm $||\Psi||$ owns the unitarily invariant property.

Now consider the norms of linear operators of the linear operator space and especially the spectral norms which own the unitarily invariant property. Here the linear operator space under study is limited to be finite-dimensional, but it may be arbitrarily large. There is one-to-one correspondence between the linear operators on an $N-$dimensional vector space (i.e., $C^N$) and the $N\times N$ matrices. All these $N \times N$ matrices form a linear space $C^{N\times N}$ of the $N \times N$ matrices [15, 16], while all these corresponding linear operators form a linear operator space too. Then the norms of the linear operators of the linear operator space correspond one-to-one to the norms of the $N \times N$ matrices of the linear space $C^{N\times N}$. It is well known that the linear space $C^{N\times N}$ of matrices with dimension $N \times N$ is isomorphic to the vector space $C^{N^2}$ with the same dimension $N \times N$ [15, 16]. These may be employed to determine the norms of the linear operators including the PDH operators $H_N^{pd}$ of the multiple-quantum operator algebra space.

Suppose that $\{Q_{kl}\}$ is any complete set of base operators of the finite-dimensional linear operator space with dimension $N \times N$. Then any operator $A$ of the linear operator space can be expanded in terms of the base operators $\{Q_{kl}\}$ :

$$A = \sum_{k,l=0}^{N-1} a_{kl} Q_{kl} \tag{4.15}$$

where $\{a_{kl}\}$ are expansional coefficients. The scalar (or inner) product $(A, B)$ of any two operators $A$ and $B$ of the linear operator space may be defined by

[1, 3, 2]

$$(A, B) = Tr\left(A^{+}B\right) \tag{4.16}$$

where the symbol $Tr$ represents the trace of operator and $A^{+}$ is the Hermitian conjugate (or adjoint) of the operator $A$. Without lost generality, below consider the representation defined by any operator $\Omega$, which is called the $\Omega$ representation. As a typical example, the $\Omega$ representation may be the energy representation which is defined by the Hamiltonian operator (i.e., the operator $\Omega$) of the quantum system under study. Suppose that in the $\Omega$ representation $\bar{A}$ and $\bar{B}$ are the representation matrices of the operators $A$ and $B$, respectively. The scalar product $\left(\bar{A}, \bar{B}\right)$ of the two matrices $\bar{A}$ and $\bar{B}$ may be expressed as

$$\left(\bar{A}, \bar{B}\right) = Tr\left(\bar{A}^{+}\bar{B}\right) = \sum_{k,l=0}^{N-1} \bar{A}_{kl}^{*}\bar{B}_{kl} \tag{4.17}$$

where $\{\bar{A}_{kl}\}$ (or $\{\bar{B}_{kl}\}$) are the elements of the $N \times N$ matrix $\bar{A}$ (or $\bar{B}$). Now the scalar product $(A, B)$ of the two operators $A$ and $B$ is represented by the scalar product $\left(\bar{A}, \bar{B}\right)$ of the two representation matrices $\bar{A}$ and $\bar{B}$. Therefore, the scalar product $(A, B)$ can be explicitly calculated by the scalar product $\left(\bar{A}, \bar{B}\right)$ of (4.17) in the $\Omega$ representation. In particular, when the matrix $\bar{A} = \bar{B}$, the scalar product $\left(\bar{A}, \bar{A}\right)$ of the matrix $\bar{A}$ is written as $\left(\bar{A}, \bar{A}\right) = Tr\left(\bar{A}^{+}\bar{A}\right)$. It is easy to find from (4.17) that the scalar product $\left(\bar{A}, \bar{A}\right)$ is non-negative. The non-negative value $\left\|\bar{A}\right\| = \sqrt{\left(\bar{A}, \bar{A}\right)}$ is called the norm of the matrix $\bar{A}$. Correspondingly the scalar product $(A, A)$ of the operator $A$ is written as $(A, A) = Tr\left(A^{+}A\right)$. The scalar product $(A, A)$ is non-negative. The non-negative value $\|A\| = \sqrt{(A, A)}$ is called the norm of the operator $A$ [1, 3]. Moreover, the norm $\|A\|$ is represented by the norm $\left\|\bar{A}\right\|$ which can be explicitly calculated by (4.17).

It is well known that the trace of any $N \times N$ matrix $\bar{A} = (\bar{A}_{kl})$ is defined as the sum of the diagonal elements of the matrix, i.e., $Tr\{\bar{A}\} = \sum_{k=0}^{N-1} \bar{A}_{kk}$. One important property for the trace of matrix is the cyclic-invariant property [2]: $Tr\left(\bar{A}\bar{B}\bar{C}\right) = Tr\left(\bar{C}\bar{A}\bar{B}\right) = Tr\left(\bar{B}\bar{C}\bar{A}\right)$, where $\bar{A}$, $\bar{B}$, and $\bar{C}$ are any three $N \times N$ matrices. With the help of the cyclic-invariant property it is easy to prove that when the matrix $\bar{A}$ undergoes any unitary transformation $\bar{U}$, its trace is invariant: $Tr\left(\bar{U}\bar{A}\bar{U}^{+}\right) = Tr\left(\bar{A}\right)$. Likewise, it is easy to prove that the norm $\left\|\bar{A}\right\|$ of the matrix $\bar{A}$ is invariant, when the matrix $\bar{A}$ undergoes any unitary transformation $\bar{U}$ (i.e., $\bar{U}\bar{A}\bar{U}^{+} = \bar{A}'$). This can be seen as follows:

$$\begin{aligned}\left\|\bar{A}'\right\|^2 &= \left(\bar{A}', \bar{A}'\right) = Tr\left(\left(\bar{A}'\right)^{+}\bar{A}'\right) = Tr\left(\left(\bar{U}\bar{A}\bar{U}^{+}\right)^{+}\left(\bar{U}\bar{A}\bar{U}^{+}\right)\right)\\ &= Tr\left(\bar{U}\bar{A}^{+}\bar{A}\bar{U}^{+}\right) = Tr\left(\bar{A}^{+}\bar{A}\right) = \left(\bar{A}, \bar{A}\right) = \left\|\bar{A}\right\|^2\end{aligned}$$

where unitarity $\bar{U}\bar{U}^{+} = \bar{U}^{+}\bar{U} = E$ (unity matrix) has been used. Note that the norm $\|A\|$ of the operator $A$ is represented by the norm $\left\|\bar{A}\right\|$ of the representation matrix $\bar{A}$.

Then the norm $\|A\|$ of the operator $A$ is invariant, when the operator $A$ undergoes any unitary transformation. Therefore, the norm $\|A\|$ owns the unitarily invariant property.

It is known above that the $N \times N$ matrix $\bar{A}$ is the representation matrix of the operator $A$ in the $\Omega$ representation. The matrix $\bar{A}$ may be expressed in the terms of the single-element matrices $\{\bar{B}_{kl}\}$ in the $\Omega$ representation. It is known from the Subsection 3.2.3 that a single-element matrix is defined as the matrix whose elements all are zero except one element equal to one. Denote $\bar{B}_{kl}$ as the single-element matrix in which only the $k-$th row and $l-$th column element is equal to one and any other elements are zero, here $0 \leq k, l \leq N-1$. Then the single-element matrix $\bar{B}_{kl}$ may be written as $\bar{B}_{kl} = \left((\bar{B}_{kl})_{ij}\right)$ whose $i-$th row and $j-$th column element $(\bar{B}_{kl})_{ij}$ is generally written as

$$\left(\bar{B}_{kl}\right)_{ij} = \delta_{ki}\delta_{lj} \text{ for } i, j = 0, 1, ..., N-1. \tag{4.18}$$

In the linear space $C^{N\times N}$ of the $N \times N$ matrices with dimension $N \times N$ there are $N^2$ orthogonal single-element matrices $\{\bar{B}_{kl}\}$ with $k, l = 0, 1, ..., N-1$. Any $N \times N$ matrix (e.g., the $N \times N$ matrix $\bar{A}$) of the linear space $C^{N\times N}$ may be expanded in terms of these $N^2$ orthogonal single-element matrices $\{\bar{B}_{kl}\}$ (in the $\Omega$ representation),

$$\bar{A} = \sum_{k,l=0}^{N-1} c_{kl}\bar{B}_{kl}, \tag{4.19}$$

where $c_{kl}$ is any expansional coefficient and moreover, it is just equal to the element $\bar{A}_{kl}$ of the matrix $\bar{A}$. Now with the help of the expression (4.19) of the matrix $\bar{A}$ in the $\Omega$ representation one may explicitly calculate the norm $\|\bar{A}\|$ of the matrix $\bar{A}$ by the formula (4.17). First of all, the scalar product $\left(\bar{A}, \bar{A}\right) = Tr\left(\bar{A}^+\bar{A}\right)$ may be expressed as

$$\left(\bar{A}, \bar{A}\right) = \left(\sum_{k,l=0}^{N-1} c_{kl}\bar{B}_{kl}, \sum_{m,n=0}^{N-1} c_{mn}\bar{B}_{mn}\right) = \sum_{k,l=0}^{N-1}\sum_{m,n=0}^{N-1} c_{kl}^* c_{mn}\left(\bar{B}_{kl}, \bar{B}_{mn}\right) \tag{4.20}$$

Here the scalar product $\left(\bar{B}_{kl}, \bar{B}_{mn}\right)$ may be explicitly calculated by using (4.17) and with the aid of (4.18). It can be found that the scalar product $\left(\bar{B}_{kl}, \bar{B}_{mn}\right)$ is given by

$$\left(\bar{B}_{kl}, \bar{B}_{mn}\right) = Tr\left(\left(\bar{B}_{kl}\right)^+ \bar{B}_{mn}\right) = \sum_{i,j=0}^{N-1} \left(\bar{B}_{kl}\right)_{ij}^* \left(\bar{B}_{mn}\right)_{ij}$$

$$= \sum_{i,j=0}^{N-1} \delta_{ki}\delta_{lj}\delta_{mi}\delta_{nj} = \delta_{km}\delta_{ln} \tag{4.21}$$

Then by substituting (4.21) into (4.20) it can prove that the square norm $\left\|\bar{A}\right\|^2 =$

$\left(\bar{A},\bar{A}\right)$ of the matrix $\bar{A}$ in the $\Omega$ representation is written as

$$\left\|\bar{A}\right\|^2 = Tr\left(\bar{A}^+\bar{A}\right) = \sum_{k,l=0}^{N-1} |c_{kl}|^2 \tag{4.22}$$

where the coefficient $c_{kl}$ is the element $\bar{A}_{kl}$ of the $N\times N$ matrix $\bar{A}$, i.e., $c_{kl} = \bar{A}_{kl}$. The formula (4.22) also can be obtained directly from (4.17).

There is one-to-one correspondence between all the linear operators of the linear operator space with dimension $N \times N$ and all the $N \times N$ matrices of the linear space $C^{N\times N}$ of matrices with dimension $N \times N$ [3,1,2]. As shown in the Subsection 3.2.3, there are the $N^2$ orthogonal single-matrix-element (SME) base operators $\{B_{kl}\}$ of the $N \times N-$dimensional linear operator space. They correspond one-to-one to the $N^2$ orthogonal single-element matrices $\{\bar{B}_{kl}\}$ with $k,l = 0,1,...,N-1$ of the linear space $C^{N\times N}$ of matrices with dimension $N \times N$. Moreover, they form a complete set $\{B_{kl}\}$ of base operators of the linear operator space. Then the operator $A$ of the linear operator space which has the representation matrix $\bar{A}$ in the $\Omega$ representation may be expanded in terms of the orthogonal SME base operators $\{B_{kl}\}$ (See the Subsection 3.2.3),

$$A = \sum_{k,l=0}^{N-1} c_{kl}B_{kl}, \tag{4.23}$$

where the expansional coefficient $c_{kl}$ is just the element $\bar{A}_{kl}$ of the matrix $\bar{A}$. The base-operator expansion (4.23) of the operator $A$ corresponds to the expansion (4.19) of the matrix $\bar{A}$. Moreover, the orthogonal SME base operators $\{B_{kl}\}$ obey the orthogonal relation:

$$(B_{kl}, B_{mn}) = Tr\left((B_{kl})^+ B_{mn}\right) = \delta_{km}\delta_{ln} \tag{4.24}$$

This orthogonal relation corresponds to that one (4.21) of the single-element matrices $\{\bar{B}_{kl}\}$. Now with the help of the orthogonal relation of (4.24) it can prove by using (4.23) that the square norm of the operator $A$ of (4.23) is written as

$$\|A\|^2 = Tr\left(A^+A\right) = \sum_{k,l=0}^{N-1} |c_{kl}|^2 \tag{4.25}$$

where the coefficient $c_{kl}$ is the element $\bar{A}_{kl}$ of the matrix $\bar{A}$, i.e., $c_{kl} = \bar{A}_{kl}$. The formula (4.25) of the operator $A$ of (4.23) corresponds to the formula (4.22) of the matrix $\bar{A}$. By the formula (4.25) one can explicitly calculate the norm $\|A\|$ of the operator $A$ of (4.23).

It is ready to calculate explicitly the norm of the PDH operator $H_N^{pd}$ of (4.3) (or (4.4)). The PDH operator $H_N^{pd}$ of (4.3) is a linear operator of the $N^2-$dimensional multiple-quantum operator algebra space (i.e., a linear operator space with dimension $N \times N$). It may be explicitly expressed as

$$H_N^{pd} = \epsilon_{pd}\left(\sum_{k=0}^{N-1}\alpha_k|k\rangle\right)\left(\sum_{l=0}^{N-1}\alpha_l|l\rangle\right)^+ = \sum_{k=0}^{N-1}\sum_{l=0}^{N-1}\epsilon_{pd}\alpha_k\alpha_l^*|k\rangle\langle l| \tag{4.26}$$

Let the base operator $B_{kl} = |k\rangle \langle l|$ and the coefficient $c_{kl} = \epsilon_{pd}\alpha_k\alpha_l^*$ for $k, l = 0, 1, ..., N-1$. Then the PDH operator $H_N^{pd}$ of (4.26) may be rewritten as

$$H_N^{pd} = \sum_{k,l=0}^{N-1} c_{kl} B_{kl} \tag{4.27}$$

where $c_{kl}$ is an expansional coefficient. Note that the $\Omega$ representation can be any representation. Without lost generality, here suppose that the base vectors $\{|k\rangle\}$ in (4.26) are the orthonormal eigenbase vectors in the $\Omega$ representation. They form a vector basis set of the $N-$dimensional vector space. Then $\{B_{kl}\} = \{|k\rangle \langle l|\}$ are the SME base operators, because their representation matrices are the single-element matrices in the $\Omega$ representation. And consequently the expansional coefficient $c_{kl}$ in (4.27) is just the element $(\bar{H}_N^{pd})_{kl}$ of the representation matrix $\bar{H}_N^{pd}$ of the PDH operator $H_N^{pd}$ in the $\Omega$ representation. Obviously, all these $N^2$ SME base operators $\{B_{kl}\}$ form a complete set of orthogonal base operators of the $N^2-$dimensional multiple-quantum operator algebra space which corresponds to the $N-$dimensional vector space.

Now according to (4.23) and (4.25) it can be found that the square norm of the PDH operator $H_N^{pd}$ of (4.27) is written as

$$\left|\left|H_N^{pd}\right|\right|^2 = Tr\left(\left(H_N^{pd}\right)^+ H_N^{pd}\right) = \sum_{k,l=0}^{N-1} |c_{kl}|^2 \tag{4.28}$$

where the coefficient $c_{kl}$ is the element $(\bar{H}_N^{pd})_{kl}$ of the representation matrix $\bar{H}_N^{pd}$, i.e., $c_{kl} = (\bar{H}_N^{pd})_{kl}$. Note that $c_{kl} = \epsilon_{pd}\alpha_k\alpha_l^*$ for $k, l = 0, 1, ..., N-1$. Then the norm of the PDH operator $H_N^{pd}$ is explicitly calculated by

$$\left|\left|H_N^{pd}\right|\right| = \sqrt{\sum_{k=0}^{N-1}\sum_{l=0}^{N-1} |\epsilon_{pd}\alpha_k\alpha_l^*|^2} = \sqrt{\left(\sum_{k=0}^{N-1} |\alpha_k|^2\right)\left(\sum_{l=0}^{N-1} |\alpha_l^*|^2\right)} = \sum_{k=0}^{N-1} |\alpha_k|^2 \tag{4.29}$$

The formula (4.28) or (4.29) is a general formula to calculate exactly the norm of the PDH operator $H_N^{pd}$ of (4.3) (or (4.4)) in the $N^2-$dimensional multiple-quantum operator algebra space.

On the other hand, the PDH operator $H_N^{pd}$ of (4.3) may be rewritten as $H_N^{pd} = \epsilon_{pd} |\Psi\rangle \langle\Psi|$, where $|\Psi\rangle = \sum_{k=0}^{N-1} \alpha_k |k\rangle$. Note that here the base vectors $\{|k\rangle\}$ are the orthonormal eigenbase vectors in the $\Omega$ representation and form a vector basis set of the $N-$dimensional vector space. Then the expansion $\sum_{k=0}^{N-1} \alpha_k |k\rangle$ in the PDH operator $H_N^{pd}$ may be separately considered as a vector $|\Psi\rangle$ of the $N-$dimensional vector space. As shown by (4.13) and (4.14), the square norm of the vector $|\Psi\rangle$ of the $N-$dimensional vector space may be given by

$$\|\Psi\|^2 = \||\Psi\rangle\|^2 = \left\|\sum_{k=0}^{N-1} \alpha_k |k\rangle\right\|^2 = \sum_{k=0}^{N-1} |\alpha_k|^2 \tag{4.30}$$

This square norm $\|\Psi\|^2$ also is the square norm of the expansion $\sum_{k=0}^{N-1} \alpha_k |k\rangle$ in the PDH operator $H_N^{pd}$. Generally the square norm $\|\Psi\|^2$ may not be equal to one. By comparing both the norms $\left\|H_N^{pd}\right\|$ of (4.29) and $\|\Psi\|^2$ of (4.30) with one another it can be found that they are really the same one: $\left\|H_N^{pd}\right\| = \|\Psi\|^2$. Therefore, the norm of the PDH operator $H_N^{pd}$ may be written as

$$\left\|H_N^{pd}\right\| = \left\|\sum_{k=0}^{N-1} \alpha_k |k\rangle\right\|^2 = \sum_{k=0}^{N-1} |\alpha_k|^2 \tag{4.31}$$

This is a general formula to calculate explicitly the norm of the PDH operator $H_N^{pd}$ of (4.3) (or (4.4)) through the $N-$dimensional vector space which corresponds to the $N^2-$dimensional multiple-quantum operator algebra space.

## 4.4. Consideration of the Hilbert-space symmetric structures

The pseudo-diagonal Hermitian operators $H_N^{pd}$ are the specific Hermitian operators of the multiple-quantum operator algebra space. They can reflect clearly the $N-$dimensional vector space (or the $N-$dimensional Hilbert space of quantum system), indicating that they are able to reflect clearly the symmetrical structure of the Hilbert space. Here the symmetrical structure of the Hilbert space of a quantum system such as an $n - spin - 1/2$ system may be specified by the direct-sum decomposition of the Hilbert space (See Ref. [7] and also the Part II and the Section A in this paper). Then the Hilbert-space symmetrical structure may be easily taken into account with the help of the PDH operators $H_N^{pd}$ in the multiple-quantum operator algebra space. Suppose that the symmetrical structure of the Hilbert space $S(N)$ of a quantum spin system under study may be characterized as follows: The whole Hilbert space $S(N)$ is divided into the $K$ direct-sum subspaces [7],

$$S(N) = S_1(N_1) \bigoplus S_2(N_2) \bigoplus ... \bigoplus S_K(N_K), \tag{4.32}$$

where $S_l(N_l)$ is the $l-$th direct-sum subspace with the dimensional size $N_l$ and the index $l = 1, 2, ..., K$, and the total dimensional size $N$ of the Hilbert space $S(N)$ is given by $N = N_1 + N_2 + ... + N_K$. These direct-sum subspaces $\{S_l(N_l)\}$ of the Hilbert space $S(N)$ are mutually orthogonal.

Now in order to match the symmetrical structure of the Hilbert space $S(N)$, i.e., the direct-sum decomposition of (4.32), the PDH operator $H_N^{pd}$ of (4.3) may be re-expressed as

$$H_N^{pd} = \epsilon_{pd}\left(\sum_{k_1}^{N_1} \alpha_{k_1} |k_1\rangle + \sum_{k_2}^{N_2} \alpha_{k_2} |k_2\rangle + ... + \sum_{k_K}^{N_K} \alpha_{k_K} |k_K\rangle\right)\langle h.c.| \tag{4.33}$$

where the expansional coefficients $\{\alpha_{k_l}\}$ can be obtained from (4.3). Note that the vector basis set $\{|k\rangle\}$ in (4.3) may stand for any complete set of orthonormal

base vectors of the Hilbert space $S(N)$. In the PDH operator $H_N^{pd}$ of (4.33) it is grouped into the $K$ vector basis subsets $\{|k_1\rangle\}$, $\{|k_2\rangle\}$, ..., and $\{|k_K\rangle\}$ (See Ref. [7] and also the Part II and the Section A),

$$\{|k\rangle\} = \{\{|k_1\rangle\}, \{|k_2\rangle\}, ..., \{|k_K\rangle\}\}. \tag{4.34}$$

Here the $l-$th vector basis subset $\{|k_l\rangle\}$ contains $N_l$ orthonormal base vectors of the complete set $\{|k\rangle\}$ in (4.3) for $l = 1, 2, ..., K$. Moreover, this vector basis subset $\{|k_l\rangle\}$ can span the $l-$th direct-sum subspace $S_l\,(N_l)$ in (4.32) with dimension $N_l$. This is a key step for the PDH operator $H_N^{pd}$ of (4.3) to take into account the symmetrical structure of the Hilbert space $S(N)$. Obviously, these $K$ vector basis subsets $\{|k_l\rangle\}$ with $l = 1, 2, ..., K$ in (4.34) correspond one-to-one to these $K$ direct-sum subspaces $\{S_l\,(N_l)\}$ in (4.32). Moreover, every subspace $S_l\,(N_l)$ is contained in the Hilbert space $S(N)$ and these $K$ subspaces $\{S_l\,(N_l)\}$ are mutually orthogonal. These indicate that the Hilbert space $S(N)$ is the direct sum of these $K$ subspaces $\{S_l\,(N_l)\}$ which are spanned by these $K$ vector basis subsets $\{|k_l\rangle\}$ with $l = 1, 2, ..., K$, respectively. And this is consistent with the direct-sum decomposition of (4.32).

For convenience denote $\{|\phi_m^{s_l}\rangle\}$ as the vector basis subset $\{|k_l\rangle\}$ of the $l-$th direct-sum subspace $S_l(N_l)$ with dimension $N_l$, where the index $m = 0, 1, ..., N_l - 1$, the superscript $s_l$ labels the $l-$th subspace $S_l(N_l)$, and the index $l = 1, 2, ..., K$. Then the vector $\sum_{k_l}^{N_l} \alpha_{k_l} |k_l\rangle$ of the $l-$th subspace $S_l(N_l)$ in (4.33) is rewritten as $\sum_{m=0}^{N_l-1} \alpha_{l,m} |\phi_m^{s_l}\rangle$ , where the coefficient $\alpha_{l,m} = \alpha_{k_l}$ if $|\phi_m^{s_l}\rangle = |k_l\rangle$ . By substituting $\sum_{k_l}^{N_l} \alpha_{k_l} |k_l\rangle = \sum_{k=0}^{N_l-1} \alpha_{l,k} |\phi_k^{s_l}\rangle$ into (4.33) the PDH operator $H_N^{pd}$ may be rewritten as

$$H_N^{pd} = \epsilon_{pd} \left( \sum_{k=0}^{N_1-1} \alpha_{1,k} |\phi_k^{s_1}\rangle + \sum_{k=0}^{N_2-1} \alpha_{2,k} |\phi_k^{s_2}\rangle + ... + \sum_{k=0}^{N_K-1} \alpha_{K,k} |\phi_k^{s_K}\rangle \right) \langle h.c.| \tag{4.35}$$

This operator $H_N^{pd}$ is the pseudo-diagonal Hermitian operator in which the symmetric structure of the Hilbert space $S\,(N)$ , i.e., the direct-sum decomposition of (4.32), is taken into account.

The PDH operator $H_N^{pd}$ of (4.35) is able to reflect the dimensional sizes of the direct-sum subspaces of the Hilbert space, if the expansional coefficients $\{\alpha_{l,k}\}$ in (4.35) are chosen suitably. If these expansional coefficients $\{\alpha_{l,k}\}$ are taken as the same value for a given subspace $S_l\,(N_l)$ , that is, $\alpha_{l,k} = \beta_l$ for $k = 0, 1, 2, ..., N_l - 1$, then the PDH operator $H_N^{pd}$ of (4.35) may be reduced to the form

$$H_N^{pd} = \epsilon_{pd} \left( \beta_1 \sum_{k=0}^{N_1-1} |\phi_k^{s_1}\rangle + \beta_2 \sum_{k=0}^{N_2-1} |\phi_k^{s_2}\rangle + ... + \beta_K \sum_{k=0}^{N_K-1} |\phi_k^{s_K}\rangle \right) \langle h.c.| \tag{4.36}$$

Obviously, here every orthonormal base vector $|\phi_k^{s_l}\rangle$ for a given subspace $S_l\,(N_l)$ has the same expansional coefficient $\beta_l$ for $l = 1, 2, ..., K$. This means that any subspace $S_l\,(N_l)$ may be characterized simply by its uniform-expansional-coefficient sum vector $\sum_{k=0}^{N_l-1} |\phi_k^{s_l}\rangle$ (or uniform-sum vector), while the only parameter that specifies the uniform-sum vector $\sum_{k=0}^{N_l-1} |\phi_k^{s_l}\rangle$ is the dimensional size $N_l$ of the subspace $S_l(N_l)$. Therefore, the uniform-sum vector $\sum_{k=0}^{N_l-1} |\phi_k^{s_l}\rangle$ of the subspace $S_l(N_l)$ in the PDH operator $H_N^{pd}$ of (4.36) may reflect conveniently the dimensional size $N_l$ of the subspace $S_l(N_l)$.

Below the full-uniform and almost-full-uniform PDH operators are simply introduced. If in the PDH operator $H_N^{pd}$ of (4.36) the expansional coefficients $\{\beta_l\}$ of all the subspaces $\{S_l(N_l)\}$ are taken as the same value $\beta$, then the PDH operator $H_N^{pd}$ is reduced to the form

$$H_N^{pd} = \frac{1}{N}\left(\sum_{k=0}^{N_1-1} |\phi_k^{s_1}\rangle + \sum_{k=0}^{N_2-1} |\phi_k^{s_2}\rangle + ... + \sum_{k=0}^{N_K-1} |\phi_k^{s_K}\rangle\right)\langle h.c.| , \tag{4.37}$$

where the sign $\epsilon_{pd}$ is taken as $\epsilon_{pd} = +1$ and $|\beta|^2$ is taken as the normalization constant $|\beta|^2 = 1/N$ with the dimensional size $N = \sum_{l=1}^{K} N_l$ of the Hilbert space $S(N)$. It can be found that the expansional coefficient of every base vector $|\phi_k^{s_l}\rangle$ has the same value in the PDH operator $H_N^{pd}$ of (4.37). Therefore, the PDH operator $H_N^{pd}$ of (4.37) is a full-uniform PDH operator.

Generally, a full-uniform PDH operator $H_N^0$ is defined as the PDH operator $H_N^{pd}$ of (4.3) in which all the expansional coefficients $\{\alpha_k\}$ are set to the same value. Then it may be written as (apart from a proportional factor)

$$H_N^0 = \frac{1}{N}\left(\sum_{k=0}^{N-1} |k\rangle\right)\langle h.c.| \tag{4.38a}$$

where the sign $\epsilon_{pd}$ is taken as $\epsilon_{pd} = +1$ and every expansional coefficient $\alpha_k = 1/\sqrt{N}$ with the dimensional size $N$ of the $N-$dimensional vector space. Obviously, the full-uniform PDH operator $H_N^0$ is defined on the $N-$dimensional vector space. The base vectors $\{|k\rangle\}$ in (4.38a) may represent any complete set of orthonormal base vectors of the $N-$dimensional vector space, for example, the complete set $\{|\phi_k^{s_l}\rangle\}$ in (4.37). By expanding the PDH operator $H_N^0$ of (4.38a) one obtains

$$H_N^0 = \frac{1}{N}\sum_{k,l=0}^{N-1} |k\rangle\langle l| \tag{4.38b}$$

where the base operators $\{|k\rangle\langle l|\}$ are just the SME base operators $\{B_{kl}\} = \{|k\rangle\langle l|\}$ (See the Subsection 3.2.3). This is the base-operator expansion of the full-uniform PDH operator $H_N^0$ in terms of the SME base operators $\{B_{kl}\}$. It can be found that the expansional coefficient of every SME base operator $B_{kl}$ in the base-operator expansion (4.38b) is equal to the same value $1/N$. This is why the PDH operator $H_N^0$ is said full-uniform.

An almost-full-uniform PDH operator may be defined as the PDH operator $H_N^{pd}$ of (4.3) in which polynomially many ($L$) expansional coefficients $\{\alpha_k\}$ take different values at most, while the other expansional coefficients $\{\alpha_k\}$ which are exponentially many $(N-L)$ take the same value. Then an almost-full-uniform PDH operator $H_N^{0,L}$ may be generally written as (apart from a constant)

$$H_N^{0,L} = \frac{1}{N}\left(\sum_{k=0}^{L-1} \alpha_k |k\rangle + \sum_{k=L}^{N-1} |k\rangle\right)\langle h.c.| \tag{4.39}$$

where the expansional coefficient $\alpha_k$ may be different from one for $k = 0, 1, ..., L-1$, the base-vector number $L = poly(\log N)$, and the dimensional size of the $N-$dimensional vector space is $N >> L$. Suppose that there is an extra constraint on these expansional coefficients $\{\alpha_k\}$ in (4.39), that is, the norm $\left\|H_N^{0,L}\right\|$ $= 1$. As shown by (4.31), the norm $\left\|H_N^{0,L}\right\|$ of the PDH operator $H_N^{0,L}$ may be given by

$$\left\|H_N^{0,L}\right\| = \left\|\frac{1}{\sqrt{N}}\sum_{k=0}^{L-1} \alpha_k |k\rangle + \frac{1}{\sqrt{N}}\sum_{k=L}^{N-1} |k\rangle\right\|^2 = 1 \tag{4.40}$$

This formula shows that the square norm $\left\|\sum_{k=0}^{L-1} \alpha_k |k\rangle\right\|^2 = \sum_{k=0}^{L-1} |\alpha_k|^2 = L$ due to that the square norm $\left\|\sum_{k=L}^{N-1} |k\rangle\right\|^2 = N-L$. Note that $N >> L$. Then $N-L >> L$. Therefore, in the PDH operator $H_N^{0,L}$ the term $\sum_{k=0}^{L-1} \alpha_k |k\rangle$ is secondary with respect to the main term $\sum_{k=L}^{N-1} |k\rangle$. This means that the almost-full-uniform PDH operator $H_N^{0,L}$ of (4.39) is much close to the full-uniform PDH operator $H_N^0$ of (4.38a), if it satisfies the constraint condition (4.40) and $N-L >> L$. In particular, when the base-vector number $L = 1$ and the expansional coefficient $\alpha_0 = 0$, the almost-full-uniform PDH operator $H_N^{0,L}$ of (4.39) is reduced to the form

$$H_N^{0,1} = \frac{1}{N}\left(\sum_{k=1}^{N-1} |k\rangle\right)\langle h.c.| \tag{4.41}$$

Below the concepts of the full-uniform and the almost-full-uniform PDH operators are employed to design some important PDH Hamiltonian operators which will be used in future work.

The full-uniform PDH operator $H_N^0$ of (4.38) and the almost-full-uniform PDH operator $H_N^{0,L}$ of (4.39) are defined on the $N-$dimensional vector space. The symmetrical structures and properties of the Hilbert space of a quantum system under study are not involved in the above discussion about the full-uniform and almost-full-uniform PDH operators. A simple scheme to consider the symmetrical structures and properties of the Hilbert space is that the $N'-$dimensional vector space $VS(N')$ on which the full-uniform PDH operator $H_{N'}^0$ or the almost-full-uniform PDH operator $H_{N'}^{0,L}$ is defined is considered as one direct-sum subspace (e.g., the subspace $S_l(N_l)$) of the Hilbert space $S(N)$ of (4.32) with dimension $N > N'$. Another equivalent scheme is that the $N-$dimensional vector space $VS(N)$ on which the full-uniform PDH operator $H_N^0$ or the almost-full-uniform PDH operator $H_N^{0,L}$ is defined is treated as one direct-sum subspace $VS(N)$ of some larger Hilbert space $S_T(N_T)$ with dimension $N_T > N$. With the second scheme a PDH operator $H_{N_T,N}^{0,L}$ on the total Hilbert space $S_T(N_T)$ is constructed as follows:

$$H_{N_T,N}^{0,L} = \left(\beta_N \left(\frac{1}{\sqrt{N}} \sum_{k=0}^{L-1} \alpha_k |k\rangle + \frac{1}{\sqrt{N}} \sum_{k=L}^{N-1} |k\rangle\right) + ...\right) \langle h.c.| \,, \qquad (4.42)$$

where for simplicity only the direct-sum subspace $VS(N)$ is explicitly written down, while any other direct-sum subspaces of the total Hilbert space $S_T(N_T)$ are hidden. Obviously, the almost-full-uniform PDH operator $H_N^{0,L}$ of (4.39) on the $N-$dimensional vector space $VS(N)$ corresponds to the PDH operator $H_{N_T,N}^{0,L}$ on the total Hilbert space $S_T(N_T)$ which contains the direct-sum subspace $VS(N)$. The PDH operator $H_{N_T,N}^{0,L}$ of (4.42) with one expansional coefficient $\alpha_k \neq 1$ at least may be called the almost-full-uniform PDH operator with respect to the direct-sum subspace $VS(N)$ of the total Hilbert space $S_T(N_T)$. When the base-vector number $L = 1$ and the expansional coefficient $\alpha_0 = 0$, the almost-full-uniform PDH operator $H_{N_T,N}^{0,L}$ of (4.42) is reduced to the form

$$H_{N_T,N}^{0,1} = \left(\beta_N \left(\frac{1}{\sqrt{N}} \sum_{k=1}^{N-1} |k\rangle\right) + ...\right) \langle h.c.| \qquad (4.43)$$

This PDH operator corresponds to the PDH operator $H_N^{0,1}$ of (4.41). The almost-full-uniform PDH operators with respect to some direct-sum subspace of the whole Hilbert space such as those of (4.42) and (4.43) will have important applications in future work.

When $L = 1$ and the expansional coefficient $\alpha_0 = 1$, the PDH operator $H_{N_T,N}^{0,L}$ of (4.42) is reduced to the form

$$H_{N_T,N}^{0} = \left(\beta_N \left(\frac{1}{\sqrt{N}} \sum_{k=0}^{N-1} |k\rangle\right) + ...\right) \langle h.c.| \qquad (4.44)$$

This is just the full-uniform PDH operator with respect to the direct-sum subspace $VS(N)$. It corresponds to the full-uniform PDH operator $H_N^0$ of (4.38a) on the $N-$dimensional vector space $VS(N)$.

The uniform-sum vector $\sum_{k=0}^{N-1} |k\rangle$ of the direct-sum subspace $VS(N)$ in the full-uniform PDH operator $H_{N_T,N}^0$ of (4.44) with respect to the direct-sum subspace $VS(N)$ may reflect the dimensional size $N$ of the direct-sum subspace $VS(N)$ of the total Hilbert space $S_T(N_T)$. Note that the almost-full-uniform PDH operator $H_N^{0,L}$ of (4.39) is much close to the full-uniform PDH operator $H_N^0$ of (4.38a), when $\left\| \sum_{k=L}^{N-1} |k\rangle \right\|^2 = N - L >> \left\| \sum_{k=0}^{L-1} \alpha_k |k\rangle \right\|^2 = \sum_{k=0}^{L-1} |\alpha_k|^2$. Then correspondingly the almost-uniform-sum vector $(\sum_{k=0}^{L-1} \alpha_k |k\rangle + \sum_{k=L}^{N-1} |k\rangle)$ of the direct-sum subspace $VS(N)$ in the almost-full-uniform PDH operator $H_{N_T,N}^{0,L}$ of (4.42) with respect to the direct-sum subspace $VS(N)$ is still able to reflect the dimensional size $N$ of the direct-sum subspace $VS(N)$ of the total Hilbert space $S_T(N_T)$.

A variety of the pseudo-diagonal Hermitian operators may have important applications in the subspace-selective unitary manipulation in the next Sections.

## 4.5. The unitary operators generated by the PDH operators

The pseudo-diagonal Hermitian operators $H_N^{pd}$ of (4.3) are Hermitian. Then the unitary propagators (or operators) $U_N^{pd}(\tau)$ may be generated directly by the PDH Hamiltonian operators $H_N^{pd}$ and can be expressed as

$$U_N^{pd}(\tau) = \exp\left(-iH_N^{pd}\tau/\hbar\right) \tag{4.45}$$

with time interval $\tau$. The unitary operator $U_N^{pd}(\tau)$ is the Pseudo-Diagonal-Hermitian-Operator-(PDHO)-generated unitary operator and may be simply called the PDHO-generated unitary operator. However, it does not have the specific form (4.3) or (4.4) of a PDH operator $H_N^{pd}$. It is a unitary exponential operator. It can be expanded in the multiple-quantum operator algebra space. Below a finite and compact operator-expansion form for the PDHO-generated unitary exponential operator $U_N^{pd}(\tau)$ is exactly derived in the multiple-quantum operator algebra space. This finite and compact operator-expansion form plays a crucial role in calculating exactly and conveniently the time-evolutional process that is governed by the PDHO-generated unitary propagator $U_N^{pd}(\tau)$.

Since the PDHO-generated unitary propagator $U_N^{pd}(\tau)$ of (4.45) is an exponential operator of the PDH operator $H_N^{pd}$, it can be generally expanded as a power series of the PDH operator $H_N^{pd}$,

$$U_N^{pd}(\tau) = \sum_{k=0}^{\infty} \frac{1}{k!}\left(-iH_N^{pd}\tau/\hbar\right)^k \tag{4.46}$$

This is an infinite power series of operators. It could not be helpful for calculating exactly and conveniently the time-evolutional process that is governed by the PDHO-generated unitary propagator $U_N^{pd}(\tau)$. Here one wants to obtain a finite and compact operator-expansion form for the PDHO-generated unitary exponential operator $U_N^{pd}(\tau)$ in the multiple-quantum operator algebra space.

First, one needs to calculate exactly the $k-$th power operator $\left(H_N^{pd}\right)^k$ of the PDH operator $H_N^{pd}$ for any integer $k \geq 0$. Here the explicit expression for the PDH operator $H_N^{pd}$ is given by (4.3). Obviously, it is trivial for the cases $k = 0, 1$. Then with the help of the expression (4.3) of the PDH operator $H_N^{pd}$ the square operator $(H_N^{pd})^2$ can be expressed as

$$\left(H_N^{pd}\right)^2 = \left(\sum_{k'=0}^{N-1} \alpha_{k'} \left|k'\right\rangle\right)\left(\sum_{l=0}^{N-1}\sum_{k=0}^{N-1} \alpha_l^* \alpha_k \left\langle l|k\right\rangle\right)\left(\sum_{l'=0}^{N-1} \alpha_{l'}^* \left\langle l'\right|\right) \tag{4.47}$$

where $\epsilon_{pd} = \pm 1$ and $(\epsilon_{pd})^2 = 1$ are already used. Note that the base vectors $\{|k\rangle\}$ are orthonormal, i.e., $\langle l|k\rangle = \delta_{lk}$. Then one has

$$\sum_{l=0}^{N-1}\sum_{k=0}^{N-1}\alpha_l^*\alpha_k\langle l|k\rangle = \sum_{l=0}^{N-1}\sum_{k=0}^{N-1}\alpha_l^*\alpha_k\delta_{lk} = \sum_{k=0}^{N-1}|\alpha_k|^2 = \left\|H_N^{pd}\right\| \tag{4.48}$$

where the last equality is obtained from (4.29). By substituting (4.48) into (4.47) the operator $(H_N^{pd})^2$ is reduced to the form

$$\left(H_N^{pd}\right)^2 = \left\|H_N^{pd}\right\|\left(\sum_{k'=0}^{N-1}\alpha_{k'}|k'\rangle\right)\left(\sum_{l'=0}^{N-1}\alpha_{l'}^*\langle l'|\right) \tag{4.49}$$

Furthermore, by using the expression (4.3) of the PDH operator $H_N^{pd}$ and noticing that $\epsilon_{pd} = \pm 1$ and $(\epsilon_{pd})^2 = 1$ it can be found from (4.49) that the operator $(H_N^{pd})^2$ may be reduced to the simple form

$$\left(H_N^{pd}\right)^2 = \left\|H_N^{pd}\right\|\left(\epsilon_{pd}H_N^{pd}\right) \tag{4.50}$$

This operator identity is basic. Below it is used repeatedly. On the basis of the operator identity (4.50) one can further obtain the recursive relations for the $k-$th power operator $(H_N^{pd})^k$ for $k = 1, 2, 3, ....$ Generally for $k \geq 1$ one has

$$\left(H_N^{pd}\right)^k = \left\|H_N^{pd}\right\|^{k-1} H_N^{pd} \text{ for } k = 1, 3, 5, ... \tag{4.51a}$$

$$\left(H_N^{pd}\right)^k = \left\|H_N^{pd}\right\|^{k-1}\left(\epsilon_{pd}H_N^{pd}\right) \text{ for } k = 2, 4, 6, ... \tag{4.51b}$$

Now by substituting these operator identities of (4.51) into the power series (4.46) of the PDHO-generated unitary propagator $U_N^{pd}(\tau)$ one obtains

$$U_N^{pd}(\tau) = \mathbf{E} + \sum_{k=1,3,5,...}^{\infty}\frac{1}{k!}\left(-iH_N^{pd}\tau/\hbar\right)^k + \sum_{k=2,4,6,...}^{\infty}\frac{1}{k!}\left(-iH_N^{pd}\tau/\hbar\right)^k$$

$$= \mathbf{E} + \frac{1}{\left\|H_N^{pd}\right\|}\left(\epsilon_{pd}H_N^{pd}\right)\left(\begin{array}{c}\sum_{k=1,3,5,...}^{\infty}\frac{1}{k!}\left(-i\epsilon_{pd}\tau/\hbar\right)^k\left\|H_N^{pd}\right\|^k \\ +\sum_{k=2,4,6,...}^{\infty}\frac{1}{k!}\left(-i\epsilon_{pd}\tau/\hbar\right)^k\left\|H_N^{pd}\right\|^k\end{array}\right)$$

where $\mathbf{E}$ is the unity operator of the multiple-quantum operator algebra space, and these relations $(\epsilon_{pd})^k = 1$ for even index $k$ and $(\epsilon_{pd})^k = \epsilon_{pd}$ for odd index $k$ are already used. Therefore, finally one has

$$U_N^{pd}(\tau) = \mathbf{E} + \frac{\left(-1 + \exp\left(-i\epsilon_{pd}\left\|H_N^{pd}\right\|\tau/\hbar\right)\right)}{\left\|H_N^{pd}\right\|}\left(\epsilon_{pd}H_N^{pd}\right) \tag{4.52}$$

where the PDH operator $H_N^{pd}$ is given by (4.3). This is a finite operator-expansion form and a compact exact expression for the PDHO-generated unitary operator $U_N^{pd}(\tau)$ of (4.45). The formula (4.52) also is an operator identity for the PDHO-generated unitary operator $U_N^{pd}(\tau)$ in the multiple-quantum operator algebra space. If $\epsilon_{pd} = +1$, then the operator identity (4.52) is reduced to the form

$$U_N^{pd}(\tau) = \mathbf{E} + \frac{\left(-1 + \exp\left(-i \left\| H_N^{pd} \right\| \tau/\hbar\right)\right)}{\left\| H_N^{pd} \right\|} H_N^{pd} \tag{4.53}$$

where the PDH operator $H_N^{pd}$ is given by (4.4).

Below the symmetrical structure of the Hilbert space of the quantum system under study such as a quantum spin system is taken into account in the PDHO-generated unitary operator $U_N^{pd}(\tau)$ of (4.52) which is generated by the PDH operator $H_N^{pd}$ of (4.3). According to the Subsection 4.4, here still suppose that the symmetric structure of the Hilbert space $S(N)$ of the spin system under study is characterized by the direct-sum decomposition (4.32) of the Hilbert space $S(N)$. When the symmetrical structure of the Hilbert space $S(N)$, i.e., the direct-sum decomposition of (4.32), is taken into account, it can be found from (4.35) that the PDH operator $H_N^{pd}$ of (4.3) may be rewritten in the form

$$\epsilon_{pd} H_N^{pd} = \left( \sum_{k=0}^{N_1-1} \alpha_{1,k} \left|\phi_k^{s_1}\right\rangle + \sum_{k=0}^{N_2-1} \alpha_{2,k} \left|\phi_k^{s_2}\right\rangle + ... + \sum_{k=0}^{N_K-1} \alpha_{K,k} \left|\phi_k^{s_K}\right\rangle \right) \langle h.c.| \tag{4.54}$$

By substituting (4.54) into (4.52) the PDHO-generated unitary propagator $U_N^{pd}(\tau)$ of (4.52) may be written in the operator-expansion form

$$U_N^{pd}(\tau) = \mathbf{E} + \frac{\left(-1 + \exp\left(-i\epsilon_{pd} \left\| H_N^{pd} \right\| \tau/\hbar\right)\right)}{\left\| H_N^{pd} \right\|}$$
$$\times \left( \sum_{k=0}^{N_1-1} \alpha_{1,k} \left|\phi_k^{s_1}\right\rangle + \sum_{k=0}^{N_2-1} \alpha_{2,k} \left|\phi_k^{s_2}\right\rangle + ... + \sum_{k=0}^{N_K-1} \alpha_{K,k} \left|\phi_k^{s_K}\right\rangle \right) \langle h.c.| \tag{4.55}$$

where the aforementioned symmetrical structure of the Hilbert space $S(N)$, i.e., the direct-sum decomposition of (4.32), is already considered explicitly. This is a general operator-expansion expression for the PDHO-generated unitary propagator $U_N^{pd}(\tau)$ of (4.45), which is generated by the PDH operator $H_N^{pd}$ of (4.3) and takes into account the symmetrical structure of the Hilbert space $S(N)$, in the multiple-quantum operator algebra space which corresponds to the Hilbert space $S(N)$.

# PART II. The subspace-selective unitary manipulation based on the Hilbert-space symmetrical structures

In quantum mechanics [3,2,1] a subspace of the Hilbert space of a quantum system generally refers to a direct-sum subspace. In the simplest case the Hilbert space is the direct sum of a subspace and its orthogonal complement subspace. Generally, in quantum mechanics [3] the Hilbert space may be the direct sum of many subspaces which are mutually orthogonal. This is the so-called direct-sum decomposition of the Hilbert space. The subspace-selective unitary manipulation [7] is directly related to the direct-sum subspaces of the Hilbert space of a quantum system. Initially it works on the Hilbert space of a quantum spin system [7]. It is closely related to the subspace-selective MQT operators and the subspace-selective multiple-quantum-transition processes between different direct-sum subspaces of the Hilbert space of the quantum spin system. It also have been used in the $HSSS$ unstructured quantum search algorithm [Ref[10]]. Theoretically it is based on the symmetrical structures of the Hilbert space of a quantum system (e.g., a quantum spin system) which are characterized by the direct-sum decomposition of the Hilbert space.

In quantum-computing and quantum-simulating research field a subspace of the Hilbert space of a composite quantum system also may refer to a tensor-product subspace of the Hilbert space. A tensor-product subspace is really the Hilbert space of a subsystem of the composite quantum system. As a typical example, a tensor-product subspace may be the Hilbert space of a spin$-1/2$ subsystem which consists of several spin$-1/2$ particles of the $n$ spin$-1/2$ particles of the composite $n - spin - 1/2$ system. The subspace-selective unitary manipulation is able to take into account a tensor-product subspace of the Hilbert space. That is, it is able to take into account the tensor-product symmetrical structure of a composite quantum system.

According to the quantum-computing speedup theory the multiple-quantum operator algebra space is the central place where the fundamental quantum-computing-speedup resources are exploited to speed up essentially quantum computing and quantum simulating. Here the fundamental quantum-computing-speedup resources are the symmetrical structures and properties of a quantum system, while the latter may be specified via the multiple-quantum operator algebra space, the density operator space, and/or the Hilbert space of the quantum system; and hence the fundamental quantum-computing-speedup resources may exist in these different kinds of basic quantum spaces for a same quantum system. Now the multiple-quantum operator algebra space is positioned as the central place where one makes use of the fundamental quantum-computing-speedup resources to speed up essentially quantum computing and quantum simulating. Then those fundamental quantum-computing-speedup resources which are original from the corresponding Hilbert space must be considered explicitly in the multiple-quantum operator algebra space. When the fundamental quantum-computing-speedup resources are employed to speed up quantum computing and quantum simulating, how can those resources which are original from the corresponding Hilbert space be explicitly taken into account in the multiple-quantum operator algebra space? This is an important problem that needs to be solved in the quantum-computing speedup theory. The subspace-selective unitary manipulation [7] is able to solve this important problem.

In quantum mechanics [1,3,2] any linear operators are defined on the Hilbert space of a quantum system (i.e., a linear vector space). Then the linear operator space of a quantum system may connect to its corresponding Hilbert space on which any linear operators of the linear operator space are defined. Consequently the symmetrical structures and properties of the Hilbert space may connect to the counterpart of the corresponding linear operator space such as the multiple-quantum operator algebra space. This results in that theoretically it becomes possible to make use of the fundamental quantum-computing-speedup resources original from the corresponding Hilbert space to speed up essentially quantum computing and quantum simulating in the multiple-quantum operator algebra space. Of course, there exist the symmetrical structures and properties of the multiple-quantum operator algebra space which are independent upon the corresponding Hilbert space. Generally a linear operator space is much more complicated than its corresponding Hilbert space.

In this paper the subspace-selective unitary manipulation aims to harness the fundamental quantum-computing-speedup resources which are original from the corresponding Hilbert space to speed up essentially quantum computing and quantum simulating in the multiple-quantum operator algebra space. Theoretically it is based on the symmetrical structures of the Hilbert space which are characterized by the direct-sum decomposition of the Hilbert space.

Theoretically the subspace-selective unitary manipulation is allowed to be applied to the chosen subspaces of the multiple-quantum operator algebra space [4], but such situation is usually much more complicated. In this paper the subspace-selective unitary manipulation is based on the chosen direct-sum subspaces of the Hilbert space [7], although it is really performed in the corresponding multiple-quantum operator algebra space. Here the symmetrical structures of the Hilbert space are still characterized via the direct-sum decomposition of the Hilbert space. Now on the one hand the subspace-selective unitary manipulation is generally related to the direct-sum subspaces of the Hilbert space. On the other hand the symmetrical structures and properties of the Hilbert space may connect to the counterpart of the corresponding the multiple-quantum operator algebra space. Then the subspace-selective unitary manipulation, which takes into account the symmetrical structures and properties of the corresponding Hilbert space, may be really performed in the multiple-quantum operator algebra space and has a real effect on the multiple-quantum operator algebra space. Consequently the fundamental quantum-computing-speedup resources which are original from the symmetrical structures and properties of the corresponding Hilbert spaces can be explicitly taken into account in the multiple-quantum operator algebra space.

The Part II in this paper consists of the Sections 5 and 6 and is devoted to investigating the subspace-selective unitary manipulation based on the Hilbert-space symmetrical structures. In addition the research work in the Section 4 also is closely related to the subspace-selective unitary manipulation. As shown in the Section 4, with the aid of the pseudo-diagonal Hermitian operators the symmetric structures and properties of a quantum system which are original from the Hilbert space can be easily taken into account explicitly in the multiple-quantum operator algebra space. Just like the PDH operators $\{H_N^{pd}\}$ of (4.3), in the Section 5 below a second kind of the Hermitian operators, i.e., the subspace-selective multiple-quantum-transition (MQT) Hermitian operators (See Subsection 5.2), are proposed to act as bridge to connect the Hilbert space to the corresponding multiple-quantum operator algebra space in the aspect of the symmetrical structures and properties. In the Section 6 below the Hilbert-space-enlarging processes are described, constructed, and realized concretely. They are one kind of the subspace-selective unitary manipulation and may be performed in the multiple-quantum operator algebra space. They are deliberately designed to make use of the fundamental quantum-computing-speedup resources original from the corresponding Hilbert space to achieve essential quantum-computing speedup in the multiple-quantum operator algebra space. As the specific kind of the subspace-selective unitary manipulation, the Hilbert-space-enlarging processes also work on the basis of the direct-sum subspaces of the Hilbert space, but they are able to take into account the tensor-product symmetrical structure of the Hilbert space of a composite quantum system under the specific (direct-sum-decomposition) Hilbert-space symmetrical structure. According to the quantum-computing speedup theory [Ref[1]] the tensor-product symmetrical structure of the Hilbert space is also the fundamental quantum-computing-speedup resource. It is expected that the Hilbert-space-enlarging processes have important applications in quantum computing and quantum simulating.

# 5. The subspace-selective Multiple Quantum Transition (MQT) operators

## 5.1. The selective and subspace-selective quantum transition operators

A $p$-order selective quantum transition operator usually refers to the spectral-line-selective (or base-vector-selective) quantum-transition operator with a given quantum-transition order $p$. In a general quantum spin system it may be defined on the basis of the representation that is defined by the total spin operator $I_z$ of the spin system (See the Section A). A subspace-selective quantum transition operator [7] with a given order $p$ may be defined in a general quantum spin system such as an $n - spin - 1/2$ system on the basis of the representation defined by the total spin operator $I_z$ and moreover, under the condition that the symmetrical structures and properties of the Hilbert space of the spin system are taken into account explicitly. Especially it is defined on the basis of the direct-sum subspaces [7] of the Hilbert space of the spin system, where the Hilbert-space symmetrical structure is specified by the direct-sum decomposition of the Hilbert space. In this paper any subspace-selective multiple-quantum-transition (MQT) operators are defined on the basis of the Hilbert-space symmetric structures and properties which are specified by the direct-sum decomposition of the Hilbert space and the representation defined by the total spin operator $I_z$.

The $p-$order selective quantum transition operators and the Hermitian selective $|p|-$quantum transition operators are basic and may be further used to construct any subspace-selective MQT operators which may own one given order $p$ or many different orders $\{p\}$. The subspace-selective MQT unitary operators may be used to realize the subspace-selective unitary manipulation [7] in the multiple-quantum operator algebra space.

### 5.1.1. Definition of the selective quantum transition operators

It is known in the Section A that for a $p-$order quantum transition operator $Q_p$ defined by (A1.5) the quantum-transition order $p$ is a relative quantity ($p = M_f - M_i$). It may be independent of any initial spin magnetic quantum number $M_i$ and hence it may be independent of any initial state vector of the spin system under study. The $p-$order quantum transitions of a spin system usually can be excited, if the spin system in any initial state vector is influenced by a $p-$order quantum transition operator $Q_p$. Such $p-$order quantum transition operator $Q_p$ is usually said non-selective (See the Section A). For example, as shown in the Section A, the non-selective multiple-quantum transition operators may include the zero-order quantum transition operators $I_k^{\pm} I_l^{\mp}$, the $\pm 2-$order quantum transition operators $I_k^{\pm} I_l^{\pm}$, and the $\pm 1-$order quantum transition operators $I_k^{\pm}$, etc., where the $k-$th and $l-$th spin particles may be any pair of the spin$-1/2$ particles of the $n - spin - 1/2$ system. The formal definition (A1.5) of a $p-$order quantum transition operator $Q_p$ may be generally used for any non-selective MQT operators, as shown in the Section A. However, it alone may not be sufficient to specify a $p-$order selective quantum-transition operator. This can be shown by the $'$non-selective$'$ MQT product operators $\{O_S\}$ of the $n - spin - 1/2$ system in the Subsection A1.4 of the Section A.

A $p-$order selective quantum-transition operator of a general spin system still may be defined on the basis of the representation that is defined by the total $z-$component spin operator $I_z$ of the total spin angular momentum operator $\mathbf{I}$ (in unit $\hbar = 1$) of the spin system. Suppose that the initial eigenvector $|\Psi_i\rangle$ is a given eigenbase vector of the total spin operator $I_z$ which obeys the eigenvalue equation:

$$I_z |\Psi_i\rangle = M_i |\Psi_i\rangle ,$$

where the eigenvalue is a value $M_i$ of the total spin magnetic quantum number. A $p-$order selective quantum transition operator $Q_p$ then may be defined by

$$|\Psi_f\rangle = Q_p |\Psi_i\rangle , \tag{5.1}$$

where the final eigenvector $|\Psi_f\rangle$ also is an eigenbase vector of the total spin operator $I_z$ and obeys the eigenvalue equation $I_z |\Psi_f\rangle = M_f |\Psi_f\rangle$ with the eigenvalue $M_f = M_i + p$. The physical meaning for the $p-$order selective quantum transition operator $Q_p$ is that the unique $p-$order selective quantum transition between the two eigenbase vectors $|\Psi_i\rangle$ and $|\Psi_f\rangle$ may be induced by the $p-$order selective quantum-transition operator $Q_p$. Obviously, the definition (5.1) can be changed to the formula $I_z Q_p |\Psi_i\rangle = (M_i + p) Q_p |\Psi_i\rangle$ and the latter is just the formal definition (A1.5) of a $p-$order quantum transition operator in the

Section A. From this point one may say the definition (5.1) is equivalent to the formal definition (A1.5). However, in the definition (5.1) the initial $|\Psi_i\rangle$ and the final eigenvector $|\Psi_f\rangle$ are limited to be the two chosen eigenbase vectors of the total spin operator $I_z$, respectively. This means that any multiple-quantum transitions of the spin system under study can not be induced by the $p-$order selective MQT operator $Q_p$ of (5.1) except the $p-$order selective quantum transition between the chosen eigenbase vectors $|\Psi_i\rangle$ and $|\Psi_f\rangle$ . The $p-$order selective MQT operator $Q_p$ tends to be related to the initial $|\Psi_i\rangle$ and the final eigenbase vector $|\Psi_f\rangle$ . In contrast, a $p-$order non-selective MQT operator $Q_p$ defined by (A1.5) is usually independent of any initial $|\Psi_i\rangle$ or final eigenvector $|\Psi_f\rangle$ , as shown in the Section A.

Generally, the definition (5.1) for a $p-$order selective MQT operator $Q_p$ also can be used for a general selective MQT operator, if the initial eigenvector $|\Psi_i\rangle$ and the final eigenvector $|\Psi_f\rangle$ are not strictly limited to the two chosen eigenbase vectors of the total spin operator $I_z$, respectively.

The $p-$order selective MQT operator $Q_p$ defined by (5.1) may not be Hermitian. A Hermitian selective $|p|-$quantum transition operator $Q_{|p|}$ may be constructed by $Q_{|p|} = \frac{1}{2}\left(Q_p + Q_p^+\right)$ or by $Q_{|p|} = \frac{1}{2i}\left(Q_p - Q_p^+\right)$ (See Ref.[7] and (A1.6) in the Section A). This is just like a Hermitian non-selective $|p|-$quantum transition operator $Q_{|p|}$ (See (A1.6)).

The $p-$order selective MQT operator $Q_p$ of (5.1) may be given a new name $Q_p^{fi}$ so as to label the chosen eigenbase vectors $|\Psi_i\rangle$ and $|\Psi_f\rangle$ and distinguish it from its non-selective counterpart. The simplest $p-$order selective MQT operator $Q_p^{fi}$ may be given by $Q_p^{fi} = |\Psi_f\rangle\langle\Psi_i|$ in accordance with the definition (5.1). Then $(Q_p^{fi})^+ = Q_{-p}^{if} = |\Psi_i\rangle\langle\Psi_f|$ . Now the Hermitian selective $|p|-$quantum transition operator $Q_{|p|}^{fi}$ may be constructed by $Q_{|p|}^{fi} = \frac{1}{2}\left(Q_p^{fi} + (Q_p^{fi})^+\right)$ or by $Q_{|p|}^{fi} = \frac{1}{2i}\left(Q_p^{fi} - (Q_p^{fi})^+\right)$ , where the $p-$order selective MQT operator $Q_p^{fi}$ induces the quantum transition from $|\Psi_i\rangle$ to $|\Psi_f\rangle$ , while the $-p-$order selective MQT operator $(Q_p^{fi})^+ = Q_{-p}^{if}$ induces the quantum transition from $|\Psi_f\rangle$ to $|\Psi_i\rangle$ . Therefore, the Hermitian selective $|p|-$quantum transition operators $Q_{|p|}^{fi}$ may be defined by [7] (See also the formal definition (A1.6) in the Section A):

$$Q_{|p|,x}^{fi} = \frac{1}{2}\left(Q_p^{fi} + Q_{-p}^{if}\right) = \frac{1}{2}\left(|\Psi_f\rangle\langle\Psi_i| + |\Psi_i\rangle\langle\Psi_f|\right) \tag{5.2a}$$

$$Q_{|p|,y}^{fi} = \frac{1}{2i}\left(Q_p^{fi} - Q_{-p}^{if}\right) = \frac{1}{2i}\left(|\Psi_f\rangle\langle\Psi_i| - |\Psi_i\rangle\langle\Psi_f|\right) \tag{5.2b}$$

The Hermitian operator $Q_{|p|,\mu}^{fi}$ with $\mu = x, y$ may be called the base-vector-selective $|p|-$quantum-transition operators, because it can cause only the selective quantum transition between the two chosen eigenbase vectors $|\Psi_f\rangle$ and $|\Psi_i\rangle$. For convenience in future applications here a Hermitian base-vector-selective diagonal operator $Q_{|p|,z}^{fi}$ is defined by

$$Q_{|p|,z}^{fi} = \frac{1}{2}\left(|\Psi_f\rangle\langle\Psi_f| - |\Psi_i\rangle\langle\Psi_i|\right) \tag{5.2c}$$

The base-vector-selective diagonal operator $Q^{fi}_{|p|,z}$ is not a multiple-quantum transition operator, but it is closely related to the Hermitian selective $|p|$-quantum transition operators $Q^{fi}_{|p|,x}$ and $Q^{fi}_{|p|,y}$. For convenience these three Hermitian operators $\{Q^{fi}_{|p|,x}, Q^{fi}_{|p|,y}, Q^{fi}_{|p|,z}\}$ of (5.2) may be called the selective-$|p|$-quantum-transition Hermitian operators $\{Q^{fi}_{|p|,\lambda}\}$ with $\lambda = x, y, z$ in unified form.

The selective-$|p|$-quantum-transition Hermitian operators $\{Q^{fi}_{|p|,\lambda}\}$ of (5.2) with $\lambda = x, y, z$ may connect to the Hermitian pseudospin operators $\{Q^{KL}_{\lambda}\}$ of (3.29) and vice versa. Actually, by comparing (5.2) with (3.29) it can be found that the Hermitian selective $|p|$ $-$quantum transition operators $Q^{fi}_{|p|,x}$ and $Q^{fi}_{|p|,y}$ are the Hermitian pseudospin operators $Q^{KL}_{x}$ and $Q^{KL}_{y}$, respectively, and the base-vector-selective diagonal operator $Q^{fi}_{|p|,z}$ is the Hermitian pseudospin operator $Q^{KL}_{z}$, if $|\Psi_f\rangle$ and $|\Psi_i\rangle$ are equal to $|K\rangle$ and $|L\rangle$ of (3.29), respectively.

The unitary exponential operator $R^{fi}_{|p|,\lambda}(\theta)$ that is generated by the selective-$|p|$-quantum-transition Hermitian operator $Q^{fi}_{|p|,\lambda}$ of (5.2) is written as

$$R^{fi}_{|p|,\lambda}(\theta) = \exp\left(-i\theta Q^{fi}_{|p|,\lambda}\right), \lambda = x, y, z \qquad (5.3a)$$

The physical meaning for the unitary operator $R^{fi}_{|p|,\mu}(\theta)$ with $\mu = x, y$ is explained as follows. For simplicity, here consider that both the spin Hamiltonian $H_s$ and the total spin operator $I_z$ of the spin system under study (e.g., an $n - spin - 1/2$ system) have the common eigenbase vectors which include $|\Psi_f\rangle$ and $|\Psi_i\rangle$. An individual selective $|p|$ $-$quantum transition between the two spin energy levels $|\Psi_f\rangle$ and $|\Psi_i\rangle$ may be excited by the unitary operator $R^{fi}_{|p|,\mu}(\theta)$, while this selective $|p|$ $-$quantum transition results in a single spectral line in the (well-resolved) magnetic resonance spectrum [11] of the $n - spin - 1/2$ system or in the atomic spectrum of an $n-$pseudospin$-1/2$ system and so on (See the Subsection 3.2.2). Therefore, the unitary operator $R^{fi}_{|p|,\mu}(\theta)$ with $\mu = x, y$ can cause the spectral-line-selective excitation of the $|p|$ $-$quantum transition. It represents a spectral-line-selective excitation pulse. Applications of the spectral-line-selective excitation-pulse experimental techniques are very popular in nuclear magnetic resonance selective-pulse experiments (See, e.g., [Ref[4]], Ref.[11]).

The unitary exponential operator $R^{fi}_{|p|,\lambda}(\theta)$ ($\lambda = x, y, z$) of (5.3a) also may be generally written in the operator-expansion form [7]:

$$R^{fi}_{|p|,\lambda}(\theta) = E + \left(-1 + \cos\frac{1}{2}\theta\right)\left(|\Psi_f\rangle\langle\Psi_f| + |\Psi_i\rangle\langle\Psi_i|\right) - i2Q^{fi}_{|p|,\lambda}\sin\frac{1}{2}\theta \qquad (5.3b)$$

This operator-expansion expression is more useful in exact calculation of the unitary transformations of the unitary operator $R^{fi}_{|p|,\lambda}(\theta)$. The basic unitary transformations in the Hilbert space for the selective$-$ $|p|$ $-$quantum-transition unitary operator $R^{fi}_{|p|,\lambda}(\theta)$ of (5.3a) with $\lambda = x, y, z$ may be easily set up with the help of the operator-expansion expressions of (5.3b). They are given respectively by

$$\exp\left(-i\theta Q_{|p|,x}^{fi}\right)|\Psi_f\rangle = |\Psi_f\rangle\cos\frac{1}{2}\theta - i\,|\Psi_i\rangle\sin\frac{1}{2}\theta \quad (5.4a)$$

$$\exp\left(-i\theta Q_{|p|,x}^{fi}\right)|\Psi_i\rangle = |\Psi_i\rangle\cos\frac{1}{2}\theta - i\,|\Psi_f\rangle\sin\frac{1}{2}\theta \quad (5.4b)$$

$$\exp\left(-i\theta Q_{|p|,y}^{fi}\right)|\Psi_f\rangle = |\Psi_f\rangle\cos\frac{1}{2}\theta + |\Psi_i\rangle\sin\frac{1}{2}\theta \quad (5.4c)$$

$$\exp\left(-i\theta Q_{|p|,y}^{fi}\right)|\Psi_i\rangle = |\Psi_i\rangle\cos\frac{1}{2}\theta - |\Psi_f\rangle\sin\frac{1}{2}\theta \quad (5.4d)$$

$$\exp\left(-i\theta Q_{|p|,z}^{fi}\right)|\Psi_f\rangle = \exp\left(-i\theta/2\right)|\Psi_f\rangle, \exp\left(-i\theta Q_{|p|,z}^{fi}\right)|\Psi_i\rangle = \exp\left(i\theta/2\right)|\Psi_i\rangle \quad (5.4e)$$

and if $|\Psi_j\rangle \neq |\Psi_f\rangle, |\Psi_i\rangle$, then there are the identical transformations:

$$\exp\left(-i\theta Q_{|p|,\lambda}^{fi}\right)|\Psi_j\rangle = |\Psi_j\rangle \text{ for } |\Psi_j\rangle \neq |\Psi_f\rangle, |\Psi_i\rangle \text{ and } \lambda = x, y, z \quad (5.4f)$$

Here $|\Psi_f\rangle$, $|\Psi_i\rangle$, and $|\Psi_j\rangle$ are any orthonormal eigenbase vectors of the total spin operator $I_z$ of the Hilbert space of the spin system under study. It can be found that these basic unitary transformations of (5.4) in the Hilbert space for the selective$-|p|-$quantum-transition unitary operators $\exp\left(-i\theta Q_{|p|,\lambda}^{fi}\right)$ ($\lambda = x, y, z$) correspond one-to-one to those basic unitary transformations of (3.37) in the Hilbert space for the pseudospin rotation operators $\exp\left(-i\theta Q_{\lambda}^{kl}\right)$ in the Subsection 3.2.2.

The subspace-selective unitary manipulation based on the Hilbert-space symmetric structure may be realized with the help of the subspace-selective multiple-quantum-transition operators [7] in the multiple-quantum operator algebra space. The subspace-selective $|p|$-quantum-transition operators may be constructed by starting from the selective-$|p|$-quantum-transition Hermitian operators $\{Q_{|p|,x}^{fi}, Q_{|p|,y}^{fi}, Q_{|p|,z}^{fi}\}$ of (5.2). Here one needs to know what is the subspace-selective concept [7]. The research works in the rest of this Subsection 5.1 are closely related to the subspace-selective concept.

The $p-$order subspace-selective quantum-transition operators are defined on the basis of the symmetrical structure of the Hilbert space of the spin system under study. As a typical example, below consider an $n-spin-1/2$ system whose Hilbert space $HS(N)$ with dimension $N = 2^n$ owns the following symmetric structure. The whole Hilbert space $HS(N)$ of the $n-spin-1/2$ system is divided into the $n+1$ direct-sum subspaces $\{S_{zq}(n/2-M)\}$ [7] (See also the Section A and the Subsection 6.1 below), each one of which is specified by one of the $n+1$ (i.e., $2I+1$) distinct values $M = -I, -I+1, ..., I-1, I$, where $M$ is the total spin magnetic quantum number and $I$ is the total spin quantum number and given by $I = n/2$ for the $n-spin-1/2$ system. These $n+1$ direct-sum subspaces $\{S_{zq}(k)\}$ with $k = n/2 - M_i = 0, 1, ..., n$ are spanned respectively by the $n+1$ vector basis subsets $\{|M_i, K_{M_i}\rangle\}$ of the orthonormal degenerate eigenbase vectors of the total spin operator $I_z$, each one of which corresponds to one of the $n+1$ different eigenvalues $M_i = -I, -I+1, ..., I-1, I$. Here the eigenvalue equation for the total spin operator $I_z$ is $I_z |M_i, K_{M_i}\rangle = M_i |M_i, K_{M_i}\rangle$ , where the eigenvalue $M_i$ is the total spin magnetic quantum number $M$ and the index $K_{M_i}$ distinguishes between different orthonormal degenerate eigenbase vectors $\{|M_i, K_{M_i}\rangle\}$. The vector basis subset $\{|M_i, K_{M_i}\rangle\}$ which spans the direct-sum subspace $S_{zq}(k)$ contains the $d(k)$ orthonormal degenerate eigenbase vectors which correspond to the same eigenvalue $M_i$, while $d(k) = \begin{pmatrix} n \\ k \end{pmatrix}$ [9, 2] with $k = n/2 - M_i = 0, 1, ..., n$ is just the dimensional size of the $k-$th direct-sum subspace $S_{zq}(k)$ of the Hilbert space $HS(N)$ of the $n-spin-1/2$ system. Therefore, these $n+1$ direct-sum subspaces $\{S_{zq}(k)\}$ of the Hilbert space $HS(N)$ have the dimensional sizes $\{d(k)\}$ with $k = 0, 1, ..., n$, respectively.

Now consider the two different direct-sum subspaces $S_{zq}(k_0)$ and $S_{zq}(n-k_0)$ with the same dimensional size $d(k_0) = d(n-k_0)$ in these $n+1$ direct-sum subspaces $\{S_{zq}(k)\}$ of the Hilbert space $HS(N)$, which are characterized by the total spin magnetic-quantum-number values $M_i = n/2 - k_0$ and $-n/2 + k_0$, respectively. These two subspaces $S_{zq}(k_0)$ and $S_{zq}(n-k_0)$ are spanned by the vector basis subsets $\{|M_i, K_{M_i}\rangle\}$ and $\{|M_f, K_{M_f}\rangle\}$, respectively, where the degenerate eigenbase vectors $\{|M_i, K_{M_i}\rangle\}$ and $\{|M_f, K_{M_f}\rangle\}$ with $K_{M_i}, K_{M_f} = 0, 1, .., d(k_0) - 1$ correspond to the eigenvalues $M_i = n/2 - k_0$ and $M_f = -n/2 + k_0$, respectively. Suppose that the chosen eigenbase vectors $|\Psi_i\rangle$ and $|\Psi_f\rangle$ of the total spin operator $I_z$ in the definition (5.1) of the $p-$order selective MQT operator $Q_p^{fi} = |\Psi_f\rangle\langle\Psi_i|$ belong to the two direct-sum subspaces $S_{zq}(k_0)$ and $S_{zq}(n-k_0)$ , respectively. Then the quantum-transition order $p$ is given by $p = M_f - M_i = -n + 2k_0 \neq 0$ for the $p-$order selective MQT operator $Q_p^{fi}$. Note that both the subspaces $S_{zq}(k_0)$ and $S_{zq}(n-k_0)$ have the same dimensional size $d(k_0) = d(n-k_0)$ . Without lost generality, let $|\Psi_i\rangle = |M_i, K_{M_i}\rangle$ and $|\Psi_f\rangle = |M_f, K_{M_f}\rangle$ with $K_{M_f} = K_{M_i}$. Then the $p-$order selective MQT operator $Q_p^{fi}$ may be simply given by $|\Psi_f\rangle\langle\Psi_i| = |M_f, K_{M_f}\rangle\langle M_i, K_{M_i}| = Q_p^{fi,l}$ with the index $l = K_{M_i} = K_{M_f} = 0, 1, .., d(k_0) - 1$. Correspondingly, according to (5.2) the Hermitian selective-$|p|$ $-$quantum-transition operators $Q_{|p|,\mu}^{fi,l}$ with $\mu = x, y$ may be respectively given by [7]

$$Q_{|p|,x}^{fi,l} = \frac{1}{2}\left(Q_p^{fi,l} + Q_{-p}^{if,l}\right) = \frac{1}{2}\left(|M_f, l\rangle\langle M_i, l| + |M_i, l\rangle\langle M_f, l|\right) \tag{5.5a}$$

$$Q_{|p|,y}^{fi,l} = \frac{1}{2i}\left(Q_p^{fi,l} - Q_{-p}^{if,l}\right) = \frac{1}{2i}\left(|M_f, l\rangle\langle M_i, l| - |M_i, l\rangle\langle M_f, l|\right) \tag{5.5b}$$

where $p = -n + 2k_0 \neq 0$ and $l = 0, 1, .., d(k_0) - 1$. There are $d(k_0)$ orthogonal Hermitian selective$-|p|-$quantum-transition operators $\{Q_{|p|,\mu}^{fi,l}\}$ with $l = 0, 1, ..., d(k_0) - 1$. Any pair of the operators $Q_{|p|,\mu}^{fi,l}$ and $Q_{|p|,\mu'}^{fi,l'}$ commute with one another: $\left[Q_{|p|,\mu}^{fi,l}, Q_{|p|,\mu'}^{fi,l'}\right] = 0$ for $l \neq l'$. The unitary exponential operator which is generated by the Hermitian selective-$|p|$ $-$quantum-transition operator

$Q^{fi,l}_{|p|,\mu}$ may be written as

$$U^{fi,l}_{|p|,\mu}(\theta) = \exp\left(-i\theta Q^{fi,l}_{|p|,\mu}\right), \ \mu = x, y \text{ and } l = 0, 1, ..., d(k_0) - 1 \tag{5.6a}$$

Obviously, any unitary operator $U^{fi,l}_{|p|,\mu}(\theta)$ of (5.6a) is the special case of the general unitary operator $R^{fi}_{|p|,\mu}(\theta)$ of (5.3a). Any pair of the unitary operators $U^{fi,l}_{|p|,\mu}(\theta)$ and $U^{fi,l'}_{|p|,\mu'}(\theta)$ for $l \neq l'$ commute with one another due to the commutation relation $\left[Q^{fi,l}_{|p|,\mu}, Q^{fi,l'}_{|p|,\mu'}\right] = 0$ for $l \neq l'$.

It can prove that the eigenbase vector $|M_i, l\rangle$ of the subspace $S_{zq}(k_0)$ can make selective quantum transition to the eigenbase vector $|M_f, l\rangle$ of the subspace $S_{zq}(n - k_0)$ and vice versa under the influence of the unitary operator $U^{fi,l}_{|p|,\mu}(\theta)$, while any other eigenbase vectors than the two eigenbase vectors $|M_i, l\rangle$ and $|M_f, l\rangle$ of the Hilbert space $HS(N)$ are not affected by the unitary operator $U^{fi,l}_{|p|,\mu}(\theta)$. The unitary operator $U^{fi,l}_{|p|,\mu}(\theta)$ $(\mu = x, y)$ may be expanded as [7]:

$$U^{fi,l}_{|p|,\mu}(\theta) = E + \left(-1 + \cos\frac{1}{2}\theta\right)(|M_i, l\rangle\langle M_i, l| + |M_f, l\rangle\langle M_f, l|)$$

$$-i2Q^{fi,l}_{|p|,\mu}\sin\frac{1}{2}\theta \tag{5.6b}$$

This is the special case of the operator-expansion expression (5.3b) of the general unitary operator $R^{fi}_{|p|,\mu}(\theta)$ of (5.3a). Without lost generality, here consider the unitary operator $U^{fi,l}_{|p|,x}(\theta)$ of (5.6) with $\mu = x$. When the eigenbase vector $|M_f, l\rangle$ (or $|M_i, l\rangle$) is acted on by the unitary operator $U^{fi,l}_{|p|,x}(\theta)$, with the help of (5.6b) and (5.5a) it is easy to find that

$$U^{fi,l}_{|p|,x}(\theta)|M_f, l\rangle = |M_f, l\rangle\cos\frac{1}{2}\theta - i|M_i, l\rangle\sin\frac{1}{2}\theta \tag{5.7a}$$

$$U^{fi,l}_{|p|,x}(\theta)|M_i, l\rangle = |M_i, l\rangle\cos\frac{1}{2}\theta - i|M_f, l\rangle\sin\frac{1}{2}\theta \tag{5.7b}$$

Actually these two unitary transformations of (5.7a) and (5.7b) are equal to the basic unitary transformations of (5.4a) and (5.4b), respectively, if in (5.4) the unitary operator $\exp\left(-i\theta Q^{fi}_{|p|,x}\right)$ is set to $U^{fi,l}_{|p|,x}(\theta)$, and $|\Psi_f\rangle$ set to $|M_f, l\rangle$ and $|\Psi_i\rangle$ set to $|M_i, l\rangle$. Then it can be found from (5.7a) and (5.7b) that when $\theta = \pi$, the eigenbase vector $|M_i, l\rangle$ can be completely converted (or make complete quantum transition) into $|M_f, l\rangle$ (up to a phase factor $-i$) and vice versa by the unitary operator $U^{fi,l}_{|p|,x}(\pi)$. Furthermore, it can prove that any other eigenbase vectors $|M_k, l'\rangle$ than the two chosen eigenbase vectors $|M_i, l\rangle$ and $|M_f, l\rangle$ in the Hilbert space $HS(N)$ are not affected by the unitary operator $U^{fi,l}_{|p|,x}(\theta)$, where $M_k \neq M_i$, $M_f$ or $l' \neq l$. Note that the complete vector basis set $\{|M, K_M\rangle\}$ of the Hilbert space $HS(N)$ is formed by the orthonormal eigenbase vectors of the total spin operator $I_z$. Then the eigenbase vectors $\{|M_k, K_{M_k}\rangle\}$ satisfy the orthonormal relations: $\langle M_k, l|M_{k'}, l'\rangle = \delta_{M_k M_{k'}}\delta_{ll'}$ where $l = K_{M_k}$ and $l' = K_{M_{k'}}$. Now with the help of the operator-expansion expression (5.6b) of the unitary operator $U^{fi,l}_{|p|,x}(\theta)$ and the orthonormal relations $\langle M_k, l|M_{k'}, l'\rangle = \delta_{M_k M_{k'}}\delta_{ll'}$ it can prove that

$$U^{fi,l}_{|p|,x}(\theta)\,|M_k, l'\rangle = |M_k, l'\rangle \text{ if } M_k \neq M_i,\ M_f \text{ or } l' \neq l \tag{5.7c}$$

This indicates that the unitary operator $U^{fi,l}_{|p|,x}(\theta)$ does not affect any eigenbase vector $|M_k, l'\rangle$ of the Hilbert space $HS(N)$ that is different from the chosen eigenbase vectors $|M_i, l\rangle$ and $|M_f, l\rangle$.

The three unitary transformations of (5.7) show clearly that the unitary operator $U^{fi,l}_{|p|,x}(\theta)$ is a base-vector-selective unitary operator in the sense that when it acts on the chosen eigenbase vector $|M_i, l\rangle$ and/or $|M_f, l\rangle$, it can cause the selective quantum transition between the two chosen eigenbase vectors $|M_i, l\rangle$ and $|M_f, l\rangle$, and when it acts on any eigenbase vector of the Hilbert space $HS(N)$ that is different from the two chosen eigenbase vectors, it does not affect the eigenbase vector. The unitary operator $U^{fi,l}_{|p|,x}(\theta)$ is also a spectral-line-selective unitary operator of the $|p|$ −quantum transition in the sense that it can cause the spectral-line-selective excitation of the $|p|$ −quantum transition between the two chosen eigenbase vectors $|M_i, l\rangle$ and $|M_f, l\rangle$. The unitary operator $U^{fi,l}_{|p|,x}(\theta)$ seems to be a subspace-selective unitary operator in the sense that the chosen eigenbase vectors $|M_i, l\rangle$ and $|M_f, l\rangle$ belong to the two direct-sum subspaces $S_{zq}(k_0)$ and $S_{zq}(n-k_0)$ of the Hilbert space $HS(N)$, respectively. However, usually it is not called a subspace-selective unitary operator, since one or several chosen eigenbase vectors are not enough to specify a chosen direct-sum subspace of the Hilbert space. Theoretically a general subspace-selective multiple-quantum-transition operator may be constructed by starting from the base-vector-selective $|p|$ −quantum-transition Hermitian operators of (5.5) or (5.2).

### 5.1.2. A subspace-selective $|p|$ −quantum transition operator

Now a subspace-selective $|p|$ −quantum transition operator $H^{M_f,M_i}_{|p|,\mu}$ with the order $p \neq 0$ may be simply constructed by the sum of the $d(k_0)$ base-vector-selective $|p|$ −quantum-transition Hermitian operators $\{Q^{fi,l}_{|p|,\mu}\}$ of (5.5) with $l = 0, 1, ..., d(k_0) - 1$, and it can be expressed as [7]

$$H^{M_f,M_i}_{|p|,\mu} = \sum_{l=0}^{d(k_0)-1} Q^{fi,l}_{|p|,\mu},\ \mu = x, y \tag{5.8}$$

where all these $d(k_0)$ Hermitian operators $\{Q^{fi,l}_{|p|,\mu}\}$ each have the same absolute quantum-transition order $|p|$. Note that $d(k_0)$ $(= d(n-k_0))$ is the dimensional size of any one of the two chosen direct-sum subspaces $S_{zq}(k_0)$ and $S_{zq}(n-k_0)$. The subspace-selective $|p|$ −quantum transition operator $H^{M_f,M_i}_{|p|,\mu}$ can cause only the selective $|p|$ −quantum transitions between the two chosen subspaces $S_{zq}(k_0)$ and $S_{zq}(n-k_0)$. Here only the two direct-sum subspaces $S_{zq}(k_0)$ and $S_{zq}(n-k_0)$ of the Hilbert space $HS(N)$ are involved in all these selective $|p|$ −quantum transitions, while any other direct-sum subspaces of the Hilbert space $HS(N)$ are not. This is the reason why the Hermitian $|p|$ −quantum-transition operator $H_{|p|,\mu}^{M_f,M_i}$ of (5.8) is said subspace-selective.

The subspace-selective $|p|$-quantum-transition unitary operator $U_{|p|,\mu}^{M_f,M_i}(\theta)$ may be generated by the subspace-selective $|p|$-quantum-transition Hermitian operator $H_{|p|,\mu}^{M_f,M_i}$ of (5.8), i.e., $U_{|p|,\mu}^{M_f,M_i}(\theta)=\exp\left(-i\theta H_{|p|,\mu}^{M_f,M_i}\right)$ with $\mu=x,y$, and with the help of (5.8) it can be expressed as [7]

$$U_{|p|,\mu}^{M_f,M_i}(\theta)=\exp\left(-i\theta\sum_{l=0}^{d(k_0)-1}Q_{|p|,\mu}^{fi,l}\right)=\prod_{l=0}^{d(k_0)-1}\exp\left(-i\theta Q_{|p|,\mu}^{fi,l}\right) \tag{5.9}$$

where the second equality holds due to the commuting relations $\left[Q_{|p|,\mu}^{fi,l},Q_{|p|,\mu'}^{fi,l'}\right]$ $=0$ for $l\neq l'$. The formulae (5.9) and (5.6a) show that the subspace-selective $|p|$-quantum-transition unitary operator $U_{|p|,\mu}^{M_f,M_i}(\theta)$ is the product of the $d(k_0)$ commuting unitary operators $\{U_{|p|,\mu}^{fi,l}(\theta)\}$ of (5.6a). The subspace-selective $|p|$-quantum-transition unitary operator $U_{|p|,\mu}^{M_f,M_i}(\theta)$ may be efficiently constructed and realized [7].

It can prove that when the subspace-selective $|p|$-quantum-transition unitary operator $U_{|p|,\mu}^{M_f,M_i}(\theta)$ is selectively applied to any one of the two chosen subspaces $S_{zq}(k_0)$ and $S_{zq}(n-k_0)$ of the Hilbert space $HS(N)$, the selective $|p|$ −quantum transitions can be excited between the two chosen subspaces $S_{zq}(k_0)$ and $S_{zq}(n-k_0)$. Without lost generality, here consider that the unitary operator $U_{|p|,x}^{M_f,M_i}(\theta)$ of (5.9) with $\mu=x$ selectively acts on the subspace $S_{zq}(k_0)$ (or $S_{zq}(n-k_0)$). Now the unitary operator $U_{|p|,x}^{M_f,M_i}(\theta)$ is applied to any eigenbase vector $|M_i,l\rangle$ of the subspace $S_{zq}(k_0)$ with $l=0,1,...,d(k_0)-1$. With the help of (5.7c) and (5.7a) it is easy to obtain [7]:

$$U_{|p|,x}^{M_f,M_i}(\theta)|M_i,l\rangle=\prod_{k=0}^{d(k_0)-1}U_{|p|,x}^{fi,k}(\theta)|M_i,l\rangle=\exp\left(-i\theta Q_{|p|,x}^{fi,l}\right)|M_i,l\rangle$$

$$=|M_i,l\rangle\cos\frac{1}{2}\theta-i|M_f,l\rangle\sin\frac{1}{2}\theta \tag{5.10a}$$

where the last equality is obtained from (5.7a) and the unitary operators $U_{|p|,x}^{fi,k}(\theta)$ of (5.6a) satisfy the commuting relations $\left[U_{|p|,x}^{fi,k}(\theta),U_{|p|,x}^{fi,k'}(\theta)\right]=0$ for $k\neq k'$. In analogous way it can prove that

$$U_{|p|,x}^{M_f,M_i}(\theta)|M_f,l\rangle=|M_f,l\rangle\cos\frac{1}{2}\theta-i|M_i,l\rangle\sin\frac{1}{2}\theta \tag{5.10b}$$

where $|M_f, l\rangle$ with $l = 0, 1, ..., d(k_0) - 1$ is any eigenbase vector of the subspace $S_{zq}(n - k_0)$. Note that $|M_i, l\rangle \in S_{zq}(k_0)$ and $|M_f, l\rangle \in S_{zq}(n - k_0)$ for $l = 0, 1, ..., d(k_0) - 1$. Then the formulae of (5.10) show that the selective $|p|$ −quantum transitions between the eigenbase vectors $\{|M_i, l\rangle\}$ of the subspace $S_{zq}(k_0)$ and the corresponding eigenbase vectors $\{|M_f, l\rangle\}$ of the subspace $S_{zq}(n - k_0)$ can be excited simultaneously by the subspace-selective $|p|$-quantum-transition unitary operator $U_{|p|,x}^{M_f,M_i}(\theta)$. It can be further found from (5.10) that when $\theta = \pi$, any eigenbase vector $|M_i, l\rangle$ $(l = 0, 1, ..., d(k_0) - 1)$ of the subspace $S_{zq}(k_0)$ can make complete quantum transition to the eigenbase vector $|M_f, l\rangle$ of the subspace $S_{zq}(n - k_0)$ (up to a phase factor $-i$) and vice versa by the subspace-selective unitary operator $U_{|p|,x}^{M_f,M_i}(\pi)$.

More generally it can prove that an arbitrary state vector of the subspace $S_{zq}(k_0)$ can make quantum transition to the state vector of the subspace $S_{zq}(n-k_0)$ and vice versa by the subspace-selective unitary operator $U_{|p|,x}^{M_f,M_i}(\theta)$. Note that the eigenbase vector basis subsets $\{|M_i, K_{M_i}\rangle\}$ and $\{|M_f, K_{M_f}\rangle\}$ span the direct-sum subspaces $S_{zq}(k_0)$ and $S_{zq}(n - k_0)$, respectively. Then an arbitrary state vector $|\Psi_i(M_i)\rangle$ of the subspace $S_{zq}(k_0)$ may be generally expressed as $|\Psi_i(M_i)\rangle = \sum_{K_{M_i}=0}^{d(k_0)-1} C_{M_i,K_{M_i}} |M_i, K_{M_i}\rangle$. Similarly, any state vector $|\Psi_f(M_f)\rangle$ of the subspace $S_{zq}(n - k_0)$ may be generally expressed as $|\Psi_f(M_f)\rangle = \sum_{K_{M_f}=0}^{d(k_0)-1} C_{M_f,K_{M_f}} |M_f, K_{M_f}\rangle$. If the subspace-selective unitary operator $U_{|p|,x}^{M_f,M_i}(\theta)$ is selectively applied to the subspace $S_{zq}(k_0)$, then an arbitrary state vector $|\Psi_i(M_i)\rangle$ of the subspace $S_{zq}(k_0)$ can make quantum transition to the subspace $S_{zq}(n - k_0)$ and this can be seen as follows:

$$U_{|p|,x}^{M_f,M_i}(\theta)|\Psi_i(M_i)\rangle = \prod_{k=0}^{d(k_0)-1} \exp\left(-i\theta Q_{|p|,x}^{fi,k}\right) \sum_{l=0}^{d(k_0)-1} C_{M_i,l}|M_i, l\rangle$$

$$= |\Psi_i(M_i)\rangle \cos\frac{1}{2}\theta - i\left|\Psi'_f(M_f)\right\rangle \sin\frac{1}{2}\theta \qquad (5.11a)$$

where the unitary transformations of (5.10a) have already been used and the state vector $\left|\Psi'_f(M_f)\right\rangle = \sum_{l=0}^{d(k_0)-1} C'_{M_f,l}|M_f, l\rangle$ with the expansional coefficient $C'_{M_f,l} = C_{M_i,l}$ belongs to the subspace $S_{zq}(n - k_0)$. In analogous way it can prove that if the subspace-selective unitary operator $U_{|p|,x}^{M_f,M_i}(\theta)$ is selectively applied to the subspace $S_{zq}(n - k_0)$, then any state vector $|\Psi_f(M_f)\rangle$ of the subspace $S_{zq}(n - k_0)$ can make quantum transition to the subspace $S_{zq}(k_0)$,

$$U_{|p|,x}^{M_f,M_i}(\theta)|\Psi_f(M_f)\rangle = |\Psi_f(M_f)\rangle \cos\frac{1}{2}\theta - i|\Psi'_i(M_i)\rangle \sin\frac{1}{2}\theta, \qquad (5.11b)$$

where the unitary transformations of (5.10b) have already been used and the state vector $|\Psi'_i(M_i)\rangle = \sum_{l=0}^{d(k_0)-1} C'_{M_i,l}|M_i, l\rangle$ with the expansional coefficient $C'_{M_i,l} = C_{M_f,l}$ belongs to the subspace $S_{zq}(k_0)$. It can be found from (5.11) that, under the influence of the subspace-selective unitary operator $U_{|p|,x}^{M_f,M_i}(\theta)$ with

$\theta = \pi$, an arbitrary state vector $|\Psi_i(M_i)\rangle$ of the subspace $S_{zq}(k_0)$ can be completely converted into the state vector $\left|\Psi'_f(M_f)\right\rangle$ of the subspace $S_{zq}(n-k_0)$ up to a phase factor $-i$ and conversely any state vector $|\Psi_f(M_f)\rangle$ of the subspace $S_{zq}(n-k_0)$ can also be completely converted into the state vector $|\Psi'_i(M_i)\rangle$ of the subspace $S_{zq}(k_0)$ up to a phase factor $-i$.

Furthermore, it can prove that an arbitrary state vector $|\Psi_k(M_k)\rangle$ of any direct-sum subspace $S_{zq}(k)$ which is not any one of the chosen subspaces $S_{zq}(k_0)$ and $S_{zq}(n-k_0)$ of the Hilbert space $HS(N)$ is not affected by the subspace-selective unitary operator $U^{M_f,M_i}_{|p|,x}(\theta)$. Any state vector $|\Psi_k(M_k)\rangle$ of the direct-sum subspace $S_{zq}(k)$ with dimension $d(k)$ may be expressed as $|\Psi_k(M_k)\rangle = \sum_{K_{M_k}=0}^{d(k)-1} C_{M_k,K_{M_k}} |M_k, K_{M_k}\rangle$, where the subspace $S_{zq}(k)$ is spanned by the eigenbase vector basis subset $\{|M_k, K_{M_k}\rangle\}$. Note that the direct-sum subspace $S_{zq}(k)$ with $k \neq k_0, n-k_0$ is orthogonal to both the chosen direct-sum subspaces $S_{zq}(k_0)$ and $S_{zq}(n-k_0)$. Then every eigenbase vector $|M_k, K_{M_k}\rangle$ of the subspace $S_{zq}(k)$ is orthogonal to every eigenbase vector $|M_i, K_{M_i}\rangle$ of the subspace $S_{zq}(k_0)$ and also every eigenbase vector $\left|M_f, K_{M_f}\right\rangle$ of the subspace $S_{zq}(n-k_0)$. Now the subspace-selective unitary operator $U^{M_f,M_i}_{|p|,x}(\theta)$ acts on any eigenbase vector $|M_k, l'\rangle$ of the subspace $S_{zq}(k)$ with $l' = 0, 1, ..., d(k)-1$. With the help of (5.7c) one can obtain

$$U^{M_f,M_i}_{|p|,x}(\theta)|M_k, l'\rangle = \prod_{l=0}^{d(k_0)-1} \exp\left(-i\theta Q^{fi,l}_{|p|,x}\right)|M_k, l'\rangle = |M_k, l'\rangle \qquad (5.12a)$$

where $l' = 0, 1, ..., d(k)-1$. Then with the help of (5.12a) it can further prove that

$$U^{M_f,M_i}_{|p|,x}(\theta)|\Psi_k(M_k)\rangle = U^{M_f,M_i}_{|p|,x}(\theta)\sum_{l'=0}^{d(k)-1} C_{M_k,l'}|M_k, l'\rangle = |\Psi_k(M_k)\rangle \quad (5.12b)$$

This unitary transformation shows that when an arbitrary state vector $|\Psi_k(M_k)\rangle$ of the subspace $S_{zq}(k)$ is acted on by the subspace-selective unitary operator $U^{M_f,M_i}_{|p|,x}(\theta)$, it keeps unchanged. This indicates that an arbitrary state vector $|\Psi_k(M_k)\rangle$ of any direct-sum subspace $S_{zq}(k)$ which is different from both the chosen subspaces $S_{zq}(k_0)$ and $S_{zq}(n-k_0)$ in the Hilbert space $HS(N)$ is not affected by the subspace-selective unitary operator $U^{M_f,M_i}_{|p|,x}(\theta)$.

With the help of (5.11a), (5.11b), and (5.12b) the physical meaning becomes clear for the subspace-selective $|p|$ −quantum-transition unitary operator $U^{M_f,M_i}_{|p|,\mu}(\theta)$ of (5.9). When the subspace-selective unitary operator $U^{M_f,M_i}_{|p|,\mu}(\theta)$ is selectively applied to any one of the two chosen direct-sum subspaces $S_{zq}(k_0)$ and $S_{zq}(n-k_0)$ of the Hilbert space $HS(N)$, there are the selective $|p|$ −quantum transitions between the two chosen subspaces. When the subspace-selective unitary operator $U^{M_f,M_i}_{|p|,\mu}(\theta)$ is selectively applied to any direct-sum subspace $S_{zq}(k)$ of the Hilbert space $HS(N)$ which is different from both the chosen direct-sum subspaces, there is nothing to occur.

As shown above, under the influence of the subspace-selective unitary operator $U_{|p|,x}^{M_f,M_i}(\pi)$, an arbitrary state vector can make complete quantum transition from one chosen subspace $S_{zq}(k_0)$ (or $S_{zq}(n-k_0)$) to another one $S_{zq}(n-k_0)$ (or $S_{zq}(k_0)$). Moreover, the subspace-selective unitary operator $U_{|p|,\mu}^{M_f,M_i}(\theta)$ may be efficiently constructed and realized [7]. Here an important point is that the two chosen direct-sum subspaces $S_{zq}(k_0)$ and $S_{zq}(n-k_0)$ on which the subspace-selective unitary operator $U_{|p|,\mu}^{M_f,M_i}(\theta)$ acts have the same dimensional size $d(k_0)$. The dimensional sizes of the direct-sum subspaces could play an important role in efficiently constructing a subspace-selective unitary operator or in constructing an efficient conversion process for an arbitrary state vector from one chosen subspace to another. Generally, an arbitrary state vector may be efficiently and completely converted from one direct-sum subspace with small dimensional size into another with large dimensional size [7]. However, the inverse process tends to be inefficient, which converts completely an arbitrary state vector from one large subspace into another small subspace, as shown in Ref. [7].[17]

There is an important physical quantity, i.e., the dimensional size $d(k)$, for a direct-sum subspace $S_{zq}(k)$ of the Hilbert space $HS(N)$. The dimensional sizes $\{d(k)\}$ of the direct-sum subspaces $\{S_{zq}(k)\}$ of the Hilbert space $HS(N)$ of the $n-spin-1/2$ system constitute one important aspect to reflect the symmetrical structures and properties of the spin system [7]. According to the quantum-computing speedup theory [Ref[1]] the symmetrical structures and properties of quantum systems are the fundamental quantum-computing-speedup resources which are responsible for essentially speeding up quantum computing and quantum simulating. Then the dimensional sizes of the direct-sum subspaces of the Hilbert space should be considered as one important resource which may be exploited to realize essential quantum-computing speedup in quantum simulating and quantum computing.

Simply speaking, an arbitrary state vector may make efficiently complete quantum transition from a small direct-sum subspace to another large one under the influence of an appropriate subspace-selective unitary operator [7]. Here the size of a direct-sum subspace of the Hilbert space of quantum system under study is measured by the dimension of the subspace. Hence the dimensional sizes of these direct-sum subspaces of the Hilbert space constitute an important physical quantity [7]. It is still thought in the present work that the direct-sum-subspace dimensional size is an important physical quantity. Rather than the dimensional size of the whole Hilbert space, the dimensional sizes of the direct-sum subspaces of the Hilbert space are really related to the symmetrical structures of the Hilbert space. The dimensional size of any direct-sum subspace is a natural number and has an infinitely high precision. As one important resource for quantum-computing speedup the dimensional sizes of the direct-sum subspaces may have important applications in quantum simulating and quantum computing.

[17] As shown in Ref. [7], the marked state (or the solution state) of an unstructured search problem, which is unknown and may be considered as an arbitary base vector before the solution to the unstructured search problem is found, may be efficiently and completely transferred from one small subspace to another large subspace, while the inverse transfer process is generally inefficient. By the way, the base-vector arrangements may be different between the conventional computational basis set and the complete set $\{|M, K_M\rangle\}$ of the eigenbase vectors of the total spin operator $I_z$, but this difference does not affect this conclusion. Of course, this difference must be taken into account, if one wants to use really the unstructured search algorithm [7] to solve the search problem. Moreover, this search algorithm is involved in the unitary manipulation on the oracle operation (or operator $C_S(\theta)$) such as $WC_S(\theta)W^+$ with the suitable unitary transformation $W$.

### 5.1.3. A subspace-selective zero-quantum transition operator

The second important instance is the subspace-selective zero-quantum transition operators [7], in addition to the first instance in the Subsection 5.1.2, i.e., the subspace-selective $|p|$ −quantum-transition operator $H_{|p|,\mu}^{M_f,M_i}$ of (5.8) with the quantum-transition order $p \neq 0$. Before the subspace-selective zero-quantum transition operators are described and constructed, a general property for any zero-order quantum transition operator is described. This general property is called the zero-quantum invariant property. According to the definition (5.1) of any zero-order selective-quantum-transition operator, when the initial eigenbase vector $|\Psi_i\rangle$ with the eigenvalue $M_i$ is acted on by any zero-order selective-quantum-transition operator $Q_0^{fi}$ with the order $p = 0$, it is changed to the final eigenbase vector $|\Psi_f\rangle = Q_0^{fi}|\Psi_i\rangle$ with the eigenvalue $M_f = M_i + p = M_i$, where $|\Psi_f\rangle \neq |\Psi_i\rangle$. This indicates that both the initial $|\Psi_i\rangle$ and the final eigenbase vector $|\Psi_f\rangle$ of the total spin operator $I_z$ correspond to the same eigenvalue $M_i$ and hence they are degenerate and in the same direct-sum subspace $S_{zq}(k_0)$ with $k_0 = n/2 - M_i$. This also means that if the initial (eigenbase) vector $|\Psi_i\rangle$ is in the subspace $S_{zq}(k_0)$, then the final (eigenbase) vector $|\Psi_f\rangle = Q_0^{fi}|\Psi_i\rangle$, which is generated by acting the zero-order selective-quantum-transition operator $Q_0^{fi}$ on the initial vector $|\Psi_i\rangle$, still remains in the same subspace $S_{zq}(k_0)$. This is a general property for any zero-order quantum transition operator which may be selective or non-selective (See below and the Section A). Here this zero-quantum invariant property is deduced from the definition (5.1) of any zero-order selective-quantum-transition operator. In other words, any zero-order selective-quantum-transition operator defined by (5.1) owns the zero-quantum invariant property.

Now it can be shown that the Hermitian selective zero-quantum-transition (ZQT) operators $Q_{|p|,\mu}^{fi}$ of (5.2a) and (5.2b) with the order $p = 0$ and $\mu = x, y$ also own the zero-quantum invariant property. For convenience here denote $Q_{zq,\lambda}^{fi}$ as the selective zero-quantum-transition (ZQT) Hermitian operator $Q_{|p|,\lambda}^{fi}$ of (5.2) with the order $p = 0$ and $\lambda = x, y, z$. Without lost generality, here consider the selective ZQT Hermitian operator $Q_{zq,x}^{fi}$ (i.e., $Q_{|p|,x}^{fi}$ of (5.2a) with the order $p = 0$). When the initial eigenbase vector $|\Psi_i\rangle$ (or $|\Psi_f\rangle$) is acted on by the selective ZQT Hermitian operator $Q_{zq,x}^{fi}$, with the help of (5.2a) it can be found that

$$Q_{zq,x}^{fi}|\Psi_i\rangle = \frac{1}{2}\left(Q_0^{fi} + Q_{-0}^{if}\right)|\Psi_i\rangle = \frac{1}{2}|\Psi_f\rangle,$$

$$Q^{fi}_{zq,x} |\Psi_f\rangle = \frac{1}{2}\left(Q^{fi}_0 + Q^{if}_{-0}\right)|\Psi_f\rangle = \frac{1}{2}|\Psi_i\rangle .$$

Here both the operators $Q^{fi}_0$ and $Q^{if}_{-0}$ are the zero-order selective-quantum-transition operators in the definition (5.1). Then both the generated eigenvectors $Q^{fi}_0 |\Psi_i\rangle$ and $Q^{if}_{-0} |\Psi_i\rangle$ each have the same eigenvalue $M_i$ as the initial eigenbase vector $|\Psi_i\rangle$ . Therefore, $Q^{fi}_{zq,x} |\Psi_i\rangle = \frac{1}{2}Q^{fi}_0 |\Psi_i\rangle + \frac{1}{2}Q^{if}_{-0} |\Psi_i\rangle$ has the same eigenvalue $M_i$ too. Actually, $Q^{fi}_0 |\Psi_i\rangle = |\Psi_f\rangle$ and $Q^{if}_{-0} |\Psi_i\rangle = 0$ due to $|\Psi_f\rangle \neq |\Psi_i\rangle$ and $\langle\Psi_f|\Psi_i\rangle = 0$. Then the final eigenbase vector $\frac{1}{2}|\Psi_f\rangle = Q^{fi}_{zq,x} |\Psi_i\rangle$ (up to a normalization constant 1/2) has the same eigenvalue $M_i$. This indicates that if the initial eigenbase vector $|\Psi_i\rangle$ is in the subspace $S_{zq}(k_0)$ , then the final eigenbase vector $|\Psi_f\rangle$ or $Q^{fi}_{zq,x} |\Psi_i\rangle$ , which is generated by acting the selective ZQT Hermitian operator $Q^{fi}_{zq,x}$ on the initial vector $|\Psi_i\rangle$, still remains in the same subspace $S_{zq}(k_0)$ [7]. Therefore, the selective ZQT Hermitian operator $Q^{fi}_{zq,x}$ owns the zero-quantum invariant property. In analogous way it can be shown that the selective ZQT Hermitian operator $Q^{fi}_{zq,y}$ owns the zero-quantum invariant property too. By the way, it can be seen that the generated eigenvector $Q^{fi}_{zq,x} |\Psi_f\rangle$ and the initial $|\Psi_i\rangle$ and the final eigenbase vector $|\Psi_f\rangle$ all are in the same subspace $S_{zq}(k_0)$ .

As shown in the Section A, any zero-order non-selective quantum transition operator also owns the zero-quantum invariant property, that is, if any state vector $|\psi\rangle$ of a direct-sum subspace $S_{zq}(k')$ is acted on by any zero-order non-selective quantum transition operator $Q_{zq}$, then the generated vector $Q_{zq}|\psi\rangle$ still remains in the same subspace $S_{zq}(k')$, where the direct-sum subspace $S_{zq}(k')$ may be any one of the direct-sum subspaces $\{S_{zq}(k)\}$ of the Hilbert space $HS(N)$.

A general conclusion therefore is obtained from the above theoretical analysis that any ZQT operator, which may be selective or non-selective and may be Hermitian or non-Hermitian, owns the zero-quantum invariant property. In other words, when any state vector $|\psi\rangle$ of a chosen direct-sum subspace $S_{zq}(k_0)$ is acted on by any zero-order quantum transition operator $Q_0$ which may be selective or non-selective, the generated vector $Q_0|\psi\rangle$ still remains in the same subspace $S_{zq}(k_0)$. (Here the special case $Q_0|\psi\rangle = 0$ is not conflict with the conclusion.)

Moreover, it is easy to deduce from the definition (5.1) or the extended definition (A1.5) (or the equation (A1.27)) in the Section A that any diagonal operator $Q_z$ also owns the same zero-quantum invariant property as any zero-order quantum transition operator $Q_0$. That is, any state vector $|\psi\rangle$ of the subspace $S_{zq}(k_0)$ remains in the same subspace $S_{zq}(k_0)$, when it is acted on by any diagonal operator $Q_z$. (Here the special case $Q_z|\psi\rangle = 0$ is not conflict with the conclusion.)

The above theoretical analysis shows that not only any zero-order quantum transition operators which may be selective or non-selective but also any diagonal operators own the zero-quantum invariant property. In other words, when any state vector $|\psi\rangle$ of a chosen direct-sum subspace $S_{zq}(k_0)$ of the Hilbert space $HS(N)$ is acted on by any diagonal operator $Q_z$ or when it is acted on by any zero-order quantum transition operator $Q_0$, the generated vector ($Q_z |\psi\rangle$ or $Q_0 |\psi\rangle$) still remains in the same direct-sum subspace $S_{zq}(k_0)$. According to quantum mechanics [3] this means that the direct-sum subspace $S_{zq}(k_0)$ is invariant under the set of the zero-quantum operators which is formed by all linearly-independent zero-order quantum transition operators $\{Q_0\}$ and all linearly-independent diagonal operators $\{Q_z\}$. This result can be proven rigorously, as shown in the Section A. It is known from Ref. [4] that the zero-quantum operator subspace is formed by all zero-order quantum transition operators and all diagonal operators of the multiple-quantum operator algebra space. Then the direct-sum subspace $S_{zq}(k_0)$ with $k_0 = n/2 - M_i = 0, 1, ..., n$ is invariant under the zero-quantum ($zq$) operator subspace in accordance with quantum mechanics [3].

A great difference between a subspace-selective (or selective) and a non-selective zero-quantum transition Hermitian operator may be reflected by their generated unitary exponential operators. If the unitary operator that is generated by a subspace-selective (or selective) ZQT Hermitian operator acts on any eigenvector $|\psi\rangle$ of the total spin operator $I_z$, then the eigenvector $|\psi\rangle$ can be affected by the unitary operator only when it belongs to the chosen direct-sum subspace and it can not be affected when it belongs to any other direct-sum subspace than the chosen subspace in the Hilbert space. In contrast, if the unitary operator that is generated by a non-selective ZQT Hermitian operator acts on the eigenvector $|\psi\rangle$, then the eigenvector $|\psi\rangle$ is affected by the unitary operator, no matter which direct-sum subspace the eigenvector $|\psi\rangle$ belongs to in the Hilbert space.

A subspace-selective zero-quantum-transition operator may be constructed as follows. As shown above, the $d(k_0)-$dimensional direct-sum subspace $S_{zq}(k_0)$ with $k_0 = n/2 - M_i$ has the $d(k_0)$ degenerate eigenbase vectors $\{|M_i, K_{M_i}\rangle\}$ of the total spin operator $I_z$ with the index $K_{M_i} = 0, 1, ..., d(k_0) - 1$, which all correspond to the same eigenvalue $M_i$. Now both the initial $|\Psi_i\rangle$ and the final eigenbase vector $|\Psi_f\rangle$ in the selective $ZQT$ Hermitian operators $\{Q_{zq,\lambda}^{fi}\}$ of (5.2) with $\lambda = x, y, z$ and the order $p = 0$ may take any pair of different eigenbase vectors $|M_i, K_{M_i}\rangle$ and $\left|M_i, K'_{M_i}\right\rangle$ of the same subspace $S_{zq}(k_0)$. Then there are $3\begin{pmatrix} d(k_0) \\ 2 \end{pmatrix} = \frac{3}{2} d(k_0)(d(k_0) - 1)$ selective $ZQT$ Hermitian operators $\{Q_{zq,\lambda}^{fi}\}$ in accordance with the formulae of (5.2). These selective $ZQT$ Hermitian operators $\{Q_{zq,\lambda}^{fi}\}$ may be further used to construct a subspace-selective $ZQT$ Hermitian operator whose unitary exponential operator is selectively applied to only the direct-sum subspace $S_{zq}(k_0)$ in the Hilbert space.

Theoretically the subspace-selective $ZQT$ operators may be generally constructed by the base-operator expansion principle (See (5.15) in the Subsection 5.2 below). Here these selective $ZQT$ Hermitian operators $\{Q_{zq,\lambda}^{fi}\}$ of (5.2) may act as the base operators in the base-operator expansion principle. Generally, a subspace-selective $ZQT$ Hermitian operator $H_{zq}^{M_i}$ may be expressed as

$$H_{zq}^{M_i} = \sum_{0=l<l'}^{d(k_0)-1} \left( \alpha_{ll'}^{x} Q_{zq,x}^{ll'} + \alpha_{ll'}^{y} Q_{zq,y}^{ll'} + \alpha_{ll'}^{z} Q_{zq,z}^{ll'} \right) \tag{5.13}$$

where the coefficients $\alpha_{ll'}^{\lambda}$ $(\lambda = x, y, z; l \neq l')$ are real and usually not all these coefficients $\{\alpha_{ll'}^{\mu}\}$ $(\mu = x, y; l \neq l')$ are zero. Evidently the operator $H_{zq}^{M_i}$ is traceless. The base operators $\{Q_{zq,\lambda}^{ll'}\}$ $(\lambda = x, y, z)$ in (5.13) are just the selective $ZQT$ Hermitian operators $\{Q_{zq,\lambda}^{fi}\}$ of (5.2), if $|\Psi_i\rangle = |M_i, l'\rangle$ with $K_{M_i} = l'$ and $|\Psi_f\rangle = |M_i, l\rangle$ with $K_{M_i} = l$ and $l \neq l'$ are substituted into (5.2). Here the Hermitian base-vector-selective diagonal operators $\{Q_{zq,z}^{ll'}\}$ are treated just like the Hermitian selective zero-quantum transition operators $Q_{zq,\mu}^{ll'}$ $(\mu = x, y)$. The subspace-selective $ZQT$ Hermitian operator $H_{zq}^{M_i}$ can cause only the selective zero-quantum transitions among the eigenbase vectors $\{|M_i, K_{M_i}\rangle\}$ of the chosen subspace $S_{zq}(k_0)$ with $k_0 = n/2 - M_i$. Therefore, here only one direct-sum subspace $S_{zq}(k_0)$ of the Hilbert space $HS(N)$ is involved in these selective zero-quantum transitions. This is different from the subspace-selective $|p|$ −quantum-transition Hermitian operator $H_{|p|,\mu}^{M_f,M_i}$ of (5.8) with the order $p \neq 0$ in the previous Subsection 5.1.2, where the two chosen direct-sum subspaces (i.e., $S_{zq}(k_0)$ and $S_{zq}(n - k_0)$) are involved in the selective $|p|$ −quantum transitions.

The subspace-selective unitary exponential operator $U_{zq}^{M_i}(\theta)$ of the zero-quantum transitions may be generated by the subspace-selective $ZQT$ Hermitian operator $H_{zq}^{M_i}$ of (5.13),

$$U_{zq}^{M_i}(\theta) = \exp\left(-i\theta H_{zq}^{M_i}\right). \tag{5.14}$$

The selective zero-quantum transitions within the chosen subspace $S_{zq}(k_0)$ can be excited, when the subspace-selective $ZQT$ unitary operator $U_{zq}^{M_i}(\theta)$ is selectively applied to the subspace $S_{zq}(k_0)$. Moreover, it can prove that an arbitrary state vector of any direct-sum subspace $S_{zq}(k)$ of the Hilbert space $HS(N)$ which is not the chosen subspace $S_{zq}(k_0)$ is not affected by the subspace-selective $ZQT$ unitary operator $U_{zq}^{M_i}(\theta)$. Therefore, the subspace-selective $ZQT$ unitary operator of (5.14) is selectively applied to only the direct-sum subspace $S_{zq}(k_0)$ and it does not affect any other direct-sum subspace $S_{zq}(k)$ with $k \neq k_0$ of the Hilbert space $HS(N)$.

## 5.2. Construction and realization of the subspace-selective MQT operators

The research work in this Subsection is one main contribution to this paper. The subspace-selective $|p|$ −quantum-transition unitary operators $U_{|p|,\mu}^{M_f,M_i}(\theta)$ of (5.9) and the subspace-selective $ZQT$ unitary operators $U_{zq}^{M_i}(\theta)$ of (5.14) are merely the simple and special ones of a number of the subspace-selective unitary operators of the multiple-quantum transitions. Generally, the subspace-selective multiple-quantum-transition (MQT) operators may be constructed by starting from the selective- $|p|$ −quantum-transition Hermitian operators $Q^{fi}_{|p|,\lambda}$ $(\lambda = x, y, z)$ of (5.2). As shown previously, the selective-$|p|$ −quantum-transition Hermitian operators $\{Q^{fi}_{|p|,x}, Q^{fi}_{|p|,y}, Q^{fi}_{|p|,z}\}$ of (5.2) correspond one-to-one to the Hermitian pseudospin operators $\{Q^{KL}_z, Q^{KL}_x, Q^{KL}_y\}$ of (3.29). Actually, if both the spin Hamiltonian $H_s$ and the total spin operator $I_z$ of the spin system under study have the common eigenbase vectors including $|\Psi_f\rangle$ and $|\Psi_i\rangle$, then both the operator sets $\{Q^{fi}_{|p|,x}, Q^{fi}_{|p|,y}, Q^{fi}_{|p|,z}\}$ and $\{Q^{KL}_z, Q^{KL}_x, Q^{KL}_y\}$ are the same one. Of course, in a general case the total spin operator $I_z$ may not commute with the spin Hamiltonian $H_s$. It is known in the Subsection 3.2.2 that all the Hermitian pseudospin operators $\{Q^{KL}_x, Q^{KL}_y, Q^{KL}_z\}$ plus the total unity operator $E$ can form at least one complete set $\{E, Q^{KL}_x, Q^{KL}_y, Q^{KL}_z\}$ of base operators of the multiple-quantum operator algebra space of the spin system. Correspondingly all these selective-$|p|$ −quantum-transition Hermitian operators $\{Q^{fi}_{|p|,x}, Q^{fi}_{|p|,y}, Q^{fi}_{|p|,z}\}$ plus the total unity operator $E$ also can constitute at least one complete set $\{E, Q^{fi}_{|p|,x}, Q^{fi}_{|p|,y}, Q^{fi}_{|p|,z}\}$ of base operators of the multiple-quantum operator algebra space.

Now any operator of the multiple-quantum operator algebra space, e.g., the spin Hamiltonian $H_s$ of the spin system such as an $n-$spin$-1/2$ system, may be expanded in terms of the complete set $\{E, Q^{fi}_{|p|,x}, Q^{fi}_{|p|,y}, Q^{fi}_{|p|,z}\}$,

$$H_s = \alpha_0 E + \sum_{f,i} \left( \alpha^x_{fi} Q^{fi}_{|p|,x} + \alpha^y_{fi} Q^{fi}_{|p|,y} + \alpha^z_{fi} Q^{fi}_{|p|,z} \right), \tag{5.15}$$

where the expansional coefficients $\{\alpha^\lambda_{fi}\}$ and $\alpha_0$ may be real, and the quantum-transition order $p = M_f - M_i$ and its absolute value $|p|$ may take $0, 1, ..., n$ for the $n - spin - 1/2$ system. The base-operator expansion (5.15) is general, just like the base-operator expansion (3.30) based on the pseudospin-operator basis set $\{E, Q^{KL}_x, Q^{KL}_y, Q^{KL}_z\}$. It also may be considered as a formal operator-expansion expression of the spin Hamiltonian $H_s$. If the symmetrical structures and properties of the $n - spin - 1/2$ system are explicitly taken into account, then the base-operator expansion (5.15) must be suitably modified or rewritten. Here the symmetrical structures are usually considered as the first priority, as shown in the Subsections 5.1.2 and 5.1.3, but the properties are often useful, as shown in the Subsection 5.1.3.

Suppose that it needs to consider the properties of the multiple-quantum transition operators in the base-operator expansion for the spin Hamiltonian $H_s$ of a quantum spin system. As an example, this consideration may simply be that the spin Hamiltonian $H_s$ is expanded in terms of the multiple-quantum transition operators (See, e.g., (5.15)) which may be classified according to different quantum-transition orders $p$ in the multiple-quantum operator algebra space (See Ref.[4], [Ref[3]]). It is known in the Subsection 3.2.1 that all the MQT product operators $\{O_S\} = \{|R\rangle \langle T|\}$ of (3.25) constitute an operator basis set of the multiple-quantum operator algebra space of the $n-$spin$-1/2$ system. Then the spin Hamiltonian $H_s$ of (5.15) of the $n - spin - 1/2$ system also may be expanded in terms of the MQT product operators $\{|R\rangle\langle T|\}$,

$$H_s = \sum_{R,T} \beta_{RT} |R\rangle\langle T| = \sum_{R} \beta_{RR} |R\rangle\langle R| + \sum_{R\neq T} \beta_{RT} |R\rangle\langle T|, \tag{5.16}$$

where the tensor-product base vectors $|R\rangle$ and $|T\rangle$ are the orthonormal eigenbase vectors of the total spin operator $I_z$ of the $n-spin-1/2$ system and may be formally given by $|R\rangle = |k_1\rangle|k_2\rangle ... |k_n\rangle$ and $|T\rangle = |l_1\rangle|l_2\rangle ... |l_n\rangle$ with $k_j, l_j = 0, 1$ and $j = 1, 2, ..., n$. In (5.16) the sum $(\sum_R)$ terms are diagonal operators and belong to the LOMSO operator subspace, while the sum $(\sum_{R\neq T})$ terms are non-diagonal operators and the multiple-quantum transition operators. The MQT operators $\sum_{R\neq T} \beta_{RT} |R\rangle\langle T|$ in (5.16) may be further classified according to different quantum-transition orders $p$ [Ref[3]] in the multiple-quantum operator algebra space and this leads to that the Hamiltonian $H_s$ of (5.16) may be written according to different quantum-transition orders $p$,

$$H_s = \sum_{R} \beta_{RR} |R\rangle\langle R| + \sum_{p=-n}^{n} H_s^{(p)}, \tag{5.17}$$

where $H_s^{(p)}$ is a component operator of the Hamiltonian $H_s$ and also a $p-$order quantum transition operator. With the help of the MQT product operators $\{|R\rangle\langle T|\}$ of (3.25) the component $p-$order quantum transition operator $H_s^{(p)}$ with the order $p = -n, -n+1, ..., n-1, n$ may be determined by

$$\sum_{p=-n}^{n} H_s^{(p)} = \sum_{R\neq T} \beta_{RT} S_1^{-k_1,+l_1} S_2^{-k_2,+l_2} ... S_n^{-k_n,+l_n} \tag{5.18}$$

where the quantum-transition order $p$ of the MQT product operator $|R\rangle\langle T| = S_1^{-k_1,+l_1} S_2^{-k_2,+l_2} ... S_n^{-k_n,+l_n}$ with $|R\rangle \neq |T\rangle$ is given by $p = \sum_{j=1}^{n} (-k_j + l_j)$ with $k_j, l_j = 0, 1$, as shown in the Subsection 3.2.1 and the Section A.

Any subspace-selective multiple-quantum transition operator may be constructed with the help of the base-operator expansion principle in the multiple-quantum operator algebra space. Generally, it may be explicitly constructed by combining the symmetrical structure of the Hilbert space, the base-operator expansion such as (5.15) or (5.16), and the definition (5.1) or (5.2) of selective quantum transition operators. This is a general method to construct explicitly any subspace-selective multiple-quantum transition operators. A typical application of the general method is the subspace-selective $ZQT$ Hermitian operator $H_{zq}^{M_i}$ of (5.13), as described in the Subsection 5.1.3. If in the base-operator expansion (5.15) the sum $\sum_{f,i}$ (for $|\Psi_i\rangle$ and $|\Psi_f\rangle$) runs over only the vector basis subset $\{|M_i, K_{M_i}\rangle\}$ of the direct-sum subspace $S_{zq}(k_0)$, then the base-operator expansion (5.15) can be reduced to the subspace-selective ZQT Hermitian operator $H_{zq}^{M_i}$ of (5.13).

Actually, the specific subspace-selective MQT Hermitian operators still may be simply constructed by using the formal definition of (5.2) or (5.1) of selective MQT operators, if the initial $|\Psi_i\rangle$ and the final eigenvector $|\Psi_f\rangle$ in (5.2) or (5.1) are not limited to the two chosen eigenbase vectors of the total spin operator $I_z$ and can be chosen suitably in accordance with the Hilbert-space symmetric structure. As a typical example, the initial $|\Psi_i\rangle$ and final eigenvector $|\Psi_f\rangle$ may be chosen as any given vectors $|\Psi_i(M_i)\rangle$ of the subspace $S_{zq}(k_0)$ and $|\Psi_f(M_f)\rangle$ of the subspace $S_{zq}(n-k_0)$, respectively, where $|\Psi_r(M_r)\rangle = \sum_{K_{M_r}=0}^{d(k_0)-1} C_{M_r,K_{M_r}} \times |M_r, K_{M_r}\rangle$ for the subscript $r = i, f$ and evidently both the vectors $|\Psi_i(M_i)\rangle$ and $|\Psi_f(M_f)\rangle$ are orthogonal to one another. Then by substituting $|\Psi_i\rangle = |\Psi_i(M_i)\rangle$ and $|\Psi_f\rangle = |\Psi_f(M_f)\rangle$ into (5.2a) and (5.2b) one can obtain the specific subspace-selective $|p|-$quantum-transition Hermitian operators $Q_{|p|,\mu}^{M_f,M_i}$ with the order $p = -n + 2k_0 \neq 0$ and $\mu = x, y$ :

$$Q_{|p|,x}^{M_f,M_i} = \frac{1}{2}\left(|\Psi_f(M_f)\rangle\langle\Psi_i(M_i)| + |\Psi_i(M_i)\rangle\langle\Psi_f(M_f)|\right)$$

$$Q_{|p|,y}^{M_f,M_i} = \frac{1}{2i}\left(|\Psi_f(M_f)\rangle\langle\Psi_i(M_i)| - |\Psi_i(M_i)\rangle\langle\Psi_f(M_f)|\right)$$

The subspace-selective $|p|-$quantum-transition Hermitian operator $Q_{|p|,\mu}^{M_f,M_i}$ can cause only the selective $|p|-$quantum transitions between the two chosen subspaces $S_{zq}(k_0)$ and $S_{zq}(n-k_0)$ in the Hilbert space $HS(N)$ (in additional to the selective zero-quantum transitions within a single chosen subspace). This method can be generalized to simply construct a more complex subspace-selective MQT Hermitian operator which may involve many different chosen direct-sum subspaces of the Hilbert space and whose order $|p|$ may take many different values. Its key point is to choose suitably the initial $|\Psi_i\rangle$ and the final vector $|\Psi_f\rangle$ in accordance with the Hilbert-space symmetric structure. This generalized method is described below.

Suppose that $S_1$ and $S_2$ are two chosen orthogonal (i.e., non-overlapping) subspaces, each one of which may contain several different direct-sum subspaces of the Hilbert space $HS(N)$ of the $n-spin-1/2$ system. Let $|\varphi_1\rangle$ and $|\varphi_2\rangle$ be any two orthonormal vectors of the Hilbert space $HS(N)$ and belong to the two chosen orthogonal subspaces $S_1$ and $S_2$, respectively. For example, $|\varphi_1\rangle = |\Psi_f\rangle$, $|\Psi_f(M_f)\rangle \in S_{zq}(n-k_0) = S_1$ and $|\varphi_2\rangle = |\Psi_i\rangle$, $|\Psi_i(M_i)\rangle \in S_{zq}(k_0) = S_2$. Then the multiple quantum transitions between the two orthonormal vectors $|\varphi_1\rangle$ and $|\varphi_2\rangle$ are said subspace-selective in the sense that they involve only the two chosen orthogonal subspaces $S_1$ and $S_2$ in the Hilbert space $HS(N)$. Obviously, these multiple quantum transitions may own many different quantum-transition orders $|p|$. Now the subspace-selective MQT Hermitian operators that can cause these multiple quantum transitions between the two chosen orthogonal subspaces $S_1$ and $S_2$ or between the two orthonormal vectors $|\varphi_1\rangle$ and $|\varphi_2\rangle$ in the Hilbert space $HS(N)$ may be formally constructed by

$$H_x^{mq} = \frac{1}{2}\left(|\varphi_1\rangle\langle\varphi_2| + |\varphi_2\rangle\langle\varphi_1|\right) \tag{5.19a}$$

$$H_y^{mq} = \frac{1}{2i}\left(|\varphi_1\rangle\langle\varphi_2| - |\varphi_2\rangle\langle\varphi_1|\right) \tag{5.19b}$$

$$H_z^{mq} = \frac{1}{2}\left(|\varphi_1\rangle\langle\varphi_1| - |\varphi_2\rangle\langle\varphi_2|\right), \tag{5.19c}$$

which constitute the definition of the subspace-selective MQT Hermitian operators, the generalized one from (5.2). Here the last Hermitian operator $H_z^{mq}$ does not cause any multiple quantum transitions between the two chosen orthogonal subspaces $S_1$ and $S_2$. However, it may not be a diagonal operator, where the Hermitian operators $B_l^{mq} = |\varphi_l\rangle\langle\varphi_l|$ with $l = 1, 2$ are the PDH operators of (4.4) (See the Section 4). It may cause those multiple quantum transitions inside a chosen orthogonal subspace $S_1$ (or $S_2$). Evidently the subspace-selective $|p|$ −quantum-transition Hermitian operator $Q_{|p|,\mu}^{M_f,M_i}$ in the last paragraph is the special case of the Hermitian operator $H_\mu^{mq}$ of (5.19) with $\mu = x, y$. These subspace-selective MQT Hermitian operators $H_x^{mq}$, $H_y^{mq}$, and $H_z^{mq}$ of (5.19) are quite simple in form. They are formally similar to the selective− $|p|$ −quantum-transition Hermitian operators $Q_{|p|,x}^{fi}$, $Q_{|p|,y}^{fi}$, and $Q_{|p|,z}^{fi}$ of (5.2), respectively.

If both the vectors $|\varphi_1\rangle$ and $|\varphi_2\rangle$ in (5.19) each are expanded in terms of the vector basis set $\{|M, K_M\rangle\}$ of the Hilbert space $HS(N)$ which are the orthonormal eigenbase vectors of the total spin operator $I_z$, then these three Hermitian operators $H_x^{mq}$, $H_y^{mq}$, and $H_z^{mq}$ of (5.19) each can be expressed as the base-operator expansion of (5.15). A concrete instance may be seen in the Subsection 6.2 below.

Just like the PDH operators $\{H_N^{pd}\}$ of (4.3), the subspace-selective MQT Hermitian operators $\{H_\lambda^{mq}\}$ of (5.19) with $\lambda = x, y, z$ are one kind of the simple Hermitian operators through which a connection may be set up conveniently between the Hilbert space and its corresponding multiple-quantum operator algebra space in the aspect of symmetrical structures and properties. These subspace-selective MQT Hermitian operators $\{H_\lambda^{mq}\}$ are independent upon any detailed vector basis set of the Hilbert space $HS(N)$. They are independent of any detailed direct-sum decomposition of the Hilbert space $HS(N)$. They are also independent upon the dimensional size $N$ of the Hilbert space $HS(N)$. These properties are important for the subspace-selective MQT Hermitian operators $\{H_\lambda^{mq}\}$ to act as bridge to connect the symmetrical structures and properties of the Hilbert space to the counterpart of the corresponding multiple-quantum operator algebra space.

The subspace-selective MQT Hermitian operators $H_x^{mq}$, $H_y^{mq}$, and $H_z^{mq}$ of (5.19) can be unitarily transformed to the selective− $|p|$ −quantum-transition Hermitian operators $Q_{|p|,x}^{fi}$, $Q_{|p|,y}^{fi}$, and $Q_{|p|,z}^{fi}$ of (5.2), respectively, and vice versa. This is an important property for these Hermitian operators $\{H_x^{mq}, H_y^{mq}, H_z^{mq}\}$. In theory this important property can be proven rigorously, as shown below. Both the orthonormal vectors $|\varphi_1\rangle$ and $|\varphi_2\rangle$ belong to the two chosen orthogonal subspaces $S_1$ and $S_2$, respectively. Then there are the orthonormal relations: $\langle\varphi_k|\varphi_l\rangle = \delta_{kl}$ for $k, l = 1, 2$. There is a famous theorem in quantum mechanics [1,2,3] that any quantum-mechanical $(QM)$ unitary transformation $V$ does not change the overlapping integral $\langle\varphi_1|\varphi_2\rangle$ between any two vectors $|\varphi_1\rangle$ and $|\varphi_2\rangle$ of the Hilbert space of quantum system under study, that is, $\langle V\varphi_k|V\varphi_l\rangle = \langle\varphi_k|\varphi_l\rangle$ for $k, l = 1, 2$. Here the overlapping integral $\langle\varphi_k|\varphi_l\rangle$ is

just the scalar (or inner) product $(\varphi_k, \varphi_l)$ (or $\langle\varphi_k|\varphi_l\rangle$) of the Hilbert space in quantum mechanics. A corollary of the theorem is: if any two vectors (e.g., $|\varphi_1\rangle$ and $|\varphi_2\rangle$) are orthogonal to one another, then any $QM$ unitary transformation $V$ does not change the orthogonal relation between the two vectors, that is, $\langle V\varphi_1|V\varphi_2\rangle = \langle\varphi_1|\varphi_2\rangle = 0$. Below this theorem and its corollary can help prove rigorously that these Hermitian operators $\{H_x^{mq}, H_y^{mq}, H_z^{mq}\}$ can be simultaneously transformed to the Hermitian operators $\{Q_{|p|,x}^{fi}, Q_{|p|,y}^{fi}, Q_{|p|,z}^{fi}\}$ by a unitary transformation, respectively.

All the orthonormal eigenbase vectors $\{|M, K_M\rangle\}$ of the total spin operator $I_z$ form a vector basis set of the $N-$dimensional Hilbert space of the spin system such as the $n-spin-1/2$ system, while $|\Psi_f\rangle$ and $|\Psi_i\rangle$ in the Hermitian operators $\{Q_{|p|,x}^{fi}, Q_{|p|,y}^{fi}, Q_{|p|,z}^{fi}\}$ of (5.2) each can take any orthonormal eigenbase vector $|M, K_M\rangle$ . Below for convenience $|K_z\rangle$ is denoted as any one of the orthonormal eigenbase vectors $\{|M, K_M\rangle\}$ and is called the Spin$-I_z$ eigenbase vector. Then the vector basis set $\{|K_z\rangle\}$ is just the vector basis set $\{|M, K_M\rangle\}$. Of course, $|K_z\rangle$ can stand for any one of the eigenbase vectors $|\Psi_f\rangle$ and $|\Psi_i\rangle$ .

According to quantum mechanics there exists a $QM$ unitary transformation $W$ on the $N-$dimensional Hilbert space of the spin system under study by which the Spin$-I_z$ eigenbase vector $|K_z\rangle$ with $0 \le K_z \le N-1$ can be changed to the vector $|\varphi_1\rangle$ and vice versa,

$$W|K_z\rangle = |\varphi_1\rangle\,,\ W^+|\varphi_1\rangle = |K_z\rangle\,. \tag{5.20}$$

If at the same time the unitary transformation $W$ also can change some Spin$-I_z$ eigenbase vector $|L_z\rangle$ ($|L_z\rangle \neq |K_z\rangle$) to the vector $|\varphi_2\rangle$, that is, $W|L_z\rangle = |\varphi_2\rangle$ and $|L_z\rangle = W^+|\varphi_2\rangle$ , then it is easy to prove that the unitary transformation $W^+$ on the operator $H_\lambda^{mq}$ of (5.19) with $\lambda = x, y, z$, i.e., $W^+H_\lambda^{mq}W$, can change the operator $H_\lambda^{mq}$ to a selective$-|p|-$quantum-transition Hermitian operator $Q_{|p|,\lambda}^{fi}$ of (5.2), that is, $W^+H_\lambda^{mq}W = Q_{|p|,\lambda}^{fi}$, where $|\Psi_f\rangle = |K_z\rangle$ and $|\Psi_i\rangle = |L_z\rangle$ . However, if at the same time the unitary transformation $W$ can not change a single Spin$-I_z$ eigenbase vector $|L_z\rangle$ to the vector $|\varphi_2\rangle$ , that is, $W|L_z\rangle \neq |\varphi_2\rangle$ and $W^+|\varphi_2\rangle \neq |L_z\rangle$ , then the theoretical proof becomes a little bit complex. Note that $\{|K_z\rangle\}$ is a complete set of the orthonormal eigenbase vectors of the $N-$dimensional Hilbert space. Then the vectors $W^+|\varphi_j\rangle$ with $j = 1, 2$ can be expanded in terms of the Spin$-I_z$ eigenbase vectors $\{|K_z\rangle\}$ :

$$W^+|\varphi_1\rangle = |K_z\rangle\,,\ W^+|\varphi_2\rangle = \sum_{K_z'=0}^{N-1} C_{K_z'}\,|K_z'\rangle \tag{5.21}$$

where $W^+|\varphi_2\rangle$ is not any single Spin$-I_z$ eigenbase vector. As pointed out above, according to quantum mechanics any $QM$ unitary transformation does not change the orthogonal relation between any two vectors of the Hilbert space. This means that the $QM$ unitary transformation $W^+$ can not change the orthogonal relation between the two orthonormal vectors $|\varphi_1\rangle$ and $|\varphi_2\rangle$ ,

$$\left\langle W^+\varphi_1|W^+\varphi_2\right\rangle = \langle\varphi_1|\varphi_2\rangle = 0. \tag{5.22}$$

By substituting the two vectors $W^{+}\left|\varphi_j\right\rangle$ with $j=1,2$ of (5.21) into (5.22) and then with the help of the orthogonal relations $\langle K_z|K_z'\rangle = \delta_{K_z K_z'}$ it can be found from (5.22) that the expansional coefficient $C_{K_z}$ in the expansion of the vector $W^{+}\left|\varphi_2\right\rangle$ in (5.21) is exactly zero, i.e., $C_{K_z}=0$. Therefore, the vector $W^{+}\left|\varphi_2\right\rangle$ in (5.21) is reduced to the form

$$W^{+}\left|\varphi_2\right\rangle = \sum_{K_z'=0, K_z'\neq K_z}^{N-1} C_{K_z'}\left|K_z'\right\rangle \tag{5.23}$$

This shows that the vector $W^{+}\left|\varphi_2\right\rangle$ belongs to the subspace $S_{\perp}^{1}$ of the $N$-dimensional Hilbert space, which is spanned by the $N-1$ Spin$-I_z$ eigenbase vectors $\{|K_z'\rangle\}$ with $0\leq K_z'\leq N-1$ and $K_z'\neq K_z$. This subspace $S_{\perp}^{1}$ is $(N-1)$ −dimensional. Obviously, the Spin$-I_z$ eigenbase vector $|K_z\rangle$ does not belong to the subspace $S_{\perp}^{1}$. Therefore, as shown by (5.20), the vector $W^{+}\left|\varphi_1\right\rangle = |K_z\rangle$ does not yet belong to the subspace $S_{\perp}^{1}$. Note that the single Spin$-I_z$ eigenbase vector $|K_z\rangle$ forms a one-dimensional subspace $S^{1}$ of the $N$-dimensional Hilbert space whose orthogonal complement subspace is just the subspace $S_{\perp}^{1}$. Therefore, the $N$-dimensional Hilbert space is the direct sum of the subspace $S^{1}$ and its orthogonal complement subspace $S_{\perp}^{1}$.

Now within the orthogonal complement subspace $S_{\perp}^{1}$ with dimension $N-1$ there exists a unitary transformation $W_1$ such that the vector $W^{+}\left|\varphi_2\right\rangle$ of (5.23) can be unitarily transformed to some single Spin$-I_z$ eigenbase vector $|L_z\rangle$ ($|L_z\rangle\neq|K_z\rangle$),

$$W_1^{+}\left(W^{+}\left|\varphi_2\right\rangle\right) = |L_z\rangle, \tag{5.24}$$

where both the vectors $W^{+}\left|\varphi_2\right\rangle$ and $|L_z\rangle$ belong to the orthogonal complement subspace $S_{\perp}^{1}$. The $QM$ unitary transformation $W_1^{+}$ in (5.24) is subspace-selective. It selectively acts on only the orthogonal complement subspace $S_{\perp}^{1}$, while it does not affect the subspace $S^{1}$ which accommodates the vector $|K_z\rangle$ (i.e., $W^{+}\left|\varphi_1\right\rangle$),

$$W_1^{+}|K_z\rangle = |K_z\rangle,\ W_1^{+}\left(W^{+}\left|\varphi_1\right\rangle\right) = W^{+}\left|\varphi_1\right\rangle. \tag{5.25}$$

One therefore finds from (5.20), (5.24), and (5.25) that

$$W_1^{+}W^{+}\left|\varphi_1\right\rangle = W^{+}\left|\varphi_1\right\rangle = |K_z\rangle,\ W_1^{+}W^{+}\left|\varphi_2\right\rangle = |L_z\rangle \tag{5.26}$$

where the Spin$-I_z$ eigenbase vectors $|K_z\rangle$ and $|L_z\rangle$ ($|L_z\rangle\neq|K_z\rangle$) belong to the subspace $S^{1}$ and its orthogonal complement subspace $S_{\perp}^{1}$, respectively. The unitary transformations of (5.26) further lead to that the operator $H_{\lambda}^{mq}$ of (5.19) can be unitarily transformed to the operator $Q_{|p|,\lambda}^{fi}$ :

$$W_1^{+}W^{+}H_{\lambda}^{mq}WW_1 = Q_{|p|,\lambda}^{fi},\ \lambda = x,y,z, \tag{5.27}$$

where $|\Psi_f\rangle = |K_z\rangle$ and $|\Psi_i\rangle = |L_z\rangle$ and $\{Q_{|p|,\lambda}^{fi}\}$ ($\lambda = x,y,z$) are the selective-$|p|$-quantum-transition Hermitian operators of (5.2). The unitary transformation (5.27) indicates that these subspace-selective MQT Hermitian operators

$\{H_x^{mq}, H_y^{mq}, H_z^{mq}\}$ of (5.19) can be unitarily transformed to the selective-$|p|$-quantum-transition Hermitian operators $\{Q_{|p|,x}^{fi}, Q_{|p|,y}^{fi}, Q_{|p|,z}^{fi}\}$ of (5.2) at the same time, respectively.

If the eigenbase vectors $|K_z\rangle$ and $|L_z\rangle$ of the total spin operator $I_z$ can be replaced with the spin energy eigenbase vectors $|K\rangle$ and $|L\rangle$ of the spin system under study, respectively, then it can prove that the unitary transformation (5.27) can be replaced with the one:

$$W_1^+ W^+ H_\lambda^{mq} W W_1 = Q_\lambda^{KL}, \ \lambda = x, y, z, \tag{5.28}$$

where $\{Q_\lambda^{KL}\}$ are the Hermitian pseudospin operators of (3.29). The unitary transformation (5.28) shows that the operators $H_\lambda^{mq}$ of (5.19) can be unitarily transformed to the corresponding pseudospin operators $Q_\lambda^{KL}$ for $\lambda = x, y, z$.

It could become simpler and more useful for the construction of the unitary transformation $W_1^+ W^+$ of (5.27) that changes the operator $H_\lambda^{mq}$ to the operator $Q_{|p|,\lambda}^{fi}$, if one takes into account the Hilbert-space symmetrical structure that both the orthonormal vectors $|\varphi_1\rangle$ and $|\varphi_2\rangle$ belong to the two chosen orthogonal subspaces $S_1$ and $S_2$ of the Hilbert space, respectively. There is the unitary transformation $W_{s_1}^+$ by which the vector $|\varphi_1\rangle$ can be changed to some Spin$-I_z$ eigenbase vector $|K_z\rangle$ within the chosen orthogonal subspace $S_1$, that is, $W_{s_1}^+ |\varphi_1\rangle = |K_z\rangle$. Similarly, there is the unitary transformation $W_{s_2}^+$ by which the vector $|\varphi_2\rangle$ is changed to some Spin$-I_z$ eigenbase vector $|L_z\rangle$ ($|L_z\rangle \neq |K_z\rangle$) within the chosen orthogonal subspace $S_2$, that is, $W_{s_2}^+ |\varphi_2\rangle = |L_z\rangle$. Here the unitary transformation $W_{s_1}^+$ is subspace-selective. It acts on selectively only the chosen orthogonal subspace $S_1$ and it does not affect any other orthogonal subspaces including the orthogonal subspace $S_2$ of the Hilbert space. Note that here both the vectors $|\varphi_2\rangle$ and $|L_z\rangle$ are in the orthogonal subspace $S_2$ but not in the chosen orthogonal subspace $S_1$. Therefore, they are not affected by the subspace-selective unitary transformation $W_{s_1}^+$, that is, $W_{s_1}^+ |\varphi_2\rangle = |\varphi_2\rangle$ and $W_{s_1}^+ |L_z\rangle = |L_z\rangle$. One therefore obtains

$$W_{s_1}^+ |\varphi_1\rangle = |K_z\rangle, \ W_{s_1}^+ |\varphi_2\rangle = |\varphi_2\rangle, \ W_{s_1}^+ |L_z\rangle = |L_z\rangle \tag{5.29}$$

The unitary transformation $W_{s_2}^+$ is subspace-selective too. It selectively acts on only the chosen orthogonal subspace $S_2$ and does not affect any other orthogonal subspaces including the orthogonal subspace $S_1$ of the Hilbert space. Since the two vectors $|\varphi_1\rangle$ and $|K_z\rangle$ are in the orthogonal subspace $S_1$ but not in the chosen orthogonal subspace $S_2$, they are not affected by the subspace-selective unitary transformation $W_{s_2}^+$, that is, $W_{s_2}^+ |\varphi_1\rangle = |\varphi_1\rangle$ and $W_{s_2}^+ |K_z\rangle = |K_z\rangle$. One therefore has

$$W_{s_2}^+ |\varphi_2\rangle = |L_z\rangle, \ W_{s_2}^+ |\varphi_1\rangle = |\varphi_1\rangle, \ W_{s_2}^+ |K_z\rangle = |K_z\rangle \tag{5.30}$$

Now by acting the subspace-selective unitary transformations $W_{s_1}^+$ and $W_{s_2}^+$ on the operator $H_\lambda^{mq}$ of (5.19), i.e., $W_{s_2}^+ W_{s_1}^+ H_\lambda^{mq} W_{s_1} W_{s_2}$, and with the help of the unitary transformations of (5.29) and (5.30) it is easy to prove that the operator $H_{\lambda}^{mq}$ can be unitarily transformed to the selective-$|p|$ $-$quantum-transition Hermitian operator $Q_{|p|,\lambda}^{fi}$,

$$W^{+}H_{\lambda}^{mq}W = Q_{|p|,\lambda}^{fi}, \ \lambda = x, y, z \tag{5.31}$$

where $|\Psi_f\rangle = |K_z\rangle$ and $|\Psi_i\rangle = |L_z\rangle$ and the unitary operator $W^{+} = W_{s_2}^{+}W_{s_1}^{+}$ or $W^{+} = W_{s_1}^{+}W_{s_2}^{+}$.

The unitary transformation $W^{+}$ of (5.31) that changes the subspace-selective MQT Hermitian operator $H_{\lambda}^{mq}$ of (5.19) to the selective-$|p|$-quantum-transition Hermitian operator $Q_{|p|,\lambda}^{fi}$ of (5.2) may be constructed by separately constructing the subspace-selective unitary operator $W_{s_1}^{+}$ within the chosen orthogonal subspace $S_1$ and the subspace-selective unitary operator $W_{s_2}^{+}$ within the chosen subspace $S_2$.

The subspace-selective MQT unitary operator (or propagator) $U_{\lambda}^{mq}(\theta)$ ($\lambda = x, y, z$) may be generated by the subspace-selective MQT Hermitian operator $H_{\lambda}^{mq}$ of (5.19),

$$U_{\lambda}^{mq}(\theta) = \exp\left(-i\theta H_{\lambda}^{mq}\right), \ \lambda = x, y, z \tag{5.32}$$

These subspace-selective MQT unitary operators $U_{\lambda}^{mq}(\theta)$ with $\lambda = x, y, z$ may induce the selective excitations of the multiple quantum transitions between the chosen orthogonal subspaces $S_1$ and $S_2$ or within any single chosen subspace ($S_1$ or $S_2$).

With the help of the unitary transformation of (5.31) the subspace-selective MQT unitary operator $U_{\lambda}^{mq}(\theta)$ with $\lambda = x, y, z$ may be unitarily transformed to the selective-$|p|$-quantum-transition unitary operator $R_{|p|,\lambda}^{fi}(\theta)$ of (5.3a) in the multiple-quantum operator algebra space,

$$W^{+}\exp\left(-i\theta H_{\lambda}^{mq}\right)W = \exp\left(-i\theta Q_{|p|,\lambda}^{fi}\right) \tag{5.33}$$

Here the unitary operator $R_{|p|,\lambda}^{fi}(\theta) = \exp\left(-i\theta Q_{|p|,\lambda}^{fi}\right)$ of (5.3a) can be easily realized, as shown in the Subsection 5.1.1. If the unitary operator $W^{+} = W_{s_2}^{+}W_{s_1}^{+}$ (or $W = W_{s_1}W_{s_2}$) in (5.31) can be efficiently realized, then the subspace-selective MQT unitary operator $U_{\lambda}^{mq}(\theta)$ of (5.32) can be efficiently realized with the help of the unitary transformation (5.33).

The unitary time-evolutional process that is governed by the subspace-selective MQT unitary propagator $U_{\lambda}^{mq}(\theta)$ of (5.32) with $\lambda = x, y, z$, where the angle $\theta$ is proportional to the time interval $\tau$, can be calculated exactly and analytically. According to the base-operator expansion principle the unitary propagator $U_{\lambda}^{mq}(\theta)$ can be expanded in terms of the complete set of base operators of the multiple-quantum operator algebra space of the $n-$spin$-1/2$ system. Without lost generality, here consider the unitary propagator $U_{y}^{mq}(\theta)$ of (5.32). First, the unitary exponential operator $U_{y}^{mq}(\theta)$ can be expanded as a power series of the Hermitian operator $H_{y}^{mq}$ of (5.19b),

$$U_{y}^{mq}(\theta) = \exp\left(-i\theta H_{y}^{mq}\right) = \sum_{k=0}^{\infty}\frac{1}{k!}\left(-i\theta H_{y}^{mq}\right)^{k} \tag{5.34a}$$

This power series may be rewritten as

$$U_y^{mq}(\theta) = \mathbf{E} + \sum_{k=1,2,...}^{\infty} \frac{1}{(2k)!}\left(-i\theta H_y^{mq}\right)^{2k} + \sum_{k=0,1,...}^{\infty} \frac{1}{(2k+1)!}\left(-i\theta H_y^{mq}\right)^{2k+1} \tag{5.34b}$$

with the unity operator $\mathbf{E}$. Note that both the vectors $|\varphi_1\rangle$ and $|\varphi_2\rangle$ are orthonormal, i.e., $\langle\varphi_k|\varphi_l\rangle = \delta_{kl}$ for $k, l = 1, 2$. Then, by using the expression (5.19b) of the operator $H_y^{mq}$ it can prove that the operators $\left(H_y^{mq}\right)^2$ and $\left(H_y^{mq}\right)^3$ may be respectively written as

$$\left(H_y^{mq}\right)^2 = \frac{1}{2^2}\left(B_1^{mq} + B_2^{mq}\right), \ \left(H_y^{mq}\right)^3 = \frac{1}{2^2} H_y^{mq}, \tag{5.35}$$

where $B_l^{mq} = |\varphi_l\rangle\langle\varphi_l|$ for $l = 1, 2$. The operator identities of (5.35) are basic and they can be used repeatedly below. With the help of the operator identities (5.35) one can set up the recursive relations for the $k-$th power operator $\left(H_y^{mq}\right)^k$ of the operator $H_y^{mq}$ for $k = 1, 2, 3, ....$ These recursive relations are written as

$$\left(H_y^{mq}\right)^{2k} = \frac{1}{2^{2k}}\left(B_1^{mq} + B_2^{mq}\right) \text{ for } k = 1, 2, 3, ... \tag{5.36a}$$

$$\left(H_y^{mq}\right)^{2k+1} = \frac{1}{2^{2k}} H_y^{mq} \text{ for } k = 1, 2, 3, ... \tag{5.36b}$$

With the aid of the recursive relations of (5.36) it can be found that

$$\sum_{k=1,2,...}^{\infty} \frac{1}{(2k)!}\left(-i\theta H_y^{mq}\right)^{2k} = \left(-1 + \cos\frac{1}{2}\theta\right)\left(B_1^{mq} + B_2^{mq}\right) \tag{5.37a}$$

$$\sum_{k=0,1,...}^{\infty} \frac{1}{(2k+1)!}\left(-i\theta H_y^{mq}\right)^{2k+1} = \left(-i2\sin\frac{1}{2}\theta\right) H_y^{mq} \tag{5.37b}$$

Now by substituting (5.37) into (5.34b) the power series (5.34b) of the unitary propagator $U_y^{mq}(\theta)$ is reduced to the simple form

$$U_y^{mq}(\theta) = \mathbf{E} + \left(-1 + \cos\frac{1}{2}\theta\right)\left(B_1^{mq} + B_2^{mq}\right) + \left(-2i\sin\frac{1}{2}\theta\right) H_y^{mq} \tag{5.38}$$

where $B_l^{mq} = |\varphi_l\rangle\langle\varphi_l|$ for $l = 1, 2$ are the PDH operators $H_N^{pd}$ of (4.4). This is the operator-expansion formula for the unitary propagator $U_y^{mq}(\theta)$ of (5.32) in the multiple-quantum operator algebra space.

More generally it can prove that the subspace-selective MQT unitary propagators $U_\lambda^{mq}(\theta)$ of (5.32) with $\lambda = x, y, z$ may be generally expanded as

$$U_\lambda^{mq}(\theta) = \mathbf{E} + \left(-1 + \cos\frac{1}{2}\theta\right)\left(B_1^{mq} + B_2^{mq}\right) + \left(-2i\sin\frac{1}{2}\theta\right) H_\lambda^{mq} \tag{5.39}$$

where the subspace-selective MQT Hermitian operators $H_\lambda^{mq}$ with $\lambda = x, y, z$ are given by (5.19), respectively, and the PDH operators $B_l^{mq} = |\varphi_l\rangle\langle\varphi_l|$ for $l = 1, 2$. This is general operator-expansion expression for the subspace-selective MQT unitary operator $U_\lambda^{mq}(\theta)$ in the multiple-quantum operator algebra space. This formal expression (5.39) is independent upon any detailed vector basis set of the Hilbert space $HS(N)$. It is independent of any detailed direct-sum decomposition of the Hilbert space $HS(N)$. It holds for any dimensional size $N$ of the Hilbert space $HS(N)$. When the unitary propagator $U_\lambda^{mq}(\theta)$ is taken as $U_y^{mq}(\theta)$, the general operator-expansion expression (5.39) is reduced to the special one of (5.38).

The operator-expansion formula (5.39) shows clearly that the subspace-selective MQT unitary operator $U_\lambda^{mq}(\theta)$ $(\lambda = x, y, z)$ is proportional to the Hermitian operator $H_\lambda^{mq}$ and the PDH operators $B_1^{mq}$ and $B_2^{mq}$. Furthermore, each one of these Hermitian operators $H_\lambda^{mq}$, $B_1^{mq}$, and $B_2^{mq}$ in (5.39) can be easily expanded in terms of the complete set of base operators of the multiple-quantum operator algebra space. Therefore, the operator-expansion formula (5.39) can be easily reduced further to the complete base-operator expansion for the subspace-selective MQT unitary operator $U_\lambda^{mq}(\theta)$ in the multiple-quantum operator algebra space. The operator-expansion formula (5.39) (or (5.38)) is usually much simpler than the complete base-operator-expansion formula. It may be more convenient to use in theory.

The operator-expansion formula (5.39) may be used to calculate exactly and analytically the unitary time-evolutional process which is governed by the subspace-selective MQT unitary propagator $U_\lambda^{mq}(\theta)$ $(\lambda = x, y, z)$. This is an exact and analytical theoretical method to simulate and calculate the unitary time-evolutional processes. General unitary propagators (or unitary operators) are complicated in quantum mechanics. The unitary time-evolutional processes which are governed by these general unitary propagators are generally hard to calculate exactly and analytically except for the simple and special cases. The unitary time-evolutional processes which are governed by the unitary propagators $U_\lambda^{mq}(\theta)$ are a novel kind of the simple and special cases that the unitary time-evolutional processes can be calculated exactly and analytically.

The subspace-selective unitary manipulation based on the subspace-selective MQT unitary operators $U_\lambda^{mq}(\theta)$ of (5.32) has important applications in future. Below the basic unitary transformations can be set up for the unitary operators $U_\lambda^{mq}(\theta)$ acting on the vectors $|\varphi_k\rangle$ in (5.19) with $k = 1, 2$ in the Hilbert space $HS(N)$ with the help of the operator-expansion formula (5.39). Here consider first the unitary operator $U_y^{mq}(\theta)$. With the help of the operator-expansion formula (5.38) the unitary transformation for the unitary operator $U_y^{mq}(\theta)$ acting on the vector $|\varphi_k\rangle$ with $k = 1, 2$ may be written as

$$U_y^{mq}(\theta)|\varphi_k\rangle = |\varphi_k\rangle + \left(-1 + \cos\frac{1}{2}\theta\right)(B_1^{mq} + B_2^{mq})|\varphi_k\rangle + \left(-2i\sin\frac{1}{2}\theta\right)H_y^{mq}|\varphi_k\rangle \quad (5.40)$$

There are the orthonormal relations $\langle\varphi_k|\varphi_l\rangle = \delta_{kl}$ for $k, l = 1, 2$. Note that $B_l^{mq} = |\varphi_l\rangle\langle\varphi_l|$ with $l = 1, 2$. Then it can be found that $B_l^{mq}|\varphi_1\rangle = \delta_{l1}|\varphi_1\rangle$ and $B_l^{mq}|\varphi_2\rangle = \delta_{l2}|\varphi_2\rangle$, and hence $(B_1^{mq}+B_2^{mq})|\varphi_l\rangle = |\varphi_l\rangle$ for $l = 1, 2$. By using the expression (5.19b) of the operator $H_y^{mq}$ and with the help of the orthonormal relation $\langle\varphi_k|\varphi_l\rangle = \delta_{kl}$ it can be found that $H_y^{mq}|\varphi_1\rangle = -\frac{1}{2i}|\varphi_2\rangle$ and $H_y^{mq}|\varphi_2\rangle = \frac{1}{2i}|\varphi_1\rangle$. Then by substituting $(B_1^{mq}+B_2^{mq})|\varphi_1\rangle = |\varphi_1\rangle$ and $H_y^{mq}|\varphi_1\rangle = -\frac{1}{2i}|\varphi_2\rangle$ into (5.40) one can obtain the basic unitary transformation in the Hilbert space:

$$U_y^{mq}(\theta)|\varphi_1\rangle = |\varphi_1\rangle\cos\frac{1}{2}\theta + |\varphi_2\rangle\sin\frac{1}{2}\theta \tag{5.41a}$$

In similar way, by substituting $(B_1^{mq}+B_2^{mq})|\varphi_2\rangle = |\varphi_2\rangle$ and $H_y^{mq}|\varphi_2\rangle = \frac{1}{2i}|\varphi_1\rangle$ into (5.40) it can prove that

$$U_y^{mq}(\theta)|\varphi_2\rangle = |\varphi_2\rangle\cos\frac{1}{2}\theta - |\varphi_1\rangle\sin\frac{1}{2}\theta \tag{5.41b}$$

In analogous way it can prove that for the unitary operator $U_x^{mq}(\theta)$ the basic unitary transformations are given by

$$U_x^{mq}(\theta)|\varphi_1\rangle = |\varphi_1\rangle\cos\frac{1}{2}\theta - i|\varphi_2\rangle\sin\frac{1}{2}\theta \tag{5.42a}$$

$$U_x^{mq}(\theta)|\varphi_2\rangle = |\varphi_2\rangle\cos\frac{1}{2}\theta - i|\varphi_1\rangle\sin\frac{1}{2}\theta \tag{5.42b}$$

and for the unitary operator $U_z^{mq}(\theta)$ the basic unitary transformations are written as

$$U_z^{mq}(\theta)|\varphi_1\rangle = \exp\left(-i\frac{1}{2}\theta\right)|\varphi_1\rangle,\ U_z^{mq}(\theta)|\varphi_2\rangle = \exp\left(i\frac{1}{2}\theta\right)|\varphi_2\rangle \tag{5.43}$$

Indeed, the unitary operator $U_z^{mq}(\theta)$ does not cause any multiple quantum transitions between the two vectors $|\varphi_1\rangle$ and $|\varphi_2\rangle$.

Generally, an arbitrary vector $|\psi\rangle$ of the Hilbert space $HS(N)$ may be different from the vectors $|\varphi_1\rangle$ and $|\varphi_2\rangle$. Then the unitary transformation $U_\lambda^{mq}(\theta)|\psi\rangle$ with $\lambda = x, y, z$ for the unitary operator $U_\lambda^{mq}(\theta)$ acting on the vector $|\psi\rangle$ in the Hilbert space may be more complex than the basic unitary transformations mentioned above. However, it still can be calculated exactly and analytically with the aid of the operator-expansion formula (5.39).

The subspace-selective property for the subspace-selective MQT unitary propagator $U_\lambda^{mq}(\theta)$ $(\lambda = x, y, z)$ of (5.32) can be confirmed with the help of the operator-expansion formula (5.39). Consider an arbitrary vector $|\psi\rangle$ of the Hilbert space $HS(N)$ which is orthogonal to both the vectors $|\varphi_1\rangle$ and $|\varphi_2\rangle$, that is, $\langle\varphi_k|\psi\rangle = 0$ for $k = 1, 2$. With the help of the operator-expansion formula (5.39) the unitary transformation $U_\lambda^{mq}(\theta)|\psi\rangle$ in the Hilbert space may be explicitly written as

$$U_\lambda^{mq}(\theta)|\psi\rangle = |\psi\rangle + \left(-1+\cos\frac{1}{2}\theta\right)(B_1^{mq}+B_2^{mq})|\psi\rangle$$

$$+\left(-2i\sin\frac{1}{2}\theta\right)H_{\lambda}^{mq}\left|\psi\right\rangle \quad (5.44)$$

Note that $B_l^{mq}=\left|\varphi_l\right\rangle\left\langle\varphi_l\right|$ with $l=1,2$. Then $B_l^{mq}\left|\psi\right\rangle=0$ due to the orthogonal relations $\left\langle\varphi_l|\psi\right\rangle=0$ for $l=1,2$. By using the explicit expressions (5.19) of the operators $H_{\lambda}^{mq}$ and employing the orthogonal relations $\left\langle\varphi_k|\psi\right\rangle=0$ for $k=1,2$ it can prove that $H_{\lambda}^{mq}\left|\psi\right\rangle=0$ for $\lambda=x,y,z$. Now by substituting $B_l^{mq}\left|\psi\right\rangle=0$ for $l=1,2$ and $H_{\lambda}^{mq}\left|\psi\right\rangle=0$ for $\lambda=x,y,z$ into (5.44) it can be found that $U_{\lambda}^{mq}(\theta)\left|\psi\right\rangle=\left|\psi\right\rangle$ for $\lambda=x,y,z$. The relations $U_{\lambda}^{mq}(\theta)\left|\psi\right\rangle=\left|\psi\right\rangle$ for $\lambda=x,y,z$ show clearly that these unitary operators $U_{\lambda}^{mq}(\theta)$ of (5.32) with $\lambda=x,y,z$ do not affect any vector $\left|\psi\right\rangle$ of the Hilbert space, if the vector $\left|\psi\right\rangle$ is orthogonal to both the vectors $\left|\varphi_1\right\rangle$ and $\left|\varphi_2\right\rangle$. Suppose now that the vector $\left|\psi\right\rangle$ belongs to any subspace which is orthogonal to (i.e., non-overlapping) both the subspace $S_1$ with the vector $\left|\varphi_1\right\rangle$ and $S_2$ with the vector $\left|\varphi_2\right\rangle$ in the Hilbert space $HS(N)$. Then the vector $\left|\psi\right\rangle$ is obviously orthogonal to both the vectors $\left|\varphi_1\right\rangle$ and $\left|\varphi_2\right\rangle$. Therefore, it is not affected by the unitary operator $U_{\lambda}^{mq}(\theta)$ and this indicates that the subspace with the vector $\left|\psi\right\rangle$ is not affected by the unitary operator $U_{\lambda}^{mq}(\theta)$.

The subspace-selective MQT Hermitian operators $\{H_{\lambda}^{mq}\}$ of (5.19) with $\lambda=x,y,z$ are able to act as bridge to connect the symmetrical structures and properties of the Hilbert space to the counterpart of the corresponding multiple-quantum operator algebra space. How to explicitly take into account the symmetrical structures and properties of the Hilbert space, with the help of the subspace-selective MQT Hermitian operators $\{H_{\lambda}^{mq}\}$, is described in the Subsection 6.2 and other Subsections below.

# 6. The Hilbert-space-enlarging processes based on the Hilbert-space symmetrical structures

The research work in this Section is one main contribution to this paper. The Hilbert-space-enlarging processes are one kind of the subspace-selective unitary manipulation. Theoretically they are based on the symmetrical structures and properties of the Hilbert space which are specified by the direct-sum decomposition of the Hilbert space. They may be performed in the multiple-quantum operator algebra space. According to the quantum-computing speedup theory the symmetrical structures and properties of a quantum system are the fundamental quantum-computing-speedup resources and may be specified via the multiple-quantum operator algebra space, the density operator space, and/or the Hilbert space of the quantum system; and hence the fundamental quantum-computing-speedup resources may exist in these basic quantum spaces of the quantum system. The subspace-selective unitary manipulation can play a crucial role in achieving concretely an essential quantum-computing speedup. It aims to harness the fundamental quantum-computing-speedup resources and especially those resources original from the corresponding Hilbert space to speed up essentially quantum computing and quantum simulating in the multiple-quantum operator algebra space. As the specific kind of the subspace-selective unitary manipulation, the Hilbert-space-enlarging processes are deliberately designed to make use of the fundamental quantum-computing-speedup resources which are original from the corresponding Hilbert space to achieve essential quantum-computing speedup in the multiple-quantum operator algebra space.

According to the quantum-computing speedup theory [Ref[1]] the tensor-product symmetrical structures of the Hilbert space of a composite quantum system are also the fundamental quantum-computing-speedup resource. A tensor-product subspace may be completely different from a direct-sum subspace in the Hilbert space of a composite quantum system. As the specific kind of the subspace-selective unitary manipulation, the Hilbert-space-enlarging processes also work on the basis of the direct-sum subspaces of the Hilbert space. Then an important problem is how the Hilbert-space-enlarging processes can explicitly take into account the tensor-product symmetrical structure of the Hilbert space of a composite quantum system. This Section is devoted to solving this important problem.

## 6.1. The specific Hilbert-space symmetrical structures

An $n-$spin$-1/2$ system is a typical composite quantum spin system which consists of $n$ individual spin$-1/2$ particles. According to quantum mechanics [2] the Hilbert space $HS(N)$ of the $n-$spin$-1/2$ system is the tensor product of the $n$ component Hilbert spaces $\{H_k(2)\}$ of the $n$ individual spin$-1/2$ particles of the spin system,

$$HS(N) = H_n(2)\bigotimes ...\bigotimes H_k(2)\bigotimes ...\bigotimes H_1(2), \qquad (6.1)$$

where without lost generality the index $k$ runs from right to left and from 1 to $n$, $H_k(2)$ stands for the two-dimensional component Hilbert space of the $k-$th spin$-1/2$ particle, and $N = 2^n$ is dimensional size of the Hilbert space $HS(N)$. The tensor-product symmetrical structure (6.1) of the Hilbert space $HS(N)$ is also the fundamental quantum-computing-speedup resource in accordance with the quantum-computing speedup theory. Beside the tensor-product symmetrical structure of (6.1) the Hilbert space $HS(N)$ owns the symmetric structures which may be characterized via the direct-sum decomposition of the Hilbert space.

The Hilbert space $HS(N)$ of the $n-spin-1/2$ system may be divided into the $n+1$ direct-sum subspaces $\{S_{zq}(k)\}$ [7] (See also the Section A),

$$HS(N) = S_{zq}(0)\bigoplus S_{zq}(1)\bigoplus ...\bigoplus S_{zq}(k)\bigoplus ...\bigoplus S_{zq}(n), \qquad (6.2)$$

where the ($k-$th) direct-sum subspace $S_{zq}(k)$ with $k = n/2 - M = 0, 1, 2, .., n$ is specified by one of the $n+1$ distinct values $M = -I, -I+1, ..., I-1, I$ with $I = n/2$ and its dimensional size is given by $d(k) = \begin{pmatrix} n \\ k \end{pmatrix}$, and here $I$ and $M$ are the total spin quantum number and the total spin magnetic quantum number of the spin system, respectively. As shown in the Section A, these $n+1$ direct-sum subspaces $S_{zq}(l)$ with $l = n/2 - M_r = 0, 1, 2, ..., n$ are spanned by the $n+1$ vector basis subsets $\{|M_r, K_{M_r}\rangle\}$, where $M_r = -I, -I+1, ..., I-1, I$ and $I = n/2$ for the $n-spin-1/2$ system, respectively. Moreover, every direct-sum subspace $S_{zq}(l)$ with $l = 0, 1, 2, ..., n$ is invariant under the zero-quantum ($zq$) operator subspace (See also the Subsection 5.1.3).

All the orthonormal eigenbase vectors $\{|M, K_M\rangle\}$ of the total spin operator $I_z$ form a complete set of the orthonormal base vectors of the Hilbert space $HS(N)$ of the $n-spin-1/2$ system. Here the eigenvalue equation for the total spin operator $I_z$ is $I_z\,|M_r, K_{M_r}\rangle = M_r\,|M_r, K_{M_r}\rangle$ , where the eigenvalue $M_r$ is the total spin magnetic quantum number $M$ and the index $K_{M_r}$ distinguishes between different orthonormal degenerate eigenbase vectors $\{|M_r, K_{M_r}\rangle\}$. The complete set $\{|M, K_M\rangle\}$ contains all the $n+1$ vector basis subsets $\{|M_r, K_{M_r}\rangle\}$ and may be expressed as [7] (See also the Section A)

$$\{|M, K_M\rangle\} = \{\{|n/2, K_0\rangle\}, \{|n/2-1, K_1\rangle\}, ...,$$
$$\{|-n/2+1, K_{n-1}\rangle\}, \{|-n/2, K_n\rangle\}\} \tag{6.3}$$

where $\{|M_r, K_{M_r}\rangle\} = \{|n/2-l, K_l\rangle\}$ with $l = 0, 1, 2, ..., n$. There is one-to-one correspondence between the direct-sum subspaces $\{S_{zq}(l)\}$ of (6.2) and the vector basis subsets $\{\{|n/2-l, K_l\rangle\}\}$ of (6.3). Moreover, the direct-sum subspace $S_{zq}(l)$ is spanned by the vector basis subset $\{|n/2-l, K_l\rangle\}$ and this may be simply expressed as $S_{zq}(l) = \{|n/2-l, K_l\rangle\}$ for $l = 0, 1, 2, ..., n$.

The complete set $\{|M, K_M\rangle\}$ of the Hilbert space $HS(N)$ of the $n-spin-1/2$ system may be set to the tensor-product base vectors $\{|\Phi_l^z\rangle\}$ (See (A1.31) in the Section A) which are the common eigenbase vectors of the commuting-operator set $\{\mathbf{I}_k^2, I_{kz}\}$ of the $n-spin-1/2$ system. Moreover, the tensor-product base vectors $\{|\Phi_l^z\rangle\}$ of (A1.31) also are the common eigenbase vectors of both the total spin operator $I_z$ and the commuting operator set $\{\mathbf{I}_k^2, I_{kz}\}$,

$$|\Phi_l^z\rangle = |I_n, m_n\rangle ... |I_j, m_j\rangle ... |I_1, m_1\rangle = |m_n\rangle ... |m_j\rangle ... |m_1\rangle\,, \tag{6.4}$$

where the eigenbase vector $|m_j\rangle \overset{Def}{\equiv} |I_j, m_j\rangle$ (here the spin quantum number $I_j = 1/2$) obeys the eigenvalue equation $I_{jz}\,|m_j\rangle = m_j\,|m_j\rangle$ with the eigenvalue $m_j = 1/2$ or $-1/2$ for $j = 1, 2, ..., n$. As shown by (A1.33) in the Section A, the eigenvalue equation for the total spin operator $I_z = \sum_{j=1}^{n} I_{jz}$ is written as $I_z\,|\Phi_l^z\rangle = M_l\,|\Phi_l^z\rangle$ with the eigenvalue $M_l = m_1 + m_2 + ... + m_n$. Let $|m_j\rangle = |1/2\rangle = |0_j\rangle$ and $|m_j\rangle = |-1/2\rangle = |1_j\rangle$ for $j = 1, 2, ..., n$. Then the tensor-product base vectors $\{|\Phi_l^z\rangle\}$ of (6.4) may be rewritten as $\{|S\rangle\}$:

$$|S\rangle = |s_n\rangle ... |s_j\rangle ... |s_2\rangle\,|s_1\rangle \tag{6.5}$$

where $s_j = 0, 1$ for $j = 1, 2, ..., n$ and $S = 0, 1, 2, ..., N-1$. The tensor-product base vectors $\{|S\rangle\}$ with $S = 0, 1, 2, ..., N-1$ may form a vector basis set of the Hilbert space $HS(N)$. This vector basis set $\{|S\rangle\}$ of the Hilbert space $HS(N)$ of the $n-spin-1/2$ system may act as the conventional computational basis set of the Hilbert space $HS(N)$ of the $n-$qubit spin$-1/2$ system in quantum computing and quantum simulating.

On the basis of the tensor-product base vectors $\{|S\rangle\}$ of (6.5) the direct-sum subspace $S_{zq}(k)$ with $k = 0, 1, ..., n$ of the Hilbert space $HS(N)$ in the direct-sum decomposition of (6.2) may be explicitly expressed as $S_{zq}(0) = \{|n/2, K_0\rangle\} = \{|0_n\rangle ... |0_k\rangle ... |0_2\rangle\, |0_1\rangle\} = \{|00...0\rangle\}$,
$S_{zq}(1) = \{|n/2-1, K_1\rangle\} = \{|10...0\rangle, |010...0\rangle, |0010...0\rangle, ..., |00...01\rangle\}$
$S_{zq}(2) = \{|n/2-2, K_2\rangle\} = \{|110...0\rangle, |0110...0\rangle, |01010...0\rangle, ..., |00...011\rangle\}$
......
$S_{zq}(n-1) = \{|-n/2+1, K_{n-1}\rangle\} = \{|01...1\rangle, |101...1\rangle, |1101...1\rangle, ..., |11...10\rangle\}$
$S_{zq}(n) = \{|-n/2, K_n\rangle\} = \{|11...1\rangle\}$

It is easy to find that the dimensional size $d(k)$ of the subspace $S_{zq}(k)$ is given by $\begin{pmatrix} n \\ k \end{pmatrix}$ for $k = 0, 1, ..., n$. It is easy to see that the direct-sum subspaces $\{S_{zq}(k)\}$ do not reflect the tensor-product subspaces of the Hilbert space $HS(N)$ of the composite $n - spin - 1/2$ system.

The tensor (or direct) product subspaces of the Hilbert space of a composite quantum system are frequently used in quantum computing and quantum simulating. A direct-product subspace is the component Hilbert space of a subsystem of the composite quantum system. It is not a direct-sum subspace of the Hilbert space. A direct-product subspace of the Hilbert space needs to be explicitly taken into account by the subspace-selective unitary manipulation in quantum computing and quantum simulating. When this is demanded, it is necessary to make a transition from the direct-product subspaces to the direct-sum subspaces in the Hilbert space of the composite quantum system in the subspace-selective unitary manipulation. It is shown below that theoretically such a transition can be easily realized for the specific Hilbert-space symmetrical structure. Therefore, the subspace-selective unitary manipulation, e.g., the Hilbert-space-enlarging processes, is able to take into account the tensor-product symmetrical structure of a composite quantum system such as that one (6.1) of the Hilbert space $HS(N)$ of the $n - spin - 1/2$ system.

Below a specific direct-sum decomposition for the Hilbert space $HS(N)$ of the composite $n - spin - 1/2$ system is constructed explicitly. It is different from that one of (6.2). It is able to reflect the specific (or selected) tensor-product subspaces of the Hilbert space $HS(N)$. Therefore, the subspace-selective unitary manipulation based on the Hilbert-space symmetrical structure which is specified by this specific direct-sum decomposition is able to take into account the selected tensor-product subspaces of the Hilbert space $HS(N)$. This specific direct-sum decomposition is described as follows. On the basis of the tensor-product base vectors $\{|S\rangle\}$ of (6.5), which form the complete set of base vectors of the Hilbert space $HS(N)$, the Hilbert space $HS(N)$ of the $n-$spin$-1/2$ system with dimension $N = 2^n$ can be divided into the $n+1$ direct-sum subspaces:

$$HS(N) = HS(1) \bigoplus SP_0(1) \bigoplus SP_1(2) \bigoplus SP_2(4) \bigoplus ... \bigoplus SP_k(2^k) \bigoplus ... \bigoplus SP_{n-1}(2^{n-1}) \quad (6.6)$$

where these direct-sum subspaces are respectively defined via the tensor-product base vectors $\{|S\rangle\}$ of (6.5) with $S = 0, 1, ..., N-1$:

$HS(1) = \{|0\rangle\}$, $HS(2) = \{|0\rangle, |1\rangle\}$, ......, $HS(N) = \{|0\rangle, |1\rangle, ..., |N-1\rangle\}$;
$SP_0(1) = \{|1\rangle\}$,

$SP_1(2) = \{|2\rangle, |3\rangle\}$,
$SP_2(4) = \{|4\rangle, |5\rangle, |6\rangle, |7\rangle\}$,
......,
$SP_k(2^k) = \{|2^k\rangle, |2^k+1\rangle, |2^k+2\rangle, ..., |2^{k+1}-1\rangle\}$,
......,
$SP_{n-1}(2^{n-1}) = \{|2^{n-1}\rangle, |2^{n-1}+1\rangle, |2^{n-1}+2\rangle, ..., |2^n-1\rangle\}$;

here the subscript $k$ and the number $L_k = 2^k$ for the $k-$th direct-sum subspace $SP_k(L_k)$ with $k = 0, 1, ..., n-1$ are used to label the first base vector $|2^k\rangle$ (or the minimum number base vector $|2^k\rangle$) and the dimensional size of the subspace $SP_k(L_k)$, respectively. For convenience the direct-sum decomposition (6.6) may be simply written as

$$HS(N) = HS(1) \bigoplus \left\{\bigoplus_{k=0}^{n-1} SP_k(2^k)\right\}. \tag{6.7}$$

There is the characteristic feature that the direct-sum decomposition (6.6) or (6.7) for the whole Hilbert space $HS(N)$ with dimension $N = 2^n$ also is available for any direct-sum subspace $SP_k(L_k)$ with dimension $L_k = 2^k$ and $0 \leq k \leq n-1$. The direct-sum decomposition (6.6) of the Hilbert space $HS(N)$ of the $n-$spin$-1/2$ system can be applied as well to the Hilbert space of an $n-$pseudospin$-1/2$ system (See the Subsection 3.2.2).

The direct-sum decomposition of (6.6) is equivalent to the one of (6.2) for the same Hilbert space $HS(N)$ in the sense that one direct-sum decomposition can be changed to another by making a suitable vector-base re-arrangement for the tensor-product base vectors $\{|S\rangle\}$ of (6.5) in decomposition.[18]

Below it is shown that the direct-sum decomposition (6.6) of the Hilbert space $HS(N)$ is able to reflect the specific (or selected) tensor-product subspaces of the Hilbert space $HS(N)$ of the composite $n-spin-1/2$ system. The tensor-product base vectors $\{|\Phi_l^{\tilde{z}}\rangle\}$ of (6.4) form a complete set of base vectors of the Hilbert space $HS(N)$ and may be rewritten as

$$|\Phi_l^{\tilde{z}}\rangle = |m_n\rangle ... |m_{n_t+1}\rangle \bigotimes |m_{n_t}\rangle ... |m_2\rangle |m_1\rangle \tag{6.8}$$

where the index $n_t$ runs from right to left and from 1 to $n$. Then the selected tensor-product subspace $TPS(n_t)$ of the Hilbert space $HS(N)$ of the composite $n - spin - 1/2$ system is defined by

$$TPS(n_t) = \{|m_{n_t}\rangle ... |m_2\rangle |m_1\rangle\} \text{ with } 1 \leq n_t \leq n \tag{6.9}$$

Obviously, the tensor-product subspace $TPS(n_t)$ is the (component) Hilbert space of the $n_t-$spin$-1/2$ subsystem ($1 \leq n_t \leq n$) which consists of the first $n_t$ spin$-1/2$ particles of the $n$ spin$-1/2$ particles of the $n - spin - 1/2$ system (from right to left). It is spanned by the complete set of the orthonormal tensor-product base vectors $\{|m_{n_t}\rangle ... |m_2\rangle |m_1\rangle\}$ of the $n_t - spin - 1/2$ subsystem of the $n - spin - 1/2$ system. However, a tensor-product subspace $TPS(n_t)$

[18] Such a vector-base re-arrangement could be realized with the help of the subspace-selective unitary manipulation in Ref. [7].

$(1 \leq n_t < n)$ is not a direct-sum subspace of the whole Hilbert space $HS(N)$. Then in the subspace-selective unitary manipulation it needs to be changed to a direct-sum subspace.

First of all, consider the tensor-product subspace $TPS(1) = \{|m_1\rangle\}$ of (6.9) with $n_t = 1$. Let $n_t = 1$, $|m_1\rangle = |1/2\rangle = |0_1\rangle$, and $|m_j\rangle = |1/2\rangle = |0_j\rangle$ for $j = 2, 3, ..., n$ in the tensor-product base vectors $\{|\Phi_l^{\tilde{z}}\rangle\}$ of (6.8). Then one obtains $|\Phi_0^{\tilde{z}}\rangle = |0_n\rangle ... |0_j\rangle ... |0_2\rangle \bigotimes |0_1\rangle$ . This tensor-product base vector $|\Phi_0^{\tilde{z}}\rangle$ can form a one-dimensional subspace $\{|\Phi_0^{\tilde{z}}\rangle\}$ of the Hilbert space $HS(N)$:

$$\{|\Phi_0^{\tilde{z}}\rangle\} = \{|0_n\rangle ... |0_j\rangle ... |0_2\rangle \bigotimes |0_1\rangle\} \tag{6.10}$$

The subspace $\{|\Phi_0^{\tilde{z}}\rangle\}$ may be briefly written as $\{|\Phi_0^{\tilde{z}}\rangle\} = \{|0\rangle\}$. Obviously, it is just the direct-sum subspace $HS(1)$ of the Hilbert space $HS(N)$ in the direct-sum decomposition (6.6). Let $n_t = 1$, $|m_1\rangle = |-1/2\rangle = |1_1\rangle$, and $|m_j\rangle = |1/2\rangle = |0_j\rangle$ for $j = 2, 3, ..., n$ in the tensor-product base vectors $\{|\Phi_l^{\tilde{z}}\rangle\}$. Then one obtains $|\Phi_1^{\tilde{z}}\rangle = |0_n\rangle ... |0_j\rangle ... |0_2\rangle \bigotimes |1_1\rangle$ . This tensor-product base vector $|\Phi_1^{\tilde{z}}\rangle$ also can form a one-dimensional subspace $\{|\Phi_1^{\tilde{z}}\rangle\}$ of the Hilbert space $HS(N)$:

$$\{|\Phi_1^{\tilde{z}}\rangle\} = \{|0_n\rangle ... |0_j\rangle ... |0_2\rangle \bigotimes |1_1\rangle\} \tag{6.11}$$

The two subspaces $\{|\Phi_0^{\tilde{z}}\rangle\}$ and $\{|\Phi_1^{\tilde{z}}\rangle\}$ evidently are mutually orthogonal, because both the component base vectors $|1_1\rangle$ and $|0_1\rangle$ of the first spin$-1/2$ particle of the $n-spin-1/2$ system are orthogonal to one another. The subspace $\{|\Phi_1^{\tilde{z}}\rangle\}$ may be briefly written as $\{|\Phi_1^{\tilde{z}}\rangle\} = \{|1\rangle\}$. Obviously, it is just the direct-sum subspace $SP_0\,(1)$ of the Hilbert space $HS(N)$ in the direct-sum decomposition (6.6).

Next, consider the tensor-product subspace $TPS(2) = \{|m_{n_t}\rangle\,|m_1\rangle\}$ with $n_t = 2$ and $|m_{n_t}\rangle = |1_2\rangle$ . Let $n_t = 2$, $|m_2\rangle = |-1/2\rangle = |1_2\rangle$, and $|m_j\rangle = |1/2\rangle = |0_j\rangle$ for $j = 3, 4, ..., n$ in the tensor-product base vectors $\{|\Phi_l^{\tilde{z}}\rangle\}$. Then one can obtain the two tensor-product base vectors $|\Phi_l^{\tilde{z}}\rangle = |0_n\rangle ... |0_j\rangle ... |0_3\rangle \bigotimes |1_2\rangle\,|m_1\rangle$ with $|m_1\rangle = |0_1\rangle\,,\ |1_1\rangle$ . These two tensor-product base vectors span a two-dimensional subspace $\{|\Phi_2^{\tilde{z}}\rangle\}$ of the Hilbert space $HS(N)$ :

$$\{|\Phi_2^{\tilde{z}}\rangle\} = \{|0_n\rangle ... |0_j\rangle ... |0_3\rangle \bigotimes |1_2\rangle\,|m_1\rangle\} \tag{6.12}$$

where $|m_1\rangle = |0_1\rangle\,,\ |1_1\rangle$ . This two-dimensional subspace $\{|\Phi_2^{\tilde{z}}\rangle\}$ may be briefly written as $\{|\Phi_2^{\tilde{z}}\rangle\} = \{|2\rangle\,,|3\rangle\}$. Obviously, it is just the direct-sum subspace $SP_1\,(2)$ in the direct-sum decomposition (6.6). It is orthogonal to every one of the previous two subspaces $\{|\Phi_0^{\tilde{z}}\rangle\}$ and $\{|\Phi_1^{\tilde{z}}\rangle\}$, because both the component base vectors $|1_2\rangle$ and $|0_2\rangle$ of the second spin$-1/2$ particle of the $n - spin - 1/2$ system are orthogonal to one another, and here the component base vector $|1_2\rangle$ is contained only in every tensor-product base vector $|\Phi_2^{\tilde{z}}\rangle$ of the subspace $\{|\Phi_2^{\tilde{z}}\rangle\}$, while the component base vector $|0_2\rangle$ is contained only in all the tensor-product base vectors of the two subspaces $\{|\Phi_0^{\tilde{z}}\rangle\}$ and $\{|\Phi_1^{\tilde{z}}\rangle\}$.

Now consider the tensor-product subspace $TPS(3) = \{|m_{n_t}\rangle\,|m_2\rangle\,|m_1\rangle\}$ with $n_t = 3$ and $|m_{n_t}\rangle = |1_3\rangle$ . Let $n_t = 3$, $|m_3\rangle = |-1/2\rangle = |1_3\rangle$, and $|m_j\rangle = |1/2\rangle = |0_j\rangle$ for $j = 4, 5, ..., n$ in the tensor-product base vectors $\{|\Phi_l^{\tilde{z}}\rangle\}$. Then

it is known from (6.8) that there are four tensor-product base vectors $|\Phi_l^z\rangle = |0_n\rangle ... |0_j\rangle ... |0_4\rangle \bigotimes |1_3\rangle |m_2\rangle |m_1\rangle$ with $|m_k\rangle = |0_k\rangle$ , $|1_k\rangle$ and $k = 1, 2$. These four tensor-product base vectors span a four-dimensional subspace $\{|\Phi_{2^2}^z\rangle\}$ of the Hilbert space $HS(N)$ :

$$\{|\Phi_{2^2}^z\rangle\} = \{|0_n\rangle ... |0_j\rangle ... |0_4\rangle \bigotimes |1_3\rangle |m_2\rangle |m_1\rangle\} \tag{6.13}$$

where $|m_2\rangle |m_1\rangle = |0_2\rangle |0_1\rangle$ , $|0_2\rangle |1_1\rangle$ , $|1_2\rangle |0_1\rangle$ , and $|1_2\rangle |1_1\rangle$ . This subspace $\{|\Phi_{2^2}^z\rangle\}$ may be briefly written as $\{|\Phi_{2^2}^z\rangle\} = \{|4\rangle, |5\rangle, |6\rangle, |7\rangle\}$, indicating that it is just the direct-sum subspace $SP_2\,(4)$ in the direct-sum decomposition (6.6). Evidently it is orthogonal to every one of the previous three subspaces $\{|\Phi_0^z\rangle\}$, $\{|\Phi_1^z\rangle\}$, and $\{|\Phi_2^z\rangle\}$.

Generally, consider the selected tensor-product subspace $TPS(k) = \{|m_{n_t}\rangle |m_{k-1}\rangle ... |m_2\rangle |m_1\rangle\}$ of (6.9) with $n_t = k \geq 2$ and $|m_{n_t}\rangle = |1_k\rangle$ . Let $n_t = k$, $|m_k\rangle = |-1/2\rangle = |1_k\rangle$, and $|m_j\rangle = |1/2\rangle = |0_j\rangle$ for $j = k+1, k+2, ..., n$ in the tensor-product base vectors $\{|\Phi_l^z\rangle\}$ of (6.8). Then it can be deduced from (6.8) that there are $2^{k-1}$ tensor-product base vectors $|\Phi_l^z\rangle = |0_n\rangle ... |0_j\rangle ... |0_{k+1}\rangle \bigotimes |1_k\rangle |m_{k-1}\rangle ... |m_2\rangle |m_1\rangle$ with $|m_l\rangle = |0_l\rangle$ , $|1_l\rangle$ and $l = 1, 2, ..., k-1$. These $2^{k-1}$ tensor-product base vectors $\{|\Phi_l^z\rangle\}$ span a $2^{k-1}-$dimensional subspace $\{|\Phi_{2^{k-1}}^z\rangle\}$ of the Hilbert space $HS(N)$ :

$$\{|\Phi_{2^{k-1}}^z\rangle\} = \{|0_n\rangle ... |0_j\rangle ... |0_{k+1}\rangle \bigotimes |1_k\rangle |m_{k-1}\rangle ... |m_2\rangle |m_1\rangle\} \tag{6.14}$$

where $k = 2, 3, ..., n$ and $|m_l\rangle = |0_l\rangle$ , $|1_l\rangle$ for $l = 1, 2, ..., k-1$. This subspace $\{|\Phi_{2^{k-1}}^z\rangle\}$ may be simply written as $\{|\Phi_{2^{k-1}}^z\rangle\} = \{|2^{k-1}\rangle, |2^{k-1}+1\rangle, ..., |2^k - 1\rangle\}$. Therefore, it is really the direct-sum subspace $SP_{k-1}\left(2^{k-1}\right)$ of the Hilbert space $HS(N)$ in the direct-sum decomposition (6.6). It is orthogonal to every one of the previous subspaces $\{|\Phi_0^z\rangle\}$, $\{|\Phi_1^z\rangle\}$, $\{|\Phi_2^z\rangle\}$, ..., and $\{|\Phi_{2^{k-2}}^z\rangle\}$. The reason for this is that both the component base vectors $|1_k\rangle$ and $|0_k\rangle$ of the $k-$th spin$-1/2$ particle of the $n - spin - 1/2$ system are orthogonal to one another, and here the component base vector $|1_k\rangle$ is contained only in every tensor-product base vector $|\Phi_{2^{k-1}}^z\rangle$ of the subspace $\{|\Phi_{2^{k-1}}^z\rangle\}$, while the component base vector $|0_k\rangle$ is contained only in all the tensor-product base vectors of the previous subspaces $\{|\Phi_0^z\rangle\}$, $\{|\Phi_1^z\rangle\}$, $\{|\Phi_2^z\rangle\}$, ..., and $\{|\Phi_{2^{k-2}}^z\rangle\}$.

In particular, the final subspace $|\Phi_{2^{n-1}}^z\rangle$ can be obtained from (6.14). Let $n_t = n$ and $|m_n\rangle = |-1/2\rangle = |1_n\rangle$. Then it can be deduced from (6.14) that the final $2^{n-1}-$dimensional subspace $\{|\Phi_{2^{n-1}}^z\rangle\}$ is given by

$$\{|\Phi_{2^{n-1}}^z\rangle\} = \{|1_n\rangle |m_{n-1}\rangle ... |m_2\rangle |m_1\rangle\} \tag{6.15}$$

where $|m_l\rangle = |0_l\rangle$ , $|1_l\rangle$ for $l = 1, 2, ..., n-1$. The $2^{n-1}-$dimensional subspace $\{|\Phi_{2^{n-1}}^z\rangle\}$ may be simply written as $\{|\Phi_{2^{n-1}}^z\rangle\} = \{|2^{n-1}\rangle, |2^{n-1}+1\rangle, ..., |2^n - 1\rangle\}$, indicating that it is really the direct-sum subspace $SP_{n-1}\left(2^{n-1}\right)$ in the direct-sum decomposition (6.6). Moreover, it is orthogonal to every one of the previous subspaces $\{|\Phi_0^z\rangle\}$, $\{|\Phi_1^z\rangle\}$, $\{|\Phi_2^z\rangle\}$, ..., and $\{|\Phi_{2^{n-2}}^z\rangle\}$.

All these $n+1$ subspaces $\{|\Phi_0^z\rangle\}$, $\{|\Phi_1^z\rangle\}$, $\{|\Phi_2^z\rangle\}$, ..., $|\Phi_{2^{k-1}}^z\rangle$ , ..., $\{|\Phi_{2^{n-1}}^z\rangle\}$ are mutually orthogonal. Sum of the dimensional sizes of all these $n+1$ subspaces is

given by $1+1+2+2^2+...+2^{n-2}+2^{n-1}=2^n$. This sum is exactly equal to the dimensional size $N=2^n$ of the whole Hilbert space $HS(N)$ of the $n-spin-1/2$ system. Therefore, the whole Hilbert space $HS(N)$ of the $n-spin-1/2$ system may be expressed as the direct sum of all these $n+1$ subspaces:

$$HS\left(N\right)=\{|\Phi_0^z\rangle\}\bigoplus\{|\Phi_1^z\rangle\}\bigoplus\{|\Phi_2^z\rangle\}\bigoplus...\bigoplus\{|\Phi_{2^{k-1}}^z\rangle\}\bigoplus...\bigoplus\{|\Phi_{2^{n-1}}^z\rangle\} \quad (6.16)$$

It can be further found that there are the relations for the direct-sum subspaces between the two direct-sum decompositions (6.16) and (6.6):

$$HS\left(1\right)=\{|\Phi_0^z\rangle\},\ SP_0\left(1\right)=\{|\Phi_1^z\rangle\},\ SP_1\left(2\right)=\{|\Phi_2^z\rangle\},...,$$

$$SP_{k-1}\left(2^{k-1}\right)=\{|\Phi_{2^{k-1}}^z\rangle\},...,\ SP_{n-1}\left(2^{n-1}\right)=\{|\Phi_{2^{n-1}}^z\rangle\}$$

Therefore, the direct-sum decomposition of (6.16) for the Hilbert space $HS(N)$ is just the previous one of (6.6).

The theoretical analysis above shows clearly that the direct-sum decomposition (6.16) or (6.6) of the Hilbert space $HS(N)$ is able to explicitly take into account the specific (or selected) tensor-product subspaces of the Hilbert space $HS(N)$ of the composite $n-spin-1/2$ system. This result is important for quantum computing and quantum simulating. According to the quantum-computing speedup theory the tensor-product symmetrical structure (6.1) of the Hilbert space $HS(N)$ is also the fundamental quantum-computing-speedup resource. This means that the fundamental quantum-computing-speedup resource may be taken into account in the direct-sum decomposition (6.6) of the Hilbert space $HS(N)$. Therefore, the subspace-selective unitary manipulation based on the symmetric structure of the Hilbert space $HS(N)$ specified by the direct-sum decomposition (6.6) is able to take into account the tensor-product symmetrical structure (6.1) of the Hilbert space $HS(N)$. Then, as the specific kind of the subspace-selective unitary manipulation, the Hilbert-space-enlarging processes are able to take into account the tensor-product symmetrical structure too.

*Remark*. Let the Hermitian operators $H_{00}^{ss}$ and $H_{10}^{ss}$ be

$$H_{00}^{ss}=|0\rangle\langle 0|=(|0_n\rangle...|0_2\rangle|\mathbf{0}_1\rangle)(|0_n\rangle...|0_2\rangle|\mathbf{0}_1\rangle)^+,$$

$$H_{10}^{ss}=|1\rangle\langle 1|=(|0_n\rangle...|0_2\rangle|\mathbf{1}_1\rangle)(|0_n\rangle...|0_2\rangle|\mathbf{1}_1\rangle)^+,$$

and generally let the Hermitian operator $H_{k,k-1}^{ss}$ for $k=2,3,...,n$ be

$$H_{k,k-1}^{ss}=(|0_n\rangle...|0_{k+1}\rangle|\mathbf{1}_k\rangle)(|0_n\rangle...|0_{k+1}\rangle|\mathbf{1}_k\rangle)^+\bigotimes H_{k-1}^s$$

Here the Hermitian operators $H_{k,k-1}^{ss}$ for $k=1,2,3,...,n$ are defined by (6.70) in the Subsection 6.4 below. As shown in the Subsection 6.4, these Hermitian operators are the subspace-selective MQT Hermitian operators. They are used to generate the first-kind subspace-selective MQT unitary operators $U_{k,k-1}^{ss}(\tau)=\exp\left(-iH_{k,k-1}^{ss}\tau/\hbar\right)$ $(1\leq k\leq n)$ and $U_{00}^{ss}(\tau)=\exp\left(-iH_{00}^{ss}\tau/\hbar\right)$. They may constitute an operator set $\{H_{00}^{ss}, H_{k,k-1}^{ss}\}$. Then it can prove that every direct-sum subspace $(HS(1),\ SP_k\left(2^k\right))$ of the Hilbert space $HS(N)$ in the direct-sum decomposition (6.6) or (6.16) is invariant under the operator set $\{H_{00}^{ss}, H_{k,k-1}^{ss}\}$ [3]. Here the specific case such as $H_{k,k-1}^{ss}\left|\Phi_{2^l}^{z}\right\rangle = 0$ is not conflict with the conclusion.

The direct-sum decomposition (6.6) or (6.16) of the Hilbert space $HS\left(N\right)$ of the $n-spin-1/2$ system also may be rewritten in the form

$$HS\left(N\right) = HS\left(N_k\right) \bigoplus SP_k\left(2^k\right) \bigoplus ... \bigoplus SP_{n-1}\left(2^{n-1}\right),\ 0 \leq k \leq n, \qquad (6.17)$$

and the direct-sum subspace $HS\left(N_k\right)$ with dimension $N_k = 2^k$ is defined by

$$HS\left(N_k\right) = HS\left(1\right) \bigoplus SP_0\left(1\right) \bigoplus SP_1\left(2\right) \bigoplus ... \bigoplus SP_{k-1}\left(2^{k-1}\right), \qquad (6.18)$$

where when $k=0$, $HS\left(N_0\right) \equiv HS\left(1\right)$ with $N_0 = 1$; when $k=1$, $HS\left(N_1\right) = HS(2) = HS\left(1\right) \bigoplus SP_0\left(1\right)$ with $N_1 = 2$; and when $k=n$, $HS(N_n) \equiv HS(N)$ with $N_n = N = 2^n$. On the basis of the direct-sum decomposition (6.6) or (6.17) of the Hilbert space $HS(N)$ the dimensional sizes of the direct-sum subspaces $HS(N_k)$ and $\{SP_k\left(2^k\right)\}$ constitute an important physical quantity and also one important aspect to reflect the symmetrical structures and properties of the Hilbert space of the $n-spin-1/2$ system [7]. They are considered as one important resource that may be exploited to realize quantum-computing speedup in quantum computing and quantum simulating. These are just like that on the basis of the direct-sum decomposition (6.2) of the Hilbert space $HS(N)$ the dimensional sizes $\{d\left(k\right)\}$ of the direct-sum subspaces $\{S_{zq}\left(k\right)\}$ constitute an important physical quantity and also one important aspect to reflect the symmetrical structures and properties of the Hilbert space $HS(N)$ and are considered as one important resource used for quantum-computing speedup (See Ref.[7] and also the Subsection 5.1).

The direct-sum decomposition (6.6) or (6.17) of the Hilbert space $HS(N)$ has important applications in the subspace-selective unitary manipulation. One of the important applications is the Hilbert-space-enlarging processes in an $n-spin-1/2$ system. The Hilbert-space-enlarging processes are able to change the dimensional sizes of the occupied direct-sum subspaces such as $HS(N_k)$ and $\{SP_k\left(2^k\right)\}$ of the Hilbert space $HS(N)$. They are described in detail later.

## 6.2. Inter-conversion between different direct-sum subspaces of the Hilbert space

The inter-conversion between different direct-sum Hilbert subspaces is basic and important in the subspace-selective unitary manipulation. It may be realized by the subspace-selective PDHO-generated unitary operators of (4.45) in the Subsection 4.5 or by the subspace-selective MQT unitary operators in the Subsection 5.2. For simplicity, here consider only the subspace-selective MQT unitary operators $U_{\lambda}^{mq}\left(\theta\right)$ $(\lambda = x, y, z)$ of (5.32) in the Subsection 5.2. As shown in the Subsection 5.2, the general operator-expansion formula (5.39) for the subspace-selective MQT unitary operators $U_{\lambda}^{mq}\left(\theta\right)$ $(\lambda = x, y, z)$ plays a key role in calculating exactly and analytically the unitary time-evolutional processes or any other unitary transformations, which are governed by the unitary propagators $U_{\lambda}^{mq}(\theta)$. The formula (5.39) is independent of any detailed vector basis set and any detailed direct-sum decomposition of the Hilbert space. It holds for any dimensional size of the Hilbert space. As shown by (5.32), the subspace-selective MQT unitary operators $U_{\lambda}^{mq}(\theta)$ $(\lambda = x, y, z)$ are directly generated by the subspace-selective MQT Hermitian operators $\{H_{\lambda}^{mq}\}$ of (5.19). The latter are able to act as bridge to connect the symmetric structures and properties of the Hilbert space to the counterpart of the corresponding multiple-quantum operator algebra space.

The symmetrical structures and properties of the Hilbert space may be characterized by the direct-sum decomposition of the Hilbert space, as shown in the preceding Subsection 6.1. Below consider an $n - spin - 1/2$ system. The vector basis set of the Hilbert space $HS(N)$ of the $n - spin - 1/2$ system may be conveniently formed by the tensor-product base vectors $\{|\Phi_l^z\rangle\}$ of (6.8) or base vectors $\{|S\rangle\}$ of (6.5) with $S = 0, 1, 2, ..., N-1$. Then the symmetrical structure of the Hilbert space $HS(N)$ may be specified by the direct-sum decomposition (6.16) or (6.6). The selective multiple quantum transitions in the $n - spin - 1/2$ system between different direct-sum subspaces $\{SP_k(L_k)\}$ of the Hilbert space $HS(N)$ with the direct-sum decomposition (6.6) or (6.7): $HS(N) = HS(1) \bigoplus \{\bigoplus_{k=0}^{n-1} SP_k(2^k)\}$ may be induced by the subspace-selective MQT unitary operators $U_{\lambda}^{mq}(\theta)$ $(\lambda = x, y)$ of (5.32). Here the subspace-selective MQT unitary operators $U_{\lambda}^{mq}(\theta)$ $(\lambda = x, y, z)$ may be explicitly constructed. As a typical example, suppose that both the vectors $|\varphi_1\rangle$ and $|\varphi_2\rangle$ in the subspace-selective MQT Hermitian operators $\{H_{\lambda}^{mq}\}$ of (5.19) belong to the two direct-sum subspaces $SP_{k_0}(2^{k_0})$ and $SP_{k_1}(2^{k_1})$, respectively, and they are simply taken as

$$|\varphi_1\rangle = \frac{1}{\sqrt{2^{k_0}}} \sum_{k=2^{k_0}}^{2^{k_0+1}-1} |k\rangle, \ \ |\varphi_2\rangle = \frac{1}{\sqrt{2^{k_1}}} \sum_{k=2^{k_1}}^{2^{k_1+1}-1} |k\rangle, \ 0 \le k_0 < k_1 \le n-1, \tag{6.19}$$

where both the vectors $|\varphi_1\rangle \in SP_{k_0}(2^{k_0})$ and $|\varphi_2\rangle \in SP_{k_1}(2^{k_1})$ are clearly orthonormal and the tensor-product base vectors $\{|k\rangle\}$ are given by (6.5). By substituting the two orthonormal vectors $|\varphi_1\rangle$ and $|\varphi_2\rangle$ of (6.19) into (5.19) one can construct the subspace-selective MQT Hermitian operators $H_{\lambda}^{mq}$ with $\lambda = x, y, z$. Then the subspace-selective MQT unitary operators $U_{\lambda}^{mq}(\theta) = \exp(-i\theta H_{\lambda}^{mq})$ can be constructed with the Hermitian operators $H_{\lambda}^{mq}$ with $\lambda = x, y, z$. It can be seen from (6.19) that for the vector $|\varphi_1\rangle$ the expansional coefficient is given by $1/\sqrt{2^{k_0}}$ for every tensor-product base vector $|k\rangle$ of (6.5) in the subspace $SP_{k_0}(2^{k_0})$, while for the vector $|\varphi_2\rangle$ the expansional coefficient is given by $1/\sqrt{2^{k_1}}$ for every tensor-product base vector $|k\rangle$ in the subspace $SP_{k_1}(2^{k_1})$. Such choice for these expansional coefficients of the vectors $|\varphi_1\rangle$ and $|\varphi_2\rangle$ can lead to that the subspace-selective MQT unitary operators $U_{\lambda}^{mq}(\theta)$ can be constructed and realized efficiently. Moreover, it can reflect well the dimensional sizes of the direct-sum subspaces $SP_{k_0}(2^{k_0})$ and $SP_{k_1}(2^{k_1})$. Such constructed subspace-selective MQT unitary operators $U_{\lambda}^{mq}(\theta)$ can be efficiently realized with the help of the unitary transformation (5.33) in the Subsection 5.2.

By substituting both the vectors $|\varphi_1\rangle$ and $|\varphi_2\rangle$ in (6.19) into (5.19) it can be shown that the subspace-selective MQT Hermitian operators $\{H_\mu^{mq}\}$ of (5.19) with $\mu = x, y$ may be explicitly written as

$$H_x^{mq} = H_x^{k_0,k_1} = \frac{1}{\sqrt{2^{k_0}}\sqrt{2^{k_1}}} \sum_{k=2^{k_0}}^{2^{k_0+1}-1} \sum_{l=2^{k_1}}^{2^{k_1+1}-1} \frac{1}{2}\left(|k\rangle\langle l| + |l\rangle\langle k|\right) \tag{6.20a}$$

$$H_y^{mq} = H_y^{k_0,k_1} = \frac{1}{\sqrt{2^{k_0}}\sqrt{2^{k_1}}} \sum_{k=2^{k_0}}^{2^{k_0+1}-1} \sum_{l=2^{k_1}}^{2^{k_1+1}-1} \frac{1}{2i}\left(|k\rangle\langle l| - |l\rangle\langle k|\right) \tag{6.20b}$$

where $Q_{|p|,x}^{kl} = \frac{1}{2}\left(|k\rangle\langle l| + |l\rangle\langle k|\right)$ and $Q_{|p|,y}^{kl} = \frac{1}{2i}\left(|k\rangle\langle l| - |l\rangle\langle k|\right)$ are just the Hermitian selective $|p|$ −quantum transition operators (i.e., $Q_{|p|,x}^{fi}$ and $Q_{|p|,y}^{fi}$ of (5.2)) with the quantum-transition order $p = M_k - M_l$, respectively, and here the total spin magnetic quantum number $M_j$ is determined by the eigenvalue equation $I_z|j\rangle = M_j|j\rangle$ for $j = k, l$. Note that here the quantum-transition order $p$ may take any value in the range from $p = -n$ to $n$ for the $n-spin-1/2$ system. By comparing (6.20) with the base-operator expansion (5.15) it is easy to see that the subspace-selective MQT Hermitian operators $H_x^{k_0,k_1}$ and $H_y^{k_0,k_1}$ each can be expanded in terms of the complete set $\{E, Q_{|p|,x}^{fi}, Q_{|p|,y}^{fi}, Q_{|p|,z}^{fi}\}$ of base operators of the multiple-quantum operator algebra space.

By applying the unitary operator $U_y^{mq}(\theta) = \exp\left(-i\theta H_y^{k_0,k_1}\right)$ with $H_y^{mq} = H_y^{k_0,k_1}$ of (6.20b) to the vector $|\varphi_1\rangle$ (or $|\varphi_2\rangle$) in (6.19) it can be obtained from the basic unitary transformations of (5.41) in the Hilbert space $HS(N)$ that

$$U_y^{mq}(\theta)|\varphi_1\rangle = \left(\frac{1}{\sqrt{2^{k_0}}} \sum_{k=2^{k_0}}^{2^{k_0+1}-1} |k\rangle\right)\cos\frac{1}{2}\theta + \left(\frac{1}{\sqrt{2^{k_1}}} \sum_{k=2^{k_1}}^{2^{k_1+1}-1} |k\rangle\right)\sin\frac{1}{2}\theta \tag{6.21a}$$

$$U_y^{mq}(\theta)|\varphi_2\rangle = \left(\frac{1}{\sqrt{2^{k_1}}} \sum_{k=2^{k_1}}^{2^{k_1+1}-1} |k\rangle\right)\cos\frac{1}{2}\theta - \left(\frac{1}{\sqrt{2^{k_0}}} \sum_{k=2^{k_0}}^{2^{k_0+1}-1} |k\rangle\right)\sin\frac{1}{2}\theta \tag{6.21b}$$

Note that $|\varphi_1\rangle \in SP_{k_0}\left(2^{k_0}\right)$ and $|\varphi_2\rangle \in SP_{k_1}\left(2^{k_1}\right)$. The basic unitary transformations of (6.21) show clearly that the selective multiple quantum transitions between the two direct-sum subspaces $SP_{k_0}\left(2^{k_0}\right)$ and $SP_{k_1}\left(2^{k_1}\right)$ may be induced by the subspace-selective MQT unitary operators $U_y^{mq}(\theta)$. When $\theta = \pi$, the unitary transformations of (6.21) may be respectively reduced to the simple forms

$$U_y^{mq}(\pi)|\varphi_1\rangle = \frac{1}{\sqrt{2^{k_1}}} \sum_{k=2^{k_1}}^{2^{k_1+1}-1} |k\rangle = |\varphi_2\rangle, \tag{6.22a}$$

$$U_y^{mq}(\pi)|\varphi_2\rangle = -\frac{1}{\sqrt{2^{k_0}}} \sum_{k=2^{k_0}}^{2^{k_0+1}-1} |k\rangle = -|\varphi_1\rangle \tag{6.22b}$$

Then the unitary transformations of (6.22) show clearly that the vector $|\varphi_1\rangle$ can make complete quantum transition to the vector $|\varphi_2\rangle$ and vice versa under the influence of the subspace-selective MQT unitary operator $U_y^{mq}(\pi)$. That is, the vector $|\varphi_1\rangle$ of the direct-sum subspace $SP_{k_0}\left(2^{k_0}\right)$ can be completely converted into the vector $|\varphi_2\rangle$ of the direct-sum subspace $SP_{k_1}\left(2^{k_1}\right)$ and vice versa, after it is acted on by the subspace-selective MQT unitary operator $U_y^{mq}(\pi)$.

Consider the special case that the subspace $SP_{k_1}\left(2^{k_1}\right)$ with the vector $|\varphi_2\rangle$ is set to the subspace $SP_{k_0+1}\left(2^{k_0+1}\right)$ (i.e., $k_1 = k_0+1$) which is nearest the subspace $SP_{k_0}\left(2^{k_0}\right)$ with the vector $|\varphi_1\rangle$. Obviously, in this case the dimensional size $2^{k_1}$ of the subspace $SP_{k_1}\left(2^{k_1}\right)$ with $k_1 = k_0 + 1$ is two times larger than the dimensional size $2^{k_0}$ of the subspace $SP_{k_0}\left(2^{k_0}\right)$. Then the unitary transformations of (6.21a) with $\theta = \pi$ and (6.21b) with $\theta = -\pi$ may be respectively reduced to the simple forms

$$U_y^{mq}(\pi)\left(\frac{1}{\sqrt{2^{k_0}}}\sum_{k=2^{k_0}}^{2^{k_0+1}-1}|k\rangle\right) = \frac{1}{\sqrt{2}}\left(\frac{1}{\sqrt{2^{k_0}}}\sum_{k=2^{k_0+1}}^{2^{k_0+2}-1}|k\rangle\right) \tag{6.23a}$$

$$U_y^{mq}(-\pi)\left(\frac{1}{\sqrt{2^{k_0+1}}}\sum_{k=2^{k_0+1}}^{2^{k_0+2}-1}|k\rangle\right) = \sqrt{2}\left(\frac{1}{\sqrt{2^{k_0+1}}}\sum_{k=2^{k_0}}^{2^{k_0+1}-1}|k\rangle\right) \tag{6.23b}$$

For the unitary transformation (6.23a) the final subspace $SP_{k_0+1}\left(2^{k_0+1}\right)$, where the $n - spin - 1/2$ system in the final vector $|\varphi_2\rangle$ occupies, is two times larger than the initial subspace $SP_{k_0}\left(2^{k_0}\right)$, where the same spin system in the initial vector $|\varphi_1\rangle$ occupies. Therefore, intuitively one may say that the unitary transformation (6.23a) is a special Hilbert-space-enlarging process. In contrast, for the unitary transformation (6.23b) the final subspace $SP_{k_0}\left(2^{k_0}\right)$, where the $n{-}spin{-}1/2$ system in the final vector $|\varphi_1\rangle$ occupies, is one half smaller than the initial subspace $SP_{k_0+1}\left(2^{k_0+1}\right)$, where the same spin system in the initial vector $|\varphi_2\rangle$ occupies. Then one may say intuitively that the unitary transformation (6.23b) is a special Hilbert-space-shrinking process. A Hilbert-space-shrinking process is the inverse of a Hilbert-space-enlarging process. General Hilbert-space-enlarging processes and their applications in the multiple-quantum operator algebra space are discussed in next Subsections.

Moreover, it can be found from (6.23) that if the initial subspace (e.g., $SP_{k_0}\left(2^{k_0}\right)$) is enlarged twofold in dimensional size, the expansional coefficient for every occupied base vector $|k\rangle$ is decreased by a fractional factor $1/\sqrt{2}$ in magnitude (as shown in (6.23a)); and in contrast, if the initial subspace (e.g., $SP_{k_0+1}\left(2^{k_0+1}\right)$) is shrunk one half in dimensional size, then the expansional coefficient for every occupied base vector $|k\rangle$ is increased by a multiplying factor $\sqrt{2}$ in magnitude (as shown in (6.23b)). This is a general rule.

Suppose now that in the unitary transformations of (6.21) the subspace $SP_{k_1}\left(2^{k_1}\right)$ with the vector $|\varphi_2\rangle$ is set to the subspace $SP_{k_0+1}\left(2^{k_0+1}\right)$ which is nearest and larger than the subspace $SP_{k_0}\left(2^{k_0}\right)$ with the vector $|\varphi_1\rangle$ and moreover, the subspace $SP_{k_0}\left(2^{k_0}\right)$ is set to $SP_0(1)$. These mean that in (6.19)

there are $|\varphi_1\rangle = |1\rangle$ and $|\varphi_2\rangle = (|2\rangle + |3\rangle)/\sqrt{2}$ and $k_0 = 0$ and $k_1 = 1$. Then it follows from (6.21) that

$$U_y^{mq}(\theta)|1\rangle = |1\rangle \cos\frac{1}{2}\theta + \frac{1}{\sqrt{2}}(|2\rangle + |3\rangle)\sin\frac{1}{2}\theta \tag{6.24a}$$

$$U_y^{mq}(\theta)\frac{1}{\sqrt{2}}(|2\rangle + |3\rangle) = \frac{1}{\sqrt{2}}(|2\rangle + |3\rangle)\cos\frac{1}{2}\theta - |1\rangle \sin\frac{1}{2}\theta \tag{6.24b}$$

In particular, when the angle $\theta$ is taken as $\theta = \pi/2$, the unitary transformations of (6.24) are respectively reduced to the simple forms

$$U_y^{mq}(\pi/2)|1\rangle = \frac{1}{\sqrt{2}}|1\rangle + \frac{1}{2}(|2\rangle + |3\rangle) \tag{6.25a}$$

$$U_y^{mq}(\pi/2)\frac{1}{\sqrt{2}}(|2\rangle + |3\rangle) = \frac{1}{2}(|2\rangle + |3\rangle) - \frac{1}{\sqrt{2}}|1\rangle \tag{6.25b}$$

where the vectors $|1\rangle$ and $(|2\rangle + |3\rangle)/\sqrt{2}$ belong to the two nearest direct-sum subspaces $SP_0(1)$ and $SP_1(2)$, respectively.

More generally, both the orthonormal vectors $|\varphi_1\rangle$ and $|\varphi_2\rangle$ in the subspace-selective MQT Hermitian operators $\{H_\lambda^{mq}\}$ of (5.19) each may encompass several direct-sum subspaces of the Hilbert space $HS(N)$ in the direct-sum decomposition (6.6) and moreover, these direct-sum subspaces occupied by the vector $|\varphi_1\rangle$ each do not overlap with those occupied by the vector $|\varphi_2\rangle$. As an example, both the vectors $|\varphi_1\rangle$ and $|\varphi_2\rangle$ are respectively taken as

$$|\varphi_1\rangle = \frac{1}{\sqrt{2^{m_1}}}\begin{pmatrix} \alpha_0|0\rangle + a_0|1\rangle + a_1(|2\rangle + |3\rangle) \\ +a_2\sum_{k=2^2}^{2^3-1}|k\rangle + ... + a_{m_1-1}\sum_{k=2^{m_1-1}}^{2^{m_1}-1}|k\rangle \end{pmatrix}; \tag{6.26a}$$

$$|\varphi_2\rangle = \frac{1}{\sqrt{2^{m_2-1}}}\sum_{k=2^{m_2-1}}^{2^{m_2}-1}|k\rangle, \ n \geq m_2 > m_1 \geq 1; \tag{6.26b}$$

where both the vectors $|\varphi_1\rangle$ and $|\varphi_2\rangle$ are orthogonal to one another, and the vector $|\varphi_2\rangle$ is obviously normalized, but the vector $|\varphi_1\rangle$ is normalized only when these expansional coefficients $\{\alpha_0, a_k\}$ satisfy the normalization condition:

$$|\alpha_0|^2 + |a_0|^2 \times 2^0 + |a_1|^2 \times 2^1 + |a_2|^2 \times 2^2 + ... + |a_{m_1-1}|^2 \times 2^{m_1-1} = 2^{m_1} \tag{6.27}$$

Obviously, the vector $|\varphi_1\rangle$ occupies the Hilbert subspace with the total dimensional size $2^{m_1}$ which contains the direct-sum subspaces $HS(1)$, $SP_0(1)$, $SP_1(2)$, ..., and $SP_{m_1-1}(2^{m_1-1})$, while the vector $|\varphi_2\rangle$ occupies only one direct-sum subspace $SP_{m_2-1}(2^{m_2-1})$ $(n \geq m_2 > m_1)$ which is different from all those direct-sum subspaces occupied by the vector $|\varphi_1\rangle$.

By substituting these two orthonormal vectors $|\varphi_1\rangle$ and $|\varphi_2\rangle$ in (6.26) into (5.19) one can construct the subspace-selective MQT Hermitian operators $H_\lambda^{mq}$ of (5.19) with $\lambda = x, y, z$ and then the subspace-selective MQT unitary operators $U_\lambda^{mq}(\theta) = \exp(-i\theta H_\lambda^{mq})$ of (5.32). By choosing suitably these expansional coefficients $\{\alpha_0, a_k\}$ in the vector $|\varphi_1\rangle$ of (6.26a) the unitary operators $U_\lambda^{mq}(\theta)$ may be efficiently realized. Moreover, these specific subspace-selective MQT unitary operators $U_\lambda^{mq}(\theta)$ could be efficiently constructed and realized by the large-scale Hilbert-space-enlarging processes in the next Subsections.

Below these subspace-selective MQT unitary operators $U_\lambda^{mq}(\theta)$ with $\lambda = x, y, z$, which are investigated and constructed in the above paragraphs in this Subsection, are employed to realize the subspace-selective unitary manipulation in the multiple-quantum operator algebra space. This is an important application for the subspace-selective MQT unitary operators $U_\lambda^{mq}(\theta)$. For simplicity, suppose in the subspace-selective unitary manipulation that the (initial) Hamiltonian operator $H_s$ of the $n - spin - 1/2$ system is simply a PDH operator $H_s = |\psi\rangle\langle\psi|$ in the multiple-quantum operator algebra space which corresponds to the Hilbert space $HS(N)$, and here the vector $|\psi\rangle$ of the Hilbert space $HS(N)$ is given by

$$|\psi\rangle = \frac{1}{\sqrt{2^{m_1}}}\begin{pmatrix} \beta_0|0\rangle + b_0|1\rangle + b_1(|2\rangle + |3\rangle) \\ +b_2\sum_{k=2^2}^{2^3-1}|k\rangle + ... + b_{m_1-1}\sum_{k=2^{m_1-1}}^{2^{m_1}-1}|k\rangle \end{pmatrix} \qquad (6.28)$$

where $\{\beta_0, b_k\}$ are expansional coefficients. This means that the vector $|\psi\rangle$ belongs to the same Hilbert subspace as the vector $|\varphi_1\rangle$ of (6.26a). Therefore, there is the orthogonal relation $\langle\varphi_2|\psi\rangle = 0$ between the vectors $|\varphi_2\rangle$ of (6.26b) and $|\psi\rangle$ of (6.28). However, the vector $|\psi\rangle$ may be different from the vector $|\varphi_1\rangle$. According to the brief expressions (4.4) and (4.36) for a PDH operator $H_N^{pd}$ the pseudo-diagonal Hermitian operator $H_s = |\psi\rangle\langle\psi|$ may be simply written as

$$H_s = \begin{pmatrix} \frac{\beta_0}{\sqrt{2^{m_1}}}|0\rangle + \frac{b_0}{\sqrt{2^{m_1}}}|1\rangle + \frac{b_1}{\sqrt{2^{m_1}}}(|2\rangle + |3\rangle) \\ +\frac{b_2}{\sqrt{2^{m_1}}}\sum_{k=2^2}^{2^3-1}|k\rangle + ... + \frac{b_{m_1-1}}{\sqrt{2^{m_1}}}\sum_{k=2^{m_1-1}}^{2^{m_1}-1}|k\rangle \end{pmatrix}\langle h.c.| \qquad (6.29)$$

where the symmetrical structure of the Hilbert space $HS(N)$, i.e., the direct-sum decomposition of (6.6), is already taken into account.

As shown in the Section 3, a unitary time-evolutional process (See (3.3)) or a unitary transformation (See (3.4)) can be performed in the multiple-quantum operator algebra space. Now the subspace-selective MQT unitary operator $U_y^{mq}(\theta) = \exp\left(-i\theta H_y^{k_0,k_1}\right)$ with $H_y^{mq} = H_y^{k_0,k_1}$ of (6.20b) is applied to the Hamiltonian operator $H_s$ of (6.29) which is treated as a dynamical variable, that is, $U_y^{mq}(\theta)H_s\left(U_y^{mq}(\theta)\right)^+$, in the multiple-quantum operator algebra space. With the help of the operator-expansion formula (5.38) of the unitary operator $U_y^{mq}(\theta)$ this unitary transformation may be written as

$$U_y^{mq}(\theta)H_s\left(U_y^{mq}(\theta)\right)^+ = \begin{pmatrix} |\psi\rangle + \left(-1 + \cos\frac{1}{2}\theta\right)\left(B_1^{mq} + B_2^{mq}\right)|\psi\rangle \\ + \left(-2i\sin\frac{1}{2}\theta\right)H_y^{mq}|\psi\rangle \end{pmatrix}\langle h.c.| \qquad (6.30)$$

This unitary transformation (6.30) may be simply explained as a unitary transformation [2,3,17] (See (3.4) in the Section 3) which acts on the dynamical variable $H_s$ of (6.29) in the multiple-quantum operator algebra space.

It is known from (5.38) that $B_l^{mq} = |\varphi_l\rangle \langle\varphi_l|$ with $l = 1, 2$. Then $B_2^{mq} |\psi\rangle = 0$ due to $\langle\varphi_2|\psi\rangle = 0$. By substituting $B_1^{mq}$, $B_2^{mq}$, and $H_y^{mq}$ of (5.19b) into (6.30) and then using the expression (6.26a) of the vector $|\varphi_1\rangle$ and the expression (6.28) of the vector $|\psi\rangle$ and noticing that $B_2^{mq} |\psi\rangle = 0$ the unitary transformation (6.30) can be exactly calculated. A detailed calculation shows that the unitary transformation (6.30) may be reduced to the form

$$U_y^{mq}(\theta) H_s \left(U_y^{mq}(\theta)\right)^+ = \begin{pmatrix} |\psi\rangle + g_1 \left(-1 + \cos\frac{1}{2}\theta\right) |\varphi_1\rangle \\ +g_1 \sin\frac{1}{2}\theta \, |\varphi_2\rangle \end{pmatrix} \langle h.c.| \tag{6.31}$$

where the factor $g_1 = \langle\varphi_1|\psi\rangle$ is given by

$$g_1 = \frac{1}{2^{m_1}} \begin{pmatrix} \alpha_0^* \beta_0 + a_0^* b_0 \times 2^0 + a_1^* b_1 \times 2^1 + a_2^* b_2 \times 2^2 \\ +a_3^* b_3 \times 2^3 + ... + a_{m_1-1}^* b_{m_1-1} \times 2^{m_1-1} \end{pmatrix} \tag{6.32}$$

In particular, when the angle $\theta = \pi$, the unitary transformation (6.31) is further reduced to the simple form

$$U_y^{mq}(\pi) H_s \left(U_y^{mq}(\pi)\right)^+ = \left(|\psi\rangle - g_1 |\varphi_1\rangle + g_1 |\varphi_2\rangle\right) \langle h.c.| \tag{6.33}$$

It can prove that the factor $g_1$ satisfies $|g_1| \leq 1$ if the vector $|\psi\rangle$ is normalized. If the vector $|\psi\rangle$ is equal to the vector $|\varphi_1\rangle$ , then the factor $g_1$ is exactly equal to one due to the normalization condition (6.27) of the vector $|\varphi_1\rangle$ . In this case it can be found from (6.33) that the vector $|\psi\rangle$ (equal to $|\varphi_1\rangle$) is completely converted into the final vector $|\varphi_2\rangle$ by the unitary transformation $U_y^{mq}(\pi)$ .

## 6.3. The Hilbert-space-enlarging processes

Generally, the so-called Hilbert-space-enlarging process is a unitary quantum dynamical process that changes one Hilbert subspace to another with a larger dimensional size. It may be performed in the multiple-quantum operator algebra space. Its inverse process is a Hilbert-space-shrinking process that changes one Hilbert subspace to another with a smaller dimensional size. Here the Hilbert subspace should refer to that Hilbert subspace that the quantum system under study really occupies. Therefore, a general Hilbert-space-enlarging process really changes a small occupied Hilbert subspace to a large one. It is the specific subspace-selective unitary manipulation. These occupied Hilbert subspaces should refer to the direct-sum subspaces of the Hilbert space. Both the occupied Hilbert subspaces before and after a Hilbert-space-enlarging process are usually contained in the same Hilbert space. Then it may be said intuitively that a Hilbert-space-enlarging process changes the occupied Hilbert subspace from a small to a large dimensional size.

A general Hilbert-space-enlarging process may be expressed as a sequence of the basic building blocks which may be conveniently chosen as the subspace-selective MQT unitary operators (or propagators). Therefore, both the Hilbert-space-enlarging processes and their inverses (i.e., the Hilbert-space-shrinking processes) constitute one kind of the subspace-selective unitary manipulation in the multiple-quantum operator algebra space. Consider the Hilbert space $HS(N)$ of an $n - spin - 1/2$ system whose dimensional size $N = 2^n$ may be very large but finite. The symmetrical structure of the Hilbert space $HS(N)$ may be characterized by the direct-sum decomposition of (6.6) or (6.7), that is, $HS\left(N\right) = HS\left(1\right) \bigoplus \{\bigoplus_{k=0}^{n-1} SP_k\left(2^k\right)\}$. Then the basic building blocks of a general Hilbert-space-enlarging process may be chosen as the subspace-selective MQT unitary operators $\{U_\lambda^{mq}\left(\theta\right)\}$ of (5.32), each one of which is selectively applied to a pair of the direct-sum subspaces (e.g., $SP_{k_0}\left(2^{k_0}\right)$ and $SP_{k_1}\left(2^{k_1}\right)$) of the Hilbert space $HS(N)$. Actually some of these basic building blocks, i.e., the subspace-selective MQT unitary operators $\{U_\lambda^{mq}\left(\theta\right)\}$, are already used to realize the special Hilbert-space-enlarging process in the previous Subsection 6.2. Suppose that the two orthonormal vectors $|\varphi_1\rangle$ and $|\varphi_2\rangle$ in the subspace-selective MQT Hermitian operators $\{H_\lambda^{mq}\}$ of (5.19) are explicitly given in (6.19) in the Subsection 6.2, respectively. They belong to the two direct-sum subspaces $SP_{k_0}\left(2^{k_0}\right)$ and $SP_{k_1}\left(2^{k_1}\right)$, respectively, that is, $|\varphi_1\rangle \in SP_{k_0}\left(2^{k_0}\right)$ and $|\varphi_2\rangle \in SP_{k_1}\left(2^{k_1}\right)$, where $0 \leq k_0 < k_1 \leq n-1$. Now both the orthonormal vectors $|\varphi_1\rangle$ and $|\varphi_2\rangle$ in (6.19) are used to construct explicitly the subspace-selective MQT Hermitian operators $H_\lambda^{mq}$ with $\lambda = x, y, z$. Then the latter are further used to generate the subspace-selective MQT unitary operators $\{U_\lambda^{mq}\left(\theta\right)\}$.

The subspace-selective MQT unitary operators $\{U_\mu^{mq}\left(\theta\right)\}$ which are generated by the subspace-selective MQT Hermitian operators $H_\mu^{mq} \equiv H_\mu^{k_0,k_1}$ of (6.20) with $\mu = x, y$ and $0 \leq k_0 < k_1 \leq n-1$ (See the Subsection 6.2) may be rewritten as

$$U_\mu^{mq}\left(\theta\right) \equiv U_\mu^{k_0,k_1}\left(\theta\right) = \exp\left(-i\theta H_\mu^{k_0,k_1}\right) \tag{6.34}$$

These subspace-selective MQT unitary operators $\{U_\mu^{k_0,k_1}\left(\theta\right)\}$ are selectively applied to only the two direct-sum subspaces $SP_{k_0}\left(2^{k_0}\right)$ and $SP_{k_1}\left(2^{k_1}\right)$ of the Hilbert space $HS(N)$. It can be shown that the subspace-selective MQT unitary operator $U_\mu^{k_0,k_1}\left(\theta\right)$ can be efficiently realized with the help of the unitary transformation (5.33) in the Subsection 5.2 for the present case that the two vectors $|\varphi_1\rangle$ and $|\varphi_2\rangle$ in the Hermitian operators $\{H_\lambda^{mq}\}$ of (5.19) are explicitly given in (6.19).

As shown by (6.21) in the Subsection 6.2, there are the basic unitary transformations in the Hilbert space for the subspace-selective MQT unitary operator $U_y^{k_0,k_1}\left(\theta\right)$ of (6.34): $U_y^{k_0,k_1}\left(\theta\right)|\varphi_1\rangle = |\varphi_1\rangle \cos\frac{1}{2}\theta + |\varphi_2\rangle \sin\frac{1}{2}\theta$, $U_y^{k_0,k_1}\left(\theta\right)|\varphi_2\rangle = |\varphi_2\rangle \cos\frac{1}{2}\theta - |\varphi_1\rangle \sin\frac{1}{2}\theta$, where the vectors $|\varphi_1\rangle$ and $|\varphi_2\rangle$ are explicitly given in (6.19). Then, as can be seen in the Subsection 3.2.2, there is a transition from the vector-basis unitary transformations in the Hilbert space to the operator-basis unitary transformations in the multiple-quantum operator algebra space,

$$\begin{aligned}
&\left(U_y^{k_0,k_1}\left(-\theta\right)\right)^+ \left(|\varphi_1\rangle + ...\right)\langle h.c.|\, U_y^{k_0,k_1}\left(-\theta\right) \\
&= U_y^{k_0,k_1}\left(\theta\right)\left(|\varphi_1\rangle + ...\right)\langle h.c.|\left(U_y^{k_0,k_1}\left(\theta\right)\right)^+ \\
&= \left(|\varphi_1\rangle \cos\frac{1}{2}\theta + |\varphi_2\rangle \sin\frac{1}{2}\theta + ...\right)\langle h.c.| \qquad (6.35a)
\end{aligned}$$

and

$$\left(U_y^{k_0,k_1}(-\theta)\right)^+ (|\varphi_2\rangle + ...) \langle h.c.| \, U_y^{k_0,k_1}(-\theta)$$
$$= U_y^{k_0,k_1}(\theta) (|\varphi_2\rangle + ...) \langle h.c.| \left(U_y^{k_0,k_1}(\theta)\right)^+$$
$$= \left(|\varphi_2\rangle \cos\frac{1}{2}\theta - |\varphi_1\rangle \sin\frac{1}{2}\theta + ...\right) \langle h.c.| \qquad (6.35b)$$

where the operators $(|\varphi_k\rangle + ...) \langle h.c.|$ $(k = 1, 2)$, *etc.*, are the PDH operators of (4.4). These are the basic unitary transformations in the multiple-quantum operator algebra space for the subspace-selective MQT unitary operator $U_y^{k_0,k_1}(\theta)$. They also may be simply represented respectively by

$$(|\varphi_1\rangle + ...) \langle h.c.| \xrightarrow{U_y^{k_0,k_1}(\theta)} \left(|\varphi_1\rangle \cos\frac{1}{2}\theta + |\varphi_2\rangle \sin\frac{1}{2}\theta + ...\right) \langle h.c.|$$

and

$$(|\varphi_2\rangle + ...) \langle h.c.| \xrightarrow{U_y^{k_0,k_1}(\theta)} \left(|\varphi_2\rangle \cos\frac{1}{2}\theta - |\varphi_1\rangle \sin\frac{1}{2}\theta + ...\right) \langle h.c.|$$

Beside the subspace-selective MQT unitary operators $\{U_\mu^{k_0,k_1}(\theta)\}$ of (6.34) there is also the subspace-selective MQT unitary operator $U_\lambda^{mq}(\theta)$ of (5.32) which is selectively applied to only the two direct-sum subspaces $HS(1)$ and $SP_{k_1}(2^{k_1})$ of the Hilbert space $HS(N)$ with $0 \leq k_1 \leq n-1$. And here for convenience such a unitary operator $U_\lambda^{mq}(\theta)$ is denoted by $U_\lambda^{00,k_1}(\theta)$ with $\lambda = x, y, z$ and $0 \leq k_1 \leq n-1$. These subspace-selective MQT unitary operators $\{U_\mu^{k_0,k_1}(\theta)\}$ and $\{U_\mu^{00,k_1}(\theta)\}$ which can be efficiently realized are the basic building blocks to construct and realize a general Hilbert-space-enlarging process.

As shown in the Section 3, a unitary time-evolutional process (See (3.3)) or a unitary transformation (See (3.4)) may be performed in the multiple-quantum operator algebra space. A general Hilbert-space-enlarging process may be performed in the multiple-quantum operator algebra space. Then it may be simply explained as a unitary time-evolutional process in the Heisenberg picture or in the Dirac picture. Here for simplicity the dynamical variable at the initial time may be taken as a PDH operator of (4.4) or (4.3). Several important Hilbert-space-enlarging processes of the $n-spin-1/2$ system are explicitly constructed below. They have important applications in the future work.

### 6.3.1. A simple Hilbert-space-enlarging process

A simple Hilbert-space-enlarging process may be intuitively described by

$$SP_0(1) \to SP_1(2) \to SP_2(2^2) \to ... \to SP_k(2^k) \qquad (6.36)$$

Here the initial Hamiltonian operator $H_s^{(0)}$ of the whole Hilbert-space-enlarging process, which is considered as a dynamical variable, is simply prepared as a diagonal operator $H_s^{(0)} = |\psi\rangle \langle\psi|$ in the multiple-quantum operator algebra space, where the initial vector $|\psi\rangle = |1\rangle$ (the tensor-product base vector of (6.5)) belongs to the direct-sum subspace $SP_0(1) = \{|1\rangle\}$ of the Hilbert space $HS(N)$

of the $n - spin - 1/2$ system. Therefore, the initial occupied subspace is the direct-sum subspace $SP_0\,(1)$ in the Hilbert space $HS(N)$. The first stage (i.e., $SP_0\,(1) \rightarrow SP_1\,(2)$) of the Hilbert-space-enlarging process then is to change the initial direct-sum subspace $SP_0\,(1)$ with dimension one to its nearest direct-sum subspace $SP_1\,(2)$ with dimension two, where the subspace $SP_1\,(2) = \{|2\rangle\,, |3\rangle\}$. Correspondingly in the first stage the initial diagonal Hamiltonian $H_s^{(0)} = |1\rangle\,\langle 1|$ is changed to the Hamiltonian $H_s^{(1)}$ by the subspace-selective unitary transformation $H_s^{(1)} = U_y^{0,1}\,(\pi)\,H_s^{(0)}\,\left(U_y^{0,1}\,(\pi)\right)^+$ in the multiple-quantum operator algebra space. Hence the first stage may be described in detail by the subspace-selective unitary transformation in the multiple-quantum operator algebra space:

$$\begin{aligned} H_s^{(1)} &= \left(U_y^{0,1}\,(-\pi)\right)^+ H_s^{(0)}\,\left(U_y^{0,1}\,(-\pi)\right) = U_y^{0,1}\,(\pi)\,H_s^{(0)}\,\left(U_y^{0,1}\,(\pi)\right)^+ \\ &= U_y^{0,1}\,(\pi)\,(|1\rangle\,\langle 1|)\,\left(U_y^{0,1}\,(\pi)\right)^+ = \left(U_y^{0,1}\,(\pi)\,|1\rangle\right)\langle h.c.| \\ &= \left(\frac{1}{\sqrt{2}}\,|2\rangle + \frac{1}{\sqrt{2}}\,|3\rangle\right)\langle h.c.| \end{aligned} \qquad (6.37a)$$

where the subspace-selective MQT unitary operator $U_y^{0,1}\,(\pi)$ is given by (6.34) with $\mu = y$ and $k_0 = 0$, $k_1 = 1$ and $\theta = \pi$, and it is selectively applied to only the two direct-sum subspaces $SP_0\,(1)$ and $SP_1\,(2)$ of the Hilbert space $HS(N)$. In (6.37a) the subspace-selective unitary transformation $U_y^{0,1}\,(\pi)\,|1\rangle\,\langle 1|\,\left(U_y^{0,1}\,(\pi)\right)^+ = \left((|2\rangle + |3\rangle)/\sqrt{2}\right)\langle h.c.|$ is obtained directly from (6.35a) where $U_y^{k_0,k_1}\,(\theta) = U_y^{0,1}\,(\pi)$ with $\theta = \pi$ and $|\varphi_1\rangle = |1\rangle\,, |\varphi_2\rangle = (|2\rangle + |3\rangle)/\sqrt{2}$. The unitary transformation (6.37a), i.e., $U_y^{0,1}\,(\pi)\,H_s^{(0)}\,\left(U_y^{0,1}\,(\pi)\right)^+$, may be simply represented by

$$H_s^{(0)} \xrightarrow{U_y^{0,1}(\pi)} H_s^{(1)} = \left(\frac{1}{\sqrt{2}}\,|2\rangle + \frac{1}{\sqrt{2}}\,|3\rangle\right)\langle h.c.| \qquad (6.37b)$$

Here the final Hamiltonian $H_s^{(1)}$ is a PDH operator of (4.4) and may be formally written as $H_s^{(1)} = |\psi\rangle\,\langle\psi|$, where the final vector $|\psi\rangle = (|2\rangle + |3\rangle)/\sqrt{2}$ occupies only the subspace $SP_1\,(2)$. This also means that at the end of the first stage the occupied subspace is the direct-sum subspace $SP_1\,(2)$ in the Hilbert space $HS(N)$.

The second stage (i.e., $SP_1\,(2) \rightarrow SP_2\,(2^2)$) of the Hilbert-space-enlarging process (6.36) is to change the occupied direct-sum subspace $SP_1\,(2) = \{|2\rangle\,, |3\rangle\}$ with dimension 2 to its nearest direct-sum subspace $SP_2\,(2^2)$ with dimension $2^2$, where the subspace $SP_2\,(2^2) = \{|4\rangle\,, |5\rangle\,, |6\rangle\,, |7\rangle\}$. Correspondingly in the second stage the initial Hamiltonian $H_s^{(1)}$, which is the final Hamiltonian of the first stage and is given in (6.37), is converted into the Hamiltonian $H_s^{(2)}$ by the subspace-selective unitary transformation $H_s^{(2)} = U_y^{1,2}\,(\pi)\,H_s^{(1)}\,\left(U_y^{1,2}\,(\pi)\right)^+$ in the multiple-quantum operator algebra space. Therefore, the second stage may be simply expressed as

$$H_s^{(1)} \xrightarrow{U_y^{1,2}(\pi)} H_s^{(2)} = \left(\frac{1}{2}\,(|4\rangle + |5\rangle + |6\rangle + |7\rangle)\right)\langle h.c.| \qquad (6.38)$$

where the subspace-selective MQT unitary operator $U_y^{1,2}(\pi)$ is given by (6.34) with $\mu = y$ and $k_0 = 1$, $k_1 = 2$ and $\theta = \pi$. In (6.38) the basic unitary transformation of (6.35a) is already used, where $U_y^{k_0,k_1}(\theta) = U_y^{1,2}(\pi)$ and $|\varphi_1\rangle = (|2\rangle + |3\rangle)/\sqrt{2}$, $|\varphi_2\rangle = (|4\rangle + |5\rangle + |6\rangle + |7\rangle)/2$. The PDH Hamiltonian operator $H_s^{(2)}$ in (6.38) may be formally written as $H_s^{(2)} = |\psi\rangle\langle\psi|$, where the final vector $|\psi\rangle = \frac{1}{2}\sum_{k=2^2}^{2^3-1}|k\rangle$ occupies only the subspace $SP_2\left(2^2\right)$, indicating that the final occupied subspace is the direct-sum subspace $SP_2\left(2^2\right)$ in the Hilbert space $HS(N)$.

Generally, the $k-$th stage (i.e., $SP_{k-1}\left(2^{k-1}\right) \rightarrow SP_k\left(2^k\right)$ for $k = 1, 2, .., n-1$) of the Hilbert-space-enlarging process (6.36) is to change the occupied direct-sum subspace $SP_{k-1}\left(2^{k-1}\right)$ with dimension $2^{k-1}$ to its nearest direct-sum subspace $SP_k\left(2^k\right)$ with dimension $2^k$ in the direct-sum decomposition (6.6) of the Hilbert space $HS(N)$, where both the subspaces $SP_{k-1}\left(2^{k-1}\right)$ and $SP_k\left(2^k\right)$ are given by $SP_{k-1}\left(2^{k-1}\right) = \{\left|2^{k-1}\right\rangle, \left|2^{k-1}+1\right\rangle, ..., \left|2^k-1\right\rangle\}$ and $SP_k\left(2^k\right) = \{\left|2^k\right\rangle, \left|2^k+1\right\rangle, ..., \left|2^{k+1}-1\right\rangle\}$, respectively. Correspondingly in the $k-$th stage the initial Hamiltonian $H_s^{(k-1)}$, which is the final Hamiltonian of the $(k-1)-$th stage and is given by $H_s^{(k-1)} = (\sum_{l=2^{k-1}}^{2^k-1}|l\rangle/\sqrt{2^{k-1}})\langle h.c.|$, is converted into the final Hamiltonian $H_s^{(k)}$ by the subspace-selective unitary transformation $H_s^{(k)} = U_y^{k-1,k}(\pi)\,H_s^{(k-1)}\left(U_y^{k-1,k}(\pi)\right)^+$ in the multiple-quantum operator algebra space. Therefore, the $k-$th stage may be simply expressed as

$$H_s^{(k-1)} \xrightarrow{U_y^{k-1,k}(\pi)} H_s^{(k)} = \left(\frac{1}{\sqrt{2^k}}\sum_{l=2^k}^{2^{k+1}-1}|l\rangle\right)\langle h.c.| \tag{6.39}$$

where the subspace-selective MQT unitary operator $U_y^{k-1,k}(\pi)$ is given by (6.34) with $\mu = y$ and $k_0 = k-1$, $k_1 = k$ and $\theta = \pi$. In (6.39) the basic unitary transformation of (6.35a) is already used, where $U_y^{k_0,k_1}(\theta) = U_y^{k-1,k}(\pi)$ and $|\varphi_1\rangle = \sum_{l=2^{k-1}}^{2^k-1}|l\rangle/\sqrt{2^{k-1}}$, $|\varphi_2\rangle = \sum_{l=2^k}^{2^{k+1}-1}|l\rangle/\sqrt{2^k}$. The PDH Hamiltonian operator $H_s^{(k)}$ in (6.39) may be formally written as $H_s^{(k)} = |\psi\rangle\langle\psi|$, where the final vector $|\psi\rangle = \sum_{l=2^k}^{2^{k+1}-1}|l\rangle/\sqrt{2^k}$ occupies only the subspace $SP_k\left(2^k\right)$ with dimension $2^k$, indicating that the final occupied subspace in the $k-$th stage is the direct-sum subspace $SP_k\left(2^k\right)$ in the Hilbert space $HS(N)$.

The theoretical analysis above shows that the occupied direct-sum subspace can be enlarged twofold in dimensional size by every stage of the Hilbert-space-enlarging process (6.36). Then the whole Hilbert-space-enlarging process (6.36) changes the initial occupied subspace $SP_0(1)$ with dimension one to the final occupied subspace $SP_k\left(2^k\right)$ with dimension $2^k$. Therefore, the whole Hilbert-space-enlarging process (6.36), which is performed in the multiple-quantum operator algebra space, may be generally expressed as a sequence of the subspace-selective unitary transformations:

$$\begin{aligned} H_s^{(k)} = &\left(U_y^{k-1,k}(-\pi)\right)^+ ... \left(U_y^{1,2}(-\pi)\right)^+ \left(U_y^{0,1}(-\pi)\right)^+ \\ &\times H_s^{(0)}\left(U_y^{0,1}(-\pi)\right)\left(U_y^{1,2}(-\pi)\right) ... \left(U_y^{k-1,k}(-\pi)\right) \end{aligned} \tag{6.40a}$$

or

$$H_s^{(k)} = U_y^{k-1,k}(\pi)\ldots U_y^{1,2}(\pi)\,U_y^{0,1}(\pi)$$

$$\times H_s^{(0)}\left(U_y^{0,1}(\pi)\right)^{+}\left(U_y^{1,2}(\pi)\right)^{+}\ldots\left(U_y^{k-1,k}(\pi)\right)^{+} \qquad (6.40\text{b})$$

where the initial Hamiltonian $H_s^{(0)}$ is the diagonal operator $H_s^{(0)} = |1\rangle\langle 1|$, while the final Hamiltonian $H_s^{(k)}$ given in (6.39) with $k = 1, 2, ..., n-1$ is a pseudo-diagonal Hermitian operator.

### 6.3.2. The small-scale Hilbert-space-enlarging processes

A small-scale Hilbert-space-enlarging process is performed in the multiple-quantum operator algebra space and may be intuitively described by

$$SP_{k_0}\left(2^{k_0}\right)\bigoplus SP_{k_0+1}\left(2^{k_0+1}\right)\bigoplus\ldots\bigoplus SP_{k_0+l}\left(2^{k_0+l}\right)$$

$$\rightarrow SP_{k_0+1}\left(2^{k_0+1}\right)\bigoplus SP_{k_0+2}\left(2^{k_0+2}\right)\bigoplus\ldots\bigoplus SP_{k_0+l+1}\left(2^{k_0+l+1}\right) \qquad (6.41)$$

with $k_0 \geq 0$ and $k_0 + l < n-1$. Here the initial Hamiltonian operator $H_s^{(0)}$ of the small-scale Hilbert-space-enlarging process, which is a dynamical variable, is still taken as a PDH operator of (4.4) in the multiple-quantum operator algebra space. Suppose that it is formally written as $H_s^{(0)} = |\psi\rangle\langle\psi|$, where the initial vector $|\psi\rangle$ of the Hilbert space $HS(N)$ may be given by

$$|\psi\rangle = \left(C_{k_0}\sum_{k=2^{k_0}}^{2^{k_0+1}-1}|k\rangle + C_{k_0+1}\sum_{k=2^{k_0+1}}^{2^{k_0+2}-1}|k\rangle + \ldots + C_{k_0+l}\sum_{k=2^{k_0+l}}^{2^{k_0+l+1}-1}|k\rangle\right) \qquad (6.42)$$

Obviously, the initial vector $|\psi\rangle$ encompasses $l+1$ occupied direct-sum subspaces $\{SP_{k_0+j}\left(2^{k_0+j}\right)\}$ in the Hilbert space $HS(N)$ for $j = 0, 1, ..., l$. Then, according to the specific form (4.4) of a PDH operator, this initial PDH Hamiltonian $H_s^{(0)}$ may be simply written as

$$H_s^{(0)} = |\psi\rangle\langle\psi| = \begin{pmatrix} C_{k_0}\sum_{k=2^{k_0}}^{2^{k_0+1}-1}|k\rangle + C_{k_0+1}\sum_{k=2^{k_0+1}}^{2^{k_0+2}-1}|k\rangle \\ +\ldots + C_{k_0+l}\sum_{k=2^{k_0+l}}^{2^{k_0+l+1}-1}|k\rangle \end{pmatrix}\langle h.c.| \qquad (6.43)$$

where the symmetrical structure of the Hilbert space $HS(N)$, i.e., the direct-sum decomposition (6.6), is already considered.

The small-scale Hilbert-space-enlarging process (6.41) with the initial Hamiltonian $H_s^{(0)}$ of (6.43) is performed in the multiple-quantum operator algebra space. The theoretical analysis and the exact calculation for the small-scale Hilbert-space-enlarging process are carried out below. It can be found that the occupied direct-sum subspace $SP_{k_0+l}\left(2^{k_0+l}\right)$ is the largest one among these $l+1$ occupied direct-sum subspaces $\{SP_{k_0+j}\left(2^{k_0+j}\right)\}$ appearing in the initial Hamiltonian $H_s^{(0)}$. Then the first stage (i.e., $SP_{k_0+l}\left(2^{k_0+l}\right) \rightarrow SP_{k_0+l+1}\left(2^{k_0+l+1}\right)$) of the small-scale Hilbert-space-enlarging process is to change selectively the largest occupied subspace $SP_{k_0+l}\left(2^{k_0+l}\right)$ in the initial Hamiltonian $H_s^{(0)}$ to its nearest, larger, and empty direct-sum subspace $SP_{k_0+l+1}\left(2^{k_0+l+1}\right)$ in the Hilbert space $HS(N)$. Correspondingly in the first stage the initial Hamiltonian $H_s^{(0)}$ is converted into the Hamiltonian $H_s^{(1)}$ by the subspace-selective unitary transformation $H_s^{(1)} = U_y^{k_0+l,k_0+l+1}(\pi) H_s^{(0)} \left(U_y^{k_0+l,k_0+l+1}(\pi)\right)^+$ in the multiple-quantum operator algebra space. Here the subspace-selective MQT unitary operator $U_y^{k_0+l,k_0+l+1}(\pi)$ is selectively applied to only the two direct-sum subspaces $SP_{k_0+l}\left(2^{k_0+l}\right)$ and $SP_{k_0+l+1}\left(2^{k_0+l+1}\right)$ and does not affect any other direct-sum subspaces of the Hilbert space $HS(N)$. It is given by (6.34) with $\mu = y$ and the index $k_0 = k_0 + l$, $k_1 = k_0 + l + 1$ and the angle $\theta = \pi$, that is, $U_y^{k_0,k_1}(\theta) = U_y^{k_0+l,k_0+l+1}(\pi)$. Its basic unitary transformation of (6.35a) in the multiple-quantum operator algebra space is explicitly represented by

$$\left(\frac{1}{\sqrt{2^{k_0+l}}}\sum_{k=2^{k_0+l}}^{2^{k_0+l+1}-1}|k\rangle + ...\right)\langle h.c.|$$

$$\xrightarrow{U_y^{k_0+l,k_0+l+1}(\pi)} \left(\frac{1}{\sqrt{2^{k_0+l+1}}}\sum_{k=2^{k_0+l+1}}^{2^{k_0+l+2}-1}|k\rangle + ...\right)\langle h.c.|, \tag{6.44}$$

where $|\varphi_1\rangle = \sum_{k=2^{k_0+l}}^{2^{k_0+l+1}-1}|k\rangle/\sqrt{2^{k_0+l}}$ and $|\varphi_2\rangle = \sum_{k=2^{k_0+l+1}}^{2^{k_0+l+2}-1}|k\rangle/\sqrt{2^{k_0+l+1}}$ in (6.35). Therefore, in the first stage the subspace-selective unitary transformation in the multiple-quantum operator algebra space may be described in detail by

$$H_s^{(1)} = \left(U_y^{k_0+l,k_0+l+1}(-\pi)\right)^+ H_s^{(0)} U_y^{k_0+l,k_0+l+1}(-\pi)$$
$$= U_y^{k_0+l,k_0+l+1}(\pi) H_s^{(0)} \left(U_y^{k_0+l,k_0+l+1}(\pi)\right)^+$$
$$= \left(\begin{array}{c} C_{k_0}U_y^{k_0+l,k_0+l+1}(\pi)\sum_{k=2^{k_0}}^{2^{k_0+1}-1}|k\rangle \\ +C_{k_0+1}U_y^{k_0+l,k_0+l+1}(\pi)\sum_{k=2^{k_0+1}}^{2^{k_0+2}-1}|k\rangle + ... \\ +C_{k_0+l}U_y^{k_0+l,k_0+l+1}(\pi)\sum_{k=2^{k_0+l}}^{2^{k_0+l+1}-1}|k\rangle \end{array}\right)\langle h.c.|$$
$$= \left(\begin{array}{c} C_{k_0}\sum_{k=2^{k_0}}^{2^{k_0+1}-1}|k\rangle + ... + C_{k_0+l-1}\sum_{k=2^{k_0+l-1}}^{2^{k_0+l}-1}|k\rangle \\ +\sqrt{2^{k_0+l}}C_{k_0+l} \times U_y^{k_0+l,k_0+l+1}(\pi)\left(\frac{1}{\sqrt{2^{k_0+l}}}\sum_{k=2^{k_0+l}}^{2^{k_0+l+1}-1}|k\rangle\right) \end{array}\right)\langle h.c.|$$
$$= \left(\begin{array}{c} C_{k_0}\sum_{k=2^{k_0}}^{2^{k_0+1}-1}|k\rangle + ... + C_{k_0+l-1}\sum_{k=2^{k_0+l-1}}^{2^{k_0+l}-1}|k\rangle \\ +\sqrt{2^{k_0+l}}C_{k_0+l}\left(\frac{1}{\sqrt{2^{k_0+l+1}}}\sum_{k=2^{k_0+l+1}}^{2^{k_0+l+2}-1}|k\rangle\right) \end{array}\right)\langle h.c.|$$
$$= \left(\begin{array}{c} C_{k_0}\sum_{k=2^{k_0}}^{2^{k_0+1}-1}|k\rangle + C_{k_0+1}\sum_{k=2^{k_0+1}}^{2^{k_0+2}-1}|k\rangle + ... \\ +C_{k_0+l-1}\sum_{k=2^{k_0+l-1}}^{2^{k_0+l}-1}|k\rangle + \frac{1}{\sqrt{2}}C_{k_0+l}\sum_{k=2^{k_0+l+1}}^{2^{k_0+l+2}-1}|k\rangle \end{array}\right)\langle h.c.|$$

Here the fourth equality holds due to the subspace-selective property of the subspace-selective MQT unitary operator $U_y^{k_0+l,k_0+l+1}(\pi)$, while the fifth equality holds due to the basic unitary transformation (6.44). The first stage may be

simply represented by

$$H_s^{(0)} \xrightarrow{U_y^{k_0+l,k_0+l+1}(\pi)} H_s^{(1)} = \begin{pmatrix} C_{k_0} \sum_{k=2^{k_0}}^{2^{k_0+1}-1} |k\rangle \\ +C_{k_0+1} \sum_{k=2^{k_0+1}}^{2^{k_0+2}-1} |k\rangle + ... \\ +C_{k_0+l-1} \sum_{k=2^{k_0+l-1}}^{2^{k_0+l}-1} |k\rangle \\ +\frac{1}{\sqrt{2}} C_{k_0+l} \sum_{k=2^{k_0+l+1}}^{2^{k_0+l+2}-1} |k\rangle \end{pmatrix} \langle h.c.| \qquad (6.45)$$

It can be found from the first stage that, after the subspace-selective unitary transformation of (6.45), only the vector $|\varphi_1\rangle = \sum_{k=2^{k_0+l}}^{2^{k_0+l+1}-1} |k\rangle / \sqrt{2^{k_0+l}}$ of the largest occupied direct-sum subspace $SP_{k_0+l}\left(2^{k_0+l}\right)$ is converted into the vector $|\varphi_2\rangle = \sum_{k=2^{k_0+l+1}}^{2^{k_0+l+2}-1} |k\rangle / \sqrt{2^{k_0+l+1}}$ of the direct-sum subspace $SP_{k_0+l+1}\left(2^{k_0+l+1}\right)$ completely, while those vectors of any smaller occupied direct-sum subspaces appearing in the initial Hamiltonian $H_s^{(0)}$ keep unchanged, and in the meantime the occupied subspace $SP_{k_0+l}\left(2^{k_0+l}\right)$ is emptied in the final Hamiltonian $H_s^{(1)}$ in (6.45).

Now the second stage (i.e., $SP_{k_0+l-1}\left(2^{k_0+l-1}\right) \rightarrow SP_{k_0+l}\left(2^{k_0+l}\right)$) of the small-scale Hilbert-space-enlarging process is to change selectively the second largest occupied subspace $SP_{k_0+l-1}\left(2^{k_0+l-1}\right)$ to its nearest, larger, and empty subspace $SP_{k_0+l}\left(2^{k_0+l}\right)$ in the initial Hamiltonian $H_s^{(1)}$ which is the final Hamiltonian of the first stage and is given in (6.45). Correspondingly in the second stage the initial Hamiltonian $H_s^{(1)}$ in (6.45) is converted into the Hamiltonian $H_s^{(2)}$ by the subspace-selective unitary transformation $H_s^{(2)} = U_y^{k_0+l-1,k_0+l}(\pi) \times H_s^{(1)} \left(U_y^{k_0+l-1,k_0+l}(\pi)\right)^+$ in the multiple-quantum operator algebra space. Here the subspace-selective MQT unitary operator $U_y^{k_0+l-1,k_0+l}(\pi)$ is selectively applied to only the two subspaces $SP_{k_0+l-1}\left(2^{k_0+l-1}\right)$ and $SP_{k_0+l}\left(2^{k_0+l}\right)$ and does not affect any other direct-sum subspaces of the Hilbert space $HS(N)$. It is given by (6.34) with $\mu = y$ and the index $k_0 = k_0 + l - 1$, $k_1 = k_0 + l$ and the angle $\theta = \pi$, that is, $U_y^{k_0,k_1}(\theta) = U_y^{k_0+l-1,k_0+l}(\pi)$. Its basic unitary transformation of (6.35a) is explicitly represented by

$$\left(\frac{1}{\sqrt{2^{k_0+l-1}}} \sum_{k=2^{k_0+l-1}}^{2^{k_0+l}-1} |k\rangle + ...\right) \langle h.c.|$$

$$\xrightarrow{U_y^{k_0+l-1,k_0+l}(\pi)} \left(\frac{1}{\sqrt{2^{k_0+l}}} \sum_{k=2^{k_0+l}}^{2^{k_0+l+1}-1} |k\rangle + ...\right) \langle h.c.| \qquad (6.46)$$

where $|\varphi_1\rangle = \sum_{k=2^{k_0+l-1}}^{2^{k_0+l}-1} |k\rangle / \sqrt{2^{k_0+l-1}}$ and $|\varphi_2\rangle = \sum_{k=2^{k_0+l}}^{2^{k_0+l+1}-1} |k\rangle / \sqrt{2^{k_0+l}}$ in (6.35). Then the subspace-selective unitary transformation in the second stage may be written as

$$H_s^{(2)} = \left(U_y^{k_0+l-1,k_0+l}(-\pi)\right)^+ H_s^{(1)} \left(U_y^{k_0+l-1,k_0+l}(-\pi)\right)$$

$$= U_y^{k_0+l-1,k_0+l}(\pi) H_s^{(1)} \left(U_y^{k_0+l-1,k_0+l}(\pi)\right)^+$$

$$= \begin{pmatrix} C_{k_0} \sum_{k=2^{k_0}}^{2^{k_0+1}-1} |k\rangle + ... + C_{k_0+l-2} \sum_{k=2^{k_0+l-2}}^{2^{k_0+l-1}-1} |k\rangle \\ +C_{k_0+l-1} \times U_y^{k_0+l-1,k_0+l}(\pi) \sum_{k=2^{k_0+l-1}}^{2^{k_0+l}-1} |k\rangle \\ +\frac{1}{\sqrt{2}} C_{k_0+l} \sum_{k=2^{k_0+l+1}}^{2^{k_0+l+2}-1} |k\rangle \end{pmatrix} \langle h.c.|$$

$$= \begin{pmatrix} C_{k_0} \sum_{k=2^{k_0}}^{2^{k_0+1}-1} |k\rangle + ... + C_{k_0+l-2} \sum_{k=2^{k_0+l-2}}^{2^{k_0+l-1}-1} |k\rangle \\ +\frac{1}{\sqrt{2}} C_{k_0+l-1} \sum_{k=2^{k_0+l}}^{2^{k_0+l+1}-1} |k\rangle + \frac{1}{\sqrt{2}} C_{k_0+l} \sum_{k=2^{k_0+l+1}}^{2^{k_0+l+2}-1} |k\rangle \end{pmatrix} \langle h.c.|$$

Here the third equality holds due to the subspace-selective property of the subspace-selective MQT unitary operator $U_y^{k_0+l-1,k_0+l}(\pi)$, while the last equality holds due to the basic unitary transformation (6.46). The second stage may be simply represented by

$$H_s^{(1)} \xrightarrow{U_y^{k_0+l-1,k_0+l}(\pi)} H_s^{(2)} = \begin{pmatrix} C_{k_0} \sum_{k=2^{k_0}}^{2^{k_0+1}-1} |k\rangle + C_{k_0+1} \sum_{k=2^{k_0+1}}^{2^{k_0+2}-1} |k\rangle \\ +... + C_{k_0+l-2} \sum_{k=2^{k_0+l-2}}^{2^{k_0+l-1}-1} |k\rangle \\ +\frac{1}{\sqrt{2}} C_{k_0+l-1} \sum_{k=2^{k_0+l}}^{2^{k_0+l+1}-1} |k\rangle \\ +\frac{1}{\sqrt{2}} C_{k_0+l} \sum_{k=2^{k_0+l+1}}^{2^{k_0+l+2}-1} |k\rangle \end{pmatrix} \langle h.c.| \tag{6.47}$$

It can be seen from the second stage that, by the subspace-selective unitary transformation (6.47), only the vector $|\varphi_1\rangle = \sum_{k=2^{k_0+l-1}}^{2^{k_0+l}-1} |k\rangle / \sqrt{2^{k_0+l-1}}$ of the occupied subspace $SP_{k_0+l-1}\left(2^{k_0+l-1}\right)$ is completely converted into the vector $|\varphi_2\rangle = \sum_{k=2^{k_0+l}}^{2^{k_0+l+1}-1} |k\rangle / \sqrt{2^{k_0+l}}$ of the subspace $SP_{k_0+l}\left(2^{k_0+l}\right)$, while those vectors of any other occupied subspaces in the initial Hamiltonian $H_s^{(1)}$ in (6.45) are not affected, and moreover, the subspace $SP_{k_0+l-1}\left(2^{k_0+l-1}\right)$ is empty in the final Hamiltonian $H_s^{(2)}$ in (6.47).

Generally, the $j-$th stage of the small-scale Hilbert-space-enlarging process, that is, $SP_{k_0+l-j+1}\left(2^{k_0+l-j+1}\right) \rightarrow SP_{k_0+l-j+2}\left(2^{k_0+l-j+2}\right)$ for $j = 1, 2, ..., l + 1$, is to selectively change the occupied subspace $SP_{k_0+l-j+1}\left(2^{k_0+l-j+1}\right)$ to its nearest, larger, and empty subspace $SP_{k_0+l-j+2}\left(2^{k_0+l-j+2}\right)$ in the initial Hamiltonian $H_s^{(j-1)}$. Correspondingly in the $j-$th stage the initial Hamiltonian $H_s^{(j-1)}$, which is the final Hamiltonian in the $(j-1)-$th stage, is converted into the Hamiltonian $H_s^{(j)}$ by the subspace-selective unitary transformation in the multiple-quantum operator algebra space:

$$H_s^{(j)} = U_y^{k_0+l-j+1,k_0+l-j+2}(\pi) H_s^{(j-1)} \left(U_y^{k_0+l-j+1,k_0+l-j+2}(\pi)\right)^+$$

Here the subspace-selective MQT unitary operator $U_y^{k_0+l-j+1,k_0+l-j+2}(\pi)$ selectively acts on only the two chosen direct-sum subspaces $SP_{k_0+l-j+1}\left(2^{k_0+l-j+1}\right)$ and $SP_{k_0+l-j+2}\left(2^{k_0+l-j+2}\right)$ and does not affect any other direct-sum subspaces of the Hilbert space $HS(N)$. It is given by (6.34) with $\mu = y$ and the index $k_0 = k_0 + l - j + 1$, $k_1 = k_0 + l - j + 2$ and the angle $\theta = \pi$, i.e., $U_y^{k_0,k_1}(\theta) = U_y^{k_0+l-j+1,k_0+l-j+2}(\pi)$. Its basic unitary transformation of (6.35a) is explicitly represented by

$$\left(\frac{1}{\sqrt{2^{k_0+l-j+1}}}\sum_{k=2^{k_0+l-j+1}}^{2^{k_0+l-j+2}-1}|k\rangle+...\right)\langle h.c.|$$

$$\xrightarrow{U_y^{k_0+l-j+1,k_0+l-j+2}(\pi)}\left(\frac{1}{\sqrt{2^{k_0+l-j+2}}}\sum_{k=2^{k_0+l-j+2}}^{2^{k_0+l-j+3}-1}|k\rangle+...\right)\langle h.c.| \tag{6.48}$$

where the two vectors $|\varphi_1\rangle$ and $|\varphi_2\rangle$ in (6.35) are given by $|\varphi_1\rangle=\sum_{k=2^{k_0+l-j+1}}^{2^{k_0+l-j+2}-1}|k\rangle/\sqrt{2^{k_0+l-j+1}}$ and $|\varphi_2\rangle=\sum_{k=2^{k_0+l-j+2}}^{2^{k_0+l-j+3}-1}|k\rangle/\sqrt{2^{k_0+l-j+2}}$, respectively.

Suppose that the initial Hamiltonian $H_s^{(j-1)}$, which is also the final Hamiltonian in the $(j-1)-$th stage, is given by

$$H_s^{(j-1)}=\begin{pmatrix} C_{k_0}\sum_{k=2^{k_0}}^{2^{k_0+1}-1}|k\rangle+C_{k_0+1}\sum_{k=2^{k_0+1}}^{2^{k_0+2}-1}|k\rangle+... \\ +C_{k_0+l-j+1}\sum_{k=2^{k_0+l-j+1}}^{2^{k_0+l-j+2}-1}|k\rangle \\ +\frac{1}{\sqrt{2}}C_{k_0+l-j+2}\sum_{k=2^{k_0+l-j+3}}^{2^{k_0+l-j+4}-1}|k\rangle+... \\ +\frac{1}{\sqrt{2}}C_{k_0+l}\sum_{k=2^{k_0+l+1}}^{2^{k_0+l+2}-1}|k\rangle \end{pmatrix}\langle h.c.|$$

Then the subspace-selective unitary transformation of the $j-$th stage for $j=1,2,...,l+1$ may be explicitly written as

$$\begin{aligned} H_s^{(j)} &= \left(U_y^{k_0+l-j+1,k_0+l-j+2}(-\pi)\right)^+ H_s^{(j-1)}\left(U_y^{k_0+l-j+1,k_0+l-j+2}(-\pi)\right) \\ &= U_y^{k_0+l-j+1,k_0+l-j+2}(\pi)\,H_s^{(j-1)}\left(U_y^{k_0+l-j+1,k_0+l-j+2}(\pi)\right)^+ \end{aligned}$$

$$=\begin{pmatrix} C_{k_0}\sum_{k=2^{k_0}}^{2^{k_0+1}-1}|k\rangle+...+C_{k_0+l-j}\sum_{k=2^{k_0+l-j}}^{2^{k_0+l-j+1}-1}|k\rangle \\ +C_{k_0+l-j+1}\times U_y^{k_0+l-j+1,k_0+l-j+2}(\pi)\sum_{k=2^{k_0+l-j+1}}^{2^{k_0+l-j+2}-1}|k\rangle \\ +\frac{1}{\sqrt{2}}C_{k_0+l-j+2}\sum_{k=2^{k_0+l-j+3}}^{2^{k_0+l-j+4}-1}|k\rangle+...+\frac{1}{\sqrt{2}}C_{k_0+l}\sum_{k=2^{k_0+l+1}}^{2^{k_0+l+2}-1}|k\rangle \end{pmatrix}\langle h.c.|$$

$$=\begin{pmatrix} C_{k_0}\sum_{k=2^{k_0}}^{2^{k_0+1}-1}|k\rangle+...+C_{k_0+l-j}\sum_{k=2^{k_0+l-j}}^{2^{k_0+l-j+1}-1}|k\rangle \\ +\frac{1}{\sqrt{2}}C_{k_0+l-j+1}\sum_{k=2^{k_0+l-j+2}}^{2^{k_0+l-j+3}-1}|k\rangle+...+\frac{1}{\sqrt{2}}C_{k_0+l}\sum_{k=2^{k_0+l+1}}^{2^{k_0+l+2}-1}|k\rangle \end{pmatrix}\langle h.c.|$$

Here the third equality holds owing to the subspace-selective property of the unitary operator $U_y^{k_0+l-j+1,k_0+l-j+2}(\pi)$, while the last equality holds due to the basic unitary transformation (6.48). Then the $j-$th stage may be simply represented by

$$H_s^{(j-1)}\xrightarrow{U_y^{k_0+l-j+1,k_0+l-j+2}(\pi)}H_s^{(j)}=\begin{pmatrix} C_{k_0}\sum_{k=2^{k_0}}^{2^{k_0+1}-1}|k\rangle+... \\ +C_{k_0+l-j}\sum_{k=2^{k_0+l-j}}^{2^{k_0+l-j+1}-1}|k\rangle \\ +\frac{1}{\sqrt{2}}C_{k_0+l-j+1}\sum_{k=2^{k_0+l-j+2}}^{2^{k_0+l-j+3}-1}|k\rangle \\ +...+\frac{1}{\sqrt{2}}C_{k_0+l}\sum_{k=2^{k_0+l+1}}^{2^{k_0+l+2}-1}|k\rangle \end{pmatrix}\langle h.c.| \tag{6.49}$$

The subspace-selective unitary transformation (6.49) of the $j-$th stage results in that only the vector $|\varphi_1\rangle=\sum_{k=2^{k_0+l-j+1}}^{2^{k_0+l-j+2}-1}|k\rangle/\sqrt{2^{k_0+l-j+1}}$ of the occupied subspace $SP_{k_0+l-j+1}\left(2^{k_0+l-j+1}\right)$ in the initial Hamiltonian $H_s^{(j-1)}$ is completely converted into the vector $|\varphi_2\rangle=\sum_{k=2^{k_0+l-j+2}}^{2^{k_0+l-j+3}-1}|k\rangle/\sqrt{2^{k_0+l-j+2}}$ of the subspace $SP_{k_0+l-j+2}\left(2^{k_0+l-j+2}\right)$ in the final Hamiltonian $H_s^{(j)}$, while those vectors of any other subspaces appearing in the initial Hamiltonian $H_s^{(j-1)}$ keep unchanged, and moreover, the subspace $SP_{k_0+l-j+1}\left(2^{k_0+l-j+1}\right)$ is empty in the final Hamiltonian $H_s^{(j)}$ in (6.49).

The above theoretical analysis and calculation in this Subsection can form evidently a complete procedure of the principle of induction in mathematics to prove rigorously that the final Hamiltonian $H_s^{(j)}$ of the $j-$th stage of the small-scale Hilbert-space-enlarging process is indeed given in (6.49) for $j=1,2,...,l+1$.

In particular, it can be deduced from (6.49) that the final $(l+1)-$th stage with $j=l+1$ (i.e., $SP_{k_0}\left(2^{k_0}\right)\rightarrow SP_{k_0+1}\left(2^{k_0+1}\right)$) of the small-scale Hilbert-space-enlarging process may be represented by

$$H_s^{(l)}\xrightarrow{U_y^{k_0,k_0+1}(\pi)}H_s^{(l+1)}=\left(\begin{array}{c}\frac{1}{\sqrt{2}}C_{k_0}\sum_{k=2^{k_0+1}}^{2^{k_0+2}-1}|k\rangle\\+\frac{1}{\sqrt{2}}C_{k_0+1}\sum_{k=2^{k_0+2}}^{2^{k_0+3}-1}|k\rangle+...\\+\frac{1}{\sqrt{2}}C_{k_0+l}\sum_{k=2^{k_0+l+1}}^{2^{k_0+l+2}-1}|k\rangle\end{array}\right)\langle h.c.| \quad (6.50)$$

It can be seen that the $(l+1)-$th stage is also the final stage of the whole small-scale Hilbert-space-enlarging process (6.41). Therefore, the Hamiltonian $H_s^{(l+1)}$ is also the final Hamiltonian of the whole small-scale Hilbert-space-enlarging process and is given by

$$H_s^{(l+1)}=\left(\frac{1}{\sqrt{2}}\left(\begin{array}{c}C_{k_0}\sum_{k=2^{k_0+1}}^{2^{k_0+2}-1}|k\rangle\\+C_{k_0+1}\sum_{k=2^{k_0+2}}^{2^{k_0+3}-1}|k\rangle+...\\+C_{k_0+l}\sum_{k=2^{k_0+l+1}}^{2^{k_0+l+2}-1}|k\rangle\end{array}\right)\right)\langle h.c.| \quad (6.51)$$

Obviously, this final Hamiltonian $H_s^{(l+1)}$ is a PDH operator of (4.4), just like the initial Hamiltonian $H_s^{(0)}$. Now by comparing the initial Hamiltonian $H_s^{(0)}$ of (6.43) with the final Hamiltonian $H_s^{(l+1)}$ of (6.51) it can be found that by the small-scale Hilbert-space-enlarging process every one of all the occupied direct-sum subspaces in the initial Hamiltonian $H_s^{(0)}$ is moved to its nearest subspace with twofold larger dimension successively. Therefore, all these occupied subspaces in the initial Hamiltonian $H_s^{(0)}$ each are enlarged twofold in dimensional size, and in the meantime the expansional coefficient of every one of all the occupied subspaces is decreased by a fractional factor $1/\sqrt{2}$. Then, after the whole small-scale Hilbert-space-enlarging process (6.41), the total dimensional size of all the occupied direct-sum subspaces together in the final Hamiltonian $H_s^{(l+1)}$ is two times that one in the initial Hamiltonian $H_s^{(0)}$.

It can be deduced from the above theoretical analysis and calculation that the small-scale Hilbert-space-enlarging process (6.41) is realized by a sequence of the subspace-selective unitary transformations in the multiple-quantum operator algebra space:

$$H_s^{(l+1)} = \left(U_y^{k_0,k_0+1}(-\pi)\right)^+ ... \left(U_y^{k_0+l-j+1,k_0+l-j+2}(-\pi)\right)^+ ...$$

$$\times \left(U_y^{k_0+l-1,k_0+l}(-\pi)\right)^+ \left(U_y^{k_0+l,k_0+l+1}(-\pi)\right)^+ H_s^{(0)} U_y^{k_0+l,k_0+l+1}(-\pi)$$

$$\times U_y^{k_0+l-1,k_0+l}(-\pi) ... U_y^{k_0+l-j+1,k_0+l-j+2}(-\pi) ... U_y^{k_0,k_0+1}(-\pi) \qquad (6.52a)$$

or

$$H_s^{(l+1)} = U_y^{k_0,k_0+1}(\pi) ... U_y^{k_0+l-j+1,k_0+l-j+2}(\pi) ... U_y^{k_0+l-1,k_0+l}(\pi)$$

$$\times U_y^{k_0+l,k_0+l+1}(\pi) H_s^{(0)} \left(U_y^{k_0+l,k_0+l+1}(\pi)\right)^+$$

$$\times \left(U_y^{k_0+l-1,k_0+l}(\pi)\right)^+ ... \left(U_y^{k_0+l-j+1,k_0+l-j+2}(\pi)\right)^+ ... \left(U_y^{k_0,k_0+1}(\pi)\right)^+ \qquad (6.52b)$$

where the initial Hamiltonian $H_s^{(0)}$ is given by (6.43), while the final Hamiltonian $H_s^{(l+1)}$ is given by (6.51). It can prove that the subspace-selective MQT unitary operator $U_y^{k_0+l-j+1,k_0+l-j+2}(\pi)$ for $j = 1, 2, ..., l+1$, which is given by (6.34), i.e., $U_y^{k_0,k_1}(\theta) = U_y^{k_0+l-j+1,k_0+l-j+2}(\pi)$, can be constructed and realized efficiently. Moreover, its basic unitary transformation of (6.35a) can be efficiently realized too. Therefore, the small-scale Hilbert-space-enlarging process (6.41) can be realized efficiently.

### 6.3.3. The large-scale Hilbert-space-enlarging processes

A large-scale Hilbert-space-enlarging process is performed in the multiple-quantum operator algebra space and may be intuitively described by

$$SP_0(1) \to SP_0(1) \bigoplus SP_1(2) \to SP_0(1) \bigoplus SP_1(2) \bigoplus SP_2(2^2) \to ...$$

$$\to SP_0(1) \bigoplus SP_1(2) \bigoplus SP_2(2^2) \bigoplus ... \bigoplus SP_{n-1}(2^{n-1}) \qquad (6.53)$$

Here initial Hamiltonian $H_s^{(0)}$ of the large-scale Hilbert-space-enlarging process, which is a dynamical variable, is simply taken as the diagonal operator $H_s^{(0)} = |1\rangle\langle 1|$ in the multiple-quantum operator algebra space, where the initial tensor-product base vector $|1\rangle$ belongs to the occupied direct-sum subspace $SP_0(1) = \{|1\rangle\}$ of the Hilbert space $HS(N)$. The large-scale Hilbert-space-enlarging process employs two different types of the subspace-selective MQT unitary operators of (6.34), one of which are $\{U_y^{k_0,k_1}(\pi)\}$ with $0 \le k_0 < k_1 \le n-1$ and another are $\{U_y^{0,1}(\theta)\}$ with the angle parameter $\theta \ne \pi$, for its efficient construction and realization in the multiple-quantum operator algebra space. Here the subspace-selective MQT unitary operator $U_y^{k_0,k_1}(\theta)$ with any angle value $\theta$ is selectively applied to only the two direct-sum subspaces $SP_{k_0}(2^{k_0})$ and $SP_{k_1}(2^{k_1})$ of the Hilbert space $HS(N)$. Its basic unitary transformations are given by (6.35) in the multiple-quantum operator algebra space and are simply represented by, when $\theta = \pi$,

$$\begin{cases} (|\varphi_1\rangle + ...) \langle h.c.| \xrightarrow{U_y^{k_0,k_1}(\pi)} (|\varphi_2\rangle + ...) \langle h.c.| \\ (|\varphi_2\rangle + ...) \langle h.c.| \xrightarrow{U_y^{k_0,k_1}(\pi)} (-|\varphi_1\rangle + ...) \langle h.c.| \end{cases} \tag{6.54}$$

where the vectors $|\varphi_1\rangle \in SP_{k_0}\left(2^{k_0}\right)$ and $|\varphi_2\rangle \in SP_{k_1}\left(2^{k_1}\right)$ are given in (6.19).

The theoretical analysis and the exact calculation for the large-scale Hilbert-space-enlarging process are carried out below. The first stage (i.e., $SP_0\,(1) \rightarrow SP_0\,(1) \bigoplus SP_1\,(2)$) of the large-scale Hilbert-space-enlarging process is to enlarge selectively the initial occupied subspace $SP_0\,(1)$ to a larger occupied subspace $SP_0\,(1) \bigoplus SP_1\,(2)$ of the Hilbert space $HS(N)$. Correspondingly in this stage the initial Hamiltonian $H_s^{(0)}$ is converted into the Hamiltonian $H_s^{(1)}$ by the subspace-selective unitary transformation $H_s^{(1)} = U_y^{0,1}\,(\theta_1)\, H_s^{(0)} \left(U_y^{0,1}\,(\theta_1)\right)^+$ in the multiple-quantum operator algebra space. Here the subspace-selective MQT unitary operator $U_y^{0,1}\,(\theta_1)$ is selectively applied only to the two subspaces $SP_0\,(1)$ and $SP_1\,(2)$ and does not affect any other subspaces of the Hilbert space $HS(N)$. It is given by (6.34) with $\mu = y$ and the index $k_0 = 0$, $k_1 = 1$ and the angle $\theta = \theta_1$, that is, $U_\mu^{k_0,k_1}\,(\theta) = U_y^{0,1}\,(\theta_1)$. Its basic unitary transformation of (6.35a) in the multiple-quantum operator algebra space is represented by

$$(|1\rangle + ...) \langle h.c.| \xrightarrow{U_y^{0,1}(\theta)} \left(|1\rangle \cos\frac{1}{2}\theta + \frac{1}{\sqrt{2}}\left(|2\rangle + |3\rangle\right) \sin\frac{1}{2}\theta + ...\right) \langle h.c.| \tag{6.55a}$$

which, for the special case $\theta = \pi/2$, is reduced to the form

$$(|1\rangle + ...) \langle h.c.| \xrightarrow{U_y^{0,1}(\pi/2)} \left(\frac{1}{\sqrt{2}}|1\rangle + \frac{1}{2}\left(|2\rangle + |3\rangle\right) + ...\right) \langle h.c.| \tag{6.55b}$$

where $|\varphi_1\rangle = |1\rangle \in SP_0\,(1)$ and $|\varphi_2\rangle = \left(|2\rangle + |3\rangle\right)/\sqrt{2} \in SP_1\,(2)$ in (6.35). The basic unitary transformations of (6.55a) and (6.55b) in the multiple-quantum operator algebra space correspond to those of (6.24a) and (6.25a) (See the Subsection 6.2) in the corresponding Hilbert space $HS(N)$, respectively.

With the help of the basic unitary transformation of (6.55a) with $\theta = \theta_1$ for the unitary operator $U_y^{0,1}\,(\theta_1)$ the subspace-selective unitary transformation of the first stage in the multiple-quantum operator algebra space is described by

$$H_s^{(1)} = \left(U_y^{0,1}\,(-\theta_1)\right)^+ H_s^{(0)} U_y^{0,1}\,(-\theta_1) = U_y^{0,1}\,(\theta_1)\, H_s^{(0)} \left(U_y^{0,1}\,(\theta_1)\right)^+$$

$$= \left(U_y^{0,1}\,(\theta_1)\,|1\rangle\right) \langle h.c.| = \left(|1\rangle \cos\frac{1}{2}\theta_1 + \frac{1}{\sqrt{2}}\left(|2\rangle + |3\rangle\right) \sin\frac{1}{2}\theta_1\right) \langle h.c.| \tag{6.56}$$

In particular, when $\theta_1 = \pi/2$, this final Hamiltonian $H_s^{(1)}$ is reduced to the form

$$H_s^{(1)} = \left(\frac{\sqrt{2}}{2}|1\rangle + \frac{1}{2}\left(|2\rangle + |3\rangle\right)\right) \langle h.c.| \tag{6.57}$$

In the final Hamiltonian $H_s^{(1)}$ the vectors $|1\rangle$ and $(|2\rangle+|3\rangle)/\sqrt{2}$ belong to the occupied subspaces $SP_0\,(1)$ and $SP_1\,(2)$, respectively, and moreover, they have different expansional coefficients. After the first stage of the large-scale Hilbert-space-enlarging process the dimensional size (one) of the occupied subspace $SP_0\,(1)$ in the initial Hamiltonian $H_s^{(0)}=|1\rangle\langle 1|$ is changed to a larger dimensional size (one+two) of the occupied subspace $SP_0\,(1)\bigoplus SP_1\,(2)$ in the final Hamiltonian $H_s^{(1)}$ of (6.56) or (6.57). Here the increased dimensional size (two) is just the dimensional size of the direct-sum subspace $SP_1\,(2)$.

The second stage (i.e., $SP_0\,(1)\bigoplus SP_1\,(2)\rightarrow SP_0\,(1)\bigoplus SP_1\,(2)\bigoplus SP_2\,(2^2)$) of the large-scale Hilbert-space-enlarging process is divided into the two time periods that the first time period is to change selectively the subspace $SP_1\,(2)$ to the subspace $SP_2\,(2^2)$ and the second is to enlarge selectively the subspace $SP_0\,(1)$ to the subspace $SP_0\,(1)\bigoplus SP_1\,(2)$. Correspondingly in the first time period of the second stage the initial Hamiltonian $H_s^{(1)}$, which is the final Hamiltonian of the first stage and is given in (6.56) (or (6.57)), is converted into the Hamiltonian $H_s^{(2,1)}$ by the subspace-selective unitary transformation $H_s^{(2,1)}=U_y^{1,2}\,(\pi)\,H_s^{(1)}\left(U_y^{1,2}\,(\pi)\right)^+$. Here the subspace-selective MQT unitary operator $U_y^{1,2}\,(\pi)$ selectively acts on only the two subspaces $SP_1\,(2)$ and $SP_2\,(2^2)$ and does not affect any other subspaces of the Hilbert space $HS(N)$. The unitary operator $U_y^{1,2}\,(\pi)$ is given by (6.34) with $\mu=y$ and $k_0=1$, $k_1=2$ and $\theta=\pi$. Its basic unitary transformation is given by (6.54) in the multiple-quantum operator algebra space,

$$\left(\frac{1}{\sqrt{2}}\,(|2\rangle+|3\rangle)\right)\langle h.c.|\xrightarrow{U_y^{1,2}(\pi)}\left(\frac{1}{2}\,(|4\rangle+|5\rangle+|6\rangle+|7\rangle)\right)\langle h.c.|\,,\tag{6.58}$$

where $|\varphi_1\rangle=(|2\rangle+|3\rangle)/\sqrt{2}\in SP_1\,(2)$ and $|\varphi_2\rangle=(|4\rangle+|5\rangle+|6\rangle+|7\rangle)/2\in SP_2\,(2^2)$, as shown by (6.19). Therefore, by starting from the initial Hamiltonian $H_s^{(1)}$ of (6.56) and with the help of the basic unitary transformation (6.58), the subspace-selective unitary transformation for the first time period of the second stage in the multiple-quantum operator algebra space may be expressed as

$$\begin{aligned}H_s^{(2,1)}&=\left(U_y^{1,2}\,(-\pi)\right)^+H_s^{(1)}U_y^{1,2}\,(-\pi)=U_y^{1,2}\,(\pi)\,H_s^{(1)}\left(U_y^{1,2}\,(\pi)\right)^+\\&=U_y^{1,2}\,(\pi)\left(|1\rangle\cos\frac{1}{2}\theta_1+\frac{1}{\sqrt{2}}\,(|2\rangle+|3\rangle)\sin\frac{1}{2}\theta_1\right)\langle h.c.|\left(U_y^{1,2}\,(\pi)\right)^+\\&=\left(|1\rangle\cos\frac{1}{2}\theta_1+U_y^{1,2}\,(\pi)\,\frac{1}{\sqrt{2}}\,(|2\rangle+|3\rangle)\sin\frac{1}{2}\theta_1\right)\langle h.c.|\\&=\left(|1\rangle\cos\frac{1}{2}\theta_1+\frac{1}{2}\,(|4\rangle+|5\rangle+|6\rangle+|7\rangle)\sin\frac{1}{2}\theta_1\right)\langle h.c.|\end{aligned}\tag{6.59}$$

Here the fourth equality holds due to the subspace-selective property of the unitary operator $U_y^{1,2}\,(\pi)$, while the last equality holds due to the basic unitary transformation (6.58). Here the vectors $|1\rangle$, $(|2\rangle+|3\rangle)/\sqrt{2}$, and $(|4\rangle+|5\rangle+|6\rangle+|7\rangle)/2$ belong to the direct-sum subspaces $SP_0(1)$, $SP_1(2)$, and $SP_2(2^2)$, respectively. It can be seen from (6.59) that, after the subspace-selective unitary transformation of the first time period, the vector $(|2\rangle+|3\rangle)/\sqrt{2}$ of the occupied subspace $SP_1(2)$ is completely converted into the vector $(|4\rangle+|5\rangle+|6\rangle+|7\rangle)/2$ of the subspace $SP_2(2^2)$ and this leads to that the subspace $SP_1(2)$ is emptied in the final Hamiltonian $H_s^{(2,1)}$.

The second time period of the second stage is to enlarge selectively the subspace $SP_0(1)$ to the subspace $SP_0(1)\bigoplus SP_1(2)$. Correspondingly in the second time period the initial Hamiltonian $H_s^{(2,1)}$, which is the final Hamiltonian (6.59) of the first time period, is converted into the Hamiltonian $H_s^{(2,2)}$ by the subspace-selective unitary transformation $H_s^{(2,2)}=U_y^{0,1}(\theta_2)H_s^{(2,1)}\left(U_y^{0,1}(\theta_2)\right)^+$. Here the subspace-selective MQT unitary operator $U_y^{0,1}(\theta_2)$ selectively acts on only the two subspaces $SP_0(1)$ and $SP_1(2)$ and does not affect any other subspaces of the Hilbert space $HS(N)$. Its basic unitary transformation is given by (6.55a) with the angle $\theta=\theta_2$ in the multiple-quantum operator algebra space. By starting from the initial Hamiltonian $H_s^{(2,1)}$ of (6.59) and with the help of the basic unitary transformation (6.55a) for the unitary operator $U_y^{0,1}(\theta_2)$ the subspace-selective unitary transformation for the second time period of the second stage in the multiple-quantum operator algebra space may be described by

$$\begin{aligned}
H_s^{(2,2)} &= \left(U_y^{0,1}(-\theta_2)\right)^+ H_s^{(2,1)} U_y^{0,1}(-\theta_2) = U_y^{0,1}(\theta_2)\, H_s^{(2,1)} \left(U_y^{0,1}(\theta_2)\right)^+ \\
&= U_y^{0,1}(\theta_2)\left(|1\rangle\cos\frac{1}{2}\theta_1+\frac{1}{2}(|4\rangle+|5\rangle+|6\rangle+|7\rangle)\sin\frac{1}{2}\theta_1\right)\langle h.c.|\left(U_y^{0,1}(\theta_2)\right)^+ \\
&= \left(U_y^{0,1}(\theta_2)\,|1\rangle\cos\frac{1}{2}\theta_1+\frac{1}{2}(|4\rangle+|5\rangle+|6\rangle+|7\rangle)\sin\frac{1}{2}\theta_1\right)\langle h.c.| \\
&= \left(\begin{array}{c}\left(|1\rangle\cos\frac{1}{2}\theta_2+\frac{1}{\sqrt{2}}(|2\rangle+|3\rangle)\sin\frac{1}{2}\theta_2\right)\cos\frac{1}{2}\theta_1 \\ +\frac{1}{2}(|4\rangle+|5\rangle+|6\rangle+|7\rangle)\sin\frac{1}{2}\theta_1\end{array}\right)\langle h.c.| \\
&= \left(\begin{array}{c}|1\rangle\cos\frac{1}{2}\theta_2\cos\frac{1}{2}\theta_1+\frac{1}{\sqrt{2}}(|2\rangle+|3\rangle)\sin\frac{1}{2}\theta_2\cos\frac{1}{2}\theta_1 \\ +\frac{1}{2}(|4\rangle+|5\rangle+|6\rangle+|7\rangle)\sin\frac{1}{2}\theta_1\end{array}\right)\langle h.c.| \qquad (6.60)
\end{aligned}$$

Here the fourth equality holds due to the subspace-selective property of the unitary operator $U_y^{0,1}(\theta_2)$ and the fifth equality holds due to the basic unitary transformation (6.55a).

The Hamiltonian $H_s^{(2,2)}$ of (6.60) is really the final Hamiltonian $H_s^{(2)}$ of the second stage of the large-scale Hilbert-space-enlarging process. By combining the first and the second time period of the second stage together it can be found that the second stage may be described by

$$\begin{aligned}
H_s^{(2)} = &\left(U_y^{0,1}(-\theta_2)\right)^+\left(U_y^{1,2}(-\pi)\right)^+\left(U_y^{0,1}(-\theta_1)\right)^+ H_s^{(0)} \\
&\times U_y^{0,1}(-\theta_1)\, U_y^{1,2}(-\pi)\, U_y^{0,1}(-\theta_2)
\end{aligned}$$

$$= U_y^{0,1}(\theta_2)\, U_y^{1,2}(\pi)\, U_y^{0,1}(\theta_1)\, H_s^{(0)}$$
$$\times \left(U_y^{0,1}(\theta_1)\right)^+ \left(U_y^{1,2}(\pi)\right)^+ \left(U_y^{0,1}(\theta_2)\right)^+ = H_s^{(2,2)} \qquad (6.61)$$

where the initial diagonal Hamiltonian $H_s^{(0)} = |1\rangle\langle 1|$ and the final Hamiltonian $H_s^{(2)}$ is given by $H_s^{(2,2)}$ of (6.60). In particular, when $\theta_2 = \theta_1 = \pi/2$, it follows from (6.60) that the final Hamiltonian $H_s^{(2)}$ of (6.61) is reduced to the form

$$H_s^{(2)} = \left(\frac{1}{2\sqrt{2}}\left(\sqrt{2}\,|1\rangle + (|2\rangle + |3\rangle) + (|4\rangle + |5\rangle + |6\rangle + |7\rangle)\right)\right)\langle h.c.| \qquad (6.62)$$

Here the uniform-sum vectors $|1\rangle$, $(|2\rangle + |3\rangle)/\sqrt{2}$, and $(|4\rangle + |5\rangle + |6\rangle + |7\rangle)/2$ in the final Hamiltonian $H_s^{(2)}$ of (6.61) or (6.62) belong to the occupied direct-sum subspaces $SP_0(1)$, $SP_1(2)$, and $SP_2(2^2)$, respectively. Their expansional coefficients which are given respectively by $\cos\frac{1}{2}\theta_2\cos\frac{1}{2}\theta_1$, $\sin\frac{1}{2}\theta_2\cos\frac{1}{2}\theta_1$, and $\sin\frac{1}{2}\theta_1$ in the final Hamiltonian of (6.60) are adjustable via the angle values $\theta_1$ and $\theta_2$. Therefore, with the help of the subspace-selective MQT unitary operators $\{U_y^{0,1}(\theta)\}$ in (6.55a) with the angle $\theta$ taking respectively different values $\theta_1$ and $\theta_2$ in the first two stages of the large-scale Hilbert-space-enlarging process, the expansional coefficients of these uniform-sum vectors of the occupied direct-sum subspaces $SP_0(1)$, $SP_1(2)$, and $SP_2(2^2)$ may be adjusted as desired in the final Hamiltonian $H_s^{(2)}$. This simple method to adjust the expansional coefficients of different direct-sum subspaces occupied by their own uniform-sum vectors may be used generally in the whole large-scale Hilbert-space-enlarging process (6.53). As an application, this simple method may be used to prepare the PDH operator $H_N^{pd}$ of (4.36) (See the Subsection 4.4), where the expansional coefficient of every direct-sum subspace occupied by its own uniform-sum vector in the Hilbert space may take a different value.

It can be found from the second stage that the dimensional size (*three*) of the occupied subspace $SP_0(1)\bigoplus SP_1(2)$ in the initial Hamiltonian $H_s^{(1)}$ of (6.57) is changed to a larger dimensional size ($three + 2^2$) of the occupied subspace $SP_0(1)\bigoplus SP_1(2)\bigoplus SP_2(2^2)$ in the final Hamiltonian $H_s^{(2)}$ of (6.62). Obviously, the increased dimensional size ($2^2$) is just the dimensional size of the direct-sum subspace $SP_2(2^2)$.

Below the subspace-selective MQT unitary operators $\{U_y^{k_0,k_1}(\pi)\}$ and $U_y^{0,1}(\theta)$ with the fixed angle $\theta = \pi/2$ are employed to construct and realize the large-scale Hilbert-space-enlarging process (6.53). The first and the second stage of the large-scale Hilbert-space-enlarging process are already analysed and calculated exactly, and their final Hamiltonians $H_s^{(1)}$ and $H_s^{(2)}$ are already given by (6.57) and (6.62), respectively. An important feature for the two Hamiltonians $H_s^{(1)}$ and $H_s^{(2)}$ is that the expansional coefficients of all the occupied tensor-product base vectors $\{|k\rangle\}$ in any one of the two Hamiltonians take the same value except the one of the occupied base vector $|1\rangle$ which is $\sqrt{2}$ times that one of any other occupied base vector $|k\rangle$ with $k > 1$. It is true that the same feature exists in every stage of the large-scale Hilbert-space-enlarging process (6.53). As an example, the third stage (i.e., $SP_0\left(1\right)\bigoplus SP_1\left(2\right)\bigoplus SP_2\left(2^2\right) \rightarrow SP_0\left(1\right)\bigoplus SP_1\left(2\right)\bigoplus SP_2\left(2^2\right)\bigoplus SP_3\left(2^3\right)$) is explicitly analyzed and calculated below. The third stage starts from the initial Hamiltonian $H_s^{(2)}$ of (6.62) which is the final Hamiltonian of the second stage. It may be divided into several time periods as follows. The first time period is to change selectively the largest occupied subspace $SP_2\left(2^2\right)$ to the empty subspace $SP_3\left(2^3\right)$ by the subspace-selective MQT unitary operator $U_y^{2,3}\left(\pi\right)$, the second time period is to change selectively the occupied subspace $SP_1\left(2\right)$ to the empty subspace $SP_2\left(2^2\right)$ by the subspace-selective MQT unitary operator $U_y^{1,2}\left(\pi\right)$, and the final time period is to enlarge selectively the occupied subspace $SP_0\left(1\right)$ to the subspace $SP_0\left(1\right)\bigoplus SP_1\left(2\right)$ by the subspace-selective MQT unitary operator $U_y^{0,1}\left(\pi/2\right)$. Therefore, the whole third stage which consists of the above three time periods may be described in detail by

$$H_s^{(2)} = \left(\frac{1}{2}\left|1\right\rangle + \frac{1}{2\sqrt{2}}\left(\left|2\right\rangle + \left|3\right\rangle\right) + \frac{1}{2\sqrt{2}}\sum_{k=2^2}^{2^3-1}\left|k\right\rangle\right)\left\langle h.c.\right| \qquad (6.63a)$$

$$\xrightarrow{U_y^{2,3}(\pi)} H_s^{(3,1)} = \left(\begin{array}{c} \frac{1}{2}\left|1\right\rangle + \frac{1}{2\sqrt{2}}\left(\left|2\right\rangle + \left|3\right\rangle\right) + 0 \times \sum_{k=2^2}^{2^3-1}\left|k\right\rangle \\ +\frac{1}{\sqrt{2}}\frac{1}{2\sqrt{2}}\sum_{k=2^3}^{2^4-1}\left|k\right\rangle \end{array}\right)\left\langle h.c.\right|$$

$$\xrightarrow{U_y^{1,2}(\pi)} H_s^{(3,2)} = \left(\begin{array}{c} \frac{1}{2}\left|1\right\rangle + 0 \times \left(\left|2\right\rangle + \left|3\right\rangle\right) + \frac{1}{\sqrt{2}}\frac{1}{2\sqrt{2}}\sum_{k=2^2}^{2^3-1}\left|k\right\rangle \\ +\frac{1}{\sqrt{2}}\frac{1}{2\sqrt{2}}\sum_{k=2^3}^{2^4-1}\left|k\right\rangle \end{array}\right)\left\langle h.c.\right|$$

$$\xrightarrow{U_y^{0,1}(\pi/2)} H_s^{(3,3)} = \left(\begin{array}{c} \frac{1}{2}\frac{1}{\sqrt{2}}\left|1\right\rangle + \frac{1}{2}\frac{1}{2}\left(\left|2\right\rangle + \left|3\right\rangle\right) + \frac{1}{\sqrt{2}}\frac{1}{2\sqrt{2}}\sum_{k=2^2}^{2^3-1}\left|k\right\rangle \\ +\frac{1}{\sqrt{2}}\frac{1}{2\sqrt{2}}\sum_{k=2^3}^{2^4-1}\left|k\right\rangle \end{array}\right)\left\langle h.c.\right|$$

$$= \left(\frac{\sqrt{2}}{4}\left|1\right\rangle + \frac{1}{4}\left(\left|2\right\rangle + \left|3\right\rangle\right) + \frac{1}{4}\sum_{k=2^2}^{2^3-1}\left|k\right\rangle + \frac{1}{4}\sum_{k=2^3}^{2^4-1}\left|k\right\rangle\right)\left\langle h.c.\right| = H_s^{(3)} \qquad (6.63b)$$

where $H_s^{(2)}$ and $H_s^{(3)} = H_s^{(3,3)}$ are the initial and the final Hamiltonian of the third stage, respectively. Indeed, in the final Hamiltonian $H_s^{(3)}$ the expansional coefficients of all the occupied tensor-product base vectors $\{\left|k\right\rangle\}$ have the same value 1/4 except the one (i.e.,$\sqrt{2}/4$) of the occupied base vector $\left|1\right\rangle$.

Note that the notation $H_1\xrightarrow{U_y^{k-1,k}(\theta)}H_2$ stands for the subspace-selective unitary transformation $U_y^{k-1,k}\left(\theta\right)$ acting on the initial Hamiltonian $H_1$, that is, $U_y^{k-1,k}\left(\theta\right)H_1\left(U_y^{k-1,k}\left(\theta\right)\right)^+ = H_2$. Then $H_s^{(2)}\xrightarrow{U_y^{2,3}(\pi)}H_s^{(3,1)}$, $H_s^{(3,1)}\xrightarrow{U_y^{1,2}(\pi)}H_s^{(3,2)}$, and $H_s^{(3,2)}\xrightarrow{U_y^{0,1}(\pi/2)}H_s^{(3,3)}$ in the third stage of (6.63) represent the subspace-selective unitary transformations of the first, the second, and the third time period of the third stage of the large-scale Hilbert-space-enlarging process (6.53) in the multiple-quantum operator algebra space, respectively. Therefore, the final Hamiltonian $H_s^{(3)}$ of the third stage of (6.63) is generated by

$$\begin{aligned} H_s^{(3)} &= \left(U_y^{0,1}(-\pi/2)\right)^+ \left(U_y^{1,2}(-\pi)\right)^+ \left(U_y^{2,3}(-\pi)\right)^+ \\ &\quad \times H_s^{(2)} U_y^{2,3}(-\pi) U_y^{1,2}(-\pi) U_y^{0,1}(-\pi/2) \\ &= U_y^{0,1}(\pi/2) U_y^{1,2}(\pi) U_y^{2,3}(\pi) H_s^{(2)} \left(U_y^{2,3}(\pi)\right)^+ \left(U_y^{1,2}(\pi)\right)^+ \left(U_y^{0,1}(\pi/2)\right)^+ \end{aligned} \tag{6.64}$$

where the initial Hamiltonian $H_s^{(2)}$ is given by (6.62) or (6.63a), while the final Hamiltonian $H_s^{(3)}$ is explicitly given by (6.63b).

It can be found that, after the third stage (6.63) of the large-scale Hilbert-space-enlarging process, the dimensional size (7) of the occupied subspace $SP_0(1) \bigoplus SP_1(2) \bigoplus SP_2(2^2)$ in the initial Hamiltonian $H_s^{(2)}$ of (6.63a) is changed to a larger dimensional size $(7+2^3)$ of the occupied subspace $SP_0(1) \bigoplus SP_1(2) \bigoplus SP_2(2^2) \bigoplus SP_3(2^3)$ in the final Hamiltonian $H_s^{(3)}$ of (6.63b). Here the increased dimensional size $(2^3)$ is just the dimensional size of the direct-sum subspace $SP_3(2^3)$.

Generally, the $(k+1)$ −th stage (i.e., $SP_0(1) \bigoplus SP_1(2) \bigoplus ... \bigoplus SP_k(2^k) \rightarrow SP_0(1) \bigoplus SP_1(2) \bigoplus ... \bigoplus SP_{k+1}(2^{k+1})$ for $k = 1, 2, ..., n-2$) of the large-scale Hilbert-space-enlarging process (6.53) may be described by

$$H_s^{(k)} = \begin{pmatrix} \frac{1}{\sqrt{2^k}}|1\rangle + \frac{1}{\sqrt{2^{k+1}}}(|2\rangle + |3\rangle) + \frac{1}{\sqrt{2^{k+1}}}\sum_{\kappa=2^2}^{2^3-1}|\kappa\rangle \\ +... + \frac{1}{\sqrt{2^{k+1}}}\sum_{\kappa=2^k}^{2^{k+1}-1}|\kappa\rangle \end{pmatrix} \langle h.c.| \tag{6.65a}$$

$$\xrightarrow{U_y^{k,k+1}(\pi)} H_s^{(k+1,1)} = \begin{pmatrix} \frac{1}{\sqrt{2^k}}|1\rangle + \frac{1}{\sqrt{2^{k+1}}}(|2\rangle + |3\rangle) \\ + \frac{1}{\sqrt{2^{k+1}}}\sum_{\kappa=2^2}^{2^3-1}|\kappa\rangle + ... + \frac{1}{\sqrt{2^{k+1}}}\sum_{\kappa=2^{k-1}}^{2^k-1}|\kappa\rangle \\ +0 \times \sum_{\kappa=2^k}^{2^{k+1}-1}|\kappa\rangle + \frac{1}{\sqrt{2}\sqrt{2^{k+1}}}\sum_{\kappa=2^{k+1}}^{2^{k+2}-1}|\kappa\rangle \end{pmatrix} \langle h.c.|$$

$$\xrightarrow{U_y^{k-1,k}(\pi)} H_s^{(k+1,2)} = \begin{pmatrix} \frac{1}{\sqrt{2^k}}|1\rangle + \frac{1}{\sqrt{2^{k+1}}}(|2\rangle + |3\rangle) \\ + \frac{1}{\sqrt{2^{k+1}}}\sum_{\kappa=2^2}^{2^3-1}|\kappa\rangle + ... + \frac{1}{\sqrt{2^{k+1}}}\sum_{\kappa=2^{k-2}}^{2^{k-1}-1}|\kappa\rangle \\ +0 \times \sum_{\kappa=2^{k-1}}^{2^k-1}|\kappa\rangle + \frac{1}{\sqrt{2}\sqrt{2^{k+1}}}\sum_{\kappa=2^k}^{2^{k+1}-1}|\kappa\rangle \\ + \frac{1}{\sqrt{2}\sqrt{2^{k+1}}}\sum_{\kappa=2^{k+1}}^{2^{k+2}-1}|\kappa\rangle \end{pmatrix} \langle h.c.|$$

......

$$\xrightarrow{U_y^{1,2}(\pi)} H_s^{(k+1,k)} = \begin{pmatrix} \frac{1}{\sqrt{2^k}}|1\rangle + 0 \times (|2\rangle + |3\rangle) + \frac{1}{\sqrt{2}\sqrt{2^{k+1}}}\sum_{\kappa=2^2}^{2^3-1}|\kappa\rangle \\ +... + \frac{1}{\sqrt{2}\sqrt{2^{k+1}}}\sum_{\kappa=2^{k+1}}^{2^{k+2}-1}|\kappa\rangle \end{pmatrix} \langle h.c.|$$

$$\xrightarrow{U_y^{0,1}(\pi/2)} H_s^{(k+1,k+1)} = \begin{pmatrix} \frac{1}{\sqrt{2}}\frac{1}{\sqrt{2^k}}|1\rangle + \frac{1}{2\sqrt{2^k}}(|2\rangle + |3\rangle) \\ + \frac{1}{\sqrt{2}\sqrt{2^{k+1}}}\sum_{\kappa=2^2}^{2^3-1}|\kappa\rangle \\ +... + \frac{1}{\sqrt{2}\sqrt{2^{k+1}}}\sum_{\kappa=2^{k+1}}^{2^{k+2}-1}|\kappa\rangle \end{pmatrix} \langle h.c.|$$

$$= \left( \frac{1}{\sqrt{2^{k+2}}} \left( \begin{array}{c} \sqrt{2}\left|1\right\rangle + \left(\left|2\right\rangle + \left|3\right\rangle\right) + \sum_{\kappa=2^2}^{2^3-1}\left|\kappa\right\rangle \\ +... + \sum_{\kappa=2^{k+1}}^{2^{k+2}-1}\left|\kappa\right\rangle \end{array} \right) \right) \left\langle h.c.\right| = H_s^{(k+1)} \quad (6.65b)$$

where $H_s^{(k)}$ and $H_s^{(k+1)} = H_s^{(k+1,k+1)}$ are the initial and the final Hamiltonian of the $(k+1)$-th stage, respectively. The $(k+1)$-th stage evidently consists of $k+1$ time periods, each one of which performs a subspace-selective unitary transformation induced by $U_y^{j-1,j}(\pi)$ and the last one induced by $U_y^{0,1}(\pi/2)$. Therefore, the final Hamiltonian $H_s^{(k+1)}$ of the $(k+1)-$th stage of the large-scale Hilbert-space-enlarging process for $k = 1, 2, ..., n-2$ may be written as

$$H_s^{(k+1)} = \left(U_y^{0,1}(-\pi/2)\right)^+ \left(U_y^{1,2}(-\pi)\right)^+ ... \left(U_y^{k-1,k}(-\pi)\right)^+ \left(U_y^{k,k+1}(-\pi)\right)^+$$

$$\times H_s^{(k)} U_y^{k,k+1}(-\pi) U_y^{k-1,k}(-\pi) ... U_y^{1,2}(-\pi) U_y^{0,1}(-\pi/2)$$

$$= U_y^{0,1}(\pi/2) U_y^{1,2}(\pi) ... U_y^{k-1,k}(\pi) U_y^{k,k+1}(\pi)$$

$$\times H_s^{(k)} \left(U_y^{k,k+1}(\pi)\right)^+ \left(U_y^{k-1,k}(\pi)\right)^+ ... \left(U_y^{1,2}(\pi)\right)^+ \left(U_y^{0,1}(\pi/2)\right)^+$$

$$= \left( \frac{1}{\sqrt{2^{k+2}}} \left( \sqrt{2}\left|1\right\rangle + \left(\left|2\right\rangle + \left|3\right\rangle\right) + \sum_{\kappa=2^2}^{2^3-1}\left|\kappa\right\rangle + ... + \sum_{\kappa=2^{k+1}}^{2^{k+2}-1}\left|\kappa\right\rangle \right) \right) \left\langle h.c.\right| \quad (6.66)$$

This is the recursive-relation-like unitary transformation between the final Hamiltonian $H_s^{(k+1)}$ of the $(k+1)-$th stage and the final Hamiltonian $H_s^{(k)}$ of the $k-$th stage for the large-scale Hilbert-space-enlarging process (6.53) for $k = 1, 2, ..., n-2$, where the Hamiltonian $H_s^{(1)}$ is given by (6.57) and it is the final Hamiltonian of the first stage ($l = 1$) which is generated by (6.56) with the angle $\theta_1 = \pi/2$. It is known above that the final Hamiltonian $H_s^{(2)}$ of the second stage ($l = 2$) is given by (6.62) and it is generated by (6.61) with $\theta_2 = \theta_1 = \pi/2$ and the final Hamiltonian $H_s^{(3)}$ of the third stage ($l = 3$) is given by (6.63b) and it is generated by (6.64). These two Hamiltonians $H_s^{(2)}$ and $H_s^{(3)}$ are the special cases of the final Hamiltonian $H_s^{(l)}$ of (6.66) with $l = k+1 = 2, 3$. The theoretical analysis and calculation above in this Subsection 6.3.3 can form evidently a complete procedure of the principle of induction in mathematics to prove rigorously that the final Hamiltonian $H_s^{(l)}$ of the $l-$th stage of the large-scale Hilbert-space-enlarging process (6.53) is indeed given in (6.66) for $l = k+1 = 2, ..., n-1$.

Obviously, the final Hamiltonian $H_s^{(1)}$ (See (6.57)) of the first stage ($l = 1$) and the final Hamiltonian $H_s^{(l)}$ (See (6.66)) of the $l-$th stage of the large-scale Hilbert-space-enlarging process for $l = 2, 3, ..., n-1$ all are the PDH operators of (4.4), indicating that the final Hamiltonian $H_s^{(l)}$ of every stage (i.e., the $l-$th stage) of the large-scale Hilbert-space-enlarging process (6.53) for $l = 1, 2, ..., n-1$ is a PDH operator of (4.4).

It can be found that, after the $(k+1)-th$ stage of (6.65) for $k = 1, 2, ..., n-2$, the dimensional size $(d_k)$ of the occupied subspace $SP_0(1) \bigoplus SP_1(2) \bigoplus ... \bigoplus$ $SP_k\left(2^k\right)$ in the initial Hamiltonian $H_s^{(k)}$ of (6.65a) is changed to a larger dimensional size $(d_k+2^{k+1})$ of the occupied subspace $SP_0\left(1\right)\bigoplus SP_1\left(2\right)\bigoplus...\bigoplus SP_{k+1}\left(2^{k+1}\right)$ in the final Hamiltonian $H_s^{(k+1)}$ of (6.65b). Here the dimensional size $d_k=1+2+2^2+...+2^k=2^{k+1}-1$ and the increased dimensional size $(2^{k+1})$ is just the dimensional size of the direct-sum subspace $SP_{k+1}\left(2^{k+1}\right)$.

Suppose that by starting from the diagonal Hamiltonian $H_s^{(0)}=|1\rangle\langle 1|$ one wants to prepare the following PDH Hamiltonian operator in the $n-spin-1/2$ system,

$$H_s^{(n_1-1)}=\left(\frac{1}{\sqrt{2^{n_1}}}\left(\begin{array}{c}\sqrt{2}\,|1\rangle+(|2\rangle+|3\rangle)+\sum_{\kappa=2^2}^{2^3-1}|\kappa\rangle\\ +...+\sum_{\kappa=2^{n_1-1}}^{2^{n_1}-1}|\kappa\rangle\end{array}\right)+...\right)\langle h.c.|\,, \quad (6.67)$$

where $1<n_1\leq n$. Then it can be deduced from the recursive-relation-like unitary transformation (6.66) that the number of the subspace-selective MQT unitary operators $\{U_y^{k_0,k_1}(\pi)\}$ $(k_1=k_0+1)$ and $U_y^{0,1}(\pi/2)$ used for preparation of the PDH Hamiltonian $H_s^{(n_1-1)}$ in the $n-spin-1/2$ system may be quadratically dependent upon the spin number $n_1$ (the worst case and the largest $n_1=n$), if the large-scale Hilbert-space-enlarging process (6.53) is employed to prepare the PDH Hamiltonian $H_s^{(n_1-1)}$. However, if a better large-scale Hilbert-space-enlarging process than the one of (6.53) is employed to prepare the PDH Hamiltonian $H_s^{(n_1-1)}$, then the number may be linearly dependent upon the spin number $n_1$. For example, the following large-scale Hilbert-space-enlarging process may be employed to prepare the PDH Hamiltonian $H_s^{(n_1-1)}$ of (6.67) with $1<n_1\leq n$ in the $n-spin-1/2$ system,

$$\begin{aligned}SP_0\left(1\right)&\rightarrow SP_0\left(1\right)\bigoplus SP_1\left(2\right)\rightarrow SP_0\left(1\right)\bigoplus SP_{n_1-1}\left(2^{n_1-1}\right)\\ &\rightarrow SP_0\left(1\right)\bigoplus SP_1\left(2\right)\bigoplus SP_{n_1-1}\left(2^{n_1-1}\right)\\ &\rightarrow SP_0\left(1\right)\bigoplus SP_{n_1-2}\left(2^{n_1-2}\right)\bigoplus SP_{n_1-1}\left(2^{n_1-1}\right)\rightarrow ...\\ &\rightarrow SP_0\left(1\right)\bigoplus SP_1\left(2\right)\bigoplus SP_2\left(2^2\right)\bigoplus...\bigoplus SP_{n_1-1}\left(2^{n_1-1}\right) \quad (6.68)\end{aligned}$$

Then the number of the subspace-selective MQT unitary operators $\{U_y^{k_0,k_1}(\pi)\}$ $(k_1\geq k_0+1)$ and $U_y^{0,1}(\pi/2)$ in (6.68) may be linearly dependent upon the number $n_1$ with $1<n_1\leq n$.

As shown in the Subsection 4.4, the concepts of the full-uniform and the almost-full-uniform pseudo-diagonal Hermitian operators can be applied to any direct-sum subspaces of the Hilbert space of a quantum system. According to the definition (4.39) of an almost-full-uniform PDH operator it can be seen that the PDH Hamiltonian operator $H_s^{(n_1-1)}$ of (6.67) with $1<n_1\leq n$ in the $n-spin-1/2$ system is an almost-full-uniform PDH operator with respect to the occupied direct-sum subspace $HS(N_{n_1})$ with dimension $N_{n_1}=2^{n_1}$ of the Hilbert space $HS(N)$ of the spin system, where the direct-sum decomposition (6.17) of the Hilbert space $HS(N)$ is simply expressed as $HS\left(N\right)=$ $HS(N_k) \bigoplus \left\{ \bigoplus_{l=k}^{n-1} SP_l\left(2^l\right) \right\}$ and the direct-sum subspace $HS(N_k)$ with dimension $N_k = 2^k$ is given by $HS(N_k) = HS(1) \bigoplus \left\{ \bigoplus_{l=0}^{k-1} SP_l\left(2^l\right) \right\}$, as shown by (6.18). Similarly it can be seen from (6.57) and (6.66) that the final PDH Hamiltonian $H_s^{(l)}$ of every stage (i.e., the $l-$th stage) of the large-scale Hilbert-space-enlarging process (6.53) for $l = k+1 = 1, 2, ..., n-1$ in the multiple-quantum operator algebra space is also an almost-full-uniform PDH operator with respect to the occupied direct-sum subspace $HS(N_{l+1})$ with dimension $N_{l+1} = 2^{l+1}$ in the Hilbert space $HS(N)$.

Generally, an almost-full-uniform PDH Hamiltonian (See (4.39)) such as the Hamiltonian $H_s^{(n_1-1)}$ of (6.67) with $1 < n_1 \leq n$ in the $n - spin - 1/2$ system may be prepared with the help of the large-scale Hilbert-space-enlarging process such as (6.53) or (6.68). The preparation may start from the initial diagonal Hamiltonian $H_s^{(0)} = |1\rangle\langle 1|$. The large-scale Hilbert-space-enlarging process may be efficiently constructed with two different types of the subspace-selective MQT unitary operators, one of which are $\{U_y^{k_0,k_1}(\pi)\}$ and another are $\{U_y^{0,1}(\theta)\}$ (the angle $\theta$ may take $\pi/2$). Therefore, the preparation may be efficiently realized in the multiple-quantum operator algebra space by combining the large-scale Hilbert-space-enlarging process with the basic unitary transformations in the Hilbert space for the selective$-|p|-$quantum-transition unitary operators $\exp\left(-i\theta Q_{|p|,\lambda}^{fi}\right)$ $(\lambda = x, y, z)$ (See (5.4)).

As shown in the Subsection 4.4, just like a full-uniform PDH operator with respect to some direct-sum subspace of the Hilbert space (e.g., the full-uniform PDH operator $H_{N_T,N}^0$ of (4.44) with respect to the direct-sum subspace $VS(N)$) an almost-full-uniform PDH operator with respect to the same direct-sum subspace (e.g., the almost-full-uniform PDH Hamiltonian $H_{N_T,N}^{0,L}$ of (4.42) with respect to the same direct-sum subspace $VS(N)$) is still able to reflect the dimensional size of the direct-sum subspace (e.g., the direct-sum subspace $VS(N)$). Then this shows that the almost-full-uniform PDH operator $H_s^{(n_1-1)}$ of (6.67) with respect to the occupied direct-sum subspace $HS(N_{n_1})$ with $1 < n_1 \leq n$ in the Hilbert space $HS(N)$ of the $n - spin - 1/2$ system is still able to reflect the dimensional size $N_{n_1} = 2^{n_1}$ of the occupied direct-sum subspace $HS(N_{n_1})$.

In the direct-sum decomposition (6.17) of the Hilbert space $HS(N)$ of an $n-spin-1/2$ system, i.e., $HS(N) = HS(N_k) \bigoplus \left\{ \bigoplus_{l=k}^{n-1} SP_l\left(2^l\right) \right\}$, the dimensional size $N_k = 2^k$ of the direct-sum subspace $HS(N_k)$ and the dimensional sizes $L_l = 2^l$ of the direct-sum subspaces $\{SP_l(L_l)\}$ with $k \leq l \leq n-1$ of the Hilbert space $HS(N)$ may constitute an important physical quantity [7] and also one important aspect to reflect the symmetrical structures and properties of the $n - spin - 1/2$ system. This is similar to that case in the direct-sum decomposition (6.2) of the Hilbert space $HS(N)$, as shown in the Subsections 5.1.2 and 6.1. The dimensional size of a direct-sum subspace may be considered as a natural number and has an infinitely high precision. According to the quantum-computing speedup theory [Ref[1]] the symmetric structures and properties of quantum systems are thought of as the fundamental quantum-computing-speedup resources which are responsible for essentially speeding up quantum computing and quantum simulating. Therefore, rather than the dimensional size $N$ of the whole Hilbert space $HS(N)$, the dimensional sizes of the direct-sum subspaces of the Hilbert space $HS(N)$ including the dimensional sizes $N_k = 2^k$ of the subspace $HS(N_k)$ and $L_l = 2^l$ of the subspaces $\{SP_l(L_l)\}$ with $k \leq l \leq n-1$ are related to the symmetrical structure of the Hilbert space $HS(N)$ in the direct-sum decomposition (6.17). They should be considered as important resource that may be used to realize quantum-computing speedup in quantum computing and quantum simulating.

The Hilbert-space-enlarging processes and their inverses (i.e., the Hilbert-space-shrinking processes) can selectively change at will the occupied subspaces of the Hilbert space of a quantum system under study. Here one important point is that the dimensional sizes of the occupied subspaces of the Hilbert space can be adjusted at will by the suitable Hilbert-space-enlarging processes and their inverses in the multiple-quantum operator algebra space. For example, in the Hamiltonian $H_s^{(n_1-1)}$ of (6.67) of the $n-spin-1/2$ system, the dimensional size $N_{n_1} = 2^{n_1}$ of the occupied subspace $HS(N_{n_1})$ in (6.67) with $1 < n_1 \leq n$ in the Hilbert space $HS(N)$ of the $n-spin-1/2$ system can be adjusted at will by the large-scale Hilbert-space-enlarging process such as (6.53) or (6.68) and its inverse. Then an important application for the Hilbert-space-enlarging process and its inverse is that one may employ them to selectively change at will the occupied subspace and hence adjust at will the dimensional size of the occupied subspace of the Hilbert space in the multiple-quantum operator algebra space. Consequently it becomes possible that the inverse of dimensional size of the occupied direct-sum subspace of the Hilbert space may act as a discrete variable which owns an infinite-high precision and can take any extreme-small discrete value (which corresponds to an exponentially-large dimensional size). As important resource for quantum-computing speedup the dimensional sizes of direct-sum subspaces and the inverse of dimensional size of the occupied direct-sum subspace which acts as a discrete variable may have important applications in quantum computing and quantum simulating in future.

## 6.4. Efficient implementations of the Hilbert-space-enlarging processes

The Hilbert-space-enlarging processes in the preceding Subsection 6.3 are performed in the multiple-quantum operator algebra space of an $n-spin-1/2$ system (or an $n-$pseudospin$-1/2$ system). For simplicity, the $n-spin-1/2$ system (or the $n-$pseudospin$-1/2$ system) is treated like a conventional $n-$qubit quantum system in this Subsection. It is known in the Subsection 6.3 that the subspace-selective MQT unitary operators $\exp(-i\theta H_\lambda^{mq})$ of (5.32), i.e., $\{U_\mu^{k_0,k_1}(\theta)\}$ of (6.34) and $\{U_\mu^{00,k_1}(\theta)\}$, may act as the basic building blocks to construct a Hilbert-space-enlarging process. Theoretically it is shown that on the basis of the direct-sum decomposition (6.6) of the Hilbert space $HS(N)$ of the $n-spin-1/2$ system the Hilbert-space-enlarging processes in the Subsection 6.3 can be efficiently constructed with these basic building blocks in the multiple-quantum operator algebra space. It can be found from the Subsections 5.2, 6.2, and 6.3 that these subspace-selective MQT unitary operators $\{U_{\mu}^{k_0,k_1}(\theta)\}$ and $\{U_{\mu}^{00,k_1}(\theta)\}$ can be efficiently realized on the basis of the symmetrical structure of the Hilbert space $HS(N)$ which is specified by the direct-sum decomposition (6.6). Therefore, the following three steps $(i)$, $(ii)$, and $(iii)$ may be used to efficiently implement the subspace-selective MQT unitary operators $\{U_{\mu}^{k_0,k_1}(\theta)\}$ and $\{U_{\mu}^{00,k_1}(\theta)\}$ :

$(i)$ The subspace-selective MQT Hermitian operators $\{H_{\lambda}^{mq}\}$ of (5.19) with $\lambda = x, y, z$ are first constructed with the two vectors $|\varphi_1\rangle$ and $|\varphi_2\rangle$ which are given in (6.19). Then they are further used to generate the subspace-selective MQT unitary operators $\exp(-i\theta H_{\lambda}^{mq})$ of (5.32), i.e., $\{U_{\lambda}^{k_0,k_1}(\theta)\}$ and $\{U_{\lambda}^{00,k_1}(\theta)\}$

$(ii)$ The subspace-selective MQT unitary operator $U_{\lambda}^{k_0,k_1}(\theta)$ (or $U_{\lambda}^{00,k_1}(\theta)$) can be unitarily transformed to the selective$-\,|p|\,-$quantum-transition unitary operator $\exp\left(-i\theta Q_{|p|,\lambda}^{fi}\right)$ of (5.3a) in the multiple-quantum operator algebra space by the unitary transformation (5.33), that is, $W^{+}\exp(-i\theta H_{\lambda}^{mq})W = \exp\left(-i\theta Q_{|p|,\lambda}^{fi}\right)$, where the unitary operator $W^{+} = W_{s_2}^{+}W_{s_1}^{+}$ (or $W = W_{s_1}W_{s_2}$) may be obtained from the unitary transformation (5.31).

$(iii)$ The efficient implementation can be carried out in the multiple-quantum operator algebra space not only for the basic unitary transformations of the selective$-\,|p|\,-$quantum-transition unitary operators $\exp\left(-i\theta Q_{|p|,\lambda}^{fi}\right)$ (See (5.4)) but also for those unitary transformations of the unitary operator $W^{+} = W_{s_2}^{+}W_{s_1}^{+}$.

The core of these three steps above is the unitary transformation (5.33) in the step $(ii)$. If the unitary transformation (5.33) can be efficiently implemented, then the step $(iii)$ can be realized and hence the subspace-selective MQT unitary operators $\{U_{\lambda}^{k_0,k_1}(\theta)\}$ and $\{U_{\lambda}^{00,k_1}(\theta)\}$ can be efficiently implemented. Consequently the Hilbert-space-enlarging processes in the Subsection 6.3 can be efficiently implemented. Here on the basis of the direct-sum decomposition (6.6) of the Hilbert space $HS(N)$ it can prove that the unitary transformation (5.33) can be efficiently implemented in the present case that both the vectors $|\varphi_1\rangle$ $(\in SP_{k_0}(2^{k_0}))$ and $|\varphi_2\rangle$ $(\in SP_{k_1}(2^{k_1}))$, which are used to construct the Hermitian operators $\{H_{\lambda}^{mq}\}$ of (5.19), are given in (6.19), respectively.

As far as any given direct-sum decomposition of the Hilbert space $HS(N)$ is concerned, the present scheme which consists of the three steps $(i)$, $(ii)$, and $(iii)$ above is general to implement efficiently the subspace-selective MQT unitary operators $\{U_{\lambda}^{k_0,k_1}(\theta)\}$ and $\{U_{\lambda}^{00,k_1}(\theta)\}$ and hence the Hilbert-space-enlarging processes in the Subsection 6.3, but precondition is that the unitary operator $W^{+} = W_{s_2}^{+}W_{s_1}^{+}$ (or $W = W_{s_1}W_{s_2}$) in the step $(ii)$ can be efficiently implemented, that is, the unitary transformation (5.33) can be efficiently implemented. As an example, it can be shown that the unitary operator $W^{+} = W_{s_2}^{+}W_{s_1}^{+}$ can be efficiently implemented on the basis of the direct-sum decomposition (6.6) of the Hilbert space $HS(N)$ and the two vectors $|\varphi_1\rangle$ and $|\varphi_2\rangle$ given in (6.19).

It is perhaps most straightforward that the Hilbert-space-enlarging processes in the Subsection 6.3 are constructed and implemented by directly making use of the symmetrical structures and properties of the Hilbert space $HS(N)$ of the $n - spin - 1/2$ system. Below consider that the direct-sum decomposition of the Hilbert space $HS(N)$ is given by (6.6) or (6.7), i.e., $HS\left(N\right) = HS\left(1\right)\bigoplus\{\bigoplus_{k=0}^{n-1}SP_k\left(2^k\right)\}$. As shown in the Subsection 6.1, the direct-sum decomposition (6.6) of the Hilbert space $HS(N)$ can reflect the specific (or selected) tensor-product subspaces $TPS(n_t)$ of (6.9) of the Hilbert space $HS(N)$ of the composite $n - spin - 1/2$ system. This point is different from the direct-sum decomposition (6.2) of the same Hilbert space $HS(N)$. The tensor-product subspaces $TPS(n_t)$ must be taken into account, when the Hilbert-space-enlarging processes are constructed and implemented on the basis of the direct-sum decomposition (6.6) of the Hilbert space $HS(N)$. In order to explicitly take into account the tensor-product subspaces $TPS(n_t)$ in the efficient construction and implementation of the Hilbert-space-enlarging processes, here the orthonormal vector basis set of the Hilbert space $HS(N)$ of the $n - spin - 1/2$ system may be set to the complete set of the tensor-product base vectors $\{|\Phi_l^z\rangle\}$ of (6.8). Then, as shown in the Subsection 6.1, the direct-sum subspaces $\{SP_k\left(2^k\right)\}$ and $HS(1)$ of the Hilbert space $HS(N)$ in the direct-sum decomposition (6.6) may be represented explicitly by the tensor-product base vectors $\{|\Phi_l^z\rangle\}$. Consequently the direct-sum decomposition (6.6) can be reduced to the direct-sum decomposition (6.16) of the Hilbert space $HS(N)$ and moreover, the direct-sum subspaces $HS\left(1\right) = \{|\Phi_0^z\rangle\}$ and $SP_{k-1}\left(2^{k-1}\right) = \{\left|\Phi_{2^{k-1}}^z\right\rangle\}$ for $k = 1, ..., n$, as shown in (6.16) in the Subsection 6.1.

There are two different kinds of the subspace-selective MQT unitary operators which may act as the basic building blocks to construct and implement a Hilbert-space-enlarging process. The first kind of the subspace-selective MQT unitary operators are defined as follows: When a subspace-selective MQT unitary operator of the first kind acts on any direct-sum subspaces of the Hilbert space $HS(N)$, it can really affect only one selected direct-sum subspace and does not affect any other direct-sum subspaces of the Hilbert space $HS(N)$. The second kind are defined as follows: When a subspace-selective MQT unitary operator of the second kind acts on any direct-sum subspaces of the Hilbert space $HS(N)$, it really acts on only two selected direct-sum subspaces and it does not affect any other direct-sum subspaces of the Hilbert space $HS(N)$. As a typical example, here consider the direct-sum decomposition (6.2) of the Hilbert space $HS(N)$ in the Subsection 6.1. As shown in the Subsection 5.1.3, the subspace-selective $ZQT$ unitary operator $U_{zq}^{M_i}\left(\theta\right)$ of (5.14) is selectively applied to only the direct-sum subspace $S_{zq}\left(k_0\right)$ of the Hilbert space $HS(N)$. Therefore, it is a subspace-selective MQT unitary operator of the first kind. As shown in the Subsection 5.1.2, the subspace-selective $|p|$ −quantum-transition unitary operator $U_{|p|,\mu}^{M_f,M_i}\left(\theta\right)$ of (5.9) with the order $p \neq 0$ is selectively applied to only the two direct-sum subspaces $S_{zq}\left(k_0\right)$ and $S_{zq}\left(n - k_0\right)$ of the Hilbert space $HS(N)$. Therefore, it is a subspace-selective MQT unitary operator of the second kind.

Below more general subspace-selective MQT unitary operators, which may or may not be equal to $U_{\lambda}^{mq}(\theta) = \exp(-i\theta H_{\lambda}^{mq})$ of (5.32), may be directly constructed on the basis of the direct-sum decomposition (6.16) of the Hilbert space $HS(N)$. Here the tensor-product subspaces $TPS(n_t)$ of (6.9) of the Hilbert space $HS(N)$ of the composite $n - spin - 1/2$ system must be considered explicitly. The selected subspace on which a subspace-selective MQT unitary operator really act may consist of one or more of these $n+1$ direct-sum subspaces of the Hilbert space $HS(N)$ in the direct-sum decomposition (6.16), i.e., $HS(1) = \{|\Phi_0^z\rangle\}$ and $SP_l(2^l) = \{|\Phi_{2^l}^z\rangle\}$ with $l = 0, 1, ..., n-1$. There are also two different kinds of the subspace-selective MQT unitary operators in accordance with the direct-sum decomposition (6.16) of the Hilbert space $HS(N)$. Here a subspace-selective MQT unitary operator of the first kind is selectively applied to only one (selected) direct-sum subspace (e.g., $SP_l(2^l) = \{|\Phi_{2^l}^z\rangle\}$) of the Hilbert space $HS(N)$ in the direct-sum decomposition (6.16). And a subspace-selective MQT unitary operator of the second kind is selectively applied to only two (selected) direct-sum subspaces (e.g., $SP_{k_0}(2^{k_0}) = \{|\Phi_{2^{k_0}}^z\rangle\}$ and $SP_{k_1}(2^{k_1}) = \{|\Phi_{2^{k_1}}^z\rangle\}$) among the $n+1$ direct-sum subspaces of the Hilbert space $HS(N)$ in the direct-sum decomposition (6.16). These two different kinds of the subspace-selective MQT unitary operators may act as the basic building blocks to construct and implement the Hilbert-space-enlarging processes including those in the Subsection 6.3. Below they are investigated in detail on the basis of the direct-sum decomposition (6.16) of the Hilbert space $HS(N)$ of the composite $n - spin - 1/2$ system.

Consider a first-kind subspace-selective MQT unitary operator (or propagator) defined by $U_{k,k-1}^{ss}(\tau) = \exp\left(-iH_{k,k-1}^{ss}\tau/\hbar\right)$. It selectively acts on the selected direct-sum subspace $SP_{k-1}\left(2^{k-1}\right) = \{|\Phi_{2^{k-1}}^z\rangle\}$ with $0 \leq k-1 \leq n-1$ and does not affect all the other $n$ direct-sum subspaces $\{|\Phi_0^z\rangle\}$ and $\{|\Phi_{2^l}^z\rangle\}$ with $l \neq k-1$ and $0 \leq l \leq n-1$ of the Hilbert space $HS(N)$ in the direct-sum decomposition (6.16). Below the subspace-selective MQT unitary operator $U_{k,k-1}^{ss}(\tau)$ of the first kind and its generating Hamiltonian $H_{k,k-1}^{ss}$ are explicitly described and constructed. First of all, examine all these $n+1$ direct-sum subspaces, i.e., $\{|\Phi_0^z\rangle\}$ and $\{|\Phi_{2^l}^z\rangle\}$ with $0 \leq l \leq n-1$, in the direct-sum decomposition (6.16) of the Hilbert space $HS(N)$ of the composite $n - spin - 1/2$ system. This examination aims to investigate where the subspace-selective property of the subspace-selective MQT unitary operator $U_{k,k-1}^{ss}(\tau)$ is originated from. According to the tensor-product base vectors $\{|\Phi_l^z\rangle\}$ of (6.8) the selected direct-sum subspace $\{|\Phi_{2^{k-1}}^z\rangle\}$ may be expressed as (6.14). Among these $n+1$ direct-sum subspaces in the direct-sum decomposition (6.16) of the Hilbert space $HS(N)$ these direct-sum subspaces $\{|\Phi_0^z\rangle\}$ and $\{|\Phi_{2^l}^z\rangle\}$ with $0 \leq l < k-1$ each have a dimension smaller than the dimension of the selected direct-sum subspace $\{|\Phi_{2^{k-1}}^z\rangle\}$, while these direct-sum subspaces $\{|\Phi_{2^l}^z\rangle\}$ with $k \leq l \leq n-1$ each have a dimension larger than the dimension of the selected direct-sum subspace $\{|\Phi_{2^{k-1}}^z\rangle\}$. As a typical example, both the direct-sum subspaces $\{|\Phi_{2^{k-2}}^z\rangle\}$ and $\{|\Phi_{2^k}^z\rangle\}$ are smaller and larger than the selected direct-sum subspace $\{|\Phi_{2^{k-1}}^z\rangle\}$, respectively, and it can be found from (6.14) that

$$\begin{cases} \left\{\left|\Phi^z_{2^{k-2}}\right\rangle\right\} = \{|0_n\rangle \dots |0_j\rangle \dots |0_{k+1}\rangle \bigotimes |0_k\rangle \, |\mathbf{1}_{k-1}\rangle \, |m_{k-2}\rangle \dots |m_2\rangle \, |m_1\rangle\} \\ \left\{\left|\Phi^z_{2^{k-1}}\right\rangle\right\} = \{|0_n\rangle \dots |0_j\rangle \dots |0_{k+1}\rangle \bigotimes |\mathbf{1}_k\rangle \, |m_{k-1}\rangle \dots |m_2\rangle \, |m_1\rangle\} \\ \left\{\left|\Phi^z_{2^k}\right\rangle\right\} = \{|0_n\rangle \dots |0_j\rangle \dots |0_{k+2}\rangle \, |\mathbf{1}_{k+1}\rangle \bigotimes |m_k\rangle \dots |m_2\rangle \, |m_1\rangle\} \end{cases} \tag{6.69}$$

where $|m_l\rangle = |0_l\rangle\,,\ |1_l\rangle$ for $l = 1, 2, ..., k$. Obviously, these three direct-sum subspaces of (6.69) are mutually orthogonal.

*Remark.* Suppose that $S_1$ and $S_2$ are any two subspaces of the Hilbert space which are mutually orthogonal. For example, both the subspaces $S_1$ and $S_2$ may be any pair of the direct-sum subspaces of the Hilbert space in any given direct-sum decomposition. Let $|\phi_1\rangle \in S_1$ and $|\phi_2\rangle \in S_2$ be any two vectors of the Hilbert space which belong to the two orthogonal subspaces $S_1$ and $S_2$, respectively. Then there is the orthogonal relation $\langle\phi_1|\phi_2\rangle = 0$ between the two vectors $|\phi_1\rangle$ and $|\phi_2\rangle$.

*Remark.* Suppose that $|R\rangle$ and $|S\rangle$ are any two orthonormal tensor-product base vectors of the vector basis set $\{|\Phi^z_l\rangle\}$ of (6.8) (or $\{|S\rangle\}$ of (6.5)) and may be formally given by $|R\rangle = |r_n\rangle\,|r_{n-1}\rangle \dots |r_1\rangle$ and $|S\rangle = |s_n\rangle\,|s_{n-1}\rangle \dots |s_1\rangle$ for $r_j, s_j = 0, 1$ and $j = 1, 2, ..., n$. Then the orthogonal relations for the tensor-product base vectors $\{|\Phi^z_l\rangle\}$ (or $\{|S\rangle\}$) may be expressed as

$$\langle R|S\rangle = \delta_{r_1 s_1}\delta_{r_2 s_2}...\delta_{r_n s_n}$$

It can be seen that one pair of different component base vectors (e.g., $|r_j\rangle \neq |s_j\rangle$ for $j = 1, 2, ..., n$) are enough to decide that both the tensor-product base vectors $|R\rangle$ and $|S\rangle$ are orthogonal to one another, i.e., $\langle R|S\rangle = 0$.

Now examine the orthogonal relations between the tensor-product base vectors in the three direct-sum subspaces of (6.69). On the one hand, by comparing the selected subspace $\left\{\left|\Phi^z_{2^{k-1}}\right\rangle\right\}$ with the smaller subspace $\left\{\left|\Phi^z_{2^{k-2}}\right\rangle\right\}$ in (6.69) it can be found that the selected direct-sum subspace $\left\{\left|\Phi^z_{2^{k-1}}\right\rangle\right\}$ is orthogonal to the smaller direct-sum subspace $\left\{\left|\Phi^z_{2^{k-2}}\right\rangle\right\}$ due to that both the component base vectors $|1_k\rangle$ and $|0_k\rangle$ of the $k-$th spin$-1/2$ particle of the composite $n - spin - 1/2$ system are orthogonal to one another, and here only the component base vector $|1_k\rangle$ is contained in every tensor-product base vector $\left|\Phi^z_{2^{k-1}}\right\rangle$ of the selected subspace $\{\left|\Phi^z_{2^{k-1}}\right\rangle\}$, while only the component base vector $|0_k\rangle$ is contained in every tensor-product base vector $\left|\Phi^z_{2^{k-2}}\right\rangle$ of the smaller subspace $\{\left|\Phi^z_{2^{k-2}}\right\rangle\}$. This result is also available for every one of the smaller direct-sum subspaces $\{|\Phi^z_0\rangle\}$ and $\{\left|\Phi^z_{2^l}\right\rangle\}$ with $0 \leq l < k - 1$ whose dimensions each are smaller than the dimension of the selected direct-sum subspace $\left\{\left|\Phi^z_{2^{k-1}}\right\rangle\right\}$. Therefore, the orthogonality that the selected direct-sum subspace $\left\{\left|\Phi^z_{2^{k-1}}\right\rangle\right\}$ is orthogonal to any smaller direct-sum subspace is achieved via both the orthonormal base vectors $|1_k\rangle$ and $|0_k\rangle$ of the $k-$th spin$-1/2$ particle of the composite $n - spin - 1/2$ system, where only the component base vector $|1_k\rangle$ is contained in every tensor-product base vector of the selected subspace $\left\{\left|\Phi^z_{2^{k-1}}\right\rangle\right\}$, while only the component base vector $|0_k\rangle$ is contained in every tensor-product base vector of any one of the smaller direct-sum subspaces $\{|\Phi^z_0\rangle\}$ and $\{\left|\Phi^z_{2^l}\right\rangle\}$ with $0 \leq l < k-1$. On the other hand, by comparing the selected subspace $\left\{\left|\Phi^z_{2^{k-1}}\right\rangle\right\}$ with the larger subspace $\left\{\left|\Phi^z_{2^k}\right\rangle\right\}$ in (6.69) it can be found that the orthonormal base vectors $|1_k\rangle$ and $|0_k\rangle$ of the $k-$th spin$-1/2$ particle can not lead to the orthogonality between the selected subspace $\left\{\left|\Phi^z_{2^{k-1}}\right\rangle\right\}$ and the larger subspace $\left\{\left|\Phi^z_{2^k}\right\rangle\right\}$, since the component base vector $|m_k\rangle$ of the $k-$th spin$-1/2$ particle in any tensor-product base vector $\left|\Phi^z_{2^k}\right\rangle$ of the larger subspace $\left\{\left|\Phi^z_{2^k}\right\rangle\right\}$ may take any one of the two base vectors $|1_k\rangle$ and $|0_k\rangle$. Instead, the orthogonality is due to that both the component base vectors $|0_{k+1}\rangle$ and $|1_{k+1}\rangle$ of the $(k+1)-$th spin$-1/2$ particle of the composite $n-spin-1/2$ system are orthogonal to one another, and here every tensor-product base vector $\left|\Phi^z_{2^{k-1}}\right\rangle$ of the selected subspace $\{\left|\Phi^z_{2^{k-1}}\right\rangle\}$ contains only the component base vector $|0_{k+1}\rangle$, while every tensor-product base vector $\left|\Phi^z_{2^k}\right\rangle$ of the larger subspace $\left\{\left|\Phi^z_{2^k}\right\rangle\right\}$ contains only the component base vector $|1_{k+1}\rangle$. This result is also available for every one of the larger direct-sum subspaces $\{\left|\Phi^z_{2^l}\right\rangle\}$ with $k \leq l \leq n-1$ whose dimensions each are larger than the dimension of the selected direct-sum subspace $\left\{\left|\Phi^z_{2^{k-1}}\right\rangle\right\}$.

It can be found from the above theoretical analysis that the component tensor-product base vector $|0_{k+1}\rangle\,|1_k\rangle$ is contained only in every tensor-product base vector $\left|\Phi^z_{2^{k-1}}\right\rangle$ of the selected subspace $\{\left|\Phi^z_{2^{k-1}}\right\rangle\}$, while it is not contained in any tensor-product base vector $\left|\Phi^z_{2^k}\right\rangle$ of the larger subspace $\left\{\left|\Phi^z_{2^k}\right\rangle\right\}$ and also any tensor-product base vector of any one of the smaller direct-sum subspaces $\{|\Phi^z_0\rangle\}$ and $\{\left|\Phi^z_{2^l}\right\rangle\}$ with $0 \leq l < k-1$. Therefore, the component tensor-product base vector $|0_{k+1}\rangle\,|1_k\rangle$ is able to control simultaneously the orthogonality between the selected subspace $\left\{\left|\Phi^z_{2^{k-1}}\right\rangle\right\}$ and the larger subspace $\left\{\left|\Phi^z_{2^k}\right\rangle\right\}$ and the orthogonality between the selected subspace $\left\{\left|\Phi^z_{2^{k-1}}\right\rangle\right\}$ and any one of the smaller direct-sum subspaces $\{|\Phi^z_0\rangle\}$ and $\{\left|\Phi^z_{2^l}\right\rangle\}$ with $0 \leq l < k-1$. In analogous way it can be shown that the component tensor-product base vector $|0_{k+2}\rangle\,|0_{k+1}\rangle\,|1_k\rangle$ which is contained only in every tensor-product base vector $\left|\Phi^z_{2^{k-1}}\right\rangle$ of the selected subspace $\{\left|\Phi^z_{2^{k-1}}\right\rangle\}$ is able to control at the same time the (first) orthogonality between the selected subspace $\left\{\left|\Phi^z_{2^{k-1}}\right\rangle\right\}$ and any one of the two larger subspaces $\left\{\left|\Phi^z_{2^k}\right\rangle\right\}$ and $\left\{\left|\Phi^z_{2^{k+1}}\right\rangle\right\}$ with $k+1 < n$ and the (second) orthogonality between the selected subspace $\left\{\left|\Phi^z_{2^{k-1}}\right\rangle\right\}$ and any one of the smaller direct-sum subspaces $\{|\Phi^z_0\rangle\}$ and $\{\left|\Phi^z_{2^l}\right\rangle\}$ with $0 \leq l < k-1$. Obviously, here the component tensor-product base vector $|0_{k+2}\rangle\,|0_{k+1}\rangle$ and the component base vector $|1_k\rangle$ of the component tensor-product base vector $|0_{k+2}\rangle\,|0_{k+1}\rangle\,|1_k\rangle$ are responsible for controlling the first and the second orthogonality, respectively. More generally, it can be shown that the component tensor-product base vector $|0_n\rangle \ldots |0_j\rangle \ldots |0_{k+2}\rangle\,|0_{k+1}\rangle\,|1_k\rangle$ which is contained only in every tensor-product base vector $\left|\Phi^z_{2^{k-1}}\right\rangle$ of the selected subspace $\{\left|\Phi^z_{2^{k-1}}\right\rangle\}$ is able to control at the same time the (first) orthogonality between the selected subspace $\left\{\left|\Phi^z_{2^{k-1}}\right\rangle\right\}$ and any one of all the larger subspaces $\{\left|\Phi^z_{2^l}\right\rangle\}$ with $k \leq l \leq n-1$ and the (second) orthogonality between the selected subspace $\left\{\left|\Phi^z_{2^{k-1}}\right\rangle\right\}$ and any one of all the smaller direct-sum subspaces $\{|\Phi^z_0\rangle\}$ and $\{\left|\Phi^z_{2^l}\right\rangle\}$ with $0 \leq l < k-1$ in the Hilbert space $HS(N)$. Here the component tensor-product base vector

$|0_n\rangle \ldots |0_j\rangle \ldots |0_{k+2}\rangle |0_{k+1}\rangle$ and the component base vector $|1_k\rangle$ in the component tensor-product base vector $|0_n\rangle \ldots |0_j\rangle \ldots |0_{k+2}\rangle |0_{k+1}\rangle |1_k\rangle$ are responsible for controlling the first and the second orthogonality, respectively.

*Remark.* Let exponential operator $\exp(-i\theta A)$ with any Hermitian operator $A$ be a unitary operator and $|\psi\rangle$ be any vector of the Hilbert space. The vector $A|\psi\rangle$ is generated by acting the Hermitian operator $A$ on the vector $|\psi\rangle$ and the vector $\exp(-i\theta A)|\psi\rangle$ is generated by acting the unitary operator $\exp(-i\theta A)$ on the vector $|\psi\rangle$. Then it can prove that if the generated vector $A|\psi\rangle = 0$, then the generated vector $\exp(-i\theta A)|\psi\rangle = |\psi\rangle$, that is, the unitary operator $\exp(-i\theta A)$ does not affect the vector $|\psi\rangle$ when it acts on the vector $|\psi\rangle$.

According to its definition, the first-kind subspace-selective MQT unitary operator $U_{k,k-1}^{ss}(\tau) = \exp\left(-iH_{k,k-1}^{ss}\tau/\hbar\right)$ $(1 \leq k \leq n)$ selectively acts on only the selected direct-sum subspace $SP_{k-1}\left(2^{k-1}\right) = \{|\Phi_{2^{k-1}}^{z}\rangle\}$ and does not affect all the other $n$ direct-sum subspaces $\{|\Phi_0^z\rangle\}$ and $\{|\Phi_{2^l}^z\rangle\}$ with $l \neq k-1$ and $0 \leq l \leq n-1$ of the Hilbert space $HS(N)$ in the direct-sum decomposition (6.16). It is generated by the subspace-selective MQT Hermitian operator $H_{k,k-1}^{ss}$. Therefore, the subspace-selective property of the unitary operator $U_{k,k-1}^{ss}(\tau)$ may be achieved via the subspace-selective MQT Hermitian operator $H_{k,k-1}^{ss}$. Now the vector $H_{k,k-1}^{ss}|\Phi_l^z\rangle$ is generated by acting the Hermitian operator $H_{k,k-1}^{ss}$ on any base vector $|\Phi_l^z\rangle$ of the complete set of the tensor-product base vectors $\{|\Phi_l^z\rangle\}$ of (6.8) and the vector $\exp\left(-iH_{k,k-1}^{ss}\tau/\hbar\right)|\Phi_l^z\rangle$ is generated by acting the unitary exponential operator $\exp\left(-iH_{k,k-1}^{ss}\tau/\hbar\right)$ on the base vector $|\Phi_l^z\rangle$. Then the unitary exponential operator $\exp\left(-iH_{k,k-1}^{ss}\tau/\hbar\right)$ does not affect the tensor-product base vector $|\Phi_l^z\rangle$ when it acts on the base vector $|\Phi_l^z\rangle$, if the generated vector $H_{k,k-1}^{ss}|\Phi_l^z\rangle = 0$. Therefore, if the generated vector $H_{k,k-1}^{ss}|\Phi_l^z\rangle = 0$, then the generated vector $U_{k,k-1}^{ss}(\tau)|\Phi_l^z\rangle$ is given by $U_{k,k-1}^{ss}(\tau)|\Phi_l^z\rangle = |\Phi_l^z\rangle$. According to its definition, the unitary operator $U_{k,k-1}^{ss}(\tau)$ does not affect any direct-sum subspaces $\{|\Phi_0^z\rangle\}$ and $\{|\Phi_{2^l}^z\rangle\}$ with $0 \leq l \leq n-1$ of the Hilbert space $HS(N)$ except the selected direct-sum subspace $\{|\Phi_{2^{k-1}}^z\rangle\}$. Then for the Hermitian operator $H_{k,k-1}^{ss}$ this requires that the generated vector $H_{k,k-1}^{ss}|\Phi_l^z\rangle = 0$ for any tensor-product base vector $|\Phi_l^z\rangle$ of the Hilbert space $HS(N)$ that does not belong to the selected subspace $\{|\Phi_{2^{k-1}}^z\rangle\}$ and in the meantime there exist the tensor-product base vectors $|\Phi_l^z\rangle$ which belong to the selected subspace $\{|\Phi_{2^{k-1}}^z\rangle\}$ such that the generated vectors $H_{k,k-1}^{ss}|\Phi_l^z\rangle \neq 0$.

The above theoretical analysis about the orthogonality between the selected direct-sum subspace $\{|\Phi_{2^{k-1}}^z\rangle\}$ and any larger direct-sum subspace or any smaller direct-sum subspace in the Hilbert space $HS(N)$ shows that the component tensor-product base vector $|0_n\rangle \ldots |0_j\rangle \ldots |0_{k+1}\rangle |\mathbf{1}_k\rangle$ of the tensor-product base vector $|\Phi_l^z\rangle$ of (6.8) is contained only in every tensor-product base vector $|\Phi_{2^{k-1}}^z\rangle$ of the selected direct-sum subspace $\{|\Phi_{2^{k-1}}^z\rangle\}$ and it is able to control at the same time the orthogonality between the selected direct-sum subspace $\{|\Phi_{2^{k-1}}^z\rangle\}$ and any other direct-sum subspaces of the Hilbert space $HS(N)$ in the direct-

sum decomposition (6.16). These results show that the subspace-selective MQT Hermitian operator $H_{k,k-1}^{ss}$ must be related to the component tensor-product base vector $|0_n\rangle ... |0_{k+1}\rangle |\mathbf{1}_k\rangle$ . Then in the simplest form the subspace-selective MQT Hermitian operator $H_{k,k-1}^{ss}$, which generates the subspace-selective MQT unitary operator $U_{k,k-1}^{ss}(\tau) = \exp\left(-iH_{k,k-1}^{ss}\tau/\hbar\right)$ of the first kind, may be chosen as such a Hermitian operator that contains the component tensor-product diagonal operator $(|0_n\rangle ... |0_{k+1}\rangle |\mathbf{1}_k\rangle)(|0_n\rangle ... |0_{k+1}\rangle |\mathbf{1}_k\rangle)^+$ . Therefore, the whole Hermitian operator $H_{k,k-1}^{ss}$ of the composite $n - spin - 1/2$ system may be generally written as

$$H_{k,k-1}^{ss} = (|0_n\rangle ... |0_{k+1}\rangle |\mathbf{1}_k\rangle)(|0_n\rangle ... |0_{k+1}\rangle |\mathbf{1}_k\rangle)^+ \bigotimes H_{k-1}^{s} \tag{6.70}$$

Here the component Hermitian operator $H_{k-1}^{s}$ with $1 \leq k-1 \leq n-1$ may be chosen as the desired component Hamiltonian and it selectively acts on only the tensor-product subspace $\{|m_{k-1}\rangle ... |m_2\rangle |m_1\rangle\}$ of the Hilbert space $HS(N)$ of the composite $n - spin - 1/2$ system. (If the index $k-1=0$ is included, then $H_0^s$ is a constant.) Now by acting the Hermitian operator $H_{k,k-1}^{ss}$ of (6.70) on any tensor-product base vector $|\Phi_l^z\rangle$ of (6.8) it can be found that the generated vector $H_{k,k-1}^{ss}|\Phi_l^z\rangle = 0$ for any tensor-product base vector $|\Phi_l^z\rangle$ of the Hilbert space $HS(N)$ that does not belong to the selected direct-sum subspace $\{|\Phi_{2^{k-1}}^z\rangle\}$. The reason for this is that any tensor-product base vector $|\Phi_l^z\rangle$ does not contain the component tensor-product base vector $|0_n\rangle ... |0_{k+1}\rangle |\mathbf{1}_k\rangle$ , if it does not belong to the selected direct-sum subspace $\{|\Phi_{2^{k-1}}^z\rangle\}$. However, the generated vector $H_{k,k-1}^{ss}|\Phi_l^z\rangle$ may not be zero, i.e., $H_{k,k-1}^{ss}|\Phi_l^z\rangle \neq 0$, if the tensor-product base vector $|\Phi_l^z\rangle$ belongs to the selected direct-sum subspace $\{|\Phi_{2^{k-1}}^z\rangle\}$. The reason for this is that $(i)$ every tensor-product base vector $|\Phi_l^z\rangle$ of the selected direct-sum subspace $\{|\Phi_{2^{k-1}}^z\rangle\}$ contains the component tensor-product base vector $|0_n\rangle ... |0_{k+1}\rangle |\mathbf{1}_k\rangle$ and $(ii)$ the component operator $H_{k-1}^{s}$ of the Hermitian operator $H_{k,k-1}^{ss}$ may be chosen as desired. Therefore, the component tensor-product diagonal operator $(|0_n\rangle ... |0_{k+1}\rangle |\mathbf{1}_k\rangle)(|0_n\rangle ... |0_{k+1}\rangle |\mathbf{1}_k\rangle)^+$ of the Hermitian operator $H_{k,k-1}^{ss}$ is responsible for the subspace-selective property of the subspace-selective MQT unitary operator (or propagator) $U_{k,k-1}^{ss}(\tau)$ of the first kind in the Hilbert space $HS(N)$.

As a typical example, if the subspace-selective MQT Hermitian operator $H_{k,k-1}^{ss}$ of (6.70) acts on any tensor-product base vectors $|\Phi_{2^{k-2}}^z\rangle$, $|\Phi_{2^{k-1}}^z\rangle$, and $|\Phi_{2^{k}}^z\rangle$ which belong to the three different direct-sum subspaces of (6.69), respectively, then it can be found that

$$\begin{cases} H_{k,k-1}^{ss}|\Phi_{2^{k-2}}^z\rangle = H_{k,k-1}^{ss}(|0_n\rangle ... |0_{k+1}\rangle \bigotimes |\mathbf{0}_k\rangle |1_{k-1}\rangle |m_{k-2}\rangle ... |m_1\rangle) = 0 \\ H_{k,k-1}^{ss}|\Phi_{2^{k-1}}^z\rangle = |0_n\rangle ... |0_{k+1}\rangle \bigotimes |\mathbf{1}_k\rangle \bigotimes H_{k-1}^{s}(|m_{k-1}\rangle ... |m_1\rangle) \neq 0 \\ H_{k,k-1}^{ss}|\Phi_{2^{k}}^z\rangle = H_{k,k-1}^{ss}(|0_n\rangle ... |0_{k+2}\rangle |\mathbf{1}_{k+1}\rangle \bigotimes |m_k\rangle |m_{k-1}\rangle ... |m_1\rangle) = 0 \end{cases} \tag{6.71}$$

These transformational equations show clearly that when the Hermitian operator $H_{k,k-1}^{ss}$ acts on these three different direct-sum subspaces in (6.69), i.e., the smaller subspace $\{|\Phi_{2^{k-2}}^z\rangle\}$, the selected subspace $\{|\Phi_{2^{k-1}}^z\rangle\}$, and the larger subspace $\{|\Phi^z_{2^k}\rangle\}$ , respectively, it can be found that $H^{ss}_{k,k-1}\{|\Phi^z_{2^{k-2}}\rangle\} = 0$, $H^{ss}_{k,k-1}\{|\Phi^z_{2^{k-1}}\rangle\} \neq 0$, and $H^{ss}_{k,k-1}\{|\Phi^z_{2^k}\rangle\} = 0$, and moreover, one has $H^{ss}_{k,k-1}\{|\Phi^z_{2^{k-1}}\rangle\} = \{|0_n\rangle ... |0_{k+1}\rangle \bigotimes |1_k\rangle \bigotimes H^s_{k-1}(|m_{k-1}\rangle ... |m_1\rangle)\}$ for the selected subspace $\{|\Phi^z_{2^{k-1}}\rangle\}$. More generally, when the Hermitian operator $H^{ss}_{k,k-1}$ acts on any direct-sum subspaces $\{|\Phi^z_0\rangle\}$ and $\{|\Phi^z_{2^l}\rangle\}$ with $l = 0, 1, ..., n-1$ of the Hilbert space $HS(N)$, it can be found that

$$\left\{\begin{array}{l} H^{ss}_{k,k-1}\{|\Phi^z_0\rangle\} = 0,\ H^{ss}_{k,k-1}\{|\Phi^z_{2^l}\rangle\} = 0 \text{ for } 0 \leq l < k-1 \\ H^{ss}_{k,k-1}\{|\Phi^z_{2^{k-1}}\rangle\} = \left\{\begin{array}{c} |0_n\rangle ... |0_{k+1}\rangle \bigotimes |\mathbf{1}_k\rangle \\ \bigotimes H^s_{k-1}(|m_{k-1}\rangle ... |m_1\rangle) \end{array}\right\} \text{ for } l = k-1 \\ H^{ss}_{k,k-1}\{|\Phi^z_{2^l}\rangle\} = 0 \text{ for } k \leq l \leq n-1 \end{array}\right. \tag{6.72}$$

Here the component operator $H^s_{k-1}$ of the Hermitian operator $H^{ss}_{k,k-1}$ of (6.70) selectively acts on only the tensor-product subspace $\{|m_{k-1}\rangle ... |m_2\rangle |m_1\rangle\}$ , when the Hermitian operator $H^{ss}_{k,k-1}$ selectively acts on the selected direct-sum subspace $\{|\Phi^z_{2^{k-1}}\rangle\}$. The transformational equations of (6.72) can directly result in that when the first-kind subspace-selective MQT unitary operator $U^{ss}_{k,k-1}(\tau) = \exp\left(-iH^{ss}_{k,k-1}\tau/\hbar\right)$ acts on any direct-sum subspaces of the Hilbert space $HS(N)$, only the selected direct-sum subspace $\{|\Phi^z_{2^{k-1}}\rangle\}$ is affected and any other direct-sum subspaces of the Hilbert space $HS(N)$ are not affected at all. This may be conveniently described by

$$\left\{\begin{array}{l} U^{ss}_{k,k-1}(\tau)\{|\Phi^z_0\rangle\} = \{|\Phi^z_0\rangle\},\ U^{ss}_{k,k-1}(\tau)\{|\Phi^z_{2^l}\rangle\} = \{|\Phi^z_{2^l}\rangle\} \text{ for } 0 \leq l < k-1 \\ U^{ss}_{k,k-1}(\tau)\{|\Phi^z_{2^{k-1}}\rangle\} = \left\{\begin{array}{c} |0_n\rangle ... |0_{k+1}\rangle \bigotimes |\mathbf{1}_k\rangle \bigotimes \\ \exp\left(-iH^s_{k-1}\tau/\hbar\right)(|m_{k-1}\rangle ... |m_1\rangle) \end{array}\right\} \text{ for } l = k-1 \\ U^{ss}_{k,k-1}(\tau)\{|\Phi^z_{2^l}\rangle\} = \{|\Phi^z_{2^l}\rangle\} \text{ for } k \leq l \leq n-1 \end{array}\right. \tag{6.73}$$

These unitary transformations show clearly the subspace-selective property of the first-kind subspace-selective MQT unitary operator $U^{ss}_{k,k-1}(\tau)$ that the unitary operator $U^{ss}_{k,k-1}(\tau)$ selectively acts on only the selected direct-sum subspace $SP_{k-1}(2^{k-1}) = \{|\Phi^z_{2^{k-1}}\rangle\}$ and it does not affect all the other $n$ direct-sum subspaces $HS(1) = \{|\Phi^z_0\rangle\}$ and $SP_l(2^l) = \{|\Phi^z_{2^l}\rangle\}$ with $l \neq k-1$ and $0 \leq l \leq n-1$ of the Hilbert space $HS(N)$. Moreover, it can be found from (6.73) that the unitary operator $\exp\left(-iH^s_{k-1}\tau/\hbar\right)$ , which is generated by the component operator $H^s_{k-1}$ of the Hermitian operator $H^{ss}_{k,k-1}$, selectively acts on only the tensor-product subspace $\{|m_{k-1}\rangle ... |m_2\rangle |m_1\rangle\}$ , when the unitary operator $U^{ss}_{k,k-1}(\tau)$ selectively acts on the selected direct-sum subspace $\{|\Phi^z_{2^{k-1}}\rangle\}$.

Below consider a subspace-selective MQT unitary operator of the second kind. When a subspace-selective MQT unitary operator of the second kind acts on any direct-sum subspaces of the Hilbert space $HS(N)$, according to its definition, it really acts on only the two selected direct-sum subspaces and it does not affect any other direct-sum subspaces of the Hilbert space $HS(N)$. For simplicity, suppose that the two selected direct-sum subspaces are the two nearest-neighborhood direct-sum subspaces $\{|\Phi^z_{2^{k-2}}\rangle\}$ and $\{|\Phi^z_{2^{k-1}}\rangle\}$ with $0 \leq k-2 < k-1 < n$ in the direct-sum decomposition (6.16) of the Hilbert space $HS(N)$. Here for convenience both the selected subspaces $\left\{\left|\Phi_{2^{k-2}}^{z}\right\rangle\right\}$ and $\left\{\left|\Phi_{2^{k-1}}^{z}\right\rangle\right\}$ are called the small and the large selected subspace, respectively. Let $\left\{\left|\Phi_{2^{k-1}}^{z}\left(\left|0_{k-1}\right\rangle\right)\right\rangle\right\}$ be a subspace of the large selected direct-sum subspace $\left\{\left|\Phi_{2^{k-1}}^{z}\right\rangle\right\}$. It is defined by

$$\{|\Phi_{2^{k-1}}^{z}(|\mathbf{0}_{k-1}\rangle)\rangle\} = \left\{ \begin{array}{c} |0_n\rangle \ldots |0_{k+1}\rangle \bigotimes |\mathbf{1}_k\rangle |\mathbf{0}_{k-1}\rangle \\ \bigotimes |m_{k-2}\rangle \ldots |m_1\rangle \end{array} \right\} \tag{6.74}$$

where $|m_{k-1}\rangle = |1/2\rangle = |0_{k-1}\rangle$. It can be found from (6.74) and (6.69) that the small selected subspace $\left\{\left|\Phi_{2^{k-2}}^{z}\right\rangle\right\}$ has the same dimensional size $(2^{k-2})$ as the subspace $\left\{\left|\Phi_{2^{k-1}}^{z}\left(\left|0_{k-1}\right\rangle\right)\right\rangle\right\}$ of the large selected subspace $\left\{\left|\Phi_{2^{k-1}}^{z}\right\rangle\right\}$.

Now suppose that the unitary operator $V_{k,k-1}^{ss}(\theta)$ is a second-kind subspace-selective MQT unitary operator which selectively acts on only the two selected direct-sum subspaces $\left\{\left|\Phi_{2^{k-2}}^{z}\right\rangle\right\}$ and $\left\{\left|\Phi_{2^{k-1}}^{z}\right\rangle\right\}$ in the Hilbert space $HS(N)$ and moreover, it is generated by $V_{k,k-1}^{ss}(\theta) = \exp\left(-i\theta Q_{k,k-1}^{ss}\right)$, where $Q_{k,k-1}^{ss}$ is a subspace-selective MQT Hermitian operator. It can prove later that the subspace-selective MQT Hermitian operator $Q_{k,k-1}^{ss}$ may be explicitly written as

$$\begin{aligned} Q_{k,k-1}^{ss} &= (|0_n\rangle \ldots |0_{k+1}\rangle)(|0_n\rangle \ldots |0_{k+1}\rangle)^{+} \bigotimes Q_{zq,y}^{k,k-1} \\ &\quad \bigotimes E_{k-2} \bigotimes E_{k-3} \bigotimes \ldots \bigotimes E_1 \\ &= (|0_n\rangle \ldots |0_{k+1}\rangle)(|0_n\rangle \ldots |0_{k+1}\rangle)^{+} \bigotimes Q_{zq,y}^{k,k-1} \end{aligned} \tag{6.75}$$

where the Hermitian zero-order quantum transition operator $Q_{zq,y}^{k,k-1}$ with $1 \leq k-1 < k \leq n$ may be explicitly given by $Q_{zq,y}^{k,k-1} = \frac{1}{2i}\left(I_k^{+} I_{k-1}^{-} - I_k^{-} I_{k-1}^{+}\right)$ and $E_l$ is the unity operator of the $l-$th spin$-1/2$ particle in the $n - spin - 1/2$ system. The ZQT Hermitian operator $Q_{zq,y}^{k,k-1}$ selectively acts on only the tensor-product subspace $\{|m_k\rangle |m_{k-1}\rangle\}$ of the $two - spin - 1/2$ subsystem which consists of the $k-$th and $(k-1)-$th spin$-1/2$ particles of the $n - spin - 1/2$ system. It is known from the Subsection 3.2.1 and the Section A that the ZQT Hermitian operator $Q_{zq,y}^{k,k-1}$ also may be written as

$$Q_{zq,y}^{k,k-1} = \frac{1}{2i}\left(|0_k\rangle |1_{k-1}\rangle \langle 0_{k-1}| \langle 1_k| - |1_k\rangle |0_{k-1}\rangle \langle 1_{k-1}| \langle 0_k|\right), \tag{6.76}$$

since $I_j^{+} = |0_j\rangle \langle 1_j|$ and $I_j^{-} = |1_j\rangle \langle 0_j|$ for $j = k,\ k-1$. Therefore, the ZQT Hermitian operator $Q_{zq,y}^{k,k-1}$ in (6.75) contains the component base vectors $|1_k\rangle$ and $|1_{k-1}\rangle$ which belong to the $k-$th and the $(k-1)-$th spin$-1/2$ particle of the $n - spin - 1/2$ system, respectively.

The above theoretical analysis about the orthogonality between the selected direct-sum subspace $\left\{\left|\Phi_{2^{k-1}}^{z}\right\rangle\right\}$ and any larger direct-sum subspace or any smaller direct-sum subspace in the Hilbert space $HS(N)$ can be applied as well to the present case. It is known above that the component base vector $|1_k\rangle$ is responsible for controlling the orthogonality between the large selected subspace $\left\{\left|\Phi_{2^{k-1}}^{z}\right\rangle\right\}$ and any one of the smaller direct-sum subspaces $\{|\Phi_0^{z}\rangle\}$ and

$\{|\Phi_{2^l}^z\rangle\}$ with $0 \leq l < k-1$. Similarly, the component base vector $|1_{k-1}\rangle$ is responsible for controlling the orthogonality between the small selected subspace $\{|\Phi_{2^{k-2}}^z\rangle\}$ and any one of the smaller direct-sum subspaces $\{|\Phi_0^z\rangle\}$ and $\{|\Phi_{2^l}^z\rangle\}$ with $0 \leq l < k-2$. Both the component base vectors $|1_k\rangle$ and $|1_{k-1}\rangle$ appear in the ZQT Hermitian operator $Q_{zq,y}^{k,k-1}$, as shown in (6.76). Then they should be responsible for controlling the orthogonality between any one of the two selected subspaces $\{|\Phi_{2^{k-2}}^z\rangle\}$ and $\{|\Phi_{2^{k-1}}^z\rangle\}$ and any one of the smaller direct-sum subspaces $\{|\Phi_0^z\rangle\}$ and $\{|\Phi_{2^l}^z\rangle\}$ with $0 \leq l < k-2$. Obviously, the component tensor-product base vector $(|0_n\rangle ... |0_{k+1}\rangle)$ in the component tensor-product diagonal operator $(|0_n\rangle ... |0_{k+1}\rangle)(|0_n\rangle ... |0_{k+1}\rangle)^+$ of the Hermitian operator $Q_{k,k-1}^{ss}$ of (6.75) should be responsible for controlling the orthogonality between any one of the two selected subspaces $\{|\Phi_{2^{k-2}}^z\rangle\}$ and $\{|\Phi_{2^{k-1}}^z\rangle\}$ and any one of the larger subspaces $\{|\Phi_{2^l}^z\rangle\}$ with $k \leq l \leq n-1$. Therefore, when the subspace-selective MQT Hermitian operator $Q_{k,k-1}^{ss}$ of (6.75) acts on any direct-sum subspaces of the Hilbert space $HS(N)$, it can be found that

$$\left\{\begin{array}{l} Q_{k,k-1}^{ss}\{|\Phi_0^z\rangle\} = 0,\ Q_{k,k-1}^{ss}\{|\Phi_{2^l}^z\rangle\} = 0 \text{ for } 0 \leq l < k-2 \\ Q_{k,k-1}^{ss}\{|\Phi_{2^{k-2}}^z\rangle\} = \left\{\begin{array}{c} |0_n\rangle ... |0_{k+1}\rangle \bigotimes Q_{zq,y}^{k,k-1}(|0_k\rangle |1_{k-1}\rangle) \\ \bigotimes (|m_{k-2}\rangle ... |m_1\rangle) \end{array}\right\} \text{ for } l = k-2 \\ Q_{k,k-1}^{ss}\{|\Phi_{2^{k-1}}^z\rangle\} = \left\{\begin{array}{c} |0_n\rangle ... |0_{k+1}\rangle \bigotimes Q_{zq,y}^{k,k-1}(|1_k\rangle |m_{k-1}\rangle) \\ \bigotimes (|m_{k-2}\rangle ... |m_1\rangle) \end{array}\right\} \text{ for } l = k-1 \\ Q_{k,k-1}^{ss}\{|\Phi_{2^l}^z\rangle\} = 0 \text{ for } k \leq l \leq n-1 \end{array}\right. \tag{6.77}$$

These transformational equations can directly lead to that when the second-kind subspace-selective MQT unitary operator $V_{k,k-1}^{ss}(\theta) = \exp\left(-i\theta Q_{k,k-1}^{ss}\right)$ acts on any direct-sum subspaces of the Hilbert space $HS(N)$, one has

$$\left\{\begin{array}{l} V_{k,k-1}^{ss}(\theta)\{|\Phi_0^z\rangle\} = \{|\Phi_0^z\rangle\},\ V_{k,k-1}^{ss}(\theta)\{|\Phi_{2^l}^z\rangle\} = \{|\Phi_{2^l}^z\rangle\} \text{ for } 0 \leq l < k-2 \\ V_{k,k-1}^{ss}(\theta)\{|\Phi_{2^{k-2}}^z\rangle\} = \left\{\begin{array}{c} |0_n\rangle ... |0_{k+1}\rangle \bigotimes \exp\left(-i\theta Q_{zq,y}^{k,k-1}\right) \\ \times (|0_k\rangle |1_{k-1}\rangle) \bigotimes (|m_{k-2}\rangle ... |m_1\rangle) \end{array}\right\} \text{for } l = k-2 \\ V_{k,k-1}^{ss}(\theta)\{|\Phi_{2^{k-1}}^z\rangle\} = \left\{\begin{array}{c} |0_n\rangle ... |0_{k+1}\rangle \bigotimes \exp\left(-i\theta Q_{zq,y}^{k,k-1}\right) \\ \times (|1_k\rangle |m_{k-1}\rangle) \bigotimes (|m_{k-2}\rangle ... |m_1\rangle) \end{array}\right\} \text{for } l = k-1 \\ V_{k,k-1}^{ss}(\theta)\{|\Phi_{2^l}^z\rangle\} = \{|\Phi_{2^l}^z\rangle\} \text{ for } k \leq l \leq n-1 \end{array}\right. \tag{6.78}$$

These unitary transformations show clearly that the second-kind subspace-selective MQT unitary operator $V_{k,k-1}^{ss}(\theta)$ selectively acts on both the selected direct-sum subspaces $\{|\Phi_{2^{k-2}}^z\rangle\}$ and $\{|\Phi_{2^{k-1}}^z\rangle\}$ and it does not affect any other direct-sum subspaces of the Hilbert space $HS(N)$. Therefore, this clearly indicates that the subspace-selective MQT Hermitian operator $Q_{k,k-1}^{ss}$ of (6.75) is responsible for the subspace-selective property of the second-kind subspace-selective MQT unitary operator $V_{k,k-1}^{ss}(\theta)$.

The ZQT unitary operator $\exp\left(-i\theta Q_{zq,y}^{k,k-1}\right)$ in the unitary transformations of (6.78) is generated by the ZQT Hermitian operator $Q_{zq,y}^{k,k-1}$. The latter selectively acts on only the tensor-product subspace $\{|m_k\rangle |m_{k-1}\rangle\}$ of the *two−spin−1/2* subsystem which consists of the $k-$th and $(k-1)-$th spin−1/2 particles of the $n-spin-1/2$ system. Then it can prove that in the tensor-product subspace $\{|m_k\rangle|m_{k-1}\rangle\} = \{|0_k\rangle|0_{k-1}\rangle, |0_k\rangle|1_{k-1}\rangle, |1_k\rangle|0_{k-1}\rangle, |1_k\rangle|1_{k-1}\rangle\}$ there are the vector-basis unitary transformations which are induced by the ZQT unitary operator $\exp\left(-i\theta Q_{zq,y}^{k,k-1}\right)$ :

$$\exp\left(-i\theta Q_{zq,y}^{k,k-1}\right)|0_k\rangle|1_{k-1}\rangle = |0_k\rangle|1_{k-1}\rangle\cos\frac{1}{2}\theta + |1_k\rangle|0_{k-1}\rangle\sin\frac{1}{2}\theta \quad (6.79a)$$

$$\exp\left(-i\theta Q_{zq,y}^{k,k-1}\right)|1_k\rangle|0_{k-1}\rangle = |1_k\rangle|0_{k-1}\rangle\cos\frac{1}{2}\theta - |0_k\rangle|1_{k-1}\rangle\sin\frac{1}{2}\theta \quad (6.79b)$$

$$\exp\left(-i\theta Q_{zq,y}^{k,k-1}\right)|0_k\rangle|0_{k-1}\rangle = |0_k\rangle|0_{k-1}\rangle \quad (6.79c)$$

$$\exp\left(-i\theta Q_{zq,y}^{k,k-1}\right)|1_k\rangle|1_{k-1}\rangle = |1_k\rangle|1_{k-1}\rangle \quad (6.79d)$$

In particular, when $\theta = \pm\pi$, one has $\exp\left(\mp i\pi Q_{zq,y}^{k,k-1}\right)|0_k\rangle|1_{k-1}\rangle = \pm|1_k\rangle|0_{k-1}\rangle$ and $\exp\left(\mp i\pi Q_{zq,y}^{k,k-1}\right)|1_k\rangle|0_{k-1}\rangle = \mp|0_k\rangle|1_{k-1}\rangle$. These vector-basis unitary transformations of (6.79) are similar to those of (5.4) or (3.37).

With the help of the second-kind subspace-selective MQT unitary operator $V_{k,k-1}^{ss}(\theta)$ it can be shown that the small selected subspace $\left\{\left|\Phi_{2^{k-2}}^z\right\rangle\right\}$ can be changed to the subspace $\left\{\left|\Phi_{2^{k-1}}^z(|0_{k-1}\rangle)\right\rangle\right\}$ of (6.74) of the large selected subspace $\left\{\left|\Phi_{2^{k-1}}^z\right\rangle\right\}$ and vice versa. This can be seen as follows:

$$V_{k,k-1}^{ss}(\pi)\{|\Phi_{2^{k-2}}^z\rangle\} = \exp\left(-i\pi Q_{k,k-1}^{ss}\right)\left\{\begin{array}{c}|0_n\rangle...|0_{k+1}\rangle \\ \bigotimes|0_k\rangle|1_{k-1}\rangle\bigotimes|m_{k-2}\rangle...|m_1\rangle\end{array}\right\}$$

$$= \left\{|0_n\rangle...|0_{k+1}\rangle\bigotimes\exp\left(-i\pi Q_{zq,y}^{k,k-1}\right)(|0_k\rangle|1_{k-1}\rangle)\bigotimes|m_{k-2}\rangle...|m_1\rangle\right\}$$

$$= \{|0_n\rangle...|0_{k+1}\rangle\bigotimes(|\mathbf{1}_k\rangle|\mathbf{0}_{k-1}\rangle)\bigotimes|m_{k-2}\rangle...|m_1\rangle\}$$

$$\equiv \{|\Phi_{2^{k-1}}^z(|0_{k-1}\rangle)\rangle\} \quad (6.80)$$

where the second equality holds due to the subspace-selective property (See (6.78)) of the unitary operator $V_{k,k-1}^{ss}(\theta)$, the third equality holds due to the unitary transformation of (6.79a) with $\theta = \pi$, and the final equality is just (6.74). Conversely, by starting from the subspace $\left\{\left|\Phi_{2^{k-1}}^z(|0_{k-1}\rangle)\right\rangle\right\}$ it can be found that

$$V_{k,k-1}^{ss}(-\pi)\{|\Phi_{2^{k-1}}^z(|0_{k-1}\rangle)\rangle\}$$

$$= \exp\left(+i\pi Q_{k,k-1}^{ss}\right)\{|0_n\rangle...|0_{k+1}\rangle\bigotimes|1_k\rangle|0_{k-1}\rangle\bigotimes|m_{k-2}\rangle...|m_1\rangle\}$$

$$= \left\{|0_n\rangle...|0_{k+1}\rangle\bigotimes\exp\left(+i\pi Q_{zq,y}^{k,k-1}\right)(|1_k\rangle|0_{k-1}\rangle)\bigotimes|m_{k-2}\rangle...|m_1\rangle\right\}$$

$$= \{|0_n\rangle...|0_{k+1}\rangle\bigotimes|0_k\rangle|\mathbf{1}_{k-1}\rangle\bigotimes|m_{k-2}\rangle...|m_1\rangle\} \equiv \{|\Phi_{2^{k-2}}^z\rangle\} \quad (6.81)$$

where the second equality holds due to the subspace-selective property of the unitary operator $V_{k,k-1}^{ss}(\theta)$ and the third equality holds due to the unitary transformation of (6.79b) with $\theta = -\pi$.

Now a Hilbert-space-enlarging process may be explicitly constructed and implemented with the aid of the subspace-selective MQT unitary operators of the first kind ($\{U_{j,j-1}^{ss}(\tau)\}$) and the second kind ($\{V_{j,j-1}^{ss}(\theta)\}$). As a typical example, a simple Hilbert-space-enlarging process is constructed such that the initial small selected direct-sum subspace $\{|\Phi^z_{2^{k-2}}\rangle\}$ with dimensional size $2^{k-2}$ is selectively changed to the large direct-sum subspace $\{|\Phi^z_{2^{k-1}}\rangle\}$ with dimensional size $2^{k-1}$. Here the small selected subspace $\{|\Phi^z_{2^{k-2}}\rangle\}$ is given in (6.69), while the large selected subspace $\{|\Phi^z_{2^{k-1}}\rangle\}$ may be written as

$$\{|\Phi^z_{2^{k-1}}\rangle\} = \left\{ \begin{array}{c} |0_n\rangle ... |0_{k+1}\rangle \bigotimes |\mathbf{1}_k\rangle (\alpha_{k-1} |0_{k-1}\rangle + \beta_{k-1} |1_{k-1}\rangle) \\ \bigotimes |m_{k-2}\rangle ... |m_1\rangle \end{array} \right\} \quad (6.82)$$

where the coefficients $\alpha_{k-1}$ and $\beta_{k-1}$ may take any values but satisfy $|\alpha_{k-1}|^2 + |\beta_{k-1}|^2 = 1$. This simple Hilbert-space enlarging process may be constructed with the help of the subspace-selective MQT unitary operators of the first-kind ($\{U^{ss}_{j,j-1}(\tau)\}$) and the second-kind ($\{V^{ss}_{j,j-1}(\theta)\}$). It is performed in the two consecutive steps such that the small selected subspace $\{|\Phi^z_{2^{k-2}}\rangle\}$ is enlarged and changed selectively to the large selected subspace $\{|\Phi^z_{2^{k-1}}\rangle\}$ and in the meantime any other direct-sum subspaces of the Hilbert space $HS(N)$ are not affected during the Hilbert-space-enlarging process. The first step is to employ the second-kind subspace-selective MQT unitary operator $V^{ss}_{k,k-1}(\theta) = \exp\left(-i\theta Q^{ss}_{k,k-1}\right)$ to change the small selected subspace $\{|\Phi^z_{2^{k-2}}\rangle\}$ to the subspace $\{|\Phi^z_{2^{k-1}}(|0_{k-1}\rangle)\rangle\}$ of (6.74) of the large selected subspace $\{|\Phi^z_{2^{k-1}}\rangle\}$. This step can be described in detail by (6.80), that is, $V^{ss}_{k,k-1}(\pi)\{|\Phi^z_{2^{k-2}}\rangle\} = \{|\Phi^z_{2^{k-1}}(|0_{k-1}\rangle)\rangle\}$. The second step is to employ the first-kind subspace-selective MQT unitary operator $U^{ss}_{k,k-1}(\tau) = \exp\left(-iH^{ss}_{k,k-1}\tau/\hbar\right)$ to enlarge the subspace $\{|\Phi^z_{2^{k-1}}(|0_{k-1}\rangle)\rangle\}$ to the large selected subspace $\{|\Phi^z_{2^{k-1}}\rangle\}$ of (6.82), where the Hamiltonian operator $H^{ss}_{k,k-1}$ of (6.70) is given by

$$H^{ss}_{k,k-1} = (|0_n\rangle ... |0_{k+1}\rangle |\mathbf{1}_k\rangle)(|0_n\rangle ... |0_{k+1}\rangle |\mathbf{1}_k\rangle)^+ \bigotimes \hbar\omega_{k-1} I_{k-1y} \quad (6.83a)$$

and the component Hamiltonian $H^s_{k-1}$ in (6.70) is explicitly given by

$$H^s_{k-1} = \hbar\omega_{k-1} I_{k-1y} \bigotimes E_{k-2} \bigotimes ... \bigotimes E_1 \quad (6.83b)$$

Therefore, the second step may be described by

$$U^{ss}_{k,k-1}(\tau)\{|\Phi^z_{2^{k-1}}(|0_{k-1}\rangle)\rangle\}$$

$$= \left\{|0_n\rangle ... |0_{k+1}\rangle \bigotimes |1_k\rangle \bigotimes (\exp\left(-iH^s_{k-1}\tau/\hbar\right)|0_{k-1}\rangle) \bigotimes |m_{k-2}\rangle ... |m_1\rangle\right\}$$

$$= \left\{ \begin{array}{c} |0_n\rangle ... |0_{k+1}\rangle \bigotimes |1_k\rangle (\alpha_{k-1} |0_{k-1}\rangle + \beta_{k-1} |1_{k-1}\rangle) \\ \bigotimes |m_{k-2}\rangle ... |m_1\rangle \end{array} \right\} \equiv \{|\Phi^z_{2^{k-1}}\rangle\} \quad (6.84)$$

where $\exp\left(-iH^s_{k-1}\tau/\hbar\right) = \exp\left(-i\theta_{k-1} I_{k-1y}\right)$ with the angle $\theta_{k-1} = \omega_{k-1}\tau$ is a spin-selective unitary operator. The angle $\theta_{k-1}$ can be determined by

$$\exp\left(-i\theta_{k-1} I_{k-1y}\right)|0_{k-1}\rangle = |0_{k-1}\rangle \cos\frac{1}{2}\theta_{k-1} + |1_{k-1}\rangle \sin\frac{1}{2}\theta_{k-1}$$

$$= \alpha_{k-1} |0_{k-1}\rangle + \beta_{k-1} |1_{k-1}\rangle \quad (6.85)$$

Therefore, the angle $\theta_{k-1}$ is obtained from $\cos\frac{1}{2}\theta_{k-1} = \alpha_{k-1}$ and $\sin\frac{1}{2}\theta_{k-1} = \beta_{k-1}$, by noting that $|\alpha_{k-1}|^2 + |\beta_{k-1}|^2 = 1$ and here the coefficients $\alpha_{k-1}$ and $\beta_{k-1}$ are real.

In the above simple Hilbert-space-enlarging process the subspace-selective MQT unitary operator $U_{k,k-1}^{ss}(\tau)$ of the first kind selectively acts on only the direct-sum subspace $\{|\Phi_{2^{k-1}}^{z}\rangle\}$, while the subspace-selective MQT unitary operator $V_{k,k-1}^{ss}(\theta)$ of the second kind selectively acts on only both the direct-sum subspaces $\{|\Phi_{2^{k-2}}^{z}\rangle\}$ and $\{|\Phi_{2^{k-1}}^{z}\rangle\}$ of the Hilbert space $HS(N)$. Therefore, this simple Hilbert-space-enlarging process is subspace-selective.

Below it is shown that the Hilbert-space-enlarging processes in the Subsection 6.3 can be efficiently constructed and implemented with the subspace-selective MQT unitary operators of the first kind $(\{U_{j,j-1}^{ss}(\tau)\})$ and the second kind $(\{V_{j,j-1}^{ss}(\theta)\})$ in the multiple-quantum operator algebra space of the $n-spin-1/2$ system. It is known that the subspace-selective MQT unitary operators $\{U_{y}^{k_0,k_1}(\pi)\}$ and $\{U_{y}^{0,1}(\theta)\}$ are the basic building blocks to construct efficiently the Hilbert-space-enlarging processes in the Subsection 6.3 which include these three Hilbert-space-enlarging processes of (6.36), (6.41), and (6.53). Below it is shown that these subspace-selective MQT unitary operators $U_{y}^{0,1}(\pi/2)$ and $U_{y}^{l-1,l}(\pi)$ with $0 \leq l-1 < l \leq n-1$, which are used to construct efficiently these three Hilbert-space-enlarging processes of (6.36), (6.41), and (6.53), can be efficiently constructed and implemented by the subspace-selective MQT unitary operators of the first kind $(\{U_{j,j-1}^{ss}(\tau)\})$ and the second kind $(\{V_{j,j-1}^{ss}(\theta)\})$ in the multiple-quantum operator algebra space.

Consider first the simple subspace-selective MQT unitary operator $U_{y}^{0,1}(\pi)$ of (6.37b). It can change selectively the initial occupied direct-sum subspace $SP_0(1) = \{|1\rangle\}$ to its nearest direct-sum subspace $SP_1(2) = \{|2\rangle, |3\rangle\}$. It is used by the first stage of the simple Hilbert-space-enlarging process (6.36) and also may be used by the small-scale Hilbert-space-enlarging process (6.41). Here the subspace-selective unitary transformation of (6.37b) may be constructed and implemented efficiently by the subspace-selective MQT unitary operators of the first kind $(\{U_{j,j-1}^{ss}(\tau)\})$ and the second kind $(\{V_{j,j-1}^{ss}(\theta)\})$. It may be explicitly constructed and implemented via the following two steps. The first step is to apply the second-kind subspace-selective MQT unitary operator $V_{2,1}^{ss}(\pi) = \exp(-i\pi Q_{2,1}^{ss})$ to the initial selected subspace $SP_0(1) = \{|\Phi_1^z\rangle\}$ of (6.11) so that the initial selected subspace $\{|\Phi_1^z\rangle\}$ is changed to the subspace $\{|\Phi_{2^1}^z(|0_1\rangle)\rangle\}$ of the selected direct-sum subspace $SP_1(2) = \{|\Phi_{2^1}^z\rangle\}$ of (6.12). Here the subspace $\{|\Phi_{2^1}^z(|0_1\rangle)\rangle\}$ is defined by (6.74) with $k=2$. Therefore, the first step is described by (6.80) with $k=2$ and is explicitly expressed as

$$V_{2,1}^{ss}(\pi)\{|\Phi_1^z\rangle\} = \exp(-i\pi Q_{2,1}^{ss})\{|0_n\rangle ... |0_3\rangle \bigotimes |0_2\rangle |1_1\rangle\}$$

$$= \{|0_n\rangle ... |0_3\rangle \bigotimes \exp(-i\pi Q_{zq,y}^{2,1})(|0_2\rangle |1_1\rangle)\}$$

$$= \{|0_n\rangle ... |0_3\rangle \bigotimes |1_2\rangle |0_1\rangle\} \equiv \{|\Phi_{2^1}^z(|0_1\rangle)\rangle\} \quad (6.86)$$

Then the second step is to apply the first-kind subspace-selective MQT unitary operator $U_{2,1}^{ss}(\tau) = \exp\left(-iH_{2,1}^{ss}\tau/\hbar\right)$ to the subspace $\left\{\left|\Phi_{2^1}^{z}(|0_1\rangle)\right\rangle\right\}$ in (6.86) so that the subspace $\left\{\left|\Phi_{2^1}^{z}(|0_1\rangle)\right\rangle\right\}$ is enlarged and changed selectively to the selected subspace $\left\{\left|\Phi_{2^1}^{z}\right\rangle\right\}$ of (6.82). Therefore, the second step is described by (6.84) with $k=2$ and is explicitly expressed as

$$U_{2,1}^{ss}(\tau)\{|\Phi_{2^1}^{z}(|0_1\rangle)\rangle\} = \{|0_n\rangle ... |0_3\rangle \bigotimes |1_2\rangle \bigotimes (\exp(-iH_1^s\tau/\hbar)|0_1\rangle)\}$$
$$= \{|0_n\rangle ... |0_3\rangle \bigotimes |1_2\rangle (\alpha_1 |0_1\rangle + \beta_1 |1_1\rangle)\} \equiv \{|\Phi_{2^1}^{z}\rangle\} \quad (6.87)$$

where the real coefficients $\alpha_1$ and $\beta_1$ as well as the real parameter $\tau$ are determined later.

Now by performing the first step of (6.86) and the second step of (6.87) in the multiple-quantum operator algebra space the subspace-selective unitary transformation of (6.37b), i.e., $H_s^{(1)} = U_y^{0,1}(\pi) H_s^{(0)} \left(U_y^{0,1}(\pi)\right)^+$, may be efficiently implemented. This may be described in detail by

$$H_s^{(0)} = |1\rangle\langle 1| = (|0_n\rangle ... |0_3\rangle \bigotimes |0_2\rangle |1_1\rangle)\langle h.c.|$$
$$\xrightarrow{V_{2,1}^{ss}(\pi)} H_{ss}^{(1,1)} = (|0_n\rangle ... |0_3\rangle \bigotimes |1_2\rangle |0_1\rangle)\langle h.c.|$$
$$\xrightarrow{U_{2,1}^{ss}(\tau)} H_{ss}^{(1,2)} = (|0_n\rangle ... |0_3\rangle \bigotimes |1_2\rangle (\alpha_1 |0_1\rangle + \beta_1 |1_1\rangle))\langle h.c.|$$
$$= \left(\frac{1}{\sqrt{2}}|0_n\rangle ... |0_3\rangle \bigotimes |1_2\rangle |0_1\rangle + \frac{1}{\sqrt{2}}|0_n\rangle ... |0_3\rangle \bigotimes |1_2\rangle |1_1\rangle\right)\langle h.c.|$$
$$= \left(\frac{1}{\sqrt{2}}|2\rangle + \frac{1}{\sqrt{2}}|3\rangle\right)\langle h.c.| = H_s^{(1)}$$

where the real coefficients $\alpha_1$ and $\beta_1$ are given by $\alpha_1 = \beta_1 = 1/\sqrt{2}$. Then it can be deduced that the subspace-selective MQT unitary operator $U_y^{0,1}(\pi)$ may be efficiently constructed and implemented by

$$U_y^{0,1}(\pi) = U_{2,1}^{ss}(\tau) V_{2,1}^{ss}(\pi) \quad (6.88)$$

Below the subspace-selective MQT unitary operators of the first kind $(U_{2,1}^{ss}(\tau))$ and the second kind $(V_{2,1}^{ss}(\pi))$ are explicitly determined. The first-kind subspace-selective MQT unitary operator $U_{2,1}^{ss}(\tau)$ is given by $U_{2,1}^{ss}(\tau) = \exp\left(-iH_{2,1}^{ss}\tau/\hbar\right)$. Here the whole Hamiltonian $H_{2,1}^{ss}$ of (6.83a) with $k=2$ is explicitly given by

$$H_{2,1}^{ss} = (|0_n\rangle ... |0_3\rangle |1_2\rangle)(|0_n\rangle ... |0_3\rangle |1_2\rangle)^+ \bigotimes H_1^s \quad (6.89)$$

where the component Hamiltonian $H_1^s = \hbar\omega_1 I_{1y}$ in accordance with (6.83b). The unitary operator $\exp(-iH_1^s\tau/\hbar)$ which is generated by $H_1^s = \hbar\omega_1 I_{1y}$ may be written as $\exp(-iH_1^s\tau/\hbar) = \exp(-i\theta_1 I_{1y})$. Then according to (6.85) the angle $\theta_1$ can be determined by $\cos\frac{1}{2}\theta_1 = \alpha_1 = 1/\sqrt{2}$ and $\sin\frac{1}{2}\theta_1 = \beta_1 = 1/\sqrt{2}$. Therefore, the angle $\theta_1 = \pi/2$ and moreover, $\theta_1 = \omega_1\tau = \pi/2$. The second-kind subspace-selective MQT unitary operator $V_{2,1}^{ss}(\pi)$ is given by $V_{2,1}^{ss}(\pi) =$ $\exp\left(-i\pi Q_{2,1}^{ss}\right)$. Here the subspace-selective MQT Hermitian operator $Q_{2,1}^{ss}$ of (6.75) with $k=2$ is explicitly given by

$$Q_{2,1}^{ss} = \left(|0_n\rangle ... |0_3\rangle\right)\left(|0_n\rangle ... |0_3\rangle\right)^+ \bigotimes Q_{zq,y}^{2,1} \tag{6.90}$$

where the ZQT Hermitian operator $Q_{zq,y}^{2,1}$ is given by $Q_{zq,y}^{2,1} = \frac{1}{2i}\left(I_2^+ I_1^- - I_2^- I_1^+\right)$. The subspace-selective MQT unitary operator of (6.88) can change selectively the initial occupied subspace $\{|\Phi_1^{\tilde{z}}\rangle\}$ to the final occupied subspace $\{|\Phi_{2^1}^{\tilde{z}}\rangle\}$ in the Hilbert space $HS(N)$.

Then consider the subspace-selective MQT unitary operator $U_y^{0,1}(\pi/2)$ of (6.55b). It can enlarge selectively the initial subspace $SP_0(1)$ to the larger subspace $SP_0(1) \bigoplus SP_1(2)$ in the Hilbert space $HS(N)$. It is employed by the first stage of the large-scale Hilbert-space-enlarging process (6.53). The subspace-selective unitary transformation of (6.55b) may be constructed and implemented efficiently by the subspace-selective MQT unitary operators of the first kind $(\{U_{j,j-1}^{ss}(\tau)\})$ and the second kind $(\{V_{j,j-1}^{ss}(\theta)\})$. It is still explicitly constructed and implemented via the two steps as follows. The first step is to enlarge the initial selected subspace $SP_0(1) = \{|\Phi_1^{\tilde{z}}\rangle\}$ to the subspace $\{|\Phi_1^{\tilde{z}}\rangle\} \bigoplus \{|\Phi_{2^1}^{\tilde{z}}(|0_1\rangle)\rangle\}$ by applying the subspace-selective MQT unitary operator of the second kind $V_{2,1}^{ss}(\theta) = \exp\left(-i\theta Q_{2,1}^{ss}\right)$ with $\theta = \pi/2$ to the initial selected subspace $\{|\Phi_1^{\tilde{z}}\rangle\}$. According to the definition (6.74), here $\{|\Phi_{2^1}^{\tilde{z}}(|0_1\rangle)\rangle\}$ is the subspace of the selected subspace $SP_1(2) = \{|\Phi_{2^1}^{\tilde{z}}\rangle\}$ of (6.12). With the help of the vector-basis unitary transformation of (6.79a), which is induced by the ZQT unitary operator $\exp\left(-i\theta Q_{zq,y}^{2,1}\right)$ with $k=2$ and $\theta = \pi/2$, it can be found that the first step is described by

$$\begin{aligned}
V_{2,1}^{ss}(\pi/2)\{|\Phi_1^{\tilde{z}}\rangle\} &= \exp\left(-i\frac{\pi}{2}Q_{2,1}^{ss}\right)\{|0_n\rangle ... |0_3\rangle \bigotimes |0_2\rangle |1_1\rangle\} \\
&= \left\{|0_n\rangle ... |0_3\rangle \bigotimes \exp\left(-i\frac{\pi}{2}Q_{zq,y}^{2,1}\right)\left(|0_2\rangle |1_1\rangle\right)\right\} \\
&= \left\{|0_n\rangle ... |0_3\rangle \bigotimes \left(\frac{1}{\sqrt{2}}|0_2\rangle |1_1\rangle + \frac{1}{\sqrt{2}}|1_2\rangle |0_1\rangle\right)\right\} \\
&= \left\{\frac{1}{\sqrt{2}}|0_n\rangle ... |0_3\rangle \bigotimes |0_2\rangle |1_1\rangle + \frac{1}{\sqrt{2}}|0_n\rangle ... |0_3\rangle \bigotimes |1_2\rangle |0_1\rangle\right\}
\end{aligned} \tag{6.91}$$

where the first tensor-product base vector belongs to the subspace $\{|\Phi_1^{\tilde{z}}\rangle\}$ and the second one belongs to the subspace $\{|\Phi_{2^1}^{\tilde{z}}(|0_1\rangle)\rangle\}$, as shown by (6.74). Then the second step is that the subspace $\{|\Phi_{2^1}^{\tilde{z}}(|0_1\rangle)\rangle\}$ is enlarged and changed selectively to the final selected subspace $\{|\Phi_{2^1}^{\tilde{z}}\rangle\}$ by the first-kind subspace-selective MQT unitary operator $U_{2,1}^{ss}(\tau) = \exp\left(-iH_{2,1}^{ss}\tau/\hbar\right)$ with the angle $\theta_1 = \omega_1\tau = \pi/2$. Therefore, the second step is still described by (6.87) with the real coefficients $\alpha_1 = \beta_1 = 1/\sqrt{2}$. In the second step only the selected subspace $\{|\Phi_{2^1}^{\tilde{z}}\rangle\}$ is affected and the subspace $\{|\Phi_1^{\tilde{z}}\rangle\}$ is not affected by the first-kind subspace-selective MQT unitary operator $U_{2,1}^{ss}(\tau)$ due to the subspace-selective property of the unitary operator $U_{2,1}^{ss}(\tau)$.

Now by performing the first step of (6.91) and the second step of (6.87) in the multiple-quantum operator algebra space the subspace-selective unitary transformation of (6.55b), i.e., $H_s^{(1)} = U_y^{0,1}(\pi/2) H_s^{(0)} \left(U_y^{0,1}(\pi/2)\right)^+$, may be efficiently implemented, where the initial Hamiltonian $H_s^{(0)} = |1\rangle\langle 1|$. This subspace-selective unitary transformation may be implemented explicitly by

$$H_s^{(0)} = |1\rangle\langle 1| = (|0_n\rangle ... |0_3\rangle \bigotimes |0_2\rangle |1_1\rangle) \langle h.c.|$$

$$\xrightarrow{V_{2,1}^{ss}(\pi/2)} \begin{pmatrix} \frac{1}{\sqrt{2}} |0_n\rangle ... |0_3\rangle \bigotimes |0_2\rangle |1_1\rangle \\ +\frac{1}{\sqrt{2}} |0_n\rangle ... |0_3\rangle \bigotimes |1_2\rangle |0_1\rangle \end{pmatrix} \langle h.c.|$$

$$\xrightarrow{U_{2,1}^{ss}(\tau)} \begin{pmatrix} \frac{1}{\sqrt{2}} |0_n\rangle ... |0_3\rangle \bigotimes |0_2\rangle |1_1\rangle \\ +\frac{1}{\sqrt{2}} |0_n\rangle ... |0_3\rangle \bigotimes |1_2\rangle (\alpha_1 |0_1\rangle + \beta_1 |1_1\rangle) \end{pmatrix} \langle h.c.|$$

$$= \begin{pmatrix} \frac{1}{\sqrt{2}} |0_n\rangle ... |0_3\rangle \bigotimes |0_2\rangle |1_1\rangle \\ +\frac{1}{2} |0_n\rangle ... |0_3\rangle \bigotimes |1_2\rangle |0_1\rangle + \frac{1}{2} |0_n\rangle ... |0_3\rangle \bigotimes |1_2\rangle |1_1\rangle \end{pmatrix} \langle h.c.|$$

$$= \left(\frac{1}{\sqrt{2}} |1\rangle + \frac{1}{2} (|2\rangle + |3\rangle)\right) \langle h.c.| = H_s^{(1)}$$

where $\alpha_1 = \beta_1 = 1/\sqrt{2}$. It is just the subspace-selective unitary transformation (6.56) with $\theta_1 = \pi/2$ of the first stage (i.e., $SP_0(1) \rightarrow SP_0(1) \bigoplus SP_1(2)$) of the large-scale Hilbert-space-enlarging process (6.53). Therefore, it can be deduced that the subspace-selective MQT unitary operator $U_y^{0,1}(\pi/2)$ may be efficiently constructed and implemented by

$$U_y^{0,1}(\pi/2) = U_{2,1}^{ss}(\tau) V_{2,1}^{ss}(\pi/2) \tag{6.92}$$

Here the subspace-selective MQT unitary operator of the first kind $U_{2,1}^{ss}(\tau) = \exp\left(-iH_{2,1}^{ss}\tau/\hbar\right)$ is generated by the Hamiltonian $H_{2,1}^{ss}$ of (6.89). Moreover, the unitary operator $\exp(-iH_1^s\tau/\hbar)$ which is generated by the component Hamiltonian $H_1^s = \hbar\omega_1 I_{1y}$ in (6.89) is given by $\exp(-iH_1^s\tau/\hbar) = \exp(-i\theta_1 I_{1y})$ with $\theta_1 = \omega_1\tau = \pi/2$. The second-kind subspace-selective MQT unitary operator $V_{2,1}^{ss}(\pi/2) = \exp\left(-i\frac{\pi}{2}Q_{2,1}^{ss}\right)$ is generated by the ZQT Hermitian operator $Q_{2,1}^{ss}$ of (6.90). The subspace-selective MQT unitary operator of (6.92) can enlarge selectively the initial occupied subspace $\{|\Phi_1^{\tilde{z}}\rangle\}$ to the final occupied subspace $\{|\Phi_1^{\tilde{z}}\rangle\}\bigoplus\{|\Phi_{2^1}^{\tilde{z}}\rangle\}$ in the Hilbert space $HS(N)$.

Next consider the subspace-selective MQT unitary operator $U_y^{1,2}(\pi)$ of (6.38). It is used by the second stage of the Hilbert-space-enlarging process (6.36) and also by the other Hilbert-space-enlarging processes. It can change selectively the occupied direct-sum subspace $SP_1(2) = \{|2\rangle, |3\rangle\}$ to its nearest direct-sum subspace $SP_2(2^2) = \{|4\rangle, |5\rangle, |6\rangle, |7\rangle\}$. Here the subspace-selective unitary transformation of (6.38) may be efficiently constructed and implemented by the subspace-selective MQT unitary operators of the first kind $(\{U_{j,j-1}^{ss}(\tau)\})$ and the second kind ($\{V^{ss}_{j,j-1}(\theta)\}$). This subspace-selective unitary transformation, $H^{(2)}_s = U^{1,2}_y(\pi)\, H^{(1)}_s \left(U^{1,2}_y(\pi)\right)^+$, in the multiple-quantum operator algebra space may be implemented in detail by

$$H^{(1)}_s = \left(\frac{1}{\sqrt{2}}|2\rangle + \frac{1}{\sqrt{2}}|3\rangle\right)\langle h.c.|$$

$$= \left(|0_n\rangle \ldots |0_4\rangle \bigotimes |0_3\rangle |1_2\rangle \bigotimes \left(\frac{1}{\sqrt{2}}|0_1\rangle + \frac{1}{\sqrt{2}}|1_1\rangle\right)\right)\langle h.c.|$$

$$\xrightarrow{V^{ss}_{3,2}(\pi)} H^{(2,1)}_{ss} = \left(|0_n\rangle \ldots |0_4\rangle \bigotimes |1_3\rangle |0_2\rangle \bigotimes \left(\frac{1}{\sqrt{2}}|0_1\rangle + \frac{1}{\sqrt{2}}|1_1\rangle\right)\right)\langle h.c.|$$

$$\xrightarrow{U^{ss}_{3,2}(\tau)} H^{(2,2)}_{ss} = \left(\begin{array}{c} |0_n\rangle \ldots |0_4\rangle \bigotimes |1_3\rangle \left(\frac{1}{\sqrt{2}}|0_2\rangle + \frac{1}{\sqrt{2}}|1_2\rangle\right) \\ \bigotimes \left(\frac{1}{\sqrt{2}}|0_1\rangle + \frac{1}{\sqrt{2}}|1_1\rangle\right) \end{array}\right)\langle h.c.|$$

$$= \left(\begin{array}{c} \frac{1}{2}|0_n\rangle \ldots |0_4\rangle \bigotimes |1_3\rangle |0_2\rangle |0_1\rangle + \frac{1}{2}|0_n\rangle \ldots |0_4\rangle \bigotimes |1_3\rangle |0_2\rangle |1_1\rangle \\ +\frac{1}{2}|0_n\rangle \ldots |0_4\rangle \bigotimes |1_3\rangle |1_2\rangle |0_1\rangle + \frac{1}{2}|0_n\rangle \ldots |0_4\rangle \bigotimes |1_3\rangle |1_2\rangle |1_1\rangle \end{array}\right)\langle h.c.|$$

$$= \left(\frac{1}{2}\left(|4\rangle + |5\rangle + |6\rangle + |7\rangle\right)\right)\langle h.c.| = H^{(2)}_s$$

Then it can be deduced that the subspace-selective MQT unitary operator $U^{1,2}_y(\pi)$ of (6.38) may be efficiently constructed and implemented by

$$U^{1,2}_y(\pi) = U^{ss}_{3,2}(\tau)\, V^{ss}_{3,2}(\pi) \tag{6.93}$$

Here the subspace-selective MQT unitary operator of the first kind $U^{ss}_{3,2}(\tau) = \exp\left(-iH^{ss}_{3,2}\tau/\hbar\right)$ is generated by the Hamiltonian $H^{ss}_{3,2}$ of (6.83a) with $k = 3$,

$$H^{ss}_{3,2} = \left(|0_n\rangle \ldots |0_4\rangle |1_3\rangle\right)\left(|0_n\rangle \ldots |0_4\rangle |1_3\rangle\right)^+ \bigotimes H^s_2 \tag{6.94}$$

where the component Hamiltonian $H^s_2 = \hbar\omega_2 I_{2y}$ in accordance with (6.83b). Moreover, the unitary operator $\exp\left(-iH^s_2\tau/\hbar\right)$ which is generated by $H^s_2 = \hbar\omega_1 I_{2y}$ with the setting $\omega_2 = \omega_1$ is given by $\exp\left(-iH^s_2\tau/\hbar\right) = \exp\left(-i\theta_1 I_{2y}\right)$ with $\theta_1 = \omega_1\tau = \pi/2$. The second-kind subspace-selective MQT unitary operator $V^{ss}_{3,2}(\pi) = \exp\left(-i\pi Q^{ss}_{3,2}\right)$ is generated by the Hermitian operator $Q^{ss}_{3,2}$ of (6.75) with $k = 3$,

$$Q^{ss}_{3,2} = \left(|0_n\rangle \ldots |0_4\rangle\right)\left(|0_n\rangle \ldots |0_4\rangle\right)^+ \bigotimes Q^{3,2}_{zq,y}, \tag{6.95}$$

where the Hermitian ZQT operator $Q^{3,2}_{zq,y}$ is given by $Q^{3,2}_{zq,y} = \frac{1}{2i}\left(I^+_3 I^-_2 - I^-_3 I^+_2\right)$. The subspace-selective MQT unitary operator of (6.93) can change selectively the initial occupied subspace $\{|\Phi^z_{2^1}\rangle\}$ to the final occupied subspace $\{|\Phi^z_{2^2}\rangle\}$ in the Hilbert space $HS(N)$.

Consider generally the subspace-selective MQT unitary operator $U^{k-1,k}_y(\pi)$ of (6.39) for $k = 1, 2, .., n-1$. It is employed by the $k-$th stage of the Hilbert-space-enlarging process (6.36). It also is used by the other Hilbert-space-enlarging processes. It can change selectively the occupied direct-sum subspace $SP_{k-1}\left(2^{k-1}\right)$ to its nearest direct-sum subspace $SP_k\left(2^k\right)$. It is shown below that the subspace-selective unitary transformation of (6.39),

$$H_s^{(k-1)}=\left(\sum_{l=2^{k-1}}^{2^k-1}|l\rangle/\sqrt{2^{k-1}}\right)\langle h.c.|\xrightarrow{U_y^{k-1,k}(\pi)}H_s^{(k)}=\left(\sum_{l=2^k}^{2^{k+1}-1}|l\rangle/\sqrt{2^k}\right)\langle h.c.|$$

for $k=1,2,...,n-1$, can be efficiently constructed and implemented by the subspace-selective MQT unitary operators of the first kind ($\{U_{j,j-1}^{ss}(\tau)\}$) and the second kind ($\{V_{j,j-1}^{ss}(\theta)\}$). This subspace-selective unitary transformation $H_s^{(k)}=U_y^{k-1,k}(\pi)\,H_s^{(k-1)}\left(U_y^{k-1,k}(\pi)\right)^+$ in the multiple-quantum operator algebra space may be explicitly described by

$$H_s^{(k-1)}=\left(\frac{1}{\sqrt{2^{k-1}}}\sum_{l=2^{k-1}}^{2^k-1}|l\rangle\right)\langle h.c.|$$

$$=\left(|0_n\rangle\ldots|0_{k+2}\rangle\bigotimes|0_{k+1}\rangle|\mathbf{1}_k\rangle\bigotimes_{l=1}^{k-1}\left(\frac{1}{\sqrt{2}}|0_l\rangle+\frac{1}{\sqrt{2}}|1_l\rangle\right)\right)\langle h.c.|$$

$$\xrightarrow{V_{k+1,k}^{ss}(\pi)}H_{ss}^{(k,1)}=\left(\begin{array}{c}|0_n\rangle\ldots|0_{k+2}\rangle\bigotimes|\mathbf{1}_{k+1}\rangle|0_k\rangle\\ \bigotimes_{l=1}^{k-1}\left(\frac{1}{\sqrt{2}}|0_l\rangle+\frac{1}{\sqrt{2}}|1_l\rangle\right)\end{array}\right)\langle h.c.|$$

$$\xrightarrow{U_{k+1,k}^{ss}(\tau)}H_{ss}^{(k,2)}=\left(\begin{array}{c}|0_n\rangle\ldots|0_{k+2}\rangle\bigotimes|\mathbf{1}_{k+1}\rangle\left(\frac{1}{\sqrt{2}}|0_k\rangle+\frac{1}{\sqrt{2}}|1_k\rangle\right)\\ \bigotimes_{l=1}^{k-1}\left(\frac{1}{\sqrt{2}}|0_l\rangle+\frac{1}{\sqrt{2}}|1_l\rangle\right)\end{array}\right)\langle h.c.|$$

$$=\left(|0_n\rangle\ldots|0_{k+2}\rangle\bigotimes|\mathbf{1}_{k+1}\rangle\bigotimes_{l=1}^{k}\left(\frac{1}{\sqrt{2}}|0_l\rangle+\frac{1}{\sqrt{2}}|1_l\rangle\right)\right)\langle h.c.|$$

$$=\left(\frac{1}{\sqrt{2^k}}\sum_{l=2^k}^{2^{k+1}-1}|l\rangle\right)\langle h.c.|=H_s^{(k)}$$

Then it can be deduced that the subspace-selective MQT unitary operator $U_y^{k-1,k}(\pi)$ of (6.39) may be efficiently constructed and implemented by

$$U_y^{k-1,k}(\pi)=U_{k+1,k}^{ss}(\tau)\,V_{k+1,k}^{ss}(\pi)\ \text{ for } k=1,2,...,n-1 \tag{6.96}$$

Here the subspace-selective MQT unitary operator of the first kind $U_{k+1,k}^{ss}(\tau)=\exp\left(-iH_{k+1,k}^{ss}\tau/\hbar\right)$ is generated by the Hamiltonian $H_{k+1,k}^{ss}$ of (6.83a),

$$H_{k+1,k}^{ss}=\left(|0_n\rangle\ldots|0_{k+2}\rangle|\mathbf{1}_{k+1}\rangle\right)\left(|0_n\rangle\ldots|0_{k+2}\rangle|\mathbf{1}_{k+1}\rangle\right)^+\bigotimes H_k^s \tag{6.97}$$

where the component Hamiltonian $H_k^s=\hbar\omega_k I_{ky}$ in accordance with (6.83b). Moreover, the unitary operator $\exp(-iH_k^s\tau/\hbar)$ which is generated by $H_k^s=\hbar\omega_1 I_{ky}$ with the setting $\omega_k=\omega_1$ is given by $\exp(-iH_k^s\tau/\hbar)=\exp(-i\theta_1 I_{ky})$ with $\theta_1=\omega_1\tau=\pi/2$. The subspace-selective MQT unitary operator of the second kind $V^{ss}_{k+1,k}(\pi) = \exp\left(-i\pi Q^{ss}_{k+1,k}\right)$ is generated by the Hermitian ZQT operator $Q^{ss}_{k+1,k}$ of (6.75),

$$Q^{ss}_{k+1,k} = (|0_n\rangle ... |0_{k+2}\rangle)(|0_n\rangle ... |0_{k+2}\rangle)^+ \bigotimes Q^{k+1,k}_{zq,y}, \tag{6.98}$$

where $Q^{k+1,k}_{zq,y} = \frac{1}{2i}\left(I^+_{k+1}I^-_k - I^-_{k+1}I^+_k\right)$ is the Hermitian ZQT operator. The subspace-selective MQT unitary operator of (6.96) can enlarge selectively the initial occupied subspace $SP_{k-1}\left(2^{k-1}\right) = \{\left|\Phi^z_{2^{k-1}}\right\rangle\}$ with dimension $2^{k-1}$ to the final occupied subspace $SP_k\left(2^k\right) = \{\left|\Phi^z_{2^k}\right\rangle\}$ with dimension $2^k$ for $k = 1, 2, ..., n-1$.

It concludes from (6.88), (6.93), and (6.96) that the subspace-selective MQT unitary operators $U^{l-1,l}_y(\pi)$ with $0 \leq l-1 < l \leq n-1$ can be efficiently constructed and implemented with the subspace-selective MQT unitary operators of the first kind ($\{U^{ss}_{j,j-1}(\tau)\}$) and the second kind ($\{V^{ss}_{j,j-1}(\theta)\}$). It concludes from (6.92) that the subspace-selective MQT unitary operator $U^{0,1}_y(\pi/2)$ also can be efficiently constructed and implemented with $\{U^{ss}_{j,j-1}(\tau)\}$ and $\{V^{ss}_{j,j-1}(\theta)\}$. Consequently the Hilbert-space-enlarging processes of (6.36), (6.41), and (6.53) in the Subsection 6.3 can be efficiently constructed and implemented with the subspace-selective MQT unitary operators of the first kind ($\{U^{ss}_{j,j-1}(\tau)\}$) and the second kind ($\{V^{ss}_{j,j-1}(\theta)\}$).

The subspace-selective MQT unitary operators $\{U^{l-1,l}_y(\pi)\}$ and $U^{0,1}_y(\pi/2)$ which are explicitly constructed and implemented above are the special ones of more general subspace-selective MQT unitary operators $U^{k_0-1,k_1-1}_\lambda(\theta)$ with $0 \leq k_0-1 < k_1-1 < n$ and $\lambda = x, y, z$. The latter still may be efficiently constructed and implemented with the subspace-selective MQT unitary operators of the first kind ($\{U^{ss}_{j,l}(\tau)\}$) and the second kind ($\{V^{ss}_{j,l}(\theta)\}$) in the multiple-quantum operator algebra space. As a typical example, it can be shown that the subspace-selective MQT unitary operators $U^{k_0-1,k_1-1}_y(\pi)$ may be efficiently constructed and implemented by

$$U^{k_0-1,k_1-1}_y(\pi) = U^{ss}_{k_1,k_1-1}(\tau) ... U^{ss}_{k_1,k_0+1}(\tau) U^{ss}_{k_1,k_0}(\tau) V^{ss}_{k_1,k_0}(\pi) \tag{6.99}$$

Here the second-kind subspace-selective MQT unitary operator $V^{ss}_{k_1,k_0}(\theta) = \exp\left(-i\theta Q^{ss}_{k_1,k_0}\right)$ is generated by the subspace-selective ZQT Hermitian operator $Q^{ss}_{k_1,k_0}$ which may be written in the direct- (or tensor-) product form

$$\begin{aligned} Q^{ss}_{k_1,k_0} &= (|0_n\rangle ... |0_{k_1+1}\rangle |0_{k_1-1}\rangle ... |0_{k_0+1}\rangle) \\ &\times (|0_n\rangle ... |0_{k_1+1}\rangle |0_{k_1-1}\rangle ... |0_{k_0+1}\rangle)^+ Q^{k_1,k_0}_{zq,y} \end{aligned} \tag{6.100}$$

where $Q^{k_1,k_0}_{zq,y} = \frac{1}{2i}\left(I^+_{k_1}I^-_{k_0} - I^-_{k_1}I^+_{k_0}\right)$ is a Hermitian ZQT operator. (Note that here the Hilbert space $HS(N)$ is written in the tensor-product form (6.1), where the index $k$ of the component Hilbert space $H_k(2)$ runs from right to left and from 1 to $n$.) The Hermitian ZQT operator $Q^{k_1,k_0}_{zq,y}$ selectively acts on only the tensor-product subspace $\{|m_{k_1}\rangle |m_{k_0}\rangle\}$ of the $two-spin-1/2$ subsystem which consists of the $k_1-$th and $k_0-$th spin$-1/2$ particles of the $n-spin-1/2$ system. The second-kind subspace-selective MQT unitary operator $V_{k_1,k_0}^{ss}(\theta)$ with $\theta=\pi$ in (6.99) selectively acts on both the selected direct-sum subspaces $SP_{k_0-1}\left(2^{k_0-1}\right)=\left\{\left|\Phi_{2^{k_0-1}}^{z}\right\rangle\right\}$ and $SP_{k_1-1}\left(2^{k_1-1}\right)=\left\{\left|\Phi_{2^{k_1-1}}^{z}\right\rangle\right\}$ in the direct-sum decomposition (6.16) of the Hilbert space $HS(N)$. Note that both the selected direct-sum subspaces $\left\{\left|\Phi_{2^{k_0-1}}^{z}\right\rangle\right\}$ and $\left\{\left|\Phi_{2^{k_1-1}}^{z}\right\rangle\right\}$ may not be two nearest-neighborhood direct-sum subspaces in the direct-sum decomposition (6.16). The first-kind subspace-selective MQT unitary operator $U_{k_1,k_0+l}^{ss}(\tau)=\exp\left(-iH_{k_1,k_0+l}^{ss}\tau/\hbar\right)$ in (6.99) with $l=0,1,2,...$, and $1\leq k_0+l<k_1\leq n$ is generated by the subspace-selective MQT Hermitian operator $H_{k_1,k_0+l}^{ss}$ which may be written in the direct- (or tensor-) product form

$$H_{k_1,k_0+l}^{ss}=\left(|0_n\rangle\,...\,|0_{k_1+1}\rangle\,|\mathbf{1}_{k_1}\rangle\right)\left(|0_n\rangle\,...\,|0_{k_1+1}\rangle\,|\mathbf{1}_{k_1}\rangle\right)^{+}\bigotimes H_{k_0+l}^{s} \qquad (6.101)$$

where the component Hamiltonian $H_{k_0+l}^{s}=\hbar\omega_1 I_{k_0+ly}$ is explicitly given by

$$H_{k_0+l}^{s}=E_{k_1-1}\bigotimes...\bigotimes E_{k_0+l+1}\bigotimes\hbar\omega_1 I_{k_0+ly}\bigotimes E_{k_0+l-1}\bigotimes...\bigotimes E_1 \qquad (6.102)$$

Moreover, the spin-selective unitary operator $\exp\left(-iH_{k_0+l}^{s}\tau/\hbar\right)$ which is generated by $H_{k_0+l}^{s}=\hbar\omega_1 I_{k_0+ly}$ is given by $\exp\left(-iH_{k_0+l}^{s}\tau/\hbar\right)=\exp\left(-i\theta_1 I_{k_0+ly}\right)$ with $\theta_1=\omega_1\tau=\pi/2$. Obviously, it selectively acts on only the $(k_0+l)-$th spin$-1/2$ particle of the $n-spin-1/2$ system.

The subspace-selective MQT unitary operator $U_y^{k_0-1,k_1-1}(\pi)$ of (6.99) can convert the initial PDH Hamiltonian $H_s^{(k_0-1)}$ into the final PDH Hamiltonian $H_s^{(k_1-1)}$ by the subspace-selective unitary transformation in the multiple-quantum operator algebra space:

$$H_s^{(k_1-1)}=U_y^{k_0-1,k_1-1}(\pi)\,H_s^{(k_0-1)}\left(U_y^{k_0-1,k_1-1}(\pi)\right)^{+}$$

where the initial $H_s^{(k_0-1)}$ and the final PDH Hamiltonian $H_s^{(k_1-1)}$ are given by $H_s^{(k_0-1)}=\left(\sum_{l=2^{k_0-1}}^{2^{k_0}-1}|l\rangle/\sqrt{2^{k_0-1}}\right)\langle h.c.|$ and $H_s^{(k_1-1)}=\left(\sum_{l=2^{k_1-1}}^{2^{k_1}-1}|l\rangle/\sqrt{2^{k_1-1}}\right)$ $\langle h.c.|$, respectively. With the help of the subspace-selective MQT unitary operator $U_y^{k_0-1,k_1-1}(\pi)$ of (6.99) this subspace-selective unitary transformation in the multiple-quantum operator algebra space may be described in detail by

$$H_s^{(k_0-1)}=\left(\frac{1}{\sqrt{2^{k_0-1}}}\sum_{l=2^{k_0-1}}^{2^{k_0}-1}|l\rangle\right)\langle h.c.|$$

$$=\left(\begin{array}{c}|0_n\rangle\,...\,|0_{k_1+1}\rangle\bigotimes|\mathbf{0}_{k_1}\rangle\,|0_{k_1-1}\rangle\,...\,|0_{k_0+1}\rangle\bigotimes|\mathbf{1}_{k_0}\rangle\\ \bigotimes_{l=1}^{k_0-1}\left(\frac{1}{\sqrt{2}}|0_l\rangle+\frac{1}{\sqrt{2}}|1_l\rangle\right)\end{array}\right)\langle h.c.|$$

$$\xrightarrow{V_{k_1,k_0}^{ss}(\pi)} H_{ss}^{(k_1-1,1)}=\left(\begin{array}{c}|0_n\rangle\,...\,|0_{k_1+1}\rangle\bigotimes|\mathbf{1}_{k_1}\rangle\,|0_{k_1-1}\rangle\,...\,|0_{k_0+1}\rangle\,|\mathbf{0}_{k_0}\rangle\\ \bigotimes_{l=1}^{k_0-1}\left(\frac{1}{\sqrt{2}}|0_l\rangle+\frac{1}{\sqrt{2}}|1_l\rangle\right)\end{array}\right)\langle h.c.|$$

$$\xrightarrow{U_{k_1,k_0}^{ss}(\tau)} H_{ss}^{(k_1-1,2)}=\left(\begin{array}{c}|0_n\rangle\,...\,|0_{k_1+1}\rangle\bigotimes|\mathbf{1}_{k_1}\rangle\,|0_{k_1-1}\rangle\,...\,|\mathbf{0}_{k_0+1}\rangle\\ \bigotimes_{l=1}^{k_0}\left(\frac{1}{\sqrt{2}}|0_l\rangle+\frac{1}{\sqrt{2}}|1_l\rangle\right)\end{array}\right)\langle h.c.|$$

$$\xrightarrow{U^{ss}_{k_1,k_0+1}(\tau)} H^{(k_1-1,3)}_{ss} = \begin{pmatrix} |0_n\rangle \ldots |0_{k_1+1}\rangle \bigotimes |\mathbf{1}_{k_1}\rangle\, |0_{k_1-1}\rangle \ldots |\mathbf{0}_{k_0+2}\rangle \\ \bigotimes_{l=1}^{k_0+1}\left(\frac{1}{\sqrt{2}}|0_l\rangle + \frac{1}{\sqrt{2}}|1_l\rangle\right) \end{pmatrix} \langle h.c.|$$

$$\ldots\ldots$$

$$\xrightarrow{U^{ss}_{k_1,k_1-1}(\tau)} H^{(k_1-1,k_1-k_0+1)}_{ss} = \begin{pmatrix} |0_n\rangle \ldots |0_{k_1+1}\rangle \bigotimes |\mathbf{1}_{k_1}\rangle \\ \bigotimes_{l=1}^{k_1-1}\left(\frac{1}{\sqrt{2}}|0_l\rangle + \frac{1}{\sqrt{2}}|1_l\rangle\right) \end{pmatrix} \langle h.c.|$$

$$= \left(\frac{1}{\sqrt{2^{k_1-1}}}\sum_{l=2^{k_1-1}}^{2^{k_1}-1}|l\rangle\right)\langle h.c.| = H^{(k_1-1)}_{s}$$

Note that $\sum_{l=2^{k_0-1}}^{2^{k_0}-1}|l\rangle/\sqrt{2^{k_0-1}} \in SP_{k_0-1}\left(2^{k_0-1}\right)$ and $\sum_{l=2^{k_1-1}}^{2^{k_1}-1}|l\rangle/\sqrt{2^{k_1-1}} \in SP_{k_1-1}\left(2^{k_1-1}\right)$. This subspace-selective unitary transformation shows clearly that the subspace-selective MQT unitary operator $U_y^{k_0-1,k_1-1}(\pi)$ of (6.99) can enlarge and change selectively the small selected subspace $SP_{k_0-1}\left(2^{k_0-1}\right)$ to the large selected subspace $SP_{k_1-1}\left(2^{k_1-1}\right)$ or the small occupied subspace $\{|\Phi^z_{2^{k_0-1}}\rangle\}$ with dimension $2^{k_0-1}$ to the large occupied subspace $\{|\Phi^z_{2^{k_1-1}}\rangle\}$ with dimension $2^{k_1-1}$. Therefore, after the subspace-selective unitary transformation in the multiple-quantum operator algebra space, a larger dimensional size (i.e., $2^{k_1-1}$) is achieved for the occupied subspace (i.e., $\{|\Phi^z_{2^{k_1-1}}\rangle\}$) in the Hilbert space $HS(N)$.

# 7. Discussion

According to the quantum-computing speedup theory [Ref[1]] the symmetrical structures and properties of quantum systems are the fundamental quantum-computing-speedup resources which are responsible for speeding up exponentially quantum computing and quantum simulating. The symmetrical structures and properties of a quantum system are the fundamental and inherent attributes of the quantum system in quantum mechanics [2,17]. Therefore, the fundamental quantum-computing-speedup resources are the fundamental and inherent attributes of the quantum system. The symmetric structures and properties of a quantum system may be characterized through these different kinds of basic quantum spaces: the multiple-quantum operator algebra space (or the Liouville operator algebra space), the density operator space, and/or the Hilbert space of the quantum system and therefore, the fundamental quantum-computing-speedup resources of the quantum system may exist in these different kinds of basic quantum spaces of the quantum system. Then one big and important problem that needs to be solved urgently in the quantum-computing speedup theory is how to concretely make use of these fundamental quantum-computing-speedup resources to speed up essentially quantum computing and quantum simulating. In this paper the author makes a great effort toward solving this important problem theoretically. The multiple-quantum operator algebra space [4] must be positioned in the central place where one makes use of the fundamental quantum-computing-speedup resources to speed up essentially quantum computing and quantum simulating. Then the fundamental quantum-computing-speedup resources which are original from the corresponding Hilbert space of the quantum system must be considered explicitly, when quantum-computing speedup is realized in the multiple-quantum operator algebra space of the quantum system. In this paper the subspace-selective unitary manipulation [7] aims to harness the fundamental quantum-computing-speedup resources which are original from the corresponding Hilbert space to essentially speed up quantum simulating and quantum computing in the multiple-quantum operator algebra space of a quantum system. Theoretically it is based on the symmetrical structures and properties of the quantum system which are original from the Hilbert space, where the symmetrical structures of the Hilbert space may be characterized via the direct-sum decomposition of the Hilbert space. The subspace-selective unitary manipulation plays a key role in realizing concretely an essential quantum-computing speedup in quantum computing and quantum simulating. One of the most important applications of the subspace-selective unitary manipulation is that it can be used to realize concretely and efficiently the search-space dynamical reduction of the $HSSS$ unstructured quantum search algorithm (See [Ref[1]] and [Ref[2]]) in the multiple-quantum operator algebra space.

Recognize that the multiple-quantum operator algebra space is the central place where the fundamental quantum-computing-speedup resources are exploited to achieve essential quantum-computing speedup. Then it becomes necessary to investigate and make use of the multiple quantum transitions and the multiple-quantum-transition operators of quantum spin systems in quantum computing and quantum simulating. Moreover, harnessing the fundamental quantum-computing-speedup resources of quantum spin systems to essentially speed up quantum computing and quantum simulating in the multiple-quantum operator algebra spaces needs to employ a variety of the multiple-quantum transition operators of the quantum spin systems. Here the symmetric structures of the multiple-quantum operator algebra spaces are characterized by the multiple-quantum transition operators. Then the properties of the multiple-quantum transition operators are also very important. One important instance is that the spin Hamiltonian of a quantum spin system may be expanded in terms of the multiple-quantum transition operators which can be classified according to different quantum-transition orders in the multiple-quantum operator algebra space (See Ref.[4], [Ref[3]], and the Subsection 5.2). Another important instance is that a direct-sum subspace of the Hilbert space under study may be invariant under the operator set of the specific subspace-selective multiple-quantum transition operators. It is described as follows. Given a direct-sum decomposition of the Hilbert space, as shown in the Subsection 6.4, any subspace-selective MQT unitary operator of the first kind can affect only one selected direct-sum subspace and does not affect any other direct-sum subspaces of the Hilbert space, when it acts on any direct-sum subspaces of the Hilbert space. Here every subspace-selective MQT unitary (exponential) operator of the first kind may be generated by a subspace-selective MQT Hermitian operator. These subspace-selective MQT Hermitian operators may constitute an operator set. Then every direct-sum subspace of the Hilbert space in the given direct-sum decomposition is invariant under the operator set. As shown in the Subsection 6.1, there may be two different direct-sum decompositions for the same Hilbert space $HS(N)$ of an $n-$spin$-1/2$ system, one is the direct-sum decomposition of (6.2) and another is the one of (6.6). For the direct-sum decomposition (6.6) the first-kind subspace-selective MQT unitary operators are the subspace-selective MQT unitary operators $U_{k,k-1}^{ss}(\tau) = \exp\left(-iH_{k,k-1}^{ss}\tau/\hbar\right)$ $(1 \leq k \leq n)$ and $U_{00}^{ss}(\tau) = \exp(-iH_{00}^{ss}\tau/\hbar)$ and here the subspace-selective MQT Hermitian operators $H_{k,k-1}^{ss}$ are given by (6.70) for $k = 1, 2, ..., n$ and $H_{00}^{ss} = |0\rangle\langle 0|$ (See the detailed theoretical analysis in the Subsections 6.4 and 6.1.) These subspace-selective MQT Hermitian operators may form an operator set $\{H_{00}^{ss}, H_{k,k-1}^{ss}\}$. Then every direct-sum subspace $(HS(1),\ SP_k\left(2^k\right))$ of the Hilbert space $HS(N)$ in the direct-sum decomposition (6.6) is invariant under the operator set $\{H_{00}^{ss}, H_{k,k-1}^{ss}\}$. For the direct-sum decomposition (6.2) the first-kind subspace-selective MQT unitary operators are the subspace-selective zero-quantum-transition (ZQT) unitary operators (See Ref.[7] and the Subsection 5.1.3.) The latter may be generated by the subspace-selective ZQT Hermitian operators, respectively. These subspace-selective ZQT Hermitian operators may form an operator set. Then every direct-sum subspace $S_{zq}(k)$ of the Hilbert space $HS(N)$ with $k = 0, 1, ..., n$ in the direct-sum decomposition (6.2) is invariant under the operator set of the subspace-selective ZQT Hermitian operators. This is a straightforward consequence of the fact that any direct-sum subspace $S_{zq}(k)$ is invariant under the zero-quantum $(zq)$ operator subspace (See the theoretical analysis in the Subsection 5.1.3 and the Section A.)

When the fundamental quantum-computing-speedup resources are exploited to speed up essentially quantum computing and quantum simulating, those resources which are original from the symmetrical structures and properties of the corresponding Hilbert spaces need to be explicitly taken into account in the multiple-quantum operator algebra spaces. Quantum-mechanically the multiple-quantum operator algebra space is a linear operator space, while the Hilbert space is a quantum-state space. However, in quantum mechanics there exists a general connection between the Hilbert space and its corresponding multiple-quantum operator algebra space. According to quantum mechanics any linear operators are defined on the Hilbert space of a quantum system which is also a linear vector space. Then the linear operator space of a quantum system such as the multiple-quantum operator algebra space may naturally connect to its corresponding Hilbert space on which any linear operators of the linear operator space are defined. Consequently theoretically the symmetrical structures and properties of the Hilbert space may connect to the counterpart of the corresponding multiple-quantum operator algebra space. Then theoretically it becomes possible to make use of the fundamental quantum-computing-speedup resources original from the corresponding Hilbert space to speed up essentially quantum computing and quantum simulating in the multiple-quantum operator algebra space. Of course, a multiple-quantum operator algebra space is independent upon its corresponding Hilbert space. Generally, the former is far more complicated than the latter.

In order to consider concretely the symmetrical structures and properties of the Hilbert space in the multiple-quantum operator algebra space, in this paper two important and different kinds of the Hermitian operators of the multiple-quantum operator algebra space are proposed to act as bridge to connect the Hilbert space to the corresponding multiple-quantum operator algebra space in the aspect of the symmetrical structures and properties. They are the pseudo-diagonal Hermitian (PDH) operators and the subspace-selective multiple-quantum-transition (MQT) Hermitian operators, respectively. Correspondingly they can generate the two different types of the unitary exponential operators, i.e., the PDHO-generated unitary operators and the subspace-selective MQT unitary operators, respectively. With the help of the PDH operators or the subspace-selective MQT Hermitian operators the symmetric structures and properties of a quantum system which are original from the Hilbert space may be easily considered explicitly in the multiple-quantum operator algebra space. In this paper these two types of the Hermitian operators and their generated unitary operators (or propagators) of the quantum spin systems such as the $n - spin - 1/2$ systems are described and investigated in detail. They are very helpful to realize concretely and efficiently the subspace-selective unitary manipulation in the multiple-quantum operator algebra space. The subspace-selective unitary manipulation therefore can be conveniently used to realize concretely and efficiently the search-space dynamical reduction of the $HSSS$ unstructured quantum search algorithm (See [Ref[1]] and [Ref[2]]) in the multiple-quantum operator algebra space.

The Hilbert-space-enlarging processes are deliberately designed to make use of the fundamental quantum-computing-speedup resources original from the Hilbert space to realize essential quantum-computing speedup in the multiple-quantum operator algebra space. The Hilbert-space-enlarging processes and their inverses (i.e., the Hilbert-space-shrinking processes) constitute the specific subspace-selective unitary manipulation in the multiple-quantum operator algebra space. They are based on the symmetrical structures and properties of quantum systems which are original from the Hilbert spaces and especially they are closely related to the direct-sum subspaces of the Hilbert spaces. Here the symmetrical structures of the Hilbert spaces are characterized via the direct-sum decomposition of the Hilbert spaces. A general Hilbert-space-enlarging process selectively changes a small occupied Hilbert subspace to a large one or it changes the occupied Hilbert subspace from a small dimensional size to a large one. It may be expressed as a sequence of the basic building blocks which may be chosen as the subspace-selective MQT unitary operators or the PDHO-generated unitary operators. In this paper the Hilbert-space-enlarging processes are described and have been investigated in detail in the $n - spin - 1/2$ systems. It is shown that the subspace-selective MQT unitary operators may be used conveniently as the basic building blocks to construct and implement efficiently the Hilbert-space-enlarging processes on the basis of the specific symmetrical structure of the Hilbert space $HS(N)$ of the $n - spin - 1/2$ system. And in the meantime the pseudo-diagonal Hermitian operators may act as the initial Hamiltonian operators (or the dynamical variables) of the Hilbert-space-enlarging processes which are performed in the multiple-quantum operator algebra space. Several important Hilbert-space-enlarging processes of the $n - spin - 1/2$ system are explicitly constructed. It can prove that they can be efficiently implemented. It is expected that they have important applications in future.

The Hilbert-space-enlarging processes may be used to harness the fundamental quantum-computing-speedup resources original from the Hilbert space to achieve essential quantum-computing speedup via the way that the Hilbert-space-enlarging processes and their inverses (i.e., the Hilbert-space-shrinking processes) can selectively change at will the occupied direct-sum subspaces of the Hilbert space in the multiple-quantum operator algebra space. This is an important application for the Hilbert-space-enlarging processes and their inverses. Here the key point is that the dimensional sizes of the occupied direct-sum subspaces of the Hilbert space can be adjusted at will by the suitable Hilbert-space-enlarging processes and their inverses. Rather than the dimensional size of the whole Hilbert space, the dimensional sizes of the direct-sum subspaces of the Hilbert space are related to the symmetrical structures of the Hilbert space. These dimensional sizes of subspaces constitute an important physical quantity [7] and also one important aspect to reflect the symmetrical structures and properties of the quantum system under study. Moreover, the dimensional size of any direct-sum subspace is a natural number and has an infinitely high precision. Since the Hilbert-space-enlarging process and its inverse can selectively change at will the occupied direct-sum subspace of the Hilbert space and are able to adjust at will the dimensional size of the occupied direct-sum subspace in the multiple-quantum operator algebra space, it becomes possible that the inverse of dimensional size of the occupied direct-sum subspace can really act as a discrete variable which owns an infinite-high precision and can take an extremely small discrete value which corresponds to the exponentially large dimensional size of the occupied direct-sum subspace. These dimensional sizes of the occupied direct-sum subspaces of the Hilbert space and especially their inverse (i.e., the discrete variable) may be considered as the important resource which may be exploited to achieve essential quantum-computing speedup in quantum computing and quantum simulating. The resource might be exploited to speed up essentially the programmable quantum simulating for the unitary time-evolutional processes [8].

As important resource used for quantum-computing speedup the discrete variable (i.e., the inverse of dimensional size of the occupied direct-sum subspace) and the dimensional sizes of the direct-sum subspaces of the Hilbert space may have important applications in quantum simulating and quantum computing in future. As an example, below consider the subspace-selective unitary manipulation in the multiple-quantum operator algebra space. Suppose in the subspace-selective unitary manipulation that the initial spin Hamiltonian $\hat{H}_{N_0}^{(i)}$ (representing a dynamical variable) of an $n - spin - 1/2$ system in the multiple-quantum operator algebra space is explicitly given by

$$\hat{H}_{N_0}^{(i)} = \left( |\Phi\rangle_a + \left(1 - \frac{1}{N_0}\right) |0\rangle_b - \frac{1}{N_0} \sum_{k=1}^{N_0-1} |k\rangle_b \right) \langle h.c.|$$

where the subscripts $a$ and $b$ in the vectors $|\Phi\rangle_a$ and $|k\rangle_b$ are used to label the occupied direct-sum subspaces $A$ and $B$, respectively. The dimensional size $N_0$ of the occupied direct-sum subspace $B$ in the Hilbert space $HS(N)$ of the $n-spin-1/2$ system is given by $N_0 = 2^{n_0}$ with $1 \leq n_0 < n$ and the inverse of the dimensional size $N_0$ is given by $x = 1/N_0$. Here the important point is that the inverse of the dimensional size $N_0$, i.e., $x = 1/N_0$, may act as a discrete variable. The discrete variable $x$ usually takes a value much smaller than one, i.e., $0 < x = 1/N_0 << 1$. The initial Hamiltonian $\hat{H}_{N_0}^{(i)}$ involves only two occupied direct-sum subspaces $A$ and $B$ in the Hilbert space $HS(N)$, where the term $|\Phi\rangle_a$ is fixed and belongs to the direct-sum subspace $A$, and those terms with the base vectors $\{|k\rangle_b\}$ for $0 \leq k < N_0$ belong to the direct-sum subspace $B$. In the subspace $A$ the term $|\Phi\rangle_a$ could hide some variable factors. In the subspace $B$ the term $(1-1/N_0)\,|0\rangle_b$ with the base vector $|0\rangle_b$ is the desired term whose amplitude coefficient $1-1/N_0 = 1-x$ is close to one, while any other terms $\{-1/N_0\,|k\rangle_b\}$ with $1 \leq k < N_0$ are the residual terms whose amplitude coefficients each are $-x = -1/N_0$ and moreover, $0 < x << 1$. Therefore, the term $(1-1/N_0)\,|0\rangle_b$ is the main term, while the sum term $-\frac{1}{N_0}\sum_{k=1}^{N_0-1}|k\rangle_b$ is the total residual term in the initial Hamiltonian $\hat{H}_{N_0}^{(i)}$.

Now by performing sequentially the specific subspace-selective unitary manipulation and especially the Hilbert-space-enlarging processes (and/or their inverses) in the multiple-quantum operator algebra space the initial Hamiltonian $\hat{H}_{N_0}^{(i)}$ may be unitarily changed to the final Hamiltonian $\hat{H}_N^{(f)}$ which, as a simple instance, may be explicitly written as

$$\hat{H}_N^{(f)} = \begin{pmatrix} |\Phi\rangle_a + \left(1 - \frac{1}{2N_0} - \frac{1}{2N_0^2}\right)|0\rangle_b \\ -\frac{1}{2}\sqrt{3}\left(\frac{1}{N_0} - \frac{1}{N_0^2}\right)|1\rangle_b - \frac{1}{\sqrt{2}}\frac{1}{N_0^2}\sum\limits_{k=2}^{2N_0-1}|k\rangle_b \end{pmatrix}\langle h.c.|$$

Here the term $|\Phi\rangle_a$ of the subspace $A$ is fixed and hence the subspace $A$ need not be further considered in the following discussion. It can be seen that both the initial $\hat{H}_N^{(i)}$ and the final Hamiltonian $\hat{H}_N^{(f)}$ are the PDH operators of (4.4) in the multiple-quantum operator algebra space. It can be found that in both the initial $\hat{H}_{N_0}^{(i)}$ and the final Hamiltonian $\hat{H}_N^{(f)}$ these nonzero amplitude coefficients of the base vectors $\{|k\rangle_b\}$ with $0 \leq k < 2N_0$ are the polynomials in the discrete variable $x = 1/N_0$. However, the final Hamiltonian $\hat{H}_N^{(f)}$ is different from the initial Hamiltonian $\hat{H}_{N_0}^{(i)}$ in $(i)$ the amplitude coefficients of the desired term with the base vector $|0\rangle_b$ and $(ii)$ the residual terms with the base vectors $\{|k\rangle_b\}$. Actually, before the final Hamiltonian $\hat{H}_N^{(f)}$ is generated, the initial Hamiltonian $\hat{H}_{N_0}^{(i)}$ may be unitarily changed to another Hamiltonian $\hat{H}_{2N_0}^{(i)}$ by a small-scale Hilbert-space-enlarging process $(U_{hse}^b)$ (See the Subsection 6.3.2) in the multiple-quantum operator algebra space:

$$\hat{H}_{2N_0}^{(i)} = U_{hse}^b \hat{H}_{N_0}^{(i)} \left(U_{hse}^b\right)^+$$

$$= \left( |\Phi\rangle_a + \left(1 - \frac{1}{N_0}\right) |0\rangle_b - 0 \times |1\rangle_b - \frac{1}{\sqrt{2}} \frac{1}{N_0} \sum_{k=2}^{2N_0-1} |k\rangle_b \right) \langle h.c.|$$

where the small-scale Hilbert-space-enlarging process ($U_{hse}^b$) affects selectively the direct-sum subspace $B$ only and does not affect the other subspace $A$ and therefore, it is subspace-selective. The small-scale Hilbert-space-enlarging process ($U_{hse}^b$) enlarges selectively the occupied direct-sum subspace $B$ from the initial dimensional size $N_0$ to a larger dimensional size $2N_0$. The initial Hamiltonian $\hat{H}_{2N_0}^{(i)}$ may be further changed unitarily to the final Hamiltonian $\hat{H}_N^{(f)}$ by the specific subspace-selective unitary manipulation (named the subspace-selective amplitude decreasing process in unpublished work).

Now by comparing the final Hamiltonian $\hat{H}_N^{(f)}$ with the initial Hamiltonian $\hat{H}_{N_0}^{(i)}$ or $\hat{H}_{2N_0}^{(i)}$ it can be found that the subspace-selective unitary manipulation that changes the initial Hamiltonian $\hat{H}_{N_0}^{(i)}$ first to the Hamiltonian $\hat{H}_{2N_0}^{(i)}$ and then to the final Hamiltonian $\hat{H}_N^{(f)}$ in the multiple-quantum operator algebra space results in that

($i$) The amplitude coefficient of the desired term with the base vector $|0\rangle_b$ is increased in magnitude from the initial amplitude coefficient $\alpha = 1 - 1/N_0$ in the initial Hamiltonian $\hat{H}_{N_0}^{(i)}$ or $\hat{H}_{2N_0}^{(i)}$ to the final amplitude coefficient $\alpha_1 = 1 - \frac{1}{2N_0} - \frac{1}{2N_0^2}$ in the final Hamiltonian $\hat{H}_N^{(f)}$. More importantly, the final amplitude coefficient $\alpha_1$ of the desired term $\alpha_1 |0\rangle_b$ in the final Hamiltonian $\hat{H}_N^{(f)}$ contains naturally the first two lower-order terms (i.e., $1 - \frac{1}{2}N_0^{-1}$) which are desired theoretically in the finite power series in the discrete variable $x = 1/N_0$ of the amplitude coefficient $\alpha_1$. This is one main achievement obtained by the present subspace-selective unitary manipulation. Here the theoretical desired term is defined as $y|0\rangle_b$ whose amplitude coefficient $y = \sqrt{1-x}$ is the power series in the variable $x$ :

$$y(x) = \sqrt{1-x} = 1 - \frac{1}{2}x - \frac{1}{8}x^2 - \frac{1}{16}x^3 - \frac{5}{128}x^4 - \frac{7}{256}x^5 + ...$$

where the variable $x$ is taken as $x = 1/N_0$ in the present case. Indeed, both the final amplitude coefficient $\alpha_1$ and the theoretical amplitude coefficient $y(x)$ with $x = 1/N_0$ share the first two lower-order terms, i.e., $1 - \frac{1}{2}x$.

($ii$) The amplitude coefficient of the residual term with $\sum_{k=2}^{2N_0-1} |k\rangle_b$ is decreased in magnitude by one order ($1/N_0$) of magnitude from the amplitude coefficient $-\frac{1}{\sqrt{2}}\frac{1}{N_0}$ in the Hamiltonian $\hat{H}_{2N_0}^{(i)}$ to the amplitude coefficient $-\frac{1}{\sqrt{2}}\frac{1}{N_0^2}$ in the final Hamiltonian $\hat{H}_N^{(f)}$. This is one desired result in the present subspace-selective unitary manipulation.

($iii$) The amplitude coefficient of the residual term with the base vector $|1\rangle_b$ is changed from zero amplitude coefficient in the Hamiltonian $\hat{H}_{2N_0}^{(i)}$ to non-zero amplitude coefficient $-\frac{1}{2}\sqrt{3}(\frac{1}{N_0} - \frac{1}{N_0^2})$ in the final Hamiltonian $\hat{H}_N^{(f)}$.

The square norm of all the residual terms in each one of these Hamiltonians $\hat{H}_{N_0}^{(i)}$, $\hat{H}_{2N_0}^{(i)}$, and $\hat{H}_N^{(f)}$ can be exactly calculated. The square norm in the initial

Hamiltonian $\hat{H}^{(i)}_{N_0}$ or $\hat{H}^{(i)}_{2N_0}$ is given by

$$\left\|\frac{-1}{N_0}\sum_{k=1}^{N_0-1}|k\rangle\right\|^2=\left\|\frac{-1}{N_0}\frac{1}{\sqrt{2}}\sum_{k=2}^{2N_0-1}|k\rangle\right\|^2=x-x^2,$$

while the square norm in the final Hamiltonian $\hat{H}^{(f)}_N$ is given by

$$\left\|-\frac{1}{2}\sqrt{3}\left(\frac{1}{N_0}-\frac{1}{N_0^2}\right)|1\rangle_b-\frac{1}{\sqrt{2}}\frac{1}{N_0^2}\sum_{k=2}^{2N_0-1}|k\rangle_b\right\|^2=\frac{3}{4}x^2-\frac{1}{2}x^3-\frac{1}{4}x^4.$$

Note that here $0<x=1/N_0<<1$. Then it can be found that the square norm of all the residual terms for the final Hamiltonian $\hat{H}^{(f)}_N$ may be much smaller than that one for the initial Hamiltonian $\hat{H}^{(i)}_{N_0}$ or $\hat{H}^{(i)}_{2N_0}$. This indicates that, after the present subspace-selective unitary manipulation (i.e., the subspace-selective amplitude decreasing process), the residual terms in the initial Hamiltonian $\hat{H}^{(i)}_{N_0}$ as a whole are decreased in magnitude, although the result (*ii*) is desired theoretically and the result (*iii*) is undesired theoretically (but its effect is limited).

In order to continue to decrease in magnitude the residual terms as a whole in the Hamiltonian $\hat{H}^{(f)}_N$ and at the same time to make the amplitude coefficient $\alpha_1$ of the desired term $\alpha_1|0\rangle_b$ more close to the theoretical amplitude coefficient $y(x)$ of the theoretical desired term $y(x)|0\rangle_b$ (i.e., both share more and more (first) lower-order terms) one needs to perform many times the subspace-selective unitary manipulation and employ many larger direct-sum subspaces. The relevant work will be reported in the future paper.

# Acknowledgements

This research project have lasted a little bit longer than five years since the author's last paper was published in arXiv in 2020 (X. Miao, arXiv.org: 2012.13250 [quant-ph] (2020)). During the research project my living cost were mainly supported by my relatives in P.R.China. I also borrowed part of my living cost from my friends. I thank my relatives for their selfless support and my friends for their kind help. I want to say this research work was finished by myself own efforts.

E-mail addresses: muyangmiao0@gmail.com, miaoxijia@yahoo.com

**Note added.** On the research subject: Applications of the Lie group and Lie algebra methods to the design of the NMR pulse sequences and the exact and analytical calculations for the unitary time-evolutional processes of the complex coupled multiple-spin systems in NMR spectroscopy. I wrote two papers at least on this research subject, one in 1990 and one in 1994. One of which had been sent to journals to publish, but unfortunately no journals would accept it to publish. I had to abandon the research work on this research subject since I finished my postdoctoral research work, largely because the papers on the subject can not be accepted to publish and the research work on the subject can not be recognized. At present some relevant papers (or manuscripts) were not at my hand. I remember that one of the papers (or manuscripts) was at my hand till the year 2002, but I found it was not at my hand in 2003, when I moved from Marney Street to Charles Street in Cambridge, MA, USA.

# References

1. J. von Neumann, *Mathematical foundations of quantum mechanics*, Translated by R. T. Beyer, Edited by N. A. Wheeler, Princeton University Press, 1955

2. L. I. Schiff, *Quantum mechanics*, 3rd, McGraw-Hill book company, New York, 1968

3. T. F. Jordan, *Linear operators for quantum mechanics*, Dover edition (2006), Originally published by John Wiley & Sons, Inc., New York, 1969

4. X. Miao, *Multiple-quantum operator algebra spaces and description for the unitary time evolution of multilevel spin systems*, Molec. Phys. 98, 625 (2000)

5. R. R. Ernst, G. Bodenhausen, and A. Wokaun, *Principles of nuclear magnetic resonance in one and two dimensions*, Oxford University Press, Oxford, 1987

6. U. Fano, *Description of states in quantum mechanics by density matrix and operator techniques*, Rev. Mod. Phys. 29, 74 (1957)

7. X. Miao, *Efficient multiple-quantum transition processes in an $n-$qubit spin system*, http://arxiv.org/abs/quant-ph/0411046 (2004)

8. X. Miao, *Universal construction of unitary transformation of quantum computation with one- and two-body interactions*, http://arxiv.org/abs/quant-ph /0003068 (2000)

9. A. R. Edmonds, *Angular momentum in quantum mechanics*, 2nd edn, Princeton University Press, Princeton, 1974

10. M. E. Rose, *Elementary theory of angular momentum*, Wiley, New York, 1957

11. R. Freeman, *Spin Choreography: Basic steps in high resolution NMR,* Oxford University Press, Oxford, 1998

12. A. Schweiger and G. Jeschke, *Principles of pulse electron paramagnetic resonance*, Oxford University Press, Oxford, 2001

13. B. W. Shore, *The theory of coherent atomic excitation*, Wiley, New York, 1990

14. R. P. Feynman, *Simulating physics in computers*, Int. J. Theor. Phys. 21, 467 (1982)

15. J. H. Wilkinson, *The algebraic eigenvalue problem*, Oxford University Press, Oxford, 1965

16. Å. Björck, *Numerical methods in matrix computations*, Subsection 1.1.7, Springer International Publishing, Switzerland, 2015

17. E. P. Wigner, *Group theory and its application to the quantum mechanics of atomic spectra*, translated by J.J. Griffin, Academic Press, New York, 1959

18. F. J. Dyson, *The radiation theories of Tomonaga, Schwinger, and Feynman,*

Phys. Rev. 75, 486 (1949)

The following references [Ref$^k$] with $k = 1, 2, ..., 14$ are given in the footnote on different pages, respectively. For example. the reference [Ref$^1$] is given in the footnote on Page One and it is simply denoted as [Ref$^1$] (Page 1). Similarly, [Ref$^2$] (Page 2); [Ref$^3$] (Page 15); [Ref$^4$] (Page 55); [Ref$^5$] (Page 175); [Ref$^6$] (Page 9); [Ref$^7$] (Page 19); [Ref$^8$] (Page 9); [Ref$^9$] (Page 53); [Ref$^{10}$] (Page 15); [Ref$^{11}$] (Page 19); [Ref$^{12}$] (Page 19); [Ref$^{13}$] (Page 19); [Ref$^{14}$] (Page 19)

# The Section A. The theoretical framework for the multiple quantum transition operators in quantum simulating

In this Section the multiple quantum transitions in quantum multiple-spin systems and the multiple quantum transition operators which can induce these multiple quantum transitions are introduced theoretically. A theoretical framework for the multiple-quantum-transition operators in quantum simulating therefore is set up. It is more suitable for applications in the research area of quantum simulating and quantum computing. Quantum computing and quantum simulating stress to be mathematical logical, strict, and quantitative. This theoretical framework meets this point. In general the multiple quantum transitions and the multiple-quantum-transition operators have been used extensively in nuclear magnetic resonance spectroscopy [5, 11].

## A1.1. Definitions of the p-order quantum transition operators

The spin energy of a spin system which may consist of many individual spin particles in an external strong static magnetic field may be characterized mainly by the Zeeman interaction in nuclear magnetic resonance spectroscopy [5]. The Zeeman interaction may be written as $H_Z = -\mu \cdot \mathbf{B}$ [2], where $\mathbf{B}$ is the external magnetic field and $\mu$ is the magnetic moment that is proportional to the total spin angular momentum $\mathbf{J}$ (i.e., $\mathbf{J} = \hbar\mathbf{I}$) of the spin system, that is, $\mu = \gamma\mathbf{J}$. For convenience, here suppose simply that the external static magnetic field $\mathbf{B}$ is oriented along the $z-$direction coordinate axis, i.e., $\mathbf{B} = B_0\mathbf{z}$. Then the Zeeman interaction may be simply written as $H_Z = -\hbar\gamma B_0 I_z$, where $\gamma$ is the gyromagnetic ratio and $B_0$ the strength of the external magnetic field $\mathbf{B}$. This indicates that the spin Hamiltonian $H_Z$ of the spin system is characterized by the $z-$component spin operator $I_z$ of the total spin angular momentum operator $\mathbf{I}$ (in unit $\hbar = 1$) of the spin system. Moreover, both the spin Hamiltonian $H_Z$ and the total $z-$component spin operator $I_z$ commute with one another and have the common eigenbase vectors. Therefore, the discrete spin energy levels of the spin system may be characterized uniquely by the total spin magnetic quantum number $M$, while the latter is the eigenvalue of the total $z-$component spin operator $I_z$, where the eigenvalue equation for the total spin operator $I_z$ may be written as $I_z |M\rangle = M |M\rangle$ with $-I \leq M \leq I$, here $I$ is the total spin quantum number of the spin system.

A quantum transition between one spin energy level and another of a spin system may be induced by an external radio-frequency electromagnetic field (RF field) in nuclear magnetic resonance spectroscopy. The spin system under study at present is a multiple-energy-level spin system whose spin Hamiltonian is simply given by the Zeeman interaction $H_Z$. As shown above, its spin energy levels are characterized uniquely by the total spin magnetic quantum number $M$. When the spin system is excited by an external RF field from the initial energy level jump to the final energy level, the total spin magnetic quantum number $M$ changes from the initial value $M_i$ to the final value $M_f$, where $M_i$ and $M_f$ correspond to the initial and the final spin energy level, respectively. Then the change $p = M_f - M_i$ of the total spin magnetic quantum number $M$ in this excitation of the RF field may characterize this quantum transition between the initial and the final spin energy level. This change value $p$ is called the order of the quantum transition, and correspondingly the quantum transition is called the $p-$order quantum transition [5]. The quantum-transition order $p$ is an integer and may be positive, zero, or negative. Because the order $p$ is a relative quantity, it may be independent of the value $M_i$ or $M_f$ but not both. When the order $p = \pm 1$, the quantum transition is a one-order quantum transition with order $p = +1$ or $-1$. In magnetic resonance spectroscopy only one-order quantum transitions with order $p = \pm 1$ are directly observable in experiment [5]. This is based on the selective rule of magnetic resonance at strong static magnetic fields: $\Delta M = M_f - M_i = \pm 1$. The quantum transitions with order $p = 0, \pm 1, \pm 2, \pm 3, ...$, may be simply called the zero-, one- (or single-), two- (or double-), three- (or triple-) quantum transitions and so on, respectively [5]. All these $p-$order quantum transitions with order $p = 0, \pm 1, \pm 2, ...$ of the spin system are called the multiple quantum transitions in unified form. Any multiple quantum transitions are usually hard to be directly observed except single-quantum transitions in magnetic resonance spectroscopy.

One theoretical basis for the present theoretical framework on the multiple quantum transitions and the multiple-quantum transition operators of quantum spin systems is the spin angular momentum theory in quantum mechanics [2] which includes the theoretical aspect for the combination (or addition) of spin angular momenta of many spin particles [9, 10]. Consider a general spin system, namely an $n - spin - I_k$ system, which consists of $n$ individual spin particles and each $(k - th)$ individual spin particle has its own spin angular momentum quantum number $I_k$ (or simply spin quantum number $I_k$). In the $n - spin - I_k$ system each ($k-$th) spin particle has its own spin angular momentum $\mathbf{I}_k$ in unit $\hbar = 1$ (or $\mathbf{J}_k = \hbar \mathbf{I}_k$). Then according to the angular momentum theory in quantum mechanics [2,9,10] the total spin angular momentum $\mathbf{I}$ (in unit $\hbar = 1$) of the $n - spin - I_k$ system is the sum of the spin angular momenta $\{\mathbf{I}_k\}$ of all the $n$ individual spin particles of the spin system,

$$\mathbf{I} = \mathbf{I}_1 + \mathbf{I}_2 + ... + \mathbf{I}_n, \tag{A1.1}$$

and correspondingly the $z-$component $I_z$ of the total spin angular momentum $\mathbf{I}$ is given by the sum:

$$I_z = I_{1z} + I_{2z} + ... + I_{nz} \tag{A1.2}$$

where $\mathbf{I}_k$ and $I_{kz}$ are the spin angular momentum and its $z-$component of the $k-$th spin particle of the $n - spin - I_k$ system with the spin quantum number $I_k$, respectively. The total spin quantum number $I$ (i.e., the largest value) is the sum of the spin quantum numbers $\{I_k\}$ of the $n$ individual spin particles of the $n - spin - I_k$ system,

$$I = I_1 + I_2 + ... + I_n, \tag{A1.3}$$

while the total spin magnetic quantum number $M$ may take any one of the values $-I, -I+1, ..., I-1, I$.

In the above discussion about the multiple quantum transitions the spin Hamiltonian of a spin system, i.e., the Zeeman interaction $H_Z = -\hbar\gamma B_0 I_z$, is really equal to the total $z-$component spin operator $I_z$ of (A1.2) of the spin system up to a constant. Generally, the spin Hamiltonian $H_s$ of a general spin system such as the $n - spin - I_k$ system may be different from the total $z-$component spin operator $I_z$. Then there are two different cases: the first case is that the spin Hamiltonian $H_s$ commutes with the total spin operator $I_z$ and the second case is that the spin Hamiltonian $H_s$ does not commute with the total spin operator $I_z$. If the spin Hamiltonian $H_s$ of the $n - spin - I_k$ system commutes with the total spin operator $I_z$, both the spin Hamiltonian $H_s$ and the total spin operator $I_z$ have the common eigenbase vectors. Then the discrete spin energy levels of the spin system may be characterized by the total spin magnetic quantum number $M$, and the latter is the eigenvalue of the total spin operator $I_z$. Therefore, a multiple quantum transition is just an energy quantum transition between a pair of the common eigenbase vectors in the spin system. There are a lot of multiple-spin systems which include the uncoupled (or non-interacting) $n-$spin$-1/2$ systems, the weakly-coupled $n-$spin$-1/2$ systems, and the strongly-coupled $n-$spin$-1/2$ systems [5], each one of which has the spin Hamiltonian $H_s$ which can commute with the total spin operator $I_z$. Then in each one of these spin systems a multiple quantum transition may be simply an energy quantum transition. Therefore, any multiple quantum transition of a general spin system still can be characterized well by the total spin magnetic quantum number $M$ or exactly by the quantum-transition order $p = M_f - M_i$, if the spin Hamiltonian $H_s$ of the spin system commutes with the total spin operator $I_z$, even though the former ($H_s$) is different from the latter ($I_z$). Later it is discussed the second case that the spin Hamiltonian $H_s$ does not commute with the total spin operator $I_z$. The second case is more complex than the first case. However, as shown later, even in the second case any multiple quantum transitions of a general spin system still can be specified by the total spin magnetic quantum number $M$.

Multiple quantum transitions of a spin system may be induced by a multiple quantum transition (MQT) operator. This is just like that energy quantum transitions of a quantum system may be induced by a (perturbation) Hamiltonian operator that is generated by an external time-varying electromagnetic field [2]. As shown above, a multiple quantum transition may be characterized via the total spin magnetic quantum number $M$, while the latter is the eigenvalue of the total $z-$component spin operator $I_z$. Then any multiple-quantum transition operator of a general spin system such as an $n-spin-I_k$ system may be formally defined on the basis of the representation that is defined by the total $z-$component spin operator $I_z$ of (A1.2) of the total spin angular momentum operator $\mathbf{I}$ of (A1.1) of the spin system.

A multiple quantum transition (MQT) operator may be formally defined as follows. Suppose that $|\Psi_i\rangle$ is an arbitrary eigenvector of the total spin operator $I_z$ of (A1.2) with the eigenvalue $M_i$, and it obeys the eigenvalue equation,

$$I_z |\Psi_i\rangle = M_i |\Psi_i\rangle\,, \tag{A1.4}$$

where the eigenvalue $M_i$, i.e., the total spin magnetic quantum number $M_i$, may take any one of the values $-I, -I+1, ..., I-1, I$ with the total spin quantum number $I$ given by (A1.3). A $p-$order multiple quantum transition operator $Q_p$ (or simply a $p-$order quantum transition operator $Q_p$) may be formally defined by ([Ref[5]][19], Ref.[8]),

$$I_z Q_p |\Psi_i\rangle = (M_i + p)\, Q_p |\Psi_i\rangle \tag{A1.5}$$

where $Q_p |\Psi_i\rangle$ also is an eigenvector of the total spin operator $I_z$ and is associated with the eigenvalue $M_i + p$. By the formal definition (A1.5) it is easy to prove that all the $p-$order quantum transition operators $\{Q_p\}$ with the same order $p$ form a linear space of operators. This means that the linear combination $\alpha_1 Q_{p1} + \alpha_2 Q_{p2}$ of two $p-$order quantum transition operators $Q_{p1}$ and $Q_{p2}$ with the same order $p$ is still a $p-$order quantum transition operator. All these $p-$order quantum transition operators with order $p = 0,\ \pm1,\ \pm2,\ ...$ of the spin system under study are called the multiple-quantum transition operators in unified form.

Let $|\Psi_f\rangle = Q_p |\Psi_i\rangle$ be the final eigenvector of the total spin operator $I_z$ with the eigenvalue $M_f = M_i + p$ which is generated by applying the $p-$order quantum transition operator $Q_p$ to arbitrary initial eigenvector $|\Psi_i\rangle$ with eigenvalue $M_i$. Then according to $|\Psi_f\rangle = Q_p |\Psi_i\rangle$ a physical explanation for the $p-$order quantum transition operator $Q_p$ in the definition (A1.5) is simply that the $p-$order quantum transitions with the same order $p = M_f - M_i$ between any initial $|\Psi_i\rangle$ and the final eigenvectors $|\Psi_f\rangle$ may be induced simultaneously by the $p-$order quantum transition operator $Q_p$.

Generally, any $p-$order quantum transition operator $Q_p$ defined by (A1.5) is not Hermitian. The only exception is any Hermitian zero-order quantum transition operator. Below it is shown that any Hermitian $|p|-$quantum transition operator $Q_{|p|}$ with the order $|p|$ may be constructed by $Q_{|p|} = \frac{1}{2}\left(Q_p + Q_p^+\right)$ or by $Q_{|p|} = \frac{1}{2i}\left(Q_p - Q_p^+\right)$. In fact, the multiple-quantum transition matrix element of the operator $Q_p$ is given by $(\Psi_f, Q_p\Psi_i)$ (or $\langle\Psi_f|Q_p|\Psi_i\rangle$) in quantum mechanics. More importantly, there is the famous identity in quantum mechanics [3, 2]: $(\Psi_f, Q_p\Psi_i) = \left(\Psi_i, Q_p^+\Psi_f\right)^*$ for any operator $Q_p$. This shows that if the operator $Q_p$ is a $p-$order quantum transition operator which induces the multiple quantum transition from $|\Psi_i\rangle$ to $|\Psi_f\rangle$ ($p = M_f - M_i$), then the operator $Q_p^+$ is a $-p-$order quantum transition operator, i.e., $Q_p^+ = Q_{-p}$, which induces the multiple quantum transition from $|\Psi_f\rangle$ to $|\Psi_i\rangle$ ($-p = M_i - M_f$). Therefore, the Hermitian $|p|-$quantum transition operators $Q_{|p|}$ may be formally defined by

$$\begin{cases} Q_{|p|} = \frac{1}{2}\left(Q_p + Q_p^+\right) = \frac{1}{2}\left(Q_p + Q_{-p}\right) \\ Q_{|p|} = \frac{1}{2i}\left(Q_p - Q_p^+\right) = \frac{1}{2i}\left(Q_p - Q_{-p}\right) \end{cases} \tag{A1.6}$$

[19] [Ref[5]] X. Miao, et al., *Application of product operator formalism to the strongly coupled spin (I=1/2) systems*, Science in China A Vol. 36, 1199 (1993) (English Ed.) and Vol. 23, 399 (1993) (Chinese Ed.); X. Miao, Ph D. thesis (in Chinese), Dalian Institute of Chemical Physics, The Chinese Academy of Sciences, 1992

One important application of the formal definition (A1.6) may be found in Ref.[7]. Now it is easy to understand why any Hermitian zero-order quantum transition operator $Q_{|p|}$ with order $p = 0$ can obey the formal definition (A1.5). Obviously, both non-Hermitian zero-order quantum transition operators $Q_p$ and $Q_{-p}$ with order $p = 0$ each satisfy exactly the definition (A1.5), that is, $I_z Q_p |\Psi_i\rangle = M_i Q_p |\Psi_i\rangle$ and $I_z Q_{-p} |\Psi_i\rangle = M_i Q_{-p} |\Psi_i\rangle$ with order $p = 0$. Then it can be easily found that the operators $(Q_p \pm Q_{-p})$ with order $p = 0$ satisfy the equation $I_z (Q_p \pm Q_{-p}) |\Psi_i\rangle = M_i (Q_p \pm Q_{-p}) |\Psi_i\rangle$ with order $p = 0$. This indicates that the Hermitian zero-order quantum transition operators $Q_{|p|}$ of (A1.6) with order $p = 0$ exactly satisfy the formal definition (A1.5), that is, $I_z Q_{|p|} |\Psi_i\rangle = M_i Q_{|p|} |\Psi_i\rangle$ with order $p = 0$. A Hermitian zero-order quantum transition operator $Q_{zq}$ may generate a unitary operator $\exp(-i\theta Q_{zq})$ which is also a zero-quantum operator ([Ref[5]], Ref.[8]).

*A*ny $p-$order quantum coherence operator $Q_p$ of the density operator space of spin system also may be formally defined by (A1.5) [Ref[5]], where the order $p$ is called the order of coherence [5]. These *m*ultiple-*q*uantum *c*oherence (MQC) operators are closely related to the density operators ($\rho$) of the density operator space of spin system. They have been extensively used to specify the quantum (mixed) states (i.e., the non-diagonal density operators) of the spin systems in magnetic resonance spectroscopy [5]. However, in this paper these multiple-quantum coherence operators are seldom discussed.

It is well known in quantum mechanics that all the orthonormal eigenbase vectors of the total spin operator $I_z$ form a complete set $\{|M\rangle\}$ of base vectors of the Hilbert space of the $n - spin - I_k$ system, where the eigenvalue equation for the total spin operator $I_z$ is simply written as $I_z |M\rangle = M |M\rangle$ with $-I \leq M \leq I$. This is the representation defined by the total spin operator $I_z$. According to quantum mechanics [2] an eigenvalue $M$ of the total spin operator $I_z$ may be degenerate and may correspond to many orthonormal eigenbase vectors $|M\rangle$, i.e., the degenerate eigenbase vectors. Then the complete set $\{|M\rangle\}$ may be rewritten as $\{|M, K_M\rangle\}$ where the index $K_M$ distinguishes between the orthonormal degenerate eigenbase vectors $|M\rangle$, and the eigenvalue equation for the total spin operator $I_z$ is rewritten as $I_z |M, K_M\rangle = M |M, K_M\rangle$ with distinct eigenvalue $M = -I, -I+1, ..., I-1, I$. Now, according to the eigenfunction expansion principle in quantum mechanics, an arbitrary vector of the Hilbert space of the spin system, e.g., an arbitrary eigenvector $|\Psi_i\rangle$ of the total spin operator $I_z$ with its own eigenvalue $M_i$ in (A1.4), may be expanded in terms of the complete set $\{|M, K_M\rangle\}$ (Ref.[7], Ref.[8], [Ref[5]]),

$$|\Psi_i\rangle = \sum_{M=-I}^{I} \sum_{K_M} A_{M,K_M} |M, K_M\rangle = \sum_{K_{M_i}} A_{M_i,K_{M_i}} |M_i, K_{M_i}\rangle \tag{A1.7}$$

where the index $K_{M_i}$ runs over all the orthonormal degenerate eigenbase vectors $\{|M_i, K_{M_i}\rangle\}$ which correspond to the same eigenvalue $M_i$. The second equality in (A1.7) holds because ($i$) any eigenvectors with different eigenvalues $M$ are mutually orthogonal and ($ii$) the eigenvector $|\Psi_i\rangle$ in (A1.4) is associated with the eigenvalue $M_i$. In analogous way the eigenvector $|\Psi_f\rangle = Q_p |\Psi_i\rangle$ in (A1.5) of the total spin operator $I_z$ with its own eigenvalue $M_f = M_i + p$ may be expanded in terms of the complete set $\{|M, K_M\rangle\}$,

$$|\Psi_f\rangle = \sum_{M=-I}^{I} \sum_{K_M} B_{M,K_M} |M, K_M\rangle = \sum_{K_{M_f}} B_{M_f,K_{M_f}} \left|M_f, K_{M_f}\right\rangle \tag{A1.8}$$

where the index $K_{M_f}$ runs over all the orthonormal degenerate eigenbase vectors $\{\left|M_f, K_{M_f}\right\rangle\}$ which correspond to the same eigenvalue $M_f = M_i + p$.

For the simple case that the spin Hamiltonian $H_s$ of the spin system commutes with the total spin operator $I_z$, the eigenbase vectors $\{|M, K_M\rangle\}$ of the total spin operator $I_z$ evidently may be the common eigenbase vectors of both the commuting operators $H_s$ and $I_z$. Then in this case a $p-$order quantum transition with the order $p = M_f - M_i$ between any pair of eigenbase vectors $\left|M_f, K_{M_f}\right\rangle$ and $|M_i, K_{M_i}\rangle$ of the total spin operator $I_z$ is just the energy quantum transition between the pair of spin energy levels $\left|M_f, K_{M_f}\right\rangle$ and $|M_i, K_{M_i}\rangle$ of the spin system.

If the spin Hamiltonian $H_s$ does not commute with the total spin operator $I_z$, then situation becomes complex. In this case there are not the common eigenbase vectors for both the spin Hamiltonian $H_s$ and the total spin operator $I_z$. Then a $p-$order quantum transition between a pair of the eigenbase vectors $|M_i, K_{M_i}\rangle$ and $\left|M_f, K_{M_f}\right\rangle$ of the total spin operator $I_z$ may not be any single energy quantum transition between a pair of spin energy levels of the spin system. However, in this case a $p-$order quantum transition operator $Q_p$ is still defined formally by (A1.5). This means that a multiple quantum transition is still characterized by the total spin magnetic quantum number $M$ (or exactly by the order $p = M_f - M_i$). This point is explained as follows. All the spin energy eigenbase vectors $\{|\Psi_{E_i}\rangle\}$ of the spin Hamiltonian $H_s$ form a complete set of the orthonormal base vectors of the Hilbert space of the spin system. Generally, according to the eigenfunction expansion principle in quantum mechanics, an arbitrary vector $|\Psi\rangle$ of the Hilbert space may be expanded in terms of the complete set $\{|\Psi_{E_i}\rangle\}$,

$$|\Psi\rangle = \sum_{E_i} C_{E_i} |\Psi_{E_i}\rangle \tag{A1.9}$$

Suppose now that the vector $|\Psi\rangle$ of (A1.9) is taken as any eigenbase vector $|M_i, K_{M_i}\rangle$ of the total spin operator $I_z$. Then it can be deduced from the eigenfunction expansion (A1.9) of the eigenbase vector $|M_i, K_{M_i}\rangle$ that any $p-$order quantum transition with the order $p = M_f - M_i$ that occurs between a pair of the eigenbase vectors $|M_i, K_{M_i}\rangle$ and $\left|M_f, K_{M_f}\right\rangle$ means that many energy quantum transitions among the spin energy levels $\{|\Psi_{E_i}\rangle\}$ occur in the spin system at the same time. Therefore, it concludes that generally a $p-$order quantum transition with the order $p = M_f - M_i$ is not a single energy quantum transition, instead it is original from many energy quantum transitions of the spin system at the same time, if the total spin operator $I_z$ does not commute with the spin Hamiltonian $H_s$.

Generally, whether or not the spin Hamiltonian $H_s$ commutes with the total spin operator $I_z$, an arbitrary vector $|\Psi\rangle$ of the Hilbert space of the spin system always can be expanded in terms of the complete set $\{|M, K_M\rangle\}$,

$$|\Psi\rangle = \sum_{M=-I}^{I} \sum_{K_M} C_{M,K_M} |M, K_M\rangle \tag{A1.10}$$

Suppose now that the vector $|\Psi\rangle$ of (A1.10) is taken as any spin energy eigenvector $|\Psi_{E_i}\rangle$. If the spin Hamiltonian $H_s$ does not commute with the total spin operator $I_z$, then the spin energy eigenvector $|\Psi_{E_i}\rangle$ is different from any eigenvector of the total spin operator $I_z$. Then it can be deduced from the eigenfunction expansion (A1.10) of the spin energy eigenvector $|\Psi_{E_i}\rangle$ that an energy quantum transition that occurs between a pair of different spin energy levels $|\Psi_{E_i}\rangle$ and $\left|\Psi_{E_f}\right\rangle$ of the spin system means that many multiple quantum transitions with various orders $p$ among the eigenbase vectors $\{|M, K_M\rangle\}$ of the total spin operator $I_z$ occur in the spin system at the same time.

On the basis of the formal definitions (A1.5) and (A1.6) for any multiple-quantum-transition operators and the spin angular momentum theory in quantum mechanics [2] which includes the theoretical aspect for the combination (or addition) of spin angular momenta of many spin particles [9, 10] a theoretical framework on the multiple quantum transitions and the multiple-quantum transition operators of quantum spin systems therefore is set up, which is more suitable for applications in the research area of quantum simulating and quantum computing. Quantum computing and quantum simulating stress to be mathematical logical, strict, and quantitative. This theoretical framework is consistent with this point. Below several important instances are discussed and analyzed theoretically on the basis of the theoretical framework.

## A1.2. Simple applications of the p-order quantum transition operators

According to the formal definition (A1.5) of a $p-$order quantum transition operator several simple examples and applications of the $p-$order quantum transition operators are described in detail below. Consider a single spin$-I$ particle, where the spin quantum number $I \geq 1/2$ may be an integer or half an odd integer [2]. According to the angular momentum theory in quantum mechanics [2,9,10] a single spin$-I$ particle may be specified by the set $\{\mathbf{I}^2, I_z\}$ of commuting operators which consists of the square of the spin angular mo-mentum operator, i.e., $\mathbf{I}^2$ and the $z-$component spin operator $I_z$ of the spin$-I$ particle. The set $\{\mathbf{I}^2, I_z\}$ of commuting operators define such a representation that both the commuting operators $\mathbf{I}^2$ and $I_z$ are diagonal at the same time. The eigenvalue equations for both the commuting operators $\mathbf{I}^2$ and $I_z$ [2] are written as $\mathbf{I}^2 |I,m\rangle = I(I+1)|I,m\rangle$ and $I_z |I,m\rangle = m|I,m\rangle$, respectively, where $|I,m\rangle$ is a common eigenbase vector of the commuting operators $\mathbf{I}^2$ and $I_z$ and the eigenvalue $m$ is the spin magnetic quantum number $m$ which is given by $m = -I, -I+1, ..., I-1, I$. For simplicity, here consider that the square of the spin angular momentum, i.e., $\mathbf{I}^2$, is a constant of motion and can be replaced by the constant $I(I+1)$ (in unit $\hbar = 1$). This is general case in magnetic resonance experiments and Stern-Gerlach experiments [2]. Then the present representation defined by both the commuting operators $\mathbf{I}^2$ and $I_z$ may be defined simply by the spin operator $I_z$, and the eigenvalue equation is simply written as $I_z|m\rangle = m|m\rangle$ with $-I \leq m \leq I$, where $|m\rangle \overset{Def}{\equiv} |I,m\rangle$. For a single spin$-1/2$ particle with $I = 1/2$, according to the eigenvalue equation $I_z|m\rangle = m|m\rangle$ with $-1/2 \leq m \leq 1/2$, there are only two (or $2I+1$) orthonormal eigenbase vectors $|\alpha\rangle = |1/2\rangle$ and $|\beta\rangle = |-1/2\rangle$ with the eigenvalues $m = 1/2$ and $m = -1/2$, respectively. Therefore, a single spin$-1/2$ particle is associated with a two-dimensional Hilbert space [2] and it is a pure two-level system.

Now it is easy to show that the raising $(I^+)$ and the lowering operator $(I^-)$ of spin angular momentum of a single spin$-I$ particle are one-order quantum transition operators with order $p = +1$ and $-1$, respectively. It is known from the angular momentum theory [2] that the spin raising operator $I^+ = I_x + iI_y$ and the lowering operator $I^- = I_x - iI_y$ acting on any eigenbase vector $|I,m\rangle$ of the spin operator $I_z$ generate respectively the following basic transformational equations:

$$I^{\pm}|I,m\rangle = \sqrt{I(I+1) - m(m\pm 1)}\,|I,m\pm 1\rangle \tag{A1.11}$$

Let $|\Psi_i\rangle = |I,m\rangle$ and $\left|\Psi_f^{\pm}\right\rangle = I^{\pm}|I,m\rangle$. On the one hand, $I_z|\Psi_i\rangle = I_z|I,m\rangle = m|I,m\rangle$ with the eigenvalue $m$. Therefore, the initial vector $|\Psi_i\rangle$ is any eigenvector of the spin operator $I_z$. On the other hand, it follows from (A1.11) and the eigenvalue equation $I_z|I,m\pm 1\rangle = (m\pm 1)|I,m\pm 1\rangle$ that

$$I_z I^{\pm}|I,m\rangle = (m\pm 1)\sqrt{I(I+1) - m(m\pm 1)}\,|I,m\pm 1\rangle \tag{A1.12}$$

Therefore, by substituting (A1.11) into (A1.12) one obtains

$$I_z I^{\pm}|I,m\rangle = (m\pm 1)\, I^{\pm}|I,m\rangle \tag{A1.13}$$

where the spin magnetic quantum number $m$ takes any one of the values $-I, -I+1, ..., I-1, I$. It can be found that the formula (A1.13) is a special case of the formal definition (A1.5) of the $p-$order quantum transition operator $Q_p$ with $Q_p = I^{\pm}$ and $p = \pm 1$. It may be further written as $I_z\left|\Psi_f^{\pm}\right\rangle = m_f^{\pm}\left|\Psi_f^{\pm}\right\rangle$ with the eigenvalue $m_f^{\pm} = m \pm 1$, indicating that $\left|\Psi_f^{\pm}\right\rangle = I^{\pm}|I,m\rangle$ is an eigenvector of the spin operator $I_z$ with the eigenvalue $m_f^{\pm} = m + p = m \pm 1$. Then the quantum transition from the initial $|\Psi_i\rangle$ to the final eigenvector $\left|\Psi_f^{\pm}\right\rangle = Q_p |\Psi_i\rangle$ has the quantum-transition order $p = m_f^{\pm} - m = \pm 1$. This indicates that the operator $Q_p = Q_{\pm 1} = I^{\pm}$ is the $\pm 1-$order quantum transition operator. There are the two special cases: $I^+ |I, I\rangle = 0$ when $m$ takes the maximum value $I$ and $I^- |I, -I\rangle = 0$ when $m$ takes the minimum value $-I$. Note that $I_z I^+ |I, I\rangle = 0 = (I+1) I^+ |I, I\rangle$ and $I_z I^- |I, -I\rangle = 0 = (-I-1) I^- |I, -I\rangle$. Both the special cases still obey the formula (A1.13) of the $\pm 1-$order quantum transition operator $Q_{\pm 1} = I^{\pm}$. In history, inspired by the formula (A1.13) of the angular momentum theory in quantum mechanics, the formal definition (A1.5) was put forward for a $p-$order quantum transition operator $Q_p$.

The $\pm 1-$order quantum transition operators $Q_{\pm 1} = I^{\pm}$ defined by (A1.13) are not Hermitian. Note that $Q_{+1}^+ = (I^+)^+ = I^- = Q_{-1}$. Then according to the definition (A1.6) a Hermitian single-quantum transition operator $Q_1$ may be constructed by $Q_1 = \frac{1}{2}\left(Q_{+1} + Q_{+1}^+\right) = \frac{1}{2}(Q_{+1} + Q_{-1}) = \frac{1}{2}(I^+ + I^-) = I_x$ or by $Q_1 = \frac{1}{2i}\left(Q_{+1} - Q_{+1}^+\right) = \frac{1}{2i}(Q_{+1} - Q_{-1}) = \frac{1}{2i}(I^+ - I^-) = I_y$. Therefore, both the spin operators $I_x$ and $I_y$ of a single spin$-I$ particle are the Hermitian single-quantum transition operators [5]. If the spin quantum number $I \geq 1$ for a single spin$-I$ particle, then beside the one-order quantum transition operators $Q_{\pm 1} = I^{\pm}$ with $|p| = 1$ there also may be higher-order quantum transition operators ($|p| > 1$) of the spin$-I$ particle [5]. For example, it can be confirmed by the definitions (A1.5) and (A1.6) that the operators $Q_{+2} = (I^+)^2$ and $Q_{-2} = (I^-)^2$ are two-order quantum transition operators with order $p = +2$ and $-2$, respectively, while the operators $Q_2 = \frac{1}{2}(Q_{+2} + Q_{-2}) = \frac{1}{2}(I^+ I^+ + I^- I^-)$ and $Q_2 = \frac{1}{2i}(Q_{+2} - Q_{-2}) = \frac{1}{2i}(I^+ I^+ - I^- I^-)$ are the Hermitian two-quantum transition operators of the spin$-I$ particle.

The $\pm 1-$order quantum transition operators $I^{\pm}$ of the single spin$-I$ particle are usually said non-selective in the sense that they are not dependent upon any eigenvector $|\Psi_i\rangle = |I, m\rangle$ in the definition (A1.13) or (A1.5) of the $p-$order quantum transition operator with the order $p = \pm 1$. This also means that when the single spin$-I$ particle in any initial state vector ($|\Psi_i\rangle$) is influenced by the $\pm 1-$order quantum transition operators $I^{\pm}$, the $\pm 1-$order quantum transitions usually can be excited in the single spin$-I$ particle. More generally, a $p-$order multiple-quantum-transition (MQT) operator $Q_p$ is usually said non-selective, if it is independent upon any eigenvector $|\Psi_i\rangle$ of the total spin operator $I_z$ in the formal definition (A1.5) of the $p-$order MQT operator $Q_p$ (this is necessary condition). This also means that when the spin system under study in any initial state vector ($|\Psi_i\rangle$) is influenced by the $p-$order MQT operator $Q_p$, the $p-$order multiple quantum transitions usually can be excited in the spin system. As a typical example, consider a $two - spin - I_k$ system which consists of the two spin$-I_j$ and spin$-I_l$ particles. It can be shown below that the $\pm 2-$order MQT operators $I_j^{\pm} I_l^{\pm}$, the zero-order MQT operators $I_j^{\pm} I_l^{\mp}$, and the $\pm 1-$order MQT operators $I_j^{\pm}$ and $I_l^{\pm}$ are usually non-selective in the two-spin$-I_k$ system, since they are independent of any eigenvector $|\Psi_i\rangle$ in the formal definition (A1.5) of a $p-$order quantum transition operator. Actually, these operators also may be considered as the product operators [5] in an $n-spin-I_k$ system with $n \geq 2$ (e.g., an $n-spin-1/2$ system). Then they are usually non-selective in the $n-spin-I_k$ system too.

It is known from the angular momentum theory in quantum mechanics [2, 9,10] that the $k-$th spin raising operator $I_k^+$ and lowering operator $I_k^-$ acting on any eigenbase vector $|I_k, m_k\rangle$ of the $k-$th spin operator $I_{kz}$ of the $k-$th spin$-I_k$ particle of the $two-spin-I_k$ system can respectively lead to the basic transformational equations:

$$I_k^{\pm} |I_k, m_k\rangle = \sqrt{I_k(I_k+1) - m_k(m_k \pm 1)}\, |I_k, m_k \pm 1\rangle \text{ for } k = j, l \qquad \text{(A1.14)}$$

where the eigenbase vector $|I_k, m_k\rangle$ obeys the eigenvalue equation $I_{kz} |I_k, m_k\rangle = m_k |I_k, m_k\rangle$ with the eigenvalue $m_k = -I_k, -I_k+1, ..., I_k-1, I_k$. Any eigenbase vector of the total spin operator $I_z = I_{jz} + I_{lz}$ may be chosen as the tensor-product base vector $|\Psi_i'\rangle = |I_j, m_j\rangle |I_l, m_l\rangle$ of the $two-spin-I_k$ system with the eigenvalue $M_i = m_j + m_l$, and it obeys the eigenvalue equation $I_z |\Psi_i'\rangle = M_i |\Psi_i'\rangle$, where the eigenvalue $M_i$ is the total spin magnetic quantum number. All the eigenbase vectors $\{|\Psi_i'\rangle\}$ form a complete set of the orthonormal base vectors of the Hilbert space of the $two-spin-I_k$ system. With the help of (A1.14) and the eigenvalue equations $I_{kz} |I_k, m_k\rangle = m_k |I_k, m_k\rangle$ with $k = j, l$ it can prove that

$$I_z \left(I_j^{\pm} I_l^{\pm}\right) |\Psi_i'\rangle = (M_i \pm 2) \left(I_j^{\pm} I_l^{\pm}\right) |\Psi_i'\rangle \qquad \text{(A1.15)}$$

$$I_z \left(I_j^{\pm} I_l^{\mp}\right) |\Psi_i'\rangle = M_i \left(I_j^{\pm} I_l^{\mp}\right) |\Psi_i'\rangle \qquad \text{(A1.16)}$$

$$I_z \left(I_k^{\pm}\right) |\Psi_i'\rangle = (M_i \pm 1) \left(I_k^{\pm}\right) |\Psi_i'\rangle, \; k = j, l \qquad \text{(A1.17)}$$

The equations (A1.15), (A1.16), and (A1.17) show that the operators $I_j^{\pm} I_l^{\pm}$ are the $\pm 2-$order MQT operators, the operators $I_j^{\pm} I_l^{\mp}$ are the zero-order MQT operators, and the operators $I_k^{\pm}$ with $k = j, l$ are the $\pm 1-$order MQT operators, respectively. It can be shown that these MQT operators $I_j^{\pm} I_l^{\pm}$, $I_j^{\pm} I_l^{\mp}$, and $I_k^{\pm}$ with $k = j, l$ are usually non-selective. Note that $|\Psi_i'\rangle$ is any eigenbase vector of the total spin operator $I_z$ with the eigenvalue $M_i = m_j + m_l$ and $\{|\Psi_i'\rangle\}$ is a complete set. Then an arbitrary eigenvector $|\Psi_i\rangle$ of the total spin operator $I_z$ of the $two-spin-I_k$ system with the eigenvalue $M_i = m_j + m_l$ may be written as [2]

$$|\Psi_i\rangle = \sum_{m_j, m_l; m_j + m_l = M_i} C_{m_j, m_l}^{M_i} |I_j, m_j\rangle |I_l, m_l\rangle \qquad \text{(A1.18a)}$$

where summations over $m_j$ from $-I_j$ to $I_j$ and over $m_l$ from $-I_l$ to $I_l$ are carried out under the constraint $m_j + m_l = M_i$. The eigenvector $|\Psi_i\rangle$ of (A1.18a) evidently obeys the eigenvalue equation $I_z |\Psi_i\rangle = M_i |\Psi_i\rangle$. Now by using arbitrary eigenvector $|\Psi_i\rangle$ of (A1.18a) and with the help of (A1.15), (A1.16), and (A1.17) it can be found that these MQT operators $I_j^{\pm} I_l^{\pm}$, $I_j^{\pm} I_l^{\mp}$, and $I_k^{\pm}$ with $k = j, l$ are usually non-selective in the $two-spin-I_k$ system. As an example, by applying the operator $I_z\left(I_j^{\pm}I_l^{\pm}\right)$ to arbitrary eigenvector $|\Psi_i\rangle$ of (A1.18a) one obtains

$$I_z\left(I_j^{\pm}I_l^{\pm}\right)|\Psi_i\rangle = \sum_{m_j,m_l;m_j+m_l=M_i} C_{m_j,m_l}^{M_i} I_z\left(I_j^{\pm}I_l^{\pm}\right)|I_j,m_j\rangle|I_l,m_l\rangle$$

$$= \sum_{m_j,m_l;m_j+m_l=M_i} C_{m_j,m_l}^{M_i}\left(m_j+m_l\pm 2\right)\left(I_j^{\pm}I_l^{\pm}\right)|I_j,m_j\rangle|I_l,m_l\rangle$$

$$= \left(M_i\pm 2\right)\left(I_j^{\pm}I_l^{\pm}\right)|\Psi_i\rangle .$$

Moreover, here the total spin magnetic quantum number $M_i$ can take any one of the values $-I, -I+1, ..., I-1, I$ with the total spin quantum number $I = I_j + I_l$. These show that the $\pm 2-$order MQT operators $I_j^{\pm}I_l^{\pm}$ are usually non-selective in the $two-spin-I_k$ system. In analogous way it can be found that the zero-order MQT operators $I_j^{\pm}I_l^{\mp}$ and the $\pm 1-$order MQT operators $I_k^{\pm}$ with $k = j, l$ are usually non-selective too.

More generally, these MQT operators $I_j^{\pm}I_l^{\pm}$, $I_j^{\pm}I_l^{\mp}$, and $I_j^{\pm}$ and $I_l^{\pm}$ for $1 \leq j, l \leq n$ may be considered as the product operators [5] of the $n-spin-I_k$ system with $n \geq 2$ (e.g., an $n-spin-1/2$ system). Then it can be shown that these product operators $I_j^{\pm}I_l^{\pm}$, $I_j^{\pm}I_l^{\mp}$, and $I_k^{\pm}$ $(k = j, l)$ are the $\pm 2-$order, *zero*$-$order, and $\pm 1-$order MQT operators, respectively, and moreover, they are usually non-selective in the $n-spin-I_k$ system. As shown later in the Subsection A1.4, an eigenbase vector of the total spin operator $I_z = \sum_{k=1}^{n} I_{kz}$ of the $n-spin-I_k$ system may be chosen as the tensor-product base vector $|\Psi_i'\rangle = |I_1,m_1\rangle|I_2,m_2\rangle...|I_n,m_n\rangle$ and it still obeys the eigenvalue equation $I_z|\Psi_i'\rangle = M_i|\Psi_i'\rangle$ with the eigenvalue $M_i$ equal to the total spin magnetic quantum number $M_i = \sum_{k=1}^{n} m_k$. Obviously, all these eigenbase vectors $\{|\Psi_i'\rangle\}$ form a complete set of the orthonormal base vectors of the Hilbert space of the $n-spin-I_k$ system. Now with the help of the equations of (A1.14) and the eigenvalue equations $I_{kz}|I_k,m_k\rangle = m_k|I_k,m_k\rangle$ with $k = 1, 2, ..., n$ it can prove that these product operators $I_j^{\pm}I_l^{\pm}$, $I_j^{\pm}I_l^{\mp}$, and $I_k^{\pm}$ with $k = j, l$ for $1 \leq j, l \leq n$ still obey the equations (A1.15), (A1.16), and (A1.17), respectively, by starting from the tensor-product base vectors $\{|\Psi_i'\rangle\}$ of the $n-spin-I_k$ system. This indicates that these product operators are the $\pm 2-$order $(I_j^{\pm}I_l^{\pm})$, the zero-order $(I_j^{\pm}I_l^{\mp})$, and the $\pm 1-$order MQT operators ($I_j^{\pm}$ and $I_l^{\pm}$), respectively, and moreover, they are usually non-selective in the $n-spin-I_k$ system. Note that $\{|\Psi_i'\rangle\}$ is a complete set. Then an arbitrary eigenvector $|\Psi_i\rangle$ of the total spin operator $I_z$ of the $n-spin-I_k$ system with the eigenvalue $M_i = \sum_{k=1}^{n} m_k$ may be generally expressed as [2]

$$|\Psi_i\rangle = \sum_{m_1,m_2,...,m_n;m_1+m_2+...+m_n=M_i} C_{m_1,m_2,...,m_n}^{M_i}|I_1,m_1\rangle|I_2,m_2\rangle...|I_n,m_n\rangle \tag{A1.18b}$$

where summations over $m_k$ from $-I_k$ to $I_k$ for $k = 1, 2, ..., n$ are carried out under the constraint $\sum_{k=1}^{n} m_k = M_i$. It is easy to verify that the eigenvector $|\Psi_i\rangle$ of (A1.18b) obeys the eigenvalue equation $I_z|\Psi_i\rangle = M_i|\Psi_i\rangle$ with the eigenvalue $M_i = \sum_{k=1}^{n} m_k$. By starting from arbitrary eigenvector $|\Psi_i\rangle$ of (A1.18b)

and with the help of (A1.15), (A1.16), and (A1.17) it can be found that these product operators $I_j^{\pm}I_l^{\pm}$, $I_j^{\pm}I_l^{\mp}$, and $I_j^{\pm}$ and $I_l^{\pm}$ with $1 \leq j, l \leq n$ are usually non-selective in the $n-spin-I_k$ system. These results obtained from the $n-spin-I_k$ system with $n \geq 2$ are a generalization of the previous results obtained from the $two-spin-I_k$ system. As an example, the spin Hamiltonians of the strongly-coupled $n-$spin$-1/2$ systems (See, e.g., [Ref[5]], Ref.[5]) in magnetic resonance spectroscopy contain the non-selective zero-order quantum transition operators $\{I_j^{\pm}I_l^{\mp}\}$.

## A1.3. The zero-order quantum transition operators

Below consider a kind of important $p-$order quantum transition operators $Q_p$ of (A1.5) with the order $p = 0$. They are the zero-order quantum transition (ZQT) operators. According to the formal definition (A1.5) any zero-order quantum transition operator $Q_0$ (or $Q_{zq}$) obeys the equation:

$$I_z Q_0 \left|\Psi_i\right\rangle = M_i Q_0 \left|\Psi_i\right\rangle \text{ with } p = 0 \tag{A1.19}$$

Here $\left|\Psi_f\right\rangle = Q_0 \left|\Psi_i\right\rangle$ ($\left|\Psi_f\right\rangle \neq \left|\Psi_i\right\rangle$) which is created by acting the ZQT operator $Q_0$ on the initial eigenvector $\left|\Psi_i\right\rangle$ is also the eigenvector of the total spin operator $I_z$ and is associated with the eigenvalue $M_f = M_i + p = M_i$. Then both the initial $\left|\Psi_i\right\rangle$ and final eigenvector $\left|\Psi_f\right\rangle$ correspond to the same eigenvalue $M_i$, and hence they are degenerate eigenvectors, by noting $\left|\Psi_f\right\rangle \neq \left|\Psi_i\right\rangle$. It follows from (A1.7) that the initial eigenvector $\left|\Psi_i\right\rangle$ can be written as

$$\left|\Psi_i\right\rangle = \sum_{K_{M_i}} A_{M_i,K_{M_i}} \left|M_i, K_{M_i}\right\rangle, \tag{A1.20a}$$

and by noticing that $M_f = M_i$ it follows from (A1.8) that the final eigenvector $\left|\Psi_f\right\rangle$ can be expressed as

$$\left|\Psi_f\right\rangle = \sum_{K_{M_f}} B_{M_f,K_{M_f}} \left|M_f, K_{M_f}\right\rangle = \sum_{K_{M_f}} B_{M_i,K_{M_f}} \left|M_i, K_{M_f}\right\rangle, \tag{A1.20b}$$

where the indices $K_{M_i}$ and $K_{M_f}$ each run over all the orthonormal degenerate eigenbase vectors $\{\left|M_i, K_i\right\rangle\}$ ($K_i = K_{M_i}$ or $K_{M_f}$) which correspond to the same eigenvalue $M_i$.

It is known from (A1.4) that the vector $\left|\Psi_i\right\rangle$ in (A1.19) is an arbitrary eigenvector of the total spin operator $I_z$ with the eigenvalue $M_i$, and the equation (A1.20a) shows that it can be expressed as a linear combination of the orthonormal degenerate eigenbase vectors $\{\left|M_i, K_{M_i}\right\rangle\}$ of the total spin operator $I_z$ which correspond to the same eigenvalue $M_i$. Then all the eigenvectors $\{\left|\Psi_i\right\rangle\}$ of the total spin operator $I_z$ with the same eigenvalue $M_i$ can form a linear vector subspace and moreover, this vector subspace is spanned by the orthonormal degenerate eigenbase vectors $\{\left|M_i, K_{M_i}\right\rangle\}$ which correspond to the same eigenvalue $M_i$ [7]. Obviously, this vector basis subset $\{\left|M_i, K_{M_i}\right\rangle\}$ is contained in the complete set $\{\left|M, K_M\right\rangle\}$, while the latter spans the whole Hilbert space of the spin system. Therefore, the subspace $\{|M_i, K_{M_i}\rangle\}$ is contained in the whole Hilbert space.

By substituting $|\Psi_f\rangle = Q_0 |\Psi_i\rangle$ into (A1.20b) it can be found that the vector $Q_0 |\Psi_i\rangle$ may be written as (Ref.[8], [Ref[5]])

$$Q_0 |\Psi_i\rangle = \sum_{K_i} B_{M_i,K_i} |M_i, K_i\rangle \tag{A1.21}$$

where the index $K_i$ runs over all the orthonormal eigenbase vectors $\{|M_i, K_i\rangle\}$ which correspond to the same eigenvalue $M_i$. The equation (A1.20b) or (A1.21) shows that the eigenvector $|\Psi_f\rangle = Q_0 |\Psi_i\rangle$ belongs to the subspace $\{|M_i, K_i\rangle\}$. Now the equations (A1.20a) and (A1.21) together show that when an arbitrary vector $|\Psi_i\rangle$ of the subspace $\{|M_i, K_i\rangle\}$ is acted on by any zero-order quantum transition operator $Q_0$, the generated vector $Q_0 |\Psi_i\rangle$ still belongs to the same subspace $\{|M_i, K_i\rangle\}$. Consequently the subspace $\{|M_i, K_i\rangle\}$ is an invariant subspace under any zero-order quantum transition operators.

The above theoretical analysis leads to the following conclusions. All these orthonormal degenerate eigenbase vectors $\{|M_r, K_{M_r}\rangle\}$ with the same eigenvalue $M_r$ evidently constitute a subset of the complete set $\{|M, K_M\rangle\}$ of base vectors of the Hilbert space of the spin system, and this subset $\{|M_r, K_{M_r}\rangle\}$ may be characterized simply by the total spin magnetic quantum number $M_r$ which takes any one of the values $-I, -I+1, ..., I-1, I$ with the total spin quantum number $I$ given by (A1.3). This subset $\{|M_r, K_{M_r}\rangle\}$ can span a direct-sum subspace of the Hilbert space of the spin system [7] and moreover, according to quantum mechanics [3] this subspace is invariant under the set of zero-quantum transition operators.

As a typical example, the direct-sum subspace $S_{zq}(k)$ with $k = n/2 - M_r = 0, 1, 2, ..., n$ of the Hilbert space $HS(N)$ of the $n-spin-1/2$ system [7] is spanned by the vector basis subset $\{|M_r, K_{M_r}\rangle\}$ where the total spin magnetic quantum number $M_r$ takes one of the $2I+1$ different values $-I, -I+1, ..., I-1, I$ with the total spin quantum number $I = n/2$. Therefore, there are $n+1$ (i.e., $2I+1$) direct-sum subspaces $\{S_{zq}(k)\}$ of the Hilbert space $HS(N)$ of the $n-spin-1/2$ system.

According to the formal definition (A1.5) all the $p-$order quantum transition operators $\{Q_p\}$ with the same order $p$ form a linear space of operators, namely the linear space of the $p-order$ quantum transition operators. The linear space of the $p-order$ quantum transition operators with order $p \neq 0$ is not closed under multiplication operation, because the product of any two $p-$order quantum transition operators with orders $p \neq 0$ does not belong to the original linear space of the $p-order$ quantum transition operators. As a typical example, in a single spin$-I$ particle with $I \geq 1$ the operators $Q_{\pm1} = I^{\pm}$ are the $\pm1-$order quantum transition operators, while the square operators $Q_{+2} = (I^+)^2$ and $Q_{-2} = (I^-)^2$ are the two-order quantum transition operators with order $p = +2$ and $-2$, respectively. Then the operators $Q_{\pm1} = I^{\pm}$ belong to the linear spaces of the $\pm1-$order quantum transition operators, respectively, while the square operators $Q_{+2} = (I^+)^2$ and $Q_{-2} = (I^-)^2$ belong to the linear spaces of the $+2-$order and $-2-$order quantum transition operators, respectively. Therefore, the square operators $Q_{\pm 2} = (I^{\pm})^2$ do not belong to the original linear spaces of the $\pm 1-$order quantum transition operators, respectively. However, an exception is the linear space of the zero-quantum operators which is formed by all the zero-order quantum transition operators with order $p = 0$ and all the diagonal operators which are not any *zero*$-$order quantum transition operators. This linear space of the zero-quantum operators is closed under multiplication operation [4, 8]. Therefore, the linear space of the zero-quantum operators may be renamed the zero-quantum operator algebra space. It is an operator subspace of the multiple-quantum operator algebra space [4]. The zero-quantum operator (algebra) subspace is formed by all the diagonal operators $\{Q_z\}$ and all the zero-order quantum transition operators $\{Q_{zq}\}$. However, all the zero-order quantum transition operators $\{Q_{zq}\}$ alone cannot form a closed operator algebra space.

A rigorous theoretical proof for the zero-quantum operator algebra subspace is described below. Suppose that $Q_{0k}$ and $Q_{0l}$ are any two zero-order quantum transition operators with orders $p = 0$. Let $|\Psi_i\rangle$ be an arbitrary eigenvector of the total spin operator $I_z$ with the eigenvalue $M_i$. By applying any one of the two ZQT operators $Q_{0k}$ and $Q_{0l}$ to the initial eigenvector $|\Psi_i\rangle$ it can be obtained from (A1.21) that

$$Q_{0\lambda}\,|\Psi_i\rangle = \sum_{K_i} B^{\lambda}_{M_i,K_i}\,|M_i,K_i\rangle\,,\ \lambda = k,l \tag{A1.22}$$

In particular, when the initial eigenvector $|\Psi_i\rangle$ is taken as any eigenbase vector $|M_i,K_r\rangle$ of the total spin operator $I_z$ with the eigenvalue $M_i$, it can be found from (A1.22) that

$$Q_{0\lambda}\,|M_i,K_r\rangle = \sum_{K_i} C^{\lambda}_{M_i,K_r,K_i}\,|M_i,K_i\rangle\,,\ \lambda = k,l \tag{A1.23}$$

where $C^{\lambda}_{M_i,K_r,K_i}$ is an expansional coefficient. If now the product of the two ZQT operators $Q_{0k}$ and $Q_{0l}$, i.e., the operator $Q_{0k}Q_{0l}$, acts on the initial eigenvector $|\Psi_i\rangle$ , then the following equation can be derived from (A1.22) and (A1.23) [8]:

$$Q_{0k}Q_{0l}\,|\Psi_i\rangle = \sum_{K_i}\sum_{K_r} B^{l}_{M_i,K_r} C^{k}_{M_i,K_r,K_i}\,|M_i,K_i\rangle\,. \tag{A1.24}$$

Let $B_{M_i,K_i} = \sum_{K_r} B^{l}_{M_i,K_r} C^{k}_{M_i,K_r,K_i}$. Then the equation (A1.24) can be reduced to the form

$$Q_{0k}Q_{0l}\,|\Psi_i\rangle = \sum_{K_i} B_{M_i,K_i}\,|M_i,K_i\rangle \tag{A1.25}$$

By comparing (A1.25) with (A1.21) and noting that $|\Psi_i\rangle$ is an arbitrary eigenvector of the total spin operator $I_z$ with the eigenvalue $M_i$ it is deduced that the operator $Q_{0k}Q_{0l}$ should be a ZQT operator $Q_0$. Actually, by applying the total spin operator $I_z$ to both sides of the equation (A1.25) and noting that there is the eigenvalue equation $I_z |M_i, K_i\rangle = M_i |M_i, K_i\rangle$ one can obtain

$$I_z (Q_{0k}Q_{0l}) |\Psi_i\rangle = M_i (Q_{0k}Q_{0l}) |\Psi_i\rangle \tag{A1.26}$$

Then by comparing the equation (A1.26) with the formal definition (A1.5) of a zero-order quantum transition operator $Q_0$ with the order $p = 0$ it can be deduced that the operator $Q_{0k}Q_{0l}$ may be a zero-order quantum transition operator $Q_{zq}$. However, there is also another possibility that the operator $Q_{0k}Q_{0l}$ may be a diagonal operator $Q_z$. This can be explained below.

Generally, it is easy to prove that an arbitrary diagonal operator $Q_z$ obeys the equation:

$$I_z Q_z |\Psi_i\rangle = M_i Q_z |\Psi_i\rangle . \tag{A1.27}$$

Any diagonal operator $Q_z$ commutes with the total spin operator $I_z$. By acting any diagonal operator $Q_z$ on both sides of the eigenvalue equation (A1.4) of the total spin operator $I_z$ one can directly obtain the equation (A1.27), that is, $Q_z I_z |\Psi_i\rangle = I_z Q_z |\Psi_i\rangle = M_i Q_z |\Psi_i\rangle$ . The eigenvalue equation (A1.27) holds generally for any diagonal operator $Q_z$ of a general spin system. Now it can be found that the definition (A1.19) for a zero-order quantum transition operator $Q_0$ is formally the same as the equation (A1.27) for a diagonal operator $Q_z$. Therefore, according to the formal definition (A1.5) alone it is impossible to distinguish a zero-order quantum transition operator $Q_{zq}$ (an off-diagonal operator) from a diagonal operator $Q_z$. In theory a diagonal operator may be treated like a zero-order quantum transition operator, and this leads to that it becomes simple for the theoretical treatment of the strongly-coupled spin$-1/2$ systems [Ref[5]] in nuclear magnetic resonance spectroscopy.

It can be easily shown that all the diagonal operators $\{Q_z\}$ can form a diagonal operator algebra space [4]. This diagonal operator algebra space is an operator subspace of the multiple-quantum operator algebra space [4] and as usual it is called the LOMSO operator subspace.

Both the definition (A1.19) for a zero-order quantum transition operator $Q_{zq}$ and the equation (A1.27) for a diagonal operator $Q_z$ need to be taken into account at the same time. This can be realized if the formal definition (A1.5) of a *zero*$-$order quantum transition operator $Q_0$ with the order $p = 0$ is extended to include the equation (A1.27) of any diagonal operator $Q_z$. Consequently the operator $Q_0$ in the extended definition (A1.5) with the order $p = 0$ may take not only any zero-order quantum transition operator $Q_{zq}$ but also any diagonal operator $Q_z$. For convenience such an operator $Q_0$ may be called the zero-quantum operator in unified form. Below it can be shown on the basis of both the definition (A1.19) and the equation (A1.27) that (*i*) the direct-sum subspace $\{|M_r, K_{M_r}\rangle\}$ is invariant under the set of the zero-order quantum transition operators $\{Q_{zq}\}$ and the diagonal operators $\{Q_z\}$ and (*ii*) all the zero-order quantum transition operators $\{Q_{zq}\}$ and all the diagonal operators $\{Q_z\}$ together can form a zero-quantum operator algebra space which is an operator subspace of the multiple-quantum operator algebra space [4].

It can be deduced that the zero-quantum operator $Q_0$ in the extended definition (A1.5) with order $p = 0$ may be a zero-order quantum transition operator $Q_{zq}$, a diagonal operator $Q_z$, or generally a linear combination of both the diagonal operators $\{Q_z\}$ and the zero-order quantum transition operators $\{Q_{zq}\}$. Therefore, generally the zero-quantum operator $Q_0$ in the extended definition (A1.5) may be expressed as

$$Q_0 = \sum_k a_k Q_z^{(k)} + \sum_k b_k Q_{zq}^{(k)} \tag{A1.28}$$

where $\{Q_{zq}^{(k)}\}$ and $\{Q_z^{(k)}\}$ are all the linearly independent zero-order quantum transition operators and all the linearly independent diagonal operators, respectively. Moreover, on the basis of the extended definition (A1.5) of a zero-quantum operator $Q_0$ it can prove that any zero-quantum operator $Q_0$ of (A1.28) acting on an arbitrary eigenvector $|\Psi_i\rangle$ of the total spin operator $I_z$ with the eigenvalue $M_i$ still can be described by (A1.21), that is, the generated vector $Q_0\,|\Psi_i\rangle$ still can be expanded in terms of the orthonormal eigenbase vectors $\{|M_i, K_i\rangle\}$ which correspond to the same eigenvalue $M_i$. Then the equations (A1.20a) and (A1.21) together show that when an arbitrary vector $|\Psi_i\rangle$ of the subspace $\{|M_i, K_i\rangle\}$ is acted on by any zero-quantum operator $Q_0$, the generated vector $Q_0\,|\Psi_i\rangle$ still belongs to the same subspace $\{|M_i, K_i\rangle\}$. This indicates that the direct-sum subspace $\{|M_i, K_i\rangle\}$ is invariant under the set of the zero-quantum operators $\{Q_0\}$ of (A1.28) which is comprised of all the zero-order quantum transition operators $\{Q_{zq}^{(k)}\}$ and all the diagonal operators $\{Q_z^{(k)}\}$.

As shown by the formal definition (A1.5) of a $p-$order quantum transition operator $Q_p$, all the $p-$order quantum transition operators $\{Q_p\}$ with the same order $p$ can form a linear space of the $p - order$ quantum transition operators. Now, on the one hand, according to the extended definition (A1.5) for a zero-quantum operator $Q_0$ it can be shown that all the *zero*−quantum operators $\{Q_0\}$ with the same order $p = 0$ can form a linear space of the *zero*−quantum operators. On the other hand, by starting from the equation (A1.21) for a zero-quantum operator $Q_0$ it can be shown that for any two zero-quantum operators $Q_{0k}$ and $Q_{0l}$ these equations (A1.22)–(A1.25) still hold; and for the product of the two zero-quantum operators, i.e., the operator $Q_{0k}Q_{0l}$, these equations (A1.24)–(A1.26) still hold. By comparing (A1.26) with the extended definition (A1.5) for a zero-quantum operator $Q_0$ with order $p = 0$ it can be found that the operator $Q_{0k}Q_{0l}$ is a zero-quantum operator $Q_0$ which can be generally written as (A1.28). Therefore, the product of any two zero-quantum operators $Q_{0k}$ and $Q_{0l}$, i.e., the zero-quantum operator $Q_{0k}Q_{0l}$, can be generally expressed as

$$Q_{0k}Q_{0l} = \sum_j a_j Q_z^{(j)} + \sum_j b_j Q_{zq}^{(j)} \tag{A1.29}$$

This equation clearly shows that the product of any two zero-quantum operators $Q_{0k}$ and $Q_{0l}$ is still a zero-quantum operator and it may be a zero-order quantum transition operator $Q_{zq}$, a diagonal operator $Q_z$, or generally a linear combination of all the linearly-independent diagonal operators $\{Q_z^{(j)}\}$ and all the linearly-independent zero-order quantum transition operators $\{Q_{zq}^{(j)}\}$. Now all the $zero-$quantum operators $\{Q_0\}$ with the same order $p = 0$ form a linear space of the $zero-$quantum operators. Moreover, the equation (A1.29) shows that this linear space of the $zero-$quantum operators is closed under the multiplication operation. Therefore, the linear space of the $zero-$quantum operators is a zero-quantum operator algebra space. The zero-quantum operator algebra space is formed by all the zero-order quantum transition operators and all the diagonal operators. It is exactly the zero-quantum operator subspace of the multiple-quantum operator algebra space [4]. Note that all the diagonal operators form the LOMSO operator subspace of the multiple-quantum operator algebra space. Therefore, the zero-quantum operator subspace contains the whole LOMSO operator subspace in the multiple-quantum operator algebra space [4].

## A1.4. The MQT product operators

The formal definition (A1.5) for a $p-$order multiple- quantum- transition (MQT) operator $Q_p$ is not dependent upon any detailed vector basis set of the Hilbert space of a quantum spin system under study, although it employs an arbitrary eigenvector $|\Psi_i\rangle$ of the total spin operator $I_z$ of the spin system. A non-selective $p-$order MQT operator $Q_p$ is even independent of any eigenvector $|\Psi_i\rangle$ in the formal definition (A1.5). As shown in the previous Subsection A1.2, these MQT operators $I_j^{\pm} I_l^{\pm}$, $I_j^{\pm} I_l^{\mp}$, and $I_k^{\pm}$ with $k = j, l$ are usually non-selective in the $n-spin-1/2$ system. They are independent of any eigenvector $|\Psi_i\rangle$ . Then a question arises why it needs to put emphasis on the complete set $\{|M, K_M\rangle\}$ (or $\{|M\rangle\}$) of the Hilbert space which is formed by all the orthonormal eigenbase vectors of the total spin operator $I_z$. A $p-$order quantum transition operator $Q_p$ is formally defined on the basis of the representation that is defined by the total spin operator $I_z$. The complete set $\{|M, K_M\rangle\}$ is a vector basis set of the Hilbert space of the spin system. There are a number of choices for the vector basis set $\{|M, K_M\rangle\}$, since in addition to the basic quantum number, i.e., the total spin magnetic quantum number $M$, the eigenbase vectors $\{|M, K_M\rangle\}$ of the total spin operator $I_z$ are also specified by the quantum number $K_M$ which distinguishes between different degenerate eigenbase vectors $\{|M, K_M\rangle\}$. As shown in the previous Subsections, a suitable vector basis set $\{|M, K_M\rangle\}$ may make the relevant theoretical treatments greatly simplified. However, a great reason for choosing suitably the vector basis set $\{|M, K_M\rangle\}$ is that the selective MQT operators and the subspace-selective MQT operators [7] are closely related to the vector basis set $\{|M, K_M\rangle\}$ of the Hilbert space of the spin system (See the Section 5). Therefore, a good choice for the vector basis set $\{|M, K_M\rangle\}$ is very important.

There are a number of choices for the complete set $\{|M, K_M\rangle\}$ of base vectors of the Hilbert space and here the total spin magnetic quantum number $M$ is basic for all these choices. As an example, if the total spin operator $I_z$ does not commute with the spin Hamiltonian $H_s$ of the spin system, then how to choose suitably the complete set $\{|M, K_M\rangle\}$? Suppose that the spin Hamiltonian $H_s$ is divided into the sum of the main term $H_s^0$ which commutes with the total spin operator $I_z$ and the perturbation term $H_s^1$ which does not commute with $I_z$, that is, $H_s = H_s^0 + H_s^1$. Then the complete set $\{|M, K_M\rangle\}$ could be conveniently chosen as the complete set of the common eigenbase vectors of both the main term $H_s^0$ and the total spin operator $I_z$.

With the help of the spin angular momentum theory in quantum mechanics [2] and especially the theoretical aspect for the combination (or addition) of spin angular momenta of many distinct spin particles $[9, 10]$ the complete sets $\{|M, K_M\rangle\}$ may be conveniently chosen in a quantum spin system. According to the spin angular momentum theory a composite multiple-spin system may be specified by a set of commuting operators which contains two or more commuting spin angular momentum operators. As a typical example, consider a $two - spin - I_k$ system. On the one hand, the set of commuting operators may consist of the square of the total spin angular momentum operator $\mathbf{I} = \mathbf{I}_1 + \mathbf{I}_2$, i.e., $\mathbf{I}^2$ and the total $z-$component spin operator $I_z = I_{1z} + I_{2z}$ of the spin system. This set is simply denoted by $\{\mathbf{I}^2, I_z\}$. On the other hand, the set of commuting operators also may consist of the square of the spin angular momentum operators, i.e., $\{\mathbf{I}_k^2\}$, and the $z-$component spin operators $\{I_{kz}\}$ of the two individual spin$-I_k$ particles of the two-spin$-I_k$ system for $k = 1, 2$. This set is simply denoted by $\{\mathbf{I}_k^2, I_{kz}\} = \{\mathbf{I}_1^2, I_{1z}; \mathbf{I}_2^2, I_{2z}\}$. On the one side, according to the angular momentum theory in quantum mechanics [2,9,10] the set $\{\mathbf{I}^2, I_z\}$ of commuting operators may define a representation which may be called the coupled representation, and the eigenvalue equations are written as $\mathbf{I}^2 |I, M\rangle = I (I + 1) |I, M\rangle$ and $I_z |I, M\rangle = M |I, M\rangle$, respectively. Here the orthonormal eigenbase vectors $\{|I, M\rangle\}$ are the common eigenbase vectors of the commuting-operator set $\{\mathbf{I}^2, I_z\}$ and they constitute a vector basis set of the Hilbert space of the two-spin$-I_k$ system. Obviously, this coupled representation is the one defined by the total spin operator $I_z$ and hence the orthonormal vector basis set $\{|I, M\rangle\}$ is naturally a complete set $\{|M, K_M\rangle\}$. On the other side, the set $\{\mathbf{I}_k^2, I_{kz}\}$ of commuting operators may define another representation which may be called the uncoupled representation, and the eigenvalue equations are written as $\mathbf{I}_k^2 |I_k, m_k\rangle = I_k (I_k + 1) |I_k, m_k\rangle$ and $I_{kz} |I_k, m_k\rangle = m_k |I_k, m_k\rangle$ for $k = 1, 2$, respectively. Then in this uncoupled representation the tensor-product base vectors $\{|I_1, m_1\rangle |I_2, m_2\rangle\}$ of the two-spin$-I_k$ system are the common eigenbase vectors of the commuting-operator set $\{\mathbf{I}_k^2, I_{kz}\}$ and constitute a vector basis set of the Hilbert space of the spin system. This orthonormal vector basis set $\{|I_1, m_1\rangle |I_2, m_2\rangle\}$ also may be chosen as a complete set $\{|M, K_M\rangle\}$, because the total spin operator $I_z = I_{1z} + I_{2z}$ commutes with every operator of the set $\{\mathbf{I}_k^2, I_{kz}\}$ of commuting operators. Here the tensor-product base vectors $\{|I_1, m_1\rangle |I_2, m_2\rangle\}$ are also the common eigenbase vectors of both the total spin operator $I_z$ and the commuting-operator set $\{\mathbf{I}_k^2, I_{kz}\}$. The above theoretical analysis for a $two-spin-I_k$ system can be generalized for a general $n-spin-I_k$ system. Generally, on the one hand, in the $n - spin - I_k$ system the coupled representation is still defined by the commuting-operator set $\{\mathbf{I}^2, I_z\}$, where the total spin angular momentum operator $\mathbf{I} = \sum_{k=1}^{n} \mathbf{I}_k$ and the total $z-$component spin operator $I_z = \sum_{k=1}^{n} I_{kz}$ are just given by (A1.1) and (A1.2), respectively. Then the orthonormal vector basis set $\{|I, M\rangle\}$ which is comprised of the common eigenbase vectors of the commuting-operator set $\{\mathbf{I}^2, I_z\}$ may be naturally chosen as a complete set $\{|M, K_M\rangle\}$. On the other hand, the uncoupled representation is defined by the commuting-operator set $\{\mathbf{I}_k^2, I_{kz}\}$ with $k = 1, 2, ..., n$. In this uncoupled representation the orthonormal vector basis set of the Hilbert space of the $n - spin - I_k$ system is formed by the tensor-product base vectors $\{|I_1, m_1\rangle |I_2, m_2\rangle ... |I_n, m_n\rangle\}$ of the spin system. Then this orthonormal vector basis set $\{|I_1, m_1\rangle |I_2, m_2\rangle ... |I_n, m_n\rangle\}$ also may be chosen as a complete set $\{|M, K_M\rangle\}$, because the total spin operator $I_z = \sum_{k=1}^{n} I_{kz}$ commutes with every operator of the commuting-operator set $\{\mathbf{I}_k^2, I_{kz}\}$ with $k = 1, 2, ..., n$. Here the tensor-product base vectors $\{|I_1, m_1\rangle |I_2, m_2\rangle ... |I_n, m_n\rangle\}$ also are the common eigenbase vectors of both the total spin operator $I_z$ and the commuting-operator set $\{\mathbf{I}_k^2, I_{kz}\}$.

With the help of the tensor product method and the angular momentum theory in quantum mechanics [2,9,10] the complete set $\{|M, K_M\rangle\}$ may be constructed in a composite $n-$spin$-I_k$ system. On the one hand, the Hilbert space of the $n-$spin$-I_k$ system may be spanned by the complete set $\{|M, K_M\rangle\}$. On the other hand, according to quantum mechanics the Hilbert space also is the tensor product of the component Hilbert spaces of the $n$ individual spin$-I_k$ particles of the composite $n-spin-I_k$ system. Suppose that $\{|\varphi_k\rangle\}$ is any complete set of the orthonormal base vectors of the component Hilbert space of the $k-$th individual spin$-I_k$ particle of the $n - spin - I_k$ system for $k = 1, 2, ..., n$. Then the tensor product of these $n$ complete sets $\{|\varphi_k\rangle\}$ of the $n$ component Hilbert spaces of the $n$ spin$-I_k$ particles of the $n - spin - I_k$ system can form a complete set $\{|\Phi_l\rangle\}$ of the orthonormal base vectors of the Hilbert space of the spin system,

$$\{|\Phi_l\rangle\} = \{|\varphi_1\rangle\} \bigotimes \{|\varphi_2\rangle\} \bigotimes ... \bigotimes \{|\varphi_n\rangle\}, \tag{A1.30a}$$

where the orthonormal base vector $|\Phi_l\rangle$ may be written as

$$|\Phi_l\rangle = |\varphi_1\rangle \bigotimes |\varphi_2\rangle \bigotimes ... \bigotimes |\varphi_n\rangle = |\varphi_1\rangle |\varphi_2\rangle ... |\varphi_n\rangle \tag{A1.30b}$$

The base vector $|\Phi_l\rangle$ is called the tensor-product base vector (or the direct-product base vector) and $\{|\Phi_l\rangle\}$ is the complete set of the tensor-product base vectors of the Hilbert space of the composite $n - spin - I_k$ system. Generally speaking, the complete set $\{|\Phi_l\rangle\}$ of the tensor-product base vectors may be different from the complete set $\{|M, K_M\rangle\}$ of the orthonormal eigenbase vectors of the total spin operator $I_z$. However, both the complete sets $\{|\Phi_l\rangle\}$ of (A1.30a) and $\{|M, K_M\rangle\}$ belong to the same Hilbert space of the $n - spin - I_k$ system. Therefore, the complete set $\{|\Phi_l\rangle\}$ can be changed to $\{|M, K_M\rangle\}$ and vice versa by a suitable unitary transformation. One important point is that it is relatively easy to construct any tensor-product base vector $|\Phi_l\rangle$ and the complete set $\{|\Phi_l\rangle\}$ and moreover, the complete set $\{|\Phi_l\rangle\}$ is relatively simple. Then by the unitary transformation the complete set $\{|M, K_M\rangle\}$ may be obtained from the complete set $\{|\Phi_l\rangle\}$.

As an application of both the tensor product method and the spin angular momentum theory in quantum mechanics, below consider a composite $n-spin-I_k$ system. Let the commuting operators $\mathbf{I}_k^2 = I_{kx}^2 + I_{ky}^2 + I_{kz}^2$ and $I_{kz}$ be the square of the spin angular momentum operator $\mathbf{I}_k$ and the $z-$component spin operator of the $k-$th individual spin$-I_k$ particle of the $n - spin - I_k$ system, respectively. According to the angular momentum theory in quantum mechanics [2, 9, 10] the eigenvalue equations for both the commuting operators $\mathbf{I}_k^2$ and $I_{kz}$ are given by $\mathbf{I}_k^2 \left|I_k, m_k\right\rangle = I_k \left(I_k+1\right) \left|I_k, m_k\right\rangle$ and $I_{kz} \left|I_k, m_k\right\rangle = m_k \left|I_k, m_k\right\rangle$ (in unit $\hbar = 1$), respectively, where $\left|I_k, m_k\right\rangle$ is any common eigenbase vector of the commuting operators $\mathbf{I}_k^2$ and $I_{kz}$ and is characterized by the spin quantum number $I_k$ and the spin magnetic quantum number $m_k$, here $m_k$ takes one of the $2I_k+1$ values $-I_k, -I_k+1, ..., I_k-1, I_k$. All the $2I_k+1$ orthonormal eigenbase vectors $\{\left|I_k, m_k\right\rangle\}$ form a complete set of the base vectors of the component Hilbert space of the $k-$th individual spin$-I_k$ particle of the composite $n - spin - I_k$ system for $k = 1, 2, ..., n$. By the tensor product of these $n$ complete sets $\{\left|I_k, m_k\right\rangle\}$ with $k = 1, 2, ..., n$ of the $n$ spin$-I_k$ particles of the $n-spin-I_k$ system one can obtain the complete set $\{\left|\Phi_l\right\rangle\}$ of the tensor-product base vectors of the Hilbert space of the composite $n-spin-I_k$ system, as shown by (A1.30a), where $\left|\varphi_k\right\rangle = \left|I_k, m_k\right\rangle$ for $k = 1, 2, ..., n$. Therefore, the complete set $\{\left|\Phi_l\right\rangle\}$ of (A1.30a) for the composite $n-spin-I_k$ system may be explicitly given by (See Refs.[2, 9, 10] and See also [Ref[5]])

$$\{\left|\Phi_l^z\right\rangle\} = \{\left|I_1, m_1\right\rangle\} \bigotimes \{\left|I_2, m_2\right\rangle\} \bigotimes ... \bigotimes \{\left|I_n, m_n\right\rangle\} \tag{A1.31}$$

where, as shown by (A1.30b), the orthonormal tensor-product base vector $\left|\Phi_l^z\right\rangle$ is explicitly written as

$$\left|\Phi_l^z\right\rangle = \left|I_1, m_1\right\rangle \bigotimes \left|I_2, m_2\right\rangle \bigotimes ... \bigotimes \left|I_n, m_n\right\rangle = \left|I_1, m_1\right\rangle \left|I_2, m_2\right\rangle ... \left|I_n, m_n\right\rangle \tag{A1.32}$$

Now with the help of the eigenvalue equations $I_{kz} \left|I_k, m_k\right\rangle = m_k \left|I_k, m_k\right\rangle$ for $k = 1, 2, ..., n$ it can be shown that any direct-product (or tensor-product) base vector $\left|\Phi_l^z\right\rangle$ of (A1.32) is an eigenbase vector of the total spin operator $I_z = \sum_{k=1}^n I_{kz}$ of the $n - spin - I_k$ system and obeys the eigenvalue equation:

$$I_z \left|\Phi_l^z\right\rangle = \left(I_{1z} + I_{2z} + ... + I_{nz}\right) \left|I_1, m_1\right\rangle \left|I_2, m_2\right\rangle ... \left|I_n, m_n\right\rangle = M_l \left|\Phi_l^z\right\rangle, \tag{A1.33}$$

where the eigenvalue $M_l = m_1 + m_2 + ... + m_n$ is the total spin magnetic quantum number of the $n-spin-I_k$ system. According to the angular momentum theory in quantum mechanics [2, 9, 10] it can be shown that the tensor-product base vectors $\{\left|\Phi_l^z\right\rangle\}$ are the common eigenbase vectors of the commuting-operator set $\{\mathbf{I}_k^2, I_{kz}\}$ with $k = 1, 2, ..., n$ in the $n-spin-I_k$ system. Note that the total spin operator $I_z$ is given by $I_z = \sum_{k=1}^n I_{kz}$ for the $n-spin-I_k$ system. Then it is easy to prove that $\left[I_z, \mathbf{I}_k^2\right] = 0$ and $\left[I_z, I_{kz}\right] = 0$ for $k = 1, 2, ..., n$, indicating that the total spin operator $I_z$ commutes with every operator of the commuting-operator set $\{\mathbf{I}_k^2, I_{kz}\}$ with $k = 1, 2, ..., n$. Therefore, it can be seen that the tensor-product base vectors $\{\left|\Phi_l^z\right\rangle\}$ are the common eigenbase vectors of both the total spin operator $I_z$ and the commuting-operator set $\{\mathbf{I}_k^2, I_{kz}\}$. As shown previously, in the $n-spin-I_k$ system the uncoupled representation is defined by the commuting-operator set $\{\mathbf{I}_k^2, I_{kz}\}$ with $k = 1, 2, ..., n$. In this uncoupled representation the tensor-product base vectors $\{|I_1, m_1\rangle\, |I_2, m_2\rangle\, ...\, |I_n, m_n\rangle\}$, which are just $\{|\Phi_l^z\rangle\}$ of (A1.31), constitute a vector basis set of the Hilbert space of the $n-spin-I_k$ system. The latter may be chosen as a complete set $\{|M, K_M\rangle\}$ and this can be confirmed by the eigenvalue equation (A1.33).

Obviously, these formulae (A1.31)-(A1.33) are available for a composite $n-spin-1/2$ system with every spin quantum number $I_k = 1/2$. Then the complete set $\{|\Phi_l^z\rangle\}$ of (A1.31) of the orthonormal tensor-product base vectors may act as a complete set $\{|M, K_M\rangle\}$ of the orthonormal eigenbase vectors of the total spin operator $I_z$ of the Hilbert space of the $n - spin - 1/2$ system ([Ref[5]], Ref.[7,8]). A large advantage to choose the complete set $\{|\Phi_l^z\rangle\}$ of (A1.31) as the complete set $\{|M, K_M\rangle\}$ is that these tensor-product base vectors $\{|\Phi_l^z\rangle\}$ are easier to construct and moreover, any tensor-product base vector $|\Phi_l^z\rangle$ of (A1.32) is simpler. There are a lot of $n - spin - 1/2$ systems which include the non-interacting $n-$spin$-1/2$ systems and the interacting $n - spin - 1/2$ systems whose interacting terms are the diagonal operators such as $2I_{kz}I_{lz}$, etc., in each one of which the spin Hamiltonian $H_s$ commutes with the total spin operator $I_z$ and moreover, the tensor-product base vectors $\{|\Phi_l^z\rangle\}$ are the common eigenbase vectors of both the spin Hamiltonian $H_s$ and the total spin operator $I_z$.

A $p-$order quantum transition operator is formally defined by (A1.5) and does not explicitly depend upon any detailed complete set $\{|M, K_M\rangle\}$. However, in the relevant theoretical treatments the eigenfunction expansions (See (A1.7) and (A1.8)) for the initial $|\Psi_i\rangle$ and the final eigenvector $Q_p\,|\Psi_i\rangle$ in the definition (A1.5) need to employ a complete set $\{|M, K_M\rangle\}$. If different complete sets $\{|M, K_M\rangle\}$ are used in the relevant theoretical treatments, what happens? Suppose that a $p-$order quantum transition operator may be constructed with the help of the formal definition (A1.5) equipped with a complete set $\{|M, K_M\rangle\}$ which takes the complete set $\{|I, M\rangle\}$ of the common eigenbase vectors of the commuting-operator set $\{\mathbf{I}^2, I_z\}$, and such constructed $p-$order quantum transition operator is denoted by $Q_{p1}$. In the meantime a $p-$order quantum transition operator also may be constructed by the definition (A1.5) equipped with another complete set $\{|M, K_M\rangle\}$ which takes the complete set $\{|\Phi_l^z\rangle\}$ of (A1.31) of the tensor-product base vectors which are the common eigenbase vectors of the commuting-operator set $\{\mathbf{I}_k^2, I_{kz}\}$, and such constructed $p-$order quantum transition operator is denoted by $Q_{p2}$. Then it can be shown that although both the $p-$order quantum transition operators $Q_{p1}$ and $Q_{p2}$ may be different from one another, they have the same quantum-transition order $p$ and belong to the same linear space of the $p-$order quantum transition operators which is comprised of all the $p-$order quantum transition operators with the same order $p$. These properties of the $p-$order quantum transition operators exist generally not only for the detailed complete sets $\{|I, M\rangle\}$ and $\{|\Phi_l^z\rangle\}$ but also for any complete set $\{|M, K_M\rangle\}$ of the eigenbase vectors of the total spin operator $I_z$.

Below the formal definition (A1.5) of a $p-$order quantum transition operator is used to analyze theoretically the MQT product operators $\{O_S\}$ of the multiple-quantum operator algebra space of an $n - spin - 1/2$ system in the Subsection 3.2.1. Here the complete set $\{|M, K_M\rangle\}$ is simply taken as the complete set $\{|\Phi_l^z\rangle\}$ of (A1.31) of the orthonormal tensor-product base vectors of the Hilbert space of the composite $n - spin - 1/2$ system. It is known from the Subsection 3.2.1 that the MQT product operators $\{O_S\}$ of (3.24) form a complete set of base operators of the multiple-quantum operator algebra space of the $n - spin - 1/2$ system and are given by

$$O_S = S_1^{-k_1,+l_1} S_2^{-k_2,+l_2} ... S_j^{-k_j,+l_j} ... S_n^{-k_n,+l_n} \tag{A1.34}$$

where the operator basis subset $\{S_j^{-k_j,+l_j}\}$ with $k_j, l_j = 0, 1$ of the $j-$th $spin-1/2$ particle of the $n - spin - 1/2$ system for $j = 1, 2, ..., n$ are given by (3.23):

$$S_j^{-0,+0} = \frac{1}{2}E_j + I_{jz},\ S_j^{-1,+1} = \frac{1}{2}E_j - I_{jz},\ S_j^{-0,+1} = I_j^+,\ S_j^{-1,+0} = I_j^- \tag{A1.35}$$

It can be deduced directly from (A1.35) that every base operator $S_j^{-k_j,+l_j}$ with $k_j = l_j = 0, 1$ is a diagonal operator and satisfies the relation $-k_j + l_j = 0$, while the off-diagonal base operator $S_j^{-k_j,+l_j}$ is a spin raising operator $I_j^+$ if it satisfies the relation $-k_j + l_j = +1$ and it is a spin lowering operator $I_j^-$ if it satisfies the relation $-k_j + l_j = -1$. Both the operators $I_j^\pm$ are the $\pm 1-$order MQT operators, respectively.

Now by applying an arbitrary base operator $S_j^{-k_j,+l_j}$ to any tensor-product base vector $|\Phi_l^z\rangle$ of (A1.32) one can set up the transformational equation for the base operator $S_j^{-k_j,+l_j}$. If the diagonal base operator $S_j^{-k_j,+l_j}$ with $k_j = l_j = 0, 1$ in (A1.35) acts on any direct-product base vector $|\Phi_l^z\rangle$, then the transformational equations for the diagonal base operators $S_j^{-k_j,+k_j}$ for $j = 1, 2, ..., n$ may be written as

$$S_j^{-k_j,+k_j} |\Phi_l^z\rangle = \left(\frac{1}{2} + (-1)^{k_j} m_j\right) |I_1, m_1\rangle ... |I_j, m_j\rangle ... |I_n, m_n\rangle \tag{A1.36}$$

where the diagonal base operator $S_j^{-k_j,+k_j}$ with $k_j = 0, 1$ is explicitly given in (A1.35), the eigenvalue equation for the spin operator $I_{jz}$ is $I_{jz} |I_j, m_j\rangle = m_j |I_j, m_j\rangle$ with $m_j = \pm 1/2$, the unity operator $E_j$ of the $j-$th spin$-1/2$ particle satisfies $E_j |I_j, m_j\rangle = |I_j, m_j\rangle$ . In analogous way the transformational equations can be set up for the non-diagonal base operators $S_j^{-k_j,+l_j}$ with $k_j, l_j = 0, 1$ and $k_j \neq l_j$ acting on any direct-product base vector $|\Phi_l^z\rangle$ and may be written as

$$\begin{aligned} S_j^{-k_j,+l_j} |\Phi_l^z\rangle &= \sqrt{I_j(I_j+1) - m_j(m_j + (-k_j + l_j))} |I_1, m_1\rangle ... \\ &\bigotimes |I_{j-1}, m_{j-1}\rangle |I_j, m_j + (-k_j + l_j)\rangle |I_{j+1}, m_{j+1}\rangle ... |I_n, m_n\rangle \end{aligned} \tag{A1.37}$$

These transformational equations can be proven as follows. As shown in (A1.35), the non-diagonal base operator $S_j^{-k_j,+l_j}$ is either the spin raising operator $I_j^+$ or the spin lowering operator $I_j^-$. According to the angular momentum theory in quantum mechanics [2,9,10] there are the basic transformational equations (A1.14) for the spin raising (or lowering) operator $I_j^+$ (or $I_j^-$), i.e., $I_j^{\pm}|I_j,m_j\rangle = \sqrt{I_j(I_j+1)-m_j(m_j\pm 1)}\,|I_j,m_j\pm 1\rangle$ for $j=1,2,...,n$. If the base operator $S_j^{-k_j,+l_j}=I_j^+$, then it can be found from (A1.35) that the index value $-k_j+l_j=+1$ and if $S_j^{-k_j,+l_j}=I_j^-$, then $-k_j+l_j=-1$. Therefore, there is one-to-one correspondence between the base operators $S_j^{-k_j,+l_j}=I_j^{\pm}$ and the index values $-k_j+l_j=\pm 1$. If now the non-diagonal base operator $S_j^{-k_j,+l_j}$ acts on the eigenbase vector $|I_j,m_j\rangle$ of the spin operator $I_{jz}$ of the $j-$th spin$-1/2$ particle, then with the help of this one-to-one correspondence it can be found that the basic transformational equations for the non-diagonal base operators $S_j^{-k_j,+l_j}=I_j^{\pm}$ may be generally written as

$$S_j^{-k_j,+l_j}|I_j,m_j\rangle = \sqrt{I_j(I_j+1)-m_j\left(m_j+(-k_j+l_j)\right)} \times |I_j,m_j+(-k_j+l_j)\rangle \qquad \text{(A1.38)}$$

where $k_j,l_j=0,1$ and $-k_j+l_j=\pm 1$ for $j=1,2,...,n$. It can be checked that the basic transformational equations (A1.38) for the non-diagonal base operators $S_j^{-k_j,+l_j}$ with $-k_j+l_j=\pm 1$ are really equal to the basic transformational equations (A1.14) for the operators $I_j^{\pm}$. The basic transformational equations (A1.38) can directly lead to the transformational equations (A1.37) for the non-diagonal base operators $S_j^{-k_j,+l_j}$ with $-k_j+l_j=\pm 1$ acting on any direct-product base vector $|\Phi_l^z\rangle$.

For convenience the transformational equations (A1.36) for the diagonal base operators $S_j^{-k_j,+l_j}$ with $k_j=l_j$ may be rewritten as

$$S_j^{-k_j,+l_j}|\Phi_l^z\rangle = \left(\frac{1}{2}+(-1)^{k_j}m_j\right)|I_1,m_1\rangle\,...\,|I_{j-1},m_{j-1}\rangle \bigotimes |I_j,m_j+(-k_j+l_j)\rangle\,|I_{j+1},m_{j+1}\rangle\,...\,|I_n,m_n\rangle \qquad \text{(A1.39)}$$

where $k_j=l_j=0,1$ and the index value $-k_j+l_j=0$ for $j=1,2,...,n$. By examining both the transformational equations (A1.37) and (A1.39) it can be found that the transformational equations for any base operators $S_j^{-k_j,+l_j}$ with $k_j,l_j=0,1$ and $j=1,2,...,n$ acting on any direct-product base vector $|\Phi_l^z\rangle$ may be written in the unified form

$$S_j^{-k_j,+l_j}|\Phi_l^z\rangle = g_j(k_j,l_j)\,|I_1,m_1\rangle\,...\,|I_{j-1},m_{j-1}\rangle \bigotimes |I_j,m_j+(-k_j+l_j)\rangle\,|I_{j+1},m_{j+1}\rangle\,...\,|I_n,m_n\rangle \qquad \text{(A1.40)}$$

where the coefficients $g_j(k_j,l_j)$ are given by

$$g_j(k_j,l_j)=g_j(k_j,k_j)=\left(\frac{1}{2}+(-1)^{k_j}m_j\right) \text{ for } k_j=l_j=0,1 \qquad \text{(A1.41a)}$$

$$g_j(k_j,l_j)=\sqrt{I_j(I_j+1)-m_j\left(m_j+(-k_j+l_j)\right)} \text{ for } k_j,l_j=0,1 \text{ and } k_j\neq l_j \qquad \text{(A1.41b)}$$

It is easy to find that the transformational equations (A1.40) are reduced to (A1.39) or (A1.36) for the diagonal base operators $S_j^{-k_j,+k_j}$ with $k_j = l_j = 0, 1$ if the coefficients $g_j(k_j, l_j)$ are given by (A1.41a) and they are reduced to (A1.37) for the non-diagonal base operators $S_j^{-k_j,+l_j}$ with $k_j \neq l_j$ and $k_j, l_j = 0, 1$ if the coefficients $g_j(k_j, l_j)$ are given by (A1.41b).

Now with the help of the transformational equations (A1.40) for any base operators $S_j^{-k_j,+l_j}$ with $k_j, l_j = 0, 1$ and $j = 1, 2, ..., n$ one can set up the transformational equations for the MQT product operator $O_S$ of (A1.34) acting on any tensor-product base vector $|\Phi_l^z\rangle$ of (A1.32),

$$\begin{aligned} O_S\,|\Phi_l^z\rangle &= S_1^{-k_1,+l_1} S_2^{-k_2,+l_2} ... S_j^{-k_j,+l_j} ... S_n^{-k_n,+l_n} \,|I_1, m_1\rangle\,|I_2, m_2\rangle\,...\,|I_n, m_n\rangle \\ &= g_1(k_1, l_1)\, g_2(k_2, l_2) ... g_j(k_j, l_j) ... g_n(k_n, l_n)\,|I_1, m_1 + (-k_1 + l_1)\rangle \\ &\bigotimes |I_2, m_2 + (-k_2 + l_2)\rangle\,...\,|I_j, m_j + (-k_j + l_j)\rangle\,...\,|I_n, m_n + (-k_n + l_n)\rangle \end{aligned} \tag{A1.42}$$

Then by applying the total spin operator $I_z = \sum_{k=1}^{n} I_{kz}$ to the generated vector $O_S\,|\Phi_l^z\rangle$ one can obtain

$$\begin{aligned} &I_z \left(S_1^{-k_1,+l_1} S_2^{-k_2,+l_2} ... S_j^{-k_j,+l_j} ... S_n^{-k_n,+l_n}\right) |I_1, m_1\rangle\,|I_2, m_2\rangle\,...\,|I_n, m_n\rangle \\ &= \left(\sum_{j=1}^{n} (m_j + (-k_j + l_j))\right) g_1(k_1, l_1)\, g_2(k_2, l_2) ... g_j(k_j, l_j) ... g_n(k_n, l_n) \\ &\quad \times |I_1, m_1 + (-k_1 + l_1)\rangle\,|I_2, m_2 + (-k_2 + l_2)\rangle\,... \\ &\quad \bigotimes |I_j, m_j + (-k_j + l_j)\rangle\,...\,|I_n, m_n + (-k_n + l_n)\rangle \end{aligned} \tag{A1.43}$$

where the eigenvalue equations $I_{jz}\,|I_j, m_j + (-k_j + l_j)\rangle = (m_j + (-k_j + l_j)) \times |I_j, m_j + (-k_j + l_j)\rangle$ are already used for $j = 1, 2, ..., n$. Furthermore, by substituting (A1.42) into (A1.43) one obtains

$$\begin{aligned} &I_z \left(S_1^{-k_1,+l_1} S_2^{-k_2,+l_2} ... S_j^{-k_j,+l_j} ... S_n^{-k_n,+l_n}\right) |I_1, m_1\rangle\,|I_2, m_2\rangle\,...\,|I_n, m_n\rangle \\ &= \left(M_l + \sum_{j=1}^{n} (-k_j + l_j)\right) \left(S_1^{-k_1,+l_1} S_2^{-k_2,+l_2} ... S_j^{-k_j,+l_j} ... S_n^{-k_n,+l_n}\right) \\ &\quad \times |I_1, m_1\rangle\,|I_2, m_2\rangle\,...\,|I_n, m_n\rangle \end{aligned} \tag{A1.44}$$

where the total spin magnetic quantum number $M_l = \sum_{j=1}^{n} m_j$. Finally, by substituting the generated vector $O_S\,|\Phi_l^z\rangle$ in (A1.42) into (A1.44) the transformational equation (A1.44) is reduced to the simple form

$$I_z O_S\,|\Phi_l^z\rangle = (M_l + p)\, O_S\,|\Phi_l^z\rangle \tag{A1.45}$$

where the quantum-transition order $p$ is given by

$$p = \sum_{j=1}^{n} (-k_j + l_j) \tag{A1.46}$$

It is known from (A1.33) that any tensor-product base vector $|\Phi_l^z\rangle$ obeys the eigenvalue equation $I_z|\Phi_l^z\rangle = M_l|\Phi_l^z\rangle$ with the eigenvalue $M_l = \sum_{j=1}^{n} m_j$. It can be deduced from (A1.45) that the generated vector $O_S|\Phi_l^z\rangle$ is also an eigenvector of the total spin operator $I_z$ with the eigenvalue $M_f = M_l + p$. Then basically it can be deduced that the MQT product operator $O_S$ of (A1.34) should be a $p-$order MQT operator $Q_p$ with the order $p$ given by (A1.46).

It is known from (A1.31) that all the tensor-product base vectors $\{|\Phi_l^z\rangle\}$ form a complete set of the orthonormal base vectors of the Hilbert space of the $n-spin-1/2$ system. Moreover, the complete set $\{|\Phi_l^z\rangle\}$ may act as a complete set $\{|M, K_M\rangle\}$ of the eigenbase vectors of the total spin operator $I_z = \sum_{k=1}^{n} I_{kz}$, where the tensor-product base vector $|\Phi_l^z\rangle$ given by (A1.32) obeys the eigenvalue equation $I_z|\Phi_l^z\rangle = M_l|\Phi_l^z\rangle$ with the eigenvalue $M_l = \sum_{j=1}^{n} m_j$, as shown by (A1.33). Then an arbitrary eigenvector $|\Psi_l\rangle$ of the total spin operator $I_z$ of the $n-spin-1/2$ system with the eigenvalue $M_l = \sum_{k=1}^{n} m_k$ may be expanded in terms of the complete set $\{|\Phi_l^z\rangle\}$ [2, 9, 10]

$$|\Psi_l\rangle = \sum_{m_1,m_2,...,m_n;m_1+m_2+...+m_n=M_l} C^{M_l}_{m_1,m_2,...,m_n} |I_1, m_1\rangle |I_2, m_2\rangle ... |I_n, m_n\rangle \tag{A1.47}$$

where summations over $m_k$ from $-1/2$ to $1/2$ for $k = 1, 2, ..., n$ are carried out under the constraint $\sum_{k=1}^{n} m_k = M_l$. It can be shown that the eigenvector $|\Psi_l\rangle$ obeys the eigenvalue equation $I_z|\Psi_l\rangle = M_l|\Psi_l\rangle$ with the eigenvalue $M_l = \sum_{k=1}^{n} m_k$. Now the MQT product operator $O_S$ of (A1.34) first and the total spin operator $I_z$ then act on an arbitrary eigenvector $|\Psi_l\rangle$ of (A1.47) of the total spin operator $I_z$. Then it can be found that

$$\begin{aligned}
I_z O_S|\Psi_l\rangle &= \sum_{m_1,m_2,...,m_n;\sum_{k=1}^{n} m_k=M_l} C^{M_l}_{m_1,m_2,...,m_n} \\
&\quad \times I_z O_S |I_1, m_1\rangle |I_2, m_2\rangle ... |I_n, m_n\rangle \\
&= \sum_{m_1,m_2,...,m_n;\sum_{k=1}^{n} m_k=M_l} C^{M_l}_{m_1,m_2,...,m_n} I_z O_S |\Phi_l^z\rangle \\
&= \sum_{m_1,m_2,...,m_n;\sum_{k=1}^{n} m_k=M_l} C^{M_l}_{m_1,m_2,...,m_n} (M_l + p) O_S |\Phi_l^z\rangle \\
&= (M_l + p) O_S \left( \sum_{m_1,m_2,...,m_n;\sum_{k=1}^{n} m_k=M_l} C^{M_l}_{m_1,m_2,...,m_n} |\Phi_l^z\rangle \right) \\
&= (M_l + p) O_S |\Psi_l\rangle
\end{aligned}$$

Here the second equality is obtained due to that $|\Phi_l^z\rangle$ is given by (A1.32), the third equality holds due to (A1.45), the fourth equality is obtained due to that the summations over $m_k$ from $-1/2$ to $1/2$ for $k = 1, 2, ..., n$ are carried out under the constraint $\sum_{k=1}^{n} m_k = M_l$ and the quantum transition order $p$ is fixed, and the final equality is obtained due to (A1.47). Therefore, the MQT product operator $O_S$ obeys the equation:

$$I_z O_S \left|\Psi_l\right\rangle = (M_l + p)\, O_S \left|\Psi_l\right\rangle \tag{A1.48}$$

where the quantum transition order $p$ is given by (A1.46). Obviously, the final vector $O_S\left|\Psi_l\right\rangle$ is also an eigenvector of the total spin operator $I_z$ with the eigenvalue $M_f = M_l + p$. Note that the initial vector $\left|\Psi_l\right\rangle$ is an arbitrary eigenvector of the total spin operator $I_z$ and obeys the eigenvalue equation $I_z\left|\Psi_l\right\rangle = M_l\left|\Psi_l\right\rangle$. By comparing (A1.48) with the formal definition (A1.5) of a $p-$order MQT operator $Q_p$ it can conclude that the MQT product operator $O_S$ of (A1.34) is a $p-$order MQT operator with the quantum-transition order $p = \sum_{j=1}^{n}(-k_j + l_j)$. Finally, according to the equation (A1.48) that is equal to the formal definition (A1.5) for the $p-$order MQT product operator $O_S$ it seems that the $p-$order MQT product operators $O_S$ of (A1.34) should be non-selective in the $n - spin - 1/2$ system. However, they may not be non-selective (See the MQT product operators $\{O_S\}$ of (3.25)).

As a typical example, below consider the MQT product operators $\{O_S\}$ of (A1.34) of a two-spin$-1/2$ system. The MQT product-operator basis set $\{O_S\}$ of the multiple-quantum operator algebra space of the two-spin$-1/2$ system is obtained by the tensor product of the two operator basis sets $\{S_j^{-k_j,+l_j}\}$ of the two spin$-1/2$ particles of the two-spin$-1/2$ system:

$$O_S = S_1^{-k_1,+l_1} \bigotimes S_2^{-k_2,+l_2} = S_1^{-k_1,+l_1} S_2^{-k_2,+l_2} \tag{A1.49}$$

where $k_1, l_1 = 0, 1$ and $k_2, l_2 = 0, 1$. This is an orthogonal operator basis set. There are only sixteen orthogonal base operators $\{S_1^{-k_1,+l_1} S_2^{-k_2,+l_2}\}$ in the product-operator basis set $\{O_S\}$ of (A1.49), because the multiple-quantum operator algebra space is sixteen-dimensional. These four diagonal operators $\{\tilde{O}_S\} = \{S_1^{-k,+k} S_2^{-l,+l}\}$ with $k, l = 0, 1$ of the MQT product operator basis set $\{O_S\}$ of (A1.49) can form an operator basis subset of the LOMSO operator subspace. They all are Hermitian. The other twelve base operators $\{S_1^{-k_1,+l_1} S_2^{-k_2,+l_2}\}$ with $k_1 \neq l_1$ and/or $k_2 \neq l_2$ in the product operator basis set $\{O_S\}$ of (A1.49) are off-diagonal operators and are the multiple-quantum transition operators. However, every one of these twelve off-diagonal base operators is not Hermitian. These twelve multiple-quantum transition operators $\{S_1^{-k_1,+l_1} S_2^{-k_2,+l_2}\}$ with $k_j, l_j = 0, 1$ and $j = 1, 2$ have the quantum-transition orders $p = \sum_{j=1}^{2}(-k_j + l_j)$ where $k_j \neq l_j$ for one individual spin$-1/2$ particle at least in the $two-$spin$-1/2$ system, respectively. Therefore, among these twelve multiple-quantum transition operators there are one $+2-$order, one $-2-$order, four $+1-$order, four $-1-$order, and two zero-order quantum transition operators, respectively. These can be confirmed later.

Generally, the MQT product-operator basis set $\{O_S\}$ of (3.24) or (A1.34) can be changed to the Cartesian product operator basis set $\{B_s\}$ of (3.10) and vice versa in the multiple-quantum operator algebra space of the $n-$spin$-1/2$ system with the help of the base-operator expansions of (3.23) or (3.16). As a typical example, below consider the two-spin$-1/2$ system whose Cartesian product operator basis sets $\{B_s\}$ and the MQT product operator basis set $\{O_S\}$ are given by (3.7) and (A1.49), respectively. First, with the help of the base-operator expansions of (3.23) any MQT product operator $S_1^{-k_1,+l_1}S_2^{-k_2,+l_2}$ of the operator basis set $\{O_S\}$ of (A1.49) can be expanded in terms of the complete set $\{B_s\}$ of the Cartesian product operators of (3.7) in the two-spin$-1/2$ system in the Subsection 3.2.1. Here the four diagonal product operators $\{\tilde{O}_S\} = \{S_1^{-k,+k}S_2^{-l,+l}\}$ with $k, l = 0, 1$ of the MQT product-operator basis set $\{O_S\}$ of (A1.49) may be respectively expressed as

$$S_1^{-0,+0}S_2^{-0,+0} = |0_1 0_2\rangle\langle 0_1 0_2| = \left(\frac{1}{2}E_1 + I_{1z}\right)\bigotimes\left(\frac{1}{2}E_2 + I_{2z}\right) \qquad \text{(A1.50a)}$$

$$S_1^{-0,+0}S_2^{-1,+1} = |0_1 1_2\rangle\langle 0_1 1_2| = \left(\frac{1}{2}E_1 + I_{1z}\right)\bigotimes\left(\frac{1}{2}E_2 - I_{2z}\right) \qquad \text{(A1.50b)}$$

$$S_1^{-1,+1}S_2^{-0,+0} = |1_1 0_2\rangle\langle 1_1 0_2| = \left(\frac{1}{2}E_1 - I_{1z}\right)\bigotimes\left(\frac{1}{2}E_2 + I_{2z}\right) \qquad \text{(A1.50c)}$$

$$S_1^{-1,+1}S_2^{-1,+1} = |1_1 1_2\rangle\langle 1_1 1_2| = \left(\frac{1}{2}E_1 - I_{1z}\right)\bigotimes\left(\frac{1}{2}E_2 - I_{2z}\right) \qquad \text{(A1.50d)}$$

It is known from (3.12a) that the complete set of base operators of the LOMSO operator subspace of the multiple-quantum operator algebra space of the two-spin$-1/2$ system may be given by $\{\tilde{B}_s\} = \{E/2, I_{1z}, I_{2z}, 2I_{1z}I_{2z}\}$. Then these expressions of (A1.50) show evidently that these four diagonal operators $\{\tilde{O}_S\}$ each can be expanded in terms of the diagonal product operators $\{\tilde{B}_s\}$ of the Cartesian product operator basis set $\{B_s\}$ of (3.7). Therefore, they belong to the LOMSO operator subspace. Moreover, these four diagonal operators $\{\tilde{O}_S\}$ are Hermitian and also form an operator basis subset of the LOMSO operator subspace. The other twelve (off-diagonal) base operators of the MQT product-operator basis set $\{O_S\}$ of (A1.49) are the multiple-quantum transition operators, each one of which can be expanded in terms of the Cartesian product operators $\{B_s\}$ of (3.7) too. Here both the base operators $S_1^{-0,+1}S_2^{-1,+0}$ and $S_1^{-1,+0}S_2^{-0,+1}$ are the zero-quantum transition operators with quantum-transition orders $p = \sum_{j=1}^{2}(-k_j + l_j) = 0$,

$$\begin{cases} S_1^{-0,+1}S_2^{-1,+0} = |0_1 1_2\rangle\langle 1_1 0_2| = (I_{1x} + iI_{1y})\bigotimes(I_{2x} - iI_{2y}) = I_1^+ I_2^- \\ S_1^{-1,+0}S_2^{-0,+1} = |1_1 0_2\rangle\langle 0_1 1_2| = (I_{1x} - iI_{1y})\bigotimes(I_{2x} + iI_{2y}) = I_1^- I_2^+ \end{cases} \qquad \text{(A1.51)}$$

where $I_k^+ = (I_{kx} + iI_{ky})$ and $I_k^- = (I_{kx} - iI_{ky})$ with $k = 1, 2$ are the raising and lowering operators of the spin angular momentum of the $k-$th spin$-1/2$ particle, respectively. And both the base operators $S_1^{-0,+1}S_2^{-0,+1}$ and $S_1^{-1,+0}S_2^{-1,+0}$ are the $\pm 2-$order quantum transition operators with the orders $p = \sum_{j=1}^{2}(-k_j + l_j)$ $= +2$ and $-2$, respectively,

$$\begin{cases} S_1^{-0+1}S_2^{-0+1} = |0_1 0_2\rangle\langle 1_1 1_2| = I_1^+ I_2^+ \text{ with } p = +2 \\ S_1^{-1+0}S_2^{-1+0} = |1_1 1_2\rangle\langle 0_1 0_2| = I_1^- I_2^- \text{ with } p = -2 \end{cases} \qquad \text{(A1.52)}$$

Now the rest eight base operators $\{S_1^{-0,+0}S_2^{-0,+1}, S_1^{-1,+1}S_2^{-0,+1}, S_1^{-0,+1}S_2^{-0,+0}, S_1^{-0,+1}S_2^{-1,+1}; S_1^{-0,+0}S_2^{-1,+0}, S_1^{-1,+1}S_2^{-1,+0}, S_1^{-1,+0}S_2^{-0,+0}, S_1^{-1,+0}S_2^{-1,+1}\}$ are the single-quantum transition operators:

$$S_1^{-0,+0}S_2^{-0,+1} = \left(\frac{1}{2}E_1 + I_{1z}\right)I_2^+,\ S_1^{-1,+1}S_2^{-0,+1} = \left(\frac{1}{2}E_1 - I_{1z}\right)I_2^+,$$

$$S_1^{-0,+1}S_2^{-0,+0} = I_1^+\left(\frac{1}{2}E_2 + I_{2z}\right),\ S_1^{-0,+1}S_2^{-1,+1} = I_1^+\left(\frac{1}{2}E_2 - I_{2z}\right);$$

$$S_1^{-0,+0}S_2^{-1,+0} = \left(\frac{1}{2}E_1 + I_{1z}\right)I_2^-,\ S_1^{-1,+1}S_2^{-1,+0} = \left(\frac{1}{2}E_1 - I_{1z}\right)I_2^-,$$

$$S_1^{-1,+0}S_2^{-0,+0} = I_1^-\left(\frac{1}{2}E_2 + I_{2z}\right),\ S_1^{-1,+0}S_2^{-1,+1} = I_1^-\left(\frac{1}{2}E_2 - I_{2z}\right)$$

where the first four base operators each have the same quantum transition order $p = \sum_{j=1}^{2}(-k_j + l_j) = +1$, while the last four base operators each have the same order $p = \sum_{j=1}^{2}(-k_j + l_j) = -1$. Note that every one of these twelve multiple-quantum transition operators is not Hermitian.

Now, the help of the MQT product-operator basis set $\{O_S\}$ of (A1.49), in the multiple-quantum operator algebra space of the two-spin$-1/2$ system $(i)$ the LOMSO operator subspace may be spanned by the diagonal operator basis subset $\{\tilde{O}_S\}$ of (A1.50):

$$\{\tilde{O}_S\} = \{S_1^{-0,+0}S_2^{-0,+0}, S_1^{-0,+0}S_2^{-1,+1}, S_1^{-1,+1}S_2^{-0,+0}, S_1^{-1,+1}S_2^{-1,+1}\}; \tag{A1.53a}$$

$(ii)$ the zero-quantum operator subspace may be spanned by the MQT product operators $\{O_S\}$:

$$\{B_s^{zq}\} = \{S_1^{-0,+0}S_2^{-0,+0}, S_1^{-0,+0}S_2^{-1,+1}, S_1^{-1,+1}S_2^{-0,+0}, S_1^{-1,+1}S_2^{-1,+1};$$
$$\frac{1}{2}\left(S_1^{-0,+1}S_2^{-1,+0} + S_1^{-1,+0}S_2^{-0,+1}\right), \frac{1}{2i}\left(S_1^{-0,+1}S_2^{-1,+0} - S_1^{-1,+0}S_2^{-0,+1}\right)\} \tag{A1.53b}$$

where the last two terms are the Hermitian zero-quantum transition operators and this can be shown by the formal definition (A1.6) of the Hermitian zero-quantum transition operators $Q_{|p|}$ with the order $p = 0$, here $Q_0 = S_1^{-0,+1}S_2^{-1,+0}$ and $Q_0^+ = S_1^{-1,+0}S_2^{-0,+1}$ which can be confirmed by (A1.51) or with the help of the operator identity $\left(S_j^{-k_j,+l_j}\right)^+ = S_j^{-l_j,+k_j}$; $(iii)$ the even-order multiple-quantum operator subspace [4,8] may be spanned by the MQT product operators $\{O_S\}$:

$$\{B_s^{emq}\} = \{S_1^{-0,+0}S_2^{-0,+0}, S_1^{-0,+0}S_2^{-1,+1}, S_1^{-1,+1}S_2^{-0,+0}, S_1^{-1,+1}S_2^{-1,+1};$$
$$\frac{1}{2}\left(S_1^{-0,+1}S_2^{-1,+0} + S_1^{-1,+0}S_2^{-0,+1}\right), \frac{1}{2i}\left(S_1^{-0,+1}S_2^{-1,+0} - S_1^{-1,+0}S_2^{-0,+1}\right);$$

$$\frac{1}{2}\left(S_1^{-0+1}S_2^{-0+1}+S_1^{-1+0}S_2^{-1+0}\right),\ \frac{1}{2i}\left(S_1^{-0+1}S_2^{-0+1}-S_1^{-1+0}S_2^{-1+0}\right)\},\tag{A1.53c}$$

where the last two terms are the Hermitian two-quantum transition operators and this can be confirmed by (A1.52) or with the help of the operator identity $\left(S_j^{-k_j,+l_j}\right)^{+}=S_j^{-l_j,+k_j}$. Obviously, the even-order multiple-quantum operator subspace $\{B_s^{emq}\}$ contains both the zero-quantum operator subspace $\{B_s^{zq}\}$ and the LOMSO operator subspace $\{\tilde{O}_S\}$, while the zero-quantum operator subspace $\{B_s^{zq}\}$ contains the LOMSO operator subspace $\{\tilde{O}_S\}$.

Conversely any base operator $B_s$ of the Cartesian product operator basis set $\{B_s\}$ of (3.7) can be expanded in terms of the MQT product-operator basis set $\{O_S\}$ of (A1.49) in the multiple-quantum operator algebra space of the *two*−spin−1/2 system. As an example, the diagonal product operators $\{\tilde{B}_s\}=\{E/2,I_{1z},I_{2z},2I_{1z}I_{2z}\}$ of the LOMSO operator subspace each may be expanded in terms of the diagonal base operators $\{\tilde{O}_S\}$ of (A1.53a) of the MQT product-operator basis set $\{O_S\}$ of (A1.49):

$$E=S_1^{-0,+0}S_2^{-0,+0}+S_1^{-0,+0}S_2^{-1,+1}+S_1^{-1,+1}S_2^{-0,+0}+S_1^{-1,+1}S_2^{-1,+1}\tag{A1.54a}$$

$$2I_{1z}=S_1^{-0,+0}S_2^{-0,+0}+S_1^{-0,+0}S_2^{-1,+1}-S_1^{-1,+1}S_2^{-0,+0}-S_1^{-1,+1}S_2^{-1,+1}\tag{A1.54b}$$

$$2I_{2z}=S_1^{-0,+0}S_2^{-0,+0}-S_1^{-0,+0}S_2^{-1,+1}+S_1^{-1,+1}S_2^{-0,+0}-S_1^{-1,+1}S_2^{-1,+1}\tag{A1.54c}$$

$$4I_{1z}I_{2z}=S_1^{-0,+0}S_2^{-0,+0}-S_1^{-0,+0}S_2^{-1,+1}-S_1^{-1,+1}S_2^{-0,+0}+S_1^{-1,+1}S_2^{-1,+1}\tag{A1.54d}$$

Beside these four diagonal product operators $\{\tilde{B}_s\}$ the other twelve (off-diagonal) base operators of the Cartesian product operator basis set $\{B_s\}$ of (3.7) each can be expanded in terms of the MQT product-operator basis set $\{O_S\}$ of (A1.49) too. Here this is not further discussed in detail.

Any operator of the multiple-quantum operator algebra space of the two-spin−1/2 system can be expanded in terms of the complete set of the MQT product operators $\{O_S\}$ of (A1.49). As the first example, the spin Hamiltonian operator $H_s$ of (3.13) of the two-spin−1/2 system in the Subsection 3.2.1 can be expanded in terms of the MQT product operators $\{O_S\}$. As shown by (3.13), $H_s=H_0+H_1+H_{02}$, where the three component Hamiltonians $H_0$, $H_1$, and $H_{02}$ are formally given by (3.14a), (3.14b), and (3.14c), respectively. Here it can be shown that the diagonal Hamiltonian $H_0$ of (3.14a) may be expressed as

$$\begin{aligned}H_0&=\beta_{00}S_1^{-0,+0}S_2^{-0,+0}+\beta_{01}S_1^{-0,+0}S_2^{-1,+1}\\&+\beta_{10}S_1^{-1,+1}S_2^{-0,+0}+\beta_{11}S_1^{-1,+1}S_2^{-1,+1}\end{aligned}\tag{A1.55a}$$

where the expansional coefficients $\beta_{kl}$ with $k,l=0,1$ are given by $\beta_{kl}=\alpha_0+\frac{1}{2}(-1)^k\,\Omega_1+\frac{1}{2}(-1)^l\,\Omega_2+\frac{1}{4}(-1)^{k+l}J_{z,z}$. The component Hamiltonian $H_1$ of (3.14b) is a linear combination of the ±1−order quantum transition operators:

$$\begin{aligned}H_1&=\xi_{00}S_1^{-1,+0}S_2^{-0,+0}+\xi_{01}S_1^{-1,+0}S_2^{-1,+1}+\eta_{00}S_1^{-0,+0}S_2^{-1,+0}\\&+\eta_{10}S_1^{-1,+1}S_2^{-1,+0}+\xi_{10}S_1^{-0,+1}S_2^{-0,+0}+\xi_{11}S_1^{-0,+1}S_2^{-1,+1}\end{aligned}$$

$$+\eta_{01}S_1^{-0,+0}S_2^{-0,+1}+\eta_{11}S_1^{-1,+1}S_2^{-0,+1} \tag{A1.55b}$$

where the expansional coefficients are given by $\xi_{kl}=\frac{1}{2}\omega_{1x}+(-1)^k\frac{1}{2}i\omega_{1y}+(-1)^l\frac{1}{4}J_{x,z}+(-1)^{k+l}\frac{1}{4}iJ_{y,z}$ and $\eta_{kl}=\frac{1}{2}\omega_{2x}+(-1)^k\frac{1}{4}J_{z,x}+(-1)^l\frac{1}{2}i\omega_{2y}+(-1)^{k+l}\frac{1}{4}iJ_{z,y}$ for $k,l=0,1$. The first four terms on the right-hand side of (A1.55b) each have the same quantum transition order $p=-1$, while the last four terms each have the order $p=+1$. Finally, the component Hamiltonian $H_{02}$ of (3.14c) is a linear combination of the zero− and ±2−order quantum transition operators:

$$H_{02}=\gamma_{00}S_1^{-0,+1}S_2^{-1,+0}+\gamma_{01}S_1^{-1,+0}S_2^{-0,+1}$$
$$+\gamma_{20}S_1^{-0,+1}S_2^{-0,+1}+\gamma_{21}S_1^{-1,+0}S_2^{-1,+0} \tag{A1.55c}$$

where the expansional coefficients are given by $\gamma_{0k}=\frac{1}{4}J_{x,x}+(-1)^k\frac{1}{4}iJ_{x,y}-(-1)^k\frac{1}{4}iJ_{y,x}+\frac{1}{4}J_{y,y}$ and $\gamma_{2k}=\frac{1}{4}J_{x,x}-(-1)^k\frac{1}{4}iJ_{x,y}-(-1)^k\frac{1}{4}iJ_{y,x}-\frac{1}{4}J_{y,y}$ for $k=0,1$. The first two terms on the right-hand side of (A1.55c) are the zero-order quantum transition operators with the order $p=0$ and the last two terms are the ±2−order quantum transition operators (the third term owns the order $p=+2$ and the fourth term owns the order $p=-2$).

As the second example, consider the specific Hermitian operator $H_4^{pd}$ of a two-spin−1/2 system which is constructed by

$$H_4^{pd}=(\alpha\left|0_10_2\right\rangle+\beta\left|0_11_2\right\rangle+\gamma\left|1_10_2\right\rangle+\delta\left|1_11_2\right\rangle)$$
$$\times(\alpha^*\left\langle 0_10_2\right|+\beta^*\left\langle 0_11_2\right|+\gamma^*\left\langle 1_10_2\right|+\delta^*\left\langle 1_11_2\right|) \tag{A1.56}$$

where the coefficients $\alpha,\beta,\gamma,\delta$ may be complex. The operator $H_4^{pd}$ is a pseudo-diagonal Hermitian (PDH) operator (See the Section 4). It can be shown that the PDH operator $H_4^{pd}$ of (A1.56) can be expanded in terms of the MQT product operators $\{O_S\}$ of (A1.49),

$$H_4^{pd}=H_0^{pd}+H_1^{pd}+H_{02}^{pd}, \tag{A1.57}$$

where

$$H_0^{pd}=|\alpha|^2S_1^{-0,+0}S_2^{-0,+0}+|\beta|^2S_1^{-0,+0}S_2^{-1,+1}$$
$$+|\gamma|^2S_1^{-1,+1}S_2^{-0,+0}+|\delta|^2S_1^{-1,+1}S_2^{-1,+1}, \tag{A1.58a}$$

$$H_1^{pd}=\alpha\beta^*S_1^{-0,+0}S_2^{-0,+1}+\alpha\gamma^*S_1^{-0,+1}S_2^{-0,+0}+\beta\delta^*S_1^{-0,+1}S_2^{-1,+1}$$
$$+\gamma\delta^*S_1^{-1,+1}S_2^{-0,+1}+\beta\alpha^*S_1^{-0,+0}S_2^{-1,+0}+\gamma\alpha^*S_1^{-1,+0}S_2^{-0,+0}$$
$$+\delta\beta^*S_1^{-1,+0}S_2^{-1,+1}+\delta\gamma^*S_1^{-1,+1}S_2^{-1,+0}, \tag{A1.58b}$$

$$H_{02}^{pd}=\beta\gamma^*S_1^{-0,+1}S_2^{-1,+0}+\gamma\beta^*S_1^{-1,+0}S_2^{-0,+1}$$
$$+\alpha\delta^*S_1^{-0,+1}S_2^{-0,+1}+\delta\alpha^*S_1^{-1,+0}S_2^{-1,+0} \tag{A1.58c}$$

Here $H_0^{pd}$ is a diagonal operator of the LOMSO operator subspace, $H_1^{pd}$ is a linear combination of the ±1−order quantum transition operators, and $H_{02}^{pd}$ is a linear combination of the zero− and ±2−order quantum transition operators. In particular, the operator $H_{02}^{pd}$ may be further written as

$$H_{02}^{pd} = H_{zq}^{pd} + H_{dq}^{pd} \tag{A1.59}$$

where

$$H_{zq}^{pd} = \beta\gamma^* S_1^{-0,+1} S_2^{-1,+0} + \gamma\beta^* S_1^{-1,+0} S_2^{-0,+1} \tag{A1.60a}$$

$$H_{dq}^{pd} = \alpha\delta^* S_1^{-0,+1} S_2^{-0,+1} + \delta\alpha^* S_1^{-1,+0} S_2^{-1,+0} \tag{A1.60b}$$

Here $H_{zq}^{pd}$ is a linear combination of the zero-order quantum transition operators ($p = 0$) and $H_{dq}^{pd}$ is a linear combination of the ±2−order quantum transition operators (the first term owns the order $p = +2$ and the second term owns the order $p = -2$). All these operators $H_0^{pd}$, $H_1^{pd}$, $H_{02}^{pd}$, $H_{zq}^{pd}$, and $H_{dq}^{pd}$ each are Hermitian.

The PDH operator $H_4^{pd}$ of (A1.57) which is expanded in terms of the MQT product operators $\{O_S\}$ of (A1.49) also can be expanded in terms of the Cartesian product operators $\{B_s\}$ of (3.7). The base-operator expansion of the PDH operator $H_4^{pd}$ in terms of the Cartesian product operators $\{B_s\}$ may be obtained as follows. Just like the spin Hamiltonian $H_s$ of (3.13), the PDH operator $H_4^{pd}$ also may be formally expressed as the base-operator expansion of (3.13) in terms of the Cartesian product operators $\{B_s\}$ of (3.7), that is,

$$H_4^{pd} = H_0 + H_1 + H_{02}, \tag{A1.61}$$

where the component Hamiltonians $H_0$, $H_1$, and $H_{02}$ are still given formally by (3.14a), (3.14b), and (3.14c) in the Subsection 3.2.1, respectively. Now one wants to determine the base-operator expansions for the component Hamiltonians $H_0$, $H_1$, and $H_{02}$ of the PDH operator $H_4^{pd}$ of (A1.61) in terms of the Cartesian product operators $\{B_s\}$. By comparing (A1.61) with (A1.57) it can be found that there are the three operator equations: $H_0 = H_0^{pd}$, $H_1 = H_1^{pd}$, and $H_{02} = H_{02}^{pd}$. With the help of these three operator identities one can determine the base-operator expansions for the component Hamiltonians $H_0$, $H_1$, and $H_{02}$ of the PDH operator $H_4^{pd}$ of (A1.61). On the one hand, these three component Hamiltonians $H_0$, $H_1$, and $H_{02}$ in (A1.61), which are formally given by the base-operator expansions (3.14a), (3.14b), and (3.14c) in terms of the Cartesian product operators $\{B_s\}$, respectively, are re-expanded in terms of the MQT product operators $\{O_S\}$ of (A1.49). This step is already done above. These three component Hamiltonians $H_0$, $H_1$, and $H_{02}$ now are given by (A1.55a), (A1.55b), and (A1.55c), respectively, which are the base-operator expansions in terms of the MQT product operators $\{O_S\}$. On the other hand, the three component Hamiltonians $H_0^{pd}$, $H_1^{pd}$, and $H_{02}^{pd}$ of the PDH operator $H_4^{pd}$ of (A1.57) are already given by (A1.58a), (A1.58b), and (A1.58c), respectively, which are also the base-operator expansions in terms of the MQT product operators $\{O_S\}$. Then the next step is to employ these operator equations: $H_0 = H_0^{pd}$, $H_1 = H_1^{pd}$, and $H_{02} = H_{02}^{pd}$ to determine the expansional coefficients of the base-operator expansions of the component Hamiltonians $H_0$, $H_1$, and $H_{02}$ of the PDH operator $H_4^{pd}$ of (A1.61).

The operator equation $H_0 = H_0^{pd}$ is employed to determine the diagonal Hamiltonian $H_0$ of the PDH operator $H_4^{pd}$ of (A1.61). Here both the diagonal Hamiltonians $H_0$ and $H_0^{pd}$ are given by (A1.55a) and (A1.58a), respectively. By using the operator equation $H_0 = H_0^{pd}$ and noticing that the MQT product operators $\{O_S\}$ are mutually orthogonal one can obtain the linear parameter equations that connect the expansional coefficients $\{\beta_{00}, \beta_{01}, \beta_{10}, \beta_{11}\}$ of the Hamiltonian $H_0$ of (A1.55a) with the expansional coefficients $\{|\alpha|^2, |\beta|^2, |\gamma|^2, |\delta|^2\}$ of the Hamiltonian $H_0^{pd}$ of (A1.58a). As shown in (A1.55a), here the expansional coefficients $\{\beta_{00}, \beta_{01}, \beta_{10}, \beta_{11}\}$ each are the linear functions of the expansional coefficients $\{\alpha_0, \Omega_1, \Omega_2, J_{z,z}\}$ of the Hamiltonian $H_0$ of (3.14a). Therefore, by solving the linear parameter equations one can determine these expansional coefficients $\{\alpha_0, \Omega_1, \Omega_2, J_{z,z}\}$ of the component Hamiltonian $H_0$ of (3.14a) of the PDH operator $H_4^{pd}$ of (A1.61). A detailed calculation shows that these expansional coefficients $\{\alpha_0, \Omega_1, \Omega_2, J_{z,z}\}$ are respectively given by

$$\alpha_0 = \frac{1}{4}\left(|\alpha|^2 + |\beta|^2 + |\gamma|^2 + |\delta|^2\right), \; \Omega_1 = \frac{1}{2}\left(|\alpha|^2 + |\beta|^2 - |\gamma|^2 - |\delta|^2\right),$$

$$\Omega_2 = \frac{1}{2}\left(|\alpha|^2 - |\beta|^2 + |\gamma|^2 - |\delta|^2\right), \; J_{z,z} = |\alpha|^2 - |\beta|^2 - |\gamma|^2 + |\delta|^2$$

Obviously, these four expansional coefficients $\{\alpha_0, \Omega_1, \Omega_2, J_{z,z}\}$ are real. In analogous way by using the operator equation $H_1 = H_1^{pd}$ to obtain a set of linear parameter equations and then by solving the set of linear parameter equations one can obtain the expansional coefficients $\{\omega_{1x}, \omega_{1y}, J_{x,z}, J_{y,z}; \omega_{2x}, \omega_{2y}, J_{z,x}, J_{z,y}\}$ of the component Hamiltonian $H_1$ of (3.14b) of the PDH operator $H_4^{pd}$ of (A1.61):

$$\omega_{1x} = \frac{1}{2}\left(\alpha\gamma^* + \gamma\alpha^* + \beta\delta^* + \delta\beta^*\right), \; \omega_{1y} = i\frac{1}{2}\left(\alpha\gamma^* - \gamma\alpha^* + \beta\delta^* - \delta\beta^*\right),$$

$$J_{x,z} = \alpha\gamma^* + \gamma\alpha^* - \beta\delta^* - \delta\beta^*, \; J_{y,z} = i\left(\alpha\gamma^* - \gamma\alpha^* - \beta\delta^* + \delta\beta^*\right),$$

$$\omega_{2x} = \frac{1}{2}\left(\alpha\beta^* + \beta\alpha^* + \gamma\delta^* + \delta\gamma^*\right), \; \omega_{2y} = i\frac{1}{2}\left(\alpha\beta^* - \beta\alpha^* + \gamma\delta^* - \delta\gamma^*\right),$$

$$J_{z,x} = \alpha\beta^* + \beta\alpha^* - \gamma\delta^* - \delta\gamma^*, \; J_{z,y} = i\left(\alpha\beta^* - \beta\alpha^* - \gamma\delta^* + \delta\gamma^*\right)$$

It can be found that all these eight expansional coefficients are real. Finally, with the help of the operator equation $H_{02} = H_{02}^{pd}$ one can obtain a set of linear parameter equations and then by solving the set of linear parameter equations one can determine the expansional coefficients $\{J_{x,x}, J_{x,y}, J_{y,x}, J_{y,y}\}$ of the component Hamiltonian $H_{02}$ of (3.14c) of the PDH operator $H_4^{pd}$ of (A1.61):

$$J_{x,x} = \alpha\delta^* + \delta\alpha^* + \beta\gamma^* + \gamma\beta^*, \; J_{y,y} = -\alpha\delta^* - \delta\alpha^* + \beta\gamma^* + \gamma\beta^*,$$

$$J_{x,y} = i\left(\alpha\delta^* - \delta\alpha^* - \beta\gamma^* + \gamma\beta^*\right), \; J_{y,x} = i\left(\alpha\delta^* - \delta\alpha^* + \beta\gamma^* - \gamma\beta^*\right),$$

where these four expansional coefficients $J_{x,x}$, $J_{x,y}$, $J_{y,x}$, and $J_{y,y}$ are real.